\documentclass[twoside,12pt,a4paper]{book}

  \usepackage[dvips]{graphics}

\usepackage{doublespace,epsfig,psfig,amssymb,ulem,longtable,float,supertabular,lscape,dcolumn,multirow,amsmath,array,rotating}

\newboolean{corrected}
\setboolean{corrected}{true}
\newcommand{\iforig}[2]{\protect\ifthenelse{\boolean{corrected}}{#2}{#1}}

\DeclareMathAlphabet{\matheuler}{U}{eus}{m}{n}
\newcommand{\mother}{\ensuremath{\matheuler{R}_{\tilde{l}}}}
\newcommand{\dEdx}{\ensuremath{\mathrm{d}E/\mathrm{d}x}}
\newlength{\ljlen}\newlength{\ljlenb}\newlength{\ljlenc}
\newcommand{\PBS}[1]{\let\temp=\\#1\let\\=\temp}
\newcommand{\ljcom}

\newcommand{\m}{\,\ensuremath{\mathrm{m}}}
\newcommand{\cm}{\,\ensuremath{\mathrm{cm}}}
\newcommand{\mm}{\,\ensuremath{\mathrm{mm}}}
\newcommand{\um}{\,\ensuremath{\mu\mathrm{m}}}

\newcommand{\us}{\,\ensuremath{\mu\mathrm{s}}}
\newcommand{\ns}{\,\ensuremath{\mathrm{ns}}}
\newcommand{\eV}{\,\ensuremath{\mathrm{eV}}}
\newcommand{\GeV}{\,\ensuremath{\mathrm{GeV}}}
\newcommand{\TeV}{\,\ensuremath{\mathrm{TeV}}}
\newcommand{\pb}{\,\ensuremath{\mathrm{pb}}}
\newcommand{\nb}{\,\ensuremath{\mathrm{nb}}}

\newcommand{\stautomultiprongtau}{$\Stau \rightarrow \tau_{multi-prong}$ }
\newcommand{\mc}{\ensuremath{m_{\chi}}}
\newcommand{\ms}{\ensuremath{m_{\tilde{l}}}}
\newcommand{\dl}{\ensuremath{d_{\tilde{l}}}}
\newcommand{\tl}{\ensuremath{\tau_{\tilde{l}}}}
\renewcommand{\S}{\ensuremath{\mathtt{S}}}
\newcommand{\Lum}{\ensuremath{\matheuler{L}}}

\newcommand{\ie}{i.e.}

\newcommand{\nt}{\rm}
\renewcommand{\$}{$}

\restylefloat{table}

\newcommand{\super}[1]{\ensuremath{\tilde{#1}}}
\newcommand{\susy}[1]{$\super{#1}$}
\renewcommand{\ss}[2]{\$\tilde{#1}_{#2}$}
\newcommand{\sss}[3]{\$\tilde{#1}_{#2}^{#3}$}

\newcommand{\sel}{\tilde{e}}
\newcommand{\selectron}{$\sel$}
\newcommand{\smu}{\tilde{\mu}}
\newcommand{\smuon}{$\smu$}
\newcommand{\Stau}{\tilde{\tau}}
\newcommand{\stau}{$\Stau$}
\newcommand{\grav}{\ensuremath{\tilde{G}}}

\newcommand{\selsel}{\ensuremath{\sel\sel}}
\newcommand{\smusmu}{\ensuremath{\smu\smu}}
\newcommand{\staustau}{\ensuremath{\Stau\Stau}}
\newcommand{\selsmu}{\ensuremath{\sel\smu}}
\newcommand{\selstau}{\ensuremath{\sel\Stau}}
\newcommand{\smustau}{\ensuremath{\smu\Stau}}

\newcommand{\slep}{\tilde{l}}
\newcommand{\slepton}{$\slep$}

\newcommand{\sferm}{\tilde{f}}
\newcommand{\sfermion}{$\sferm$}

\newcommand{\neut}{\tilde{\chi}}
\newcommand{\neutralino}{$\neut$}

\newcommand{\Mmess}{\ensuremath{\mathrm{M_{mess}}}}
\newcommand{\Mgrav}{\ensuremath{\mathrm{M_{\tilde{G}}}}}
\newcommand{\Lam}{\ensuremath{\Lambda}}
\newcommand{\tanb}{\ensuremath{\tan\beta}}
\newcommand{\Nfive}{\ensuremath{\mathrm{N5}}}
\newcommand{\signmu}{\ensuremath{\mathrm{sign(\mu)}}}
\newcommand{\rootF}{\ensuremath{\mathrm{\sqrt{F}}}}

\newcommand{\HS}[1][1.]{\hspace*{\stretch{#1}}}

\newcommand{\mr}{\multirow}
\newcolumntype{m}{>{$}c<{$}}

\newcommand{\ar}{2}
\newlength{\xa}
\newlength{\ya}
\newlength{\xb}
\newlength{\yb}
\newlength{\dx}
\newlength{\dy}
\newcommand{\setp}[4]
{
\setlength{\xa}{#1}
\setlength{\ya}{#2}
\setlength{\dx}{#3}
\renewcommand{\ar}{#4}
\addtolength{\xa}{-0.5\dx}
\setlength{\dy}{\ar\dx}
\setlength{\xb}{\xa}
\addtolength{\xb}{\dx}
\setlength{\yb}{\ya}
\addtolength{\yb}{\dy}
}
\newcommand{\setq}[4]
{
\setlength{\xa}{#1}
\setlength{\ya}{#2}
\setlength{\dx}{#3}
\renewcommand{\ar}{#4}
\setlength{\dy}{\ar\dx}
\setlength{\xb}{\xa}
\addtolength{\xb}{\dx}
\setlength{\yb}{\ya}
\addtolength{\yb}{\dy}
}
\newcommand{\setr}[4]
{
\setlength{\xa}{#1}
\setlength{\ya}{#2}
\setlength{\dx}{#3}
\renewcommand{\ar}{#4}
\addtolength{\xa}{-0.5\dx}
\setlength{\dy}{\ar\dx}
\addtolength{\ya}{-0.5\dy}
\setlength{\xb}{\xa}
\addtolength{\xb}{\dx}
\setlength{\yb}{\ya}
\addtolength{\yb}{\dy}
}

\begin{document}
\pagenumbering{arabic}

\newcommand{\lscap}[2]{\protect\ifthenelse{\value{thisistoc} = 0}{#1}{#2}}
\newlength{\aspace}
\newcommand{\ladd}[1]{\protect\ifthenelse{\value{thisistoc} = 0}{#1}{\settowidth{\aspace}{ }\hspace{-\aspace}}}
\newcommand{\backspace}{\settowidth{\aspace}{ }\hspace{-\aspace}}
\newcounter{thisistoc}

\newcommand{\lcurl}{\{\,}
\newcommand{\rcurl}{\,\}}

\newcommand{\dzero}{\ensuremath{\mathrm{d}_0}}
\newcommand{\zzero}{\ensuremath{\mathrm{z}_0}}
\newcommand{\phizero}{\ensuremath{\phi_0}}

\begin{singlespace}
\begin{titlepage}
\begin{center}
\centering
\LARGE\bf\mdseries\scshape
\iforig{\framebox{\parbox{\textwidth}{\centering $\times\times\times\times\times\times$ \\ \texttt{$\times\times\times$ UNCORRECTED VERSION! $\times\times\times$} \\ $\times\times\times\times\times\times$}} \\ \vspace{1cm} }
{\vspace*{2cm}}
A Search for GMSB Sleptons with Lifetime at ALEPH \\
\vspace{3cm}
Luke Timothy Jones \\
\vspace{4cm}
\large
Department of Physics, \\
Royal Holloway, \\
University of London. \\
\vspace{5cm}
\normalsize
A thesis submitted to the University of London \\
for the degree of Doctor of Philosophy. \\ 
September, 2001.
\end{center}
\end{titlepage}
\setcounter{page}{2}

\newpage
\Large
\begin{center}

{\bf Abstract}\\
\end{center}
\normalsize
\begin{spacing}{1.5}
\vspace{1.0cm}
A search for slepton production via the decay of pair-produced
neutralinos has been performed under the assumption that the sleptons
have observable lifetime in the detector before each
\iforig{decaying}{decays} to a lepton and a gravitino. Sleptons,
neutralinos and gravitinos are particles predicted by the theory of
supersymmetry, and are the supersymmetric partners of the Standard
Model leptons, neutral bosons and of the graviton respectively. The
search was performed in 628\pb$^{-1}$ of data taken by the ALEPH
detector at LEP centre-of-mass energies from 189 to 208\GeV. It was
motivated by general predictions of Gauge-Mediated Supersymmetry
Breaking (GMSB) models in which the lightest supersymmetric particle
is always the gravitino. No evidence of the process was
found. Model-independent cross-section limits are quoted as a function
of neutralino mass, slepton mass and slepton lifetime in the case that
the neutralino branching ratios to each slepton are equal at
$\frac{1}{3}$ (the so-called slepton co-NLSP scenario\iforig{}{, where
NLSP stands for `Next-to-Lightest Supersymmetric Particle'}) and in
the case that the neutralino decays exclusively to the stau (the
stau-NLSP scenario). Excluded regions in the neutralino-stau mass
plane are shown for four gravitino masses under model-specific
assumptions.
\end{spacing}
\end{singlespace}

\newpage
\centerline{\Large {\bf Acknowledgements}}
\vspace{1.5cm}
\noindent I would like to thank...\\
\begin{spacing}{1.2}\renewcommand{\arraystretch}{1.3}
\noindent\begin{tabular}{>{$\hspace*{-2mm}\cdot$}c<{\hspace*{-2mm}}p{14cm}}
& Grahame Blair and Terry Medcalf for being my supervisors.
\\
& Jon Chambers and Julian Von Wimmersburg for holding my hand during my
first tottering steps into the worlds of Unix, Hbook and Paw.
\\
& Jeremy Coles, Nick Robertson, John Kennedy, David Smith and all the
other \mbox{friends} I made at CERN for advice, support and
friendship, sorry I can't name you all -- you know who you are!
\\
& Clemens Mannert for kindly providing me with the modified versions of
GALEPH and other packages that were absolutely necessary for my search
and which would have taken me months to do myself, plus the
plain-English documentation of how to use them.
\\
& Simon George and all the technical support staff in the group, without
whom our computers would be expensive paperweights.
\\
& Glen Cowan for help on all matters statistical.
\\
& Aran Garcia-Bellido, especially for his patience and assistance while
I debugged the exclusion routine.
\\
& Grahame Blair (again), Mike Green and Glen Cowan (again) for reading
my thesis and giving me invaluable feedback.
\\
& Anybody who has lent me, or offered to lend me, money during the
financial wilderness of this last year (especially my parents), or
anyone who lent me a TV (I suppose that's just you then Bodger).
\\
& I would also like to thank every single one of the friends I have
known over the past four years by name, all of whom have provided much
needed (and frequently \mbox{pub-oriented}) distractions from particle
physics and life as a Royal Holloway student, but I'd inevitably
forget some people so I'll just say -- if you've been a mate, thanks.
\end{tabular}
\end{spacing}

\setcounter{thisistoc}{1}
\newpage
\vspace*{7cm}
\centerline{\Large {\bf Declaration}}
\vspace{1cm}

The copyright of this thesis rests with the author and no quotation
from it or information derived from it may be published without the
prior written consent of the author.
\\
\vspace{3cm}
\hspace{10cm}
Luke Jones.

\tableofcontents
\listoffigures
\listoftables


\setcounter{thisistoc}{0}
\chapter{Introduction}

This thesis documents a search for evidence of supersymmetry, which is
a theory relating fermions and bosons. Supersymmetry predicts that
every Standard Model particle should have a partner which is identical
in every way apart from its spin, which should differ by
$\frac{1}{2}$. Thus, under supersymmetry, the existing particle
spectrum must be at least doubled in size to accommodate the new
supersymmetric particles. There are many motivations for taking
supersymmetry as a serious candidate for physics beyond the Standard
Model, but a complete and viable model of supersymmetry must include a
mechanism for its breaking, since if the supersymmetric particles had
the same masses as their Standard Model counterparts they would have
been discovered some time ago. GMSB (Gauge-Mediated Supersymmetry
Breaking) is such a model. A key prediction that makes its
phenomenology different from other models is that the lightest
supersymmetric particle should be the `gravitino' (the supersymmetric
partner of the graviton -- the quantum of the gravitational
interaction), and that it is very light.
\par
This thesis documents a search for the production of sleptons (the
supersymmetric partners of leptons) via the decay of neutralinos (the
supersymmetric partners of the Standard Model neutral bosons) that in
turn are pair-produced in electron-positron collisions. It is assumed
that the sleptons travel an observable distance in the detector before
decaying to their Standard Model counterparts and a gravitino. The
primary distinguishing feature of these events in a detector would
then be the presence of particles whose trajectories do not originate
at the primary $e^+e^-$ interaction point. This process can have an
advantage over direct slepton pair-production for the prospect of
discovering supersymmetry since GMSB models can predict the neutralino
production cross-section to be higher than that of the slepton.
\par
Chapter~\ref{susy} will introduce the theory of supersymmetry in more
detail, beginning with how it can provide answers to some of the
unanswered questions of the Standard Model, and ending with how GMSB
could manifest itself in an $e^+e^-$ collider.  Chapter~\ref{detector}
describes the ALEPH detector. Chapter~\ref{search} discusses the
simulated data sets used to develop the search techniques, and the
real data set in which the search was
performed. Chapter~\ref{analysis} will describe the techniques used to
analyse particle trajectories found not to originate at the primary
interaction point in order to test whether they conform either to the
hypothesis that they originated from certain known background
processes, or from slepton decay. Chapter~\ref{selections} gives the
cut-based selections on event variables that are used as the final
discrimination between signal and background. Finally, 
\iforig{Chapter 7}{Chapter~\ref{results}} gives the search outcome 
(that no evidence for the process was found)\iforig{,}{, and}
then proceeds to discuss methods for quantifying this result in terms
of supersymmetric parameter space, before giving model-independent
cross-section limits for the process and some model-dependent
parameter space exclusion plots.

\chapter{Theory}
\label{susy}

\section{The Standard Model}

The experimental discoveries and theoretical developments that have
led to our current understanding of the physical world are presently
described within a single theoretical framework -- the Standard Model
of particle physics. It is the culmination of a history of division
and of unification, both driven by discovery. The division has been in
the understanding of matter, as entities once thought fundamental have
successively been shown to be systems of even smaller entities. The
atom (supposedly indivisible by definition) was found to be a system
of negatively charged electrons with a positively charged nucleus. The
nucleus was found to consist of positively charged `protons' and
neutral `neutrons', and both these have been found to be comprised of
yet smaller particles called quarks. At the same time understanding of
the behaviour and interactions of these particles has been steadily
consolidating. Maxwell showed that electricity and magnetism,
previously thought to be entirely different phenomena, were in fact
different manifestations of a single force subsequently dubbed
electromagnetism. His theory also tied visible light and all other
forms of electromagnetic radiation from radio waves to X-rays into a
tight theoretical framework. Einstein provided the theories to link
the worlds of the very fast and very large with the everyday world of
Newton; and the other pioneers of quantum physics did the same for the
very small.
\par
The principal discovery tools in particle physics this century have
been particle accelerators/colliders. These have allowed the study of
particle collisions at high energy, and unveiled a rich particle
spectrum beyond that from which the everyday world is made up. These
particles have been categorised by their properties. A principal
distinction that can be drawn is based on particle spin. Spin is the
`internal' angular momentum that a particle possesses, not by virtue
of any motion it may have, but as an intrinsic property tied up in its
very identity. All particles have spin in integer units of
$\frac{1}{2}$ (in natural units). Fermions possess an odd multiple,
bosons an even multiple. As a consequence, fermions and bosons have
drastically different behaviour since fermions are subjected to the
Pauli exclusion principle while bosons are not. Fermions are further
split into leptons and quarks. The distinction between these two
groups is based on the forces they experience.
\par
There are four known forces which, in order of weakest to strongest,
are gravity, the weak force, electromagnetism, and the strong
force. All particles experience gravity, but since it is so weak at
the microscopic level it has no observable effect on the interactions
of individual particles. All fermions also experience the weak
force. Particles will experience the electromagnetic force if they
have electric charge. The distinction between leptons and quarks is
that quarks experience the strong force whereas leptons do not. The
`charge' of the strong force is called colour. Particles can be red,
green, blue or colourless. The nature of the strong force means that
particles can only exist in isolation of others if they are colourless,
\ie\ if they contain equal amounts of red, green and blue quarks, a 
quark and an anti-quark of the same colour, or no quarks at all. Both
quarks and leptons can be further split into `up-type' and
`down-type'. All leptons/quarks of the same type have the same
charge. They can also be split into three `generations'. The first
generation consists of the particles that make up everyday matter. The
second and third are very similar to the first but heavier and
unstable. The three down-type leptons are the electron, muon and
tau. The three up-type are their associated neutrinos. The three
down-type quarks are the down, strange and bottom, while the up-type
are the up, charm and top. Each fermion has an associated
anti-particle possessing the same mass and spin, but with opposite
charge and colour (the anti-particle of the charge $-\frac{1}{3}$ red
up quark is the charge $+\frac{1}{3}$ anti-red up anti-quark; note
that \mbox{red + anti-red = colourless}).
\par
Since the earliest days of physics as a subject in its own right, it
was understood that net amounts of certain quantities could neither be
created nor destroyed -- they are conserved. This was first observed
in dynamical quantities such as energy, momentum and angular
momentum. But as research into matter continued, new conserved
quantities were found that did not relate to the motion of particles,
but to `internal' properties that appeared bound up in the particle's
identity, such as electric charge. Initially these conservation laws
were just that -- laws in their own right which had to be inserted
into the description of the world by hand. However, it was realised
that conservation laws can actually be inescapable results of
symmetries. That is, if the universe possesses a given symmetry, then
any system should remain invariant under the respective transformation
of the symmetry. Once this is inserted into the description of
particles, a new prediction can appear which requires an associated
quantity to be conserved. For example, invariance of a system under
translation leads to momentum conservation, invariance under
displacement in time leads to energy conservation, and invariance
under a change in the phase of the wave-function leads to charge
conservation. As opposed to a new conservation law which involves the
introduction of a new arbitrary rule in need of further explanation, a
new symmetry simply says that something is not important in our description
of matter (it is a `gauge') and so can be considered a simplification.
\par
This principle was extended further to great success. It was shown
that considering the phase of the wave-function of a charged particle
to be invariant under local transformation necessitates the existence
of the electromagnetic force. The associated transformations form the
group, U(1), and the associated eigenstate is called the
$B$\label{Bdef}. Similarly, invariance under a change from up-type to down-type
fermions necessitates the weak force. These transformations form the
group SU(2), and the weak eigenstates are the $W^1$, $W^2$ and
$W^3$. And invariance under colour rotation necessitates the strong
force, whose associated group is SU(3) and whose associated
eigenstates are the gluons (of which there are eight corresponding to
different permutations of colour-anticolour). Thus these three forces
have been shown to be the results of symmetries, or `gauge
invariance'.\footnote{Gravity is also a result of gauge invariance,
specifically invariance under local coordinate transformation. But
this is not yet incorporated in the Standard Model.}

\par
Gauge invariance predicts the gauge quanta to be massless. This is
fine in the case of the electromagnetic force since its range is known
to be infinite. For a force to be transmitted over an arbitrarily long
distance, the force mediators (the gauge quanta) have to be able to
exist for an arbitrarily long period of time, and so by the Heisenberg
uncertainty principle, $\Delta E\Delta t \lesssim
\hbar$, have to have an arbitrarily small amount of energy, and this can 
only be achieved if the quanta are massless. It is also fine for the
strong force, because although its range is only of the order of
$10^{-15}$\m, its quanta (the gluons) possess colour and so feel the
force that they mediate, which can explain their restricted range. But
while the quanta of the weak force can also interact with each other,
the nature of their interaction cannot explain their very short range
($\sim$$10^{-17}$\m). This implies that the quanta are heavy, and so
could just be fixed by introducing masses for the $W^{1,2,3}$ by hand,
but this destroys the gauge invariance and leads to unresolvable
technical problems. The solution is provided by the Higgs
mechanism. Higgs predicted the existence of a new spin-0 (or `scalar')
field which couples to the Standard Model particles. Its coupling to
the SU(2) gauge quanta makes their symmetric states unstable, causing
a `spontaneous' breaking of the symmetry as it is realised in nature,
and observed states in which the symmetry is not manifest. This
implies the existence of new, massless bosons called Goldstone
bosons. But rather than new physical states, these can be shown to
manifest themselves as longitudinal degrees of freedom for the gauge
bosons, giving the bosons mass in a manner consistent with gauge
invariance. The coupling of the Higgs field to the fermions also
accounts for their masses. A by-product of the Higgs mechanism is the
(massive) Higgs boson, which has yet to be discovered despite the best
efforts of experimental physicists.
\newcommand{\frst}{1$^{st}$}\newcommand{\scnd}{2$^{nd}$}\newcommand{\thrd}{3$^{rd}$}
\newlength{\wfrst}\newlength{\wscnd}\newlength{\wthrd}
\settowidth{\wfrst}{\frst}\settowidth{\wscnd}{\scnd}\settowidth{\wthrd}{\thrd}
\begin{table}[htpb]
\renewcommand{\arraystretch}{1.3}
\centerline{
\begin{tabular}{|c|c|m|cmmm|} \hline
 &Generation & $\scriptsize{$+$1=up-type}$ & Name & $Symbol$ & $Mass$ & $Charge$ \\
 &           & $\scriptsize{$-$1=down-type}$ &      &          &$(MeV)$& \\ \hline
\mr{6}{2ex}{\rotatebox{90}{Leptons}}
 &\mr{2}{\wfrst}{\frst}& -1 & electron	      & e	 & 0.511            & -1 \\
 &                     & +1 & $e$-neutrino    & \nu_e	 & <3\times10^{-6}  & 0 \\ \cline{2-7}
 &\mr{2}{\wscnd}{\scnd}& -1 & muon	      & \mu	 & 106              & -1 \\
 &                     & +1 & $\mu$-neutrino  & \nu_\mu  & <0.19            & 0 \\ \cline{2-7}
 &\mr{2}{\wthrd}{\thrd}& -1 & tau	      & \tau     & 1780             & -1 \\
 &                     & +1 & $\tau$-neutrino & \nu_\tau & <18.2            & 0 \\ \hline
\mr{6}{2ex}{\rotatebox{90}{Quarks}}	      
 &\mr{2}{\wfrst}{\frst}& -1 & down	      & d	 & 3-9       	    & -\frac{1}{3} \\
 &                     & +1 & up	      & u	 & 1-5	            & +\frac{2}{3} \\ \cline{2-7}
 &\mr{2}{\wscnd}{\scnd}& -1 & strange	      & s	 & 75-170	    & -\frac{1}{3} \\
 &                     & +1 & charm	      & c	 & 1150-1350	    & +\frac{2}{3} \\ \cline{2-7}
 &\mr{2}{\wthrd}{\thrd}& -1 & bottom	      & b	 & 4000-4400 	    & -\frac{1}{3} \\
 &                     & +1 & top	      & t	 & 174000\pm5000    & +\frac{2}{3} \\ \hline
\end{tabular}
}
\caption{Fermion summary table. \ladd{Only upper limits exist on the masses of the neutrinos. Whether they have any mass at all is still under investigation. The masses of the quarks are not well defined because, due to the nature of the strong force, they cannot be isolated. All these values are taken from \cite{pdg}.}}
\label{fermions}
\end{table}
\begin{table}
\renewcommand{\arraystretch}{1.3}
\centerline{
\begin{tabular}{|cmcmc|} \hline
Name     & $Symbol$ & Mass   & $Charge$ & Spin \\
         &          & (GeV)  &          &      \\ \hline
Photon   & \gamma   & 0      & 0        & 1    \\
Z-boson  & Z        & 91.2   & 0        & 1    \\
W-boson  & W^\pm    & 80.4   & \pm1     & 1    \\
Gluon    & g        & 0      & 0        & 1    \\
Higgs    & H        & $>114.1$ & 0        & 0    \\
Graviton & G        & 0      & 0        & 2    \\ \hline
\end{tabular}
}
\caption{Boson summary table. \ladd{The upper limit on the Higgs mass is from the combined LEP search for the Standard Model Higgs boson \cite{higgslimit}. The graviton has not yet been directly observed.}}
\label{bosons}
\end{table}
\par
The mass eigenstates of the gauge bosons (the photon, $\gamma$, the $Z$
and the $W^\pm$) do not end up in a one-to-one correspondence with the
weak eigenstates, but as mixtures of them --
\begin{gather*}
\gamma=\cos\theta_WB+\sin\theta_WW^3 \ ,\\
Z=-\sin\theta_WB+\cos\theta_WW^3 \ ,\\
W^\pm=(W^1\mp iW^2)/\sqrt{2} \ ,
\end{gather*}
where $\theta_W$ is known as the weak mixing angle and is
$\approx$\,28$^\circ$. The masses of the weak gauge bosons, and the
expected mass of the Higgs boson, are of the order of 100\GeV, and
this is referred to as the `weak scale'. The properties of all the
particles of the Standard Model are summarised in
Tables~\ref{fermions} and
\ref{bosons}. The charges are given as multiples of that on the
positron. Also shown is the graviton, which is the hypothetical
quantum of the gravitational force, which is not yet encompassed by
the Standard Model and has not been directly observed.

\section{The hierarchy problem}
The Standard Model of particle physics has been an acclaimed success
in describing with predictive power many aspects of the known particle
spectrum. Despite this it is not regarded as the `whole story'. It
contains many free parameters which can only be determined by
experiment, and it offers no explanation for many of the most
intriguing aspects of the particle spectrum. Such as, why are there
three generations? What are the origins of the values of the free
parameters? Why is electric charge quantised in units of one third
that on the electron? It certainly will not be able to describe
physics at the Planck scale where quantum gravitational effects become
important. And it is considered unlikely that in the many orders of
magnitude between this scale and the weak scale there will be no
intermediate new physics. Although questions such as these call into
doubt the completeness of the Standard Model, they do not indicate any
problems with the predictions that the Model does make. However, there
is one area in which the Standard model is considered to have a
flaw. This is in the Higgs sector.
\par
	The Standard Model Higgs field is a complex scalar, $H$,
with a potential given by
\[
	V=\mu^2 |H|^2 + \lambda |H|^4 \ ,
\]
\par
where $\mu$ is the Higgs mass parameter and $\mu^2<0$ is required for
electroweak symmetry breaking. Then the Vacuum Expectation Value (VEV)
for the field is given by $\langle
H\rangle=\sqrt{-\mu^2/2\lambda}$. It is known from electroweak
measurements that $\langle H \rangle=246$\GeV. Thus if the Standard
Model Higgs mechanism is responsible for electroweak symmetry
breaking, $\mu^2$ must be of the order of $-$(100\GeV)$^2$. But
$\mu^2$ is subject to loop corrections from every particle which
couples to the Higgs field. These loop corrections are proportional to
the square of the energy scale at which the loop integral is cut off,
which would be the scale at which some new physics enters the picture
to prevent further corrections. In general, this scale could be
anywhere between the weak and Planck scale. To assume that it is low
enough to prevent disastrously large corrections to $\mu^2$, but high
enough to have had no measurable consequences given the energies of
today's accelerators, is to ask it to lie in an extremely narrow
region of the complete allowable scale. Without a model to explain why
this should be so, it would seem like a very unnatural assumption.
\par
It is possible to remove this divergence by dimensional regularisation
\cite{susyintro}, but one is still left with corrections proportional
to the squares of the loop-particle masses. So the largest correction
will be of the order of the (mass)$^2$ of the heaviest particle that
couples to the Higgs field, even if this is only indirectly via other
interactions \cite{martin}. This means that to preserve $\mu^2$ at
around $-$(100\GeV)$^2$, there would have to be no new particles that
couple, directly or indirectly, to the Higgs field in a very large
energy range up to and including the Planck scale. This is considered
equally unnatural.
\par
One could assume that the various terms that contribute corrections to
$\mu^2$ manage \iforig{to almost cancel such as}{almost to cancel such
as to} leave the right amount. But for the reasons given above it
would be expected that the largest of these terms would be very large
indeed, very possibly of the order of the Planck mass. It would seem
very unlikely that terms this large would cancel to leave just a few
hundred GeV. For this to be the origin of the value of $\mu^2$ would
mean that the parameters of\iforig{}{ the} theory would have to be
fine-tuned to many significant figures.
\par
A physically justifiable model would be one in which the correct size
for $\mu^2$ falls out naturally, without requiring any parameter
conspiracies or a veto on new physics at higher scales. It is the
failure of the Standard Model to provide this that is referred to as
the ``naturalness'', ``fine-tuning'' or ``hierarchy'' problem.
\section{Supersymmetry as a solution}

Supersymmetry is a symmetry that relates fermions and bosons. Because
of the relative minus sign between fermion and boson loop
contributions to the Higgs mass, such a symmetry offers an immediate
hope for a natural explanation of cancellations between
loop-correction terms. Indeed, it is considered the primary attraction
of supersymmetry (or `SUSY') that, as an extension to the Standard
Model and at least in its unbroken form, these cancellations are
required to be exact.
\par
A SUSY transformation alters the spin of a state by one half unit,
turning a bosonic state into a fermionic one and vice-versa:
\begin{center}
$Q|boson\rangle=|fermion\rangle$;\\ $Q|fermion\rangle=|boson\rangle$.
\end{center}
The required operators, $Q$ and $Q^\dagger$ (the hermitian conjugate of
$Q$), must be anti-commuting spinors. Their algebra can be expressed in
terms of commutation and anti-commutation relations;
\begin{center}
$\{Q,Q^\dagger\}=P^\mu$; \\
$\{Q,Q\}=\{Q^\dagger,Q^\dagger\}=0$; \\
$[P^\mu,Q]=[P^\mu,Q^\dagger]=0$.
\end{center}

SUSY eigenstates which can be transformed into one another by the SUSY
generators are called ``superpartners''. The collections of single
particle states which form the irreducible representations of the SUSY
algebra are called ``supermultiplets'', and consist of pairs of
superpartners. Since the (mass)$^2$ operator ($-P^2$) and the gauge
group generators commute with $Q$ and $Q^\dagger$, it follows that
superpartner pairs have the same masses, gauge quantum numbers and
couplings as each other.  It also follows (though not so simply) that
each supermultiplet must contain an equal number of fermionic and
bosonic degrees of freedom ($n_B=n_F$). It is then guaranteed that all
the loop contributions to the Higgs mass from a single supermultiplet
must cancel exactly, and since under SUSY every particle is in a
supermultiplet this forms a very neat solution to the hierarchy
problem.

\subsection{SUSY and grand unification}
Supersymmetry also offers a solution to another problem of the
Standard Model, that of grand unification, or lack thereof.
\par
Using the renormalisation group equations the Standard Model gauge
couplings can be run up to higher scales, and it is found that they
almost converge on a single value.  This would be very desirable since
it would be a clear indication of more fundamental physics at a higher
scale.  But measurements have shown that `almost' is as good as it
gets, they do not fully converge \cite{granduni}. The evolutions of
the couplings depend on the particle content of the theory under
consideration however, and so the additional particle content that
results from supersymmetry changes the evolution. It just so happens
that if no more supersymmetric particles than are absolutely necessary
are included, the couplings evolve such as to unify at around
$10^{16}$\GeV\ to within current experimental errors (see
Figure~\ref{gut}\footnote{Thanks to Aran Garcia-Bellido for providing
me with this figure.}). This could be a numerical coincidence, but it
is certainly another motivation for supersymmetry.
\begin{figure}[tbh]
\centerline{
\epsfig{figure=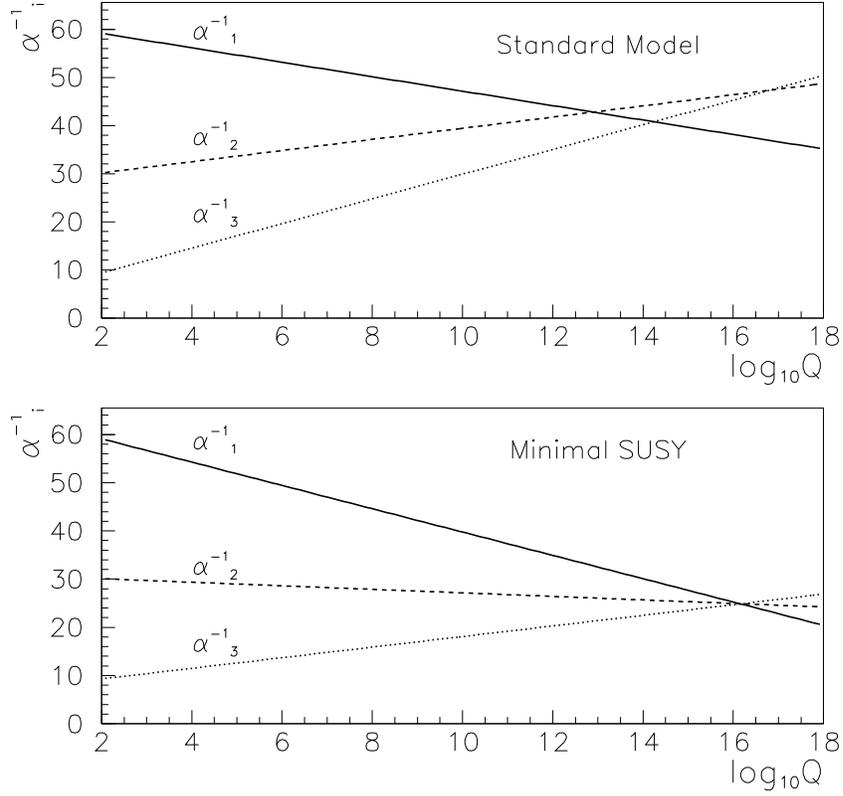,height=12cm}
}
\caption{\lscap{The evolution of the gauge coupling constants with energy, as predicted by both the Standard Model particle content alone (upper), and with the minimum additional content that could come from supersymmetry (lower). The energy scale, Q, is in units of GeV. Supersymmetry appears to allow grand unification.}{The evolution of the gauge coupling constants with and without minimal supersymmetry}}
\label{gut}
\end{figure}
\par
It should be stated at this point that there are many different models
for the implementation of supersymmetry in nature, some more exotic
than others. All require that the existing particle spectrum be at
least doubled in size. Only the Minimal SuperSymmetric Model (or MSSM)
is considered here. This is the model which introduces as few as
possible new particles whilst still providing a consistent,
phenomenologically viable model. As such it has been the most
intensely investigated of all supersymmetric theories.

\section{The SUSY eigenstates}

How then, do the Standard Model particles fit into supermultiplets?
Let's start with the fermions. Since $n_B=n_F$, a supermultiplet
containing a single massless fermion ($n_F=2$) should also contain
either a single massless spin-1 boson ($n_B=2$) or two scalar bosons
(2$\times [n_B=1]$). In fact, for the theory to be renormalisable, the
only spin-1 bosons allowed are the gauge bosons.
Since no gauge boson shares the same quantum numbers as any fermion we
are forced to propose a completely new set of scalar particles for the
supersymmetric partners of the fermions. The two mass degenerate
scalars associated with each fermion are the supersymmetric partners
of the left-handed and right-handed states of the fermion. The naming
convention for these new scalar particles is simply to append an `s'
(for scalar) to the front of the name of their Standard Model
counterpart. The notation convention for all the new supersymmetric
particles is that of a tilde above the letter used to denote the
Standard Model equivalent. Thus we may refer for example to the
``selectron'' (\selectron), ``smuon'' (\smuon) and ``sneutrino''
(\susy{\nu}), or more generically to a ``slepton''
(\slepton). Alternatively there is the ``stop'' (\susy{t}) and the
``sbottom'' (\susy{b}), both examples of a ``squark''
(\susy{q}). These are all ``sfermions'' (\sfermion 's). With the
exception of the sneutrinos, there are two of each corresponding to
the left-handed and right-handed states of the respective Standard
Model particle. Thus the difference between a left-handed and a
right-handed sfermion is nothing to do with spin states of the
sfermions themselves (they are scalar), but instead refers to the
spin-states of their Standard Model superpartners.
\par
Each gauge boson ($n_B=2$ before electroweak symmetry breaking) must
be in a supermultiplet with a completely new massless
spin-$\frac{1}{2}$ fermion (again, a spin-$\frac{3}{2}$ fermion would
lead to an unrenormalisable theory). The naming convention for the
fermionic partners of bosons is to add `ino' to the end of the name of
the Standard Model counterpart, replacing `on' where it exists. Thus
we have the ``photino'' (\susy{\gamma}), ``zino'' (\susy{Z}), ``wino''
(\susy{W^\pm}) and ``gluino'' (\susy{g}), all of which are
``gauginos''.
\par
In supersymmetric theories it turns out that one Higgs boson is not
enough. In order that all the Standard Model fermions gain mass, and
in order to avoid triangle gauge anomalies, there must be at least two
Higgs doublets \cite{susyhiggs}. This doubles the number of real
degrees of freedom to eight. After electroweak symmetry breaking, as
in the Standard Model, three of these become the longitudinal modes of
the $Z$ and $W^\pm$\iforig{. A}{, a}lthough now five as opposed to
just one are left over to form new scalar bosons. Thus supersymmetry
dictates that there should be at least five Higgs bosons. Of these,
three are neutral, one charge positive and one charge negative. Their
spin-$\frac{1}{2}$ superpartners are referred to as the
``Higgsinos''. All the supersymmetric partners of the Standard Model
particles are collectively known as ``sparticles''.

\section{R-parity}

Supersymmetry opens the possibility of Lagrangian terms which violate
lepton and baryon number conservation. Although allowed they generally
give rise to counterfactual predictions such as fast proton
decay. Thus strong experimental limits exist on the couplings; they
must either be very small or be forbidden by some conservation
law. There are problems associated with the introduction of baryon and
lepton number conservation as fundamental conservation laws in their
own right. In the Standard Model they are satisfied `accidentally'
since there are no renormalisable terms which can violate them. In the
MSSM they are not and their introduction as basic postulates would be
a step backwards. Also they are known to be violated by
non-perturbative electroweak effects. Instead, a new property known as
``R-parity'' can be introduced. This is defined for a (s)particle with
baryon number, B, lepton number, L, and spin, S, as
\[\mathrm{P_R=(-1)^{3(B-L)+2S}}\ . \]
It is +1 for all particles, and $-$1 for all sparticles, and there are
no technical problems associated with its introduction as a
fundamental, multiplicatively conserved quantum number. Thus it
can be assumed that Lagrangian terms are only allowed if the product
of the R-parities of the involved fields is +1. This would rule out
all baryon and lepton number violating terms. It would also have a
huge impact on the phenomenology of SUSY. It would mean that
sparticles can only be produced in pairs, and that all subsequent
supersymmetric decay chains must end in the production of the lightest
supersymmetric particle (LSP), which remains stable. Thus the LSP may
be a good dark-matter candidate, and R-parity conserving models are
subject to cosmological constraint.
\par
R-parity conserving and violating models of SUSY thus form two very
contrasting groups from the experimentalist's viewpoint. The search
documented in this thesis assumes conservation of R-parity and so
R-parity violation will not be discussed further.

\section{Supersymmetry breaking}
As previously mentioned, supersymmetry dictates that sparticles should
be degenerate in mass with their corresponding particles. But if this
were true they would have been discovered a long time ago. So if we
are to continue with the hypothesis that supersymmetry is realised in
nature, we must accept that it is a broken symmetry resulting in
sparticle masses significantly greater than their corresponding
particles. However, the mass-splitting cannot be too large without
unacceptably large loop corrections creeping back into the Higgs mass
parameter. So if we are to retain supersymmetry as a solution to the
hierarchy problem we find that
\[ m_{sparticle} - m_{particle} \lesssim {\mathcal{O}} (1\TeV) \ ,\]
leaving the sparticle spectrum still accessible to experiment. This is
called `soft' supersymmetry breaking, and it means we can write the
effective Lagrangian of the MSSM as the Lagrangian of unbroken
supersymmetry plus a Lagrangian containing the SUSY violating terms
(the ``soft'' terms),
\[ \mathcal{L}_{MSSM} = \mathcal{L}_{susy} + \mathcal{L}_{soft}\ .\]
We do not need to make any assumptions about the origin of
$\mathcal{L}_{soft}$, our ignorance of the mechanism by which
supersymmetry is broken can be parameterised by introducing every
allowable soft term \cite{softbreaking}. This, however, is
costly. While the unbroken MSSM introduces only one new, free,
non-trivial parameter (the ratio of the VEVs of the two Higgs
doublets, \tanb), the breaking of the MSSM in its most general form
introduces 105 \cite{count}. Thus what was a highly predictive theory
becomes so flexible it is almost untestable by experiment. It's clear
then, that a model for the SUSY-breaking mechanism would be very
desirable.
\par
If any random set of values were chosen for the 105 new soft
parameters, one would almost certainly end up with a model
incompatible with experimental observation. Loop diagrams involving
sparticles would lead to decays such as $\mu \rightarrow e \gamma$ and
$ b \rightarrow s \gamma$ with branching ratios far in excess of limits
set by searches for flavour changing neutral currents (FCNC's). And mixing
between down and strange squarks would give extra contributions to
$K^0-\bar{K}^0$ mixing incompatible with experimental measurements of
the CP violating parameters. A desirable model would be one which
naturally forbids all these counterfactual predictions.
\par
SUSY-breaking by renormalisable interactions at tree level does not do
this. It also runs into technical problems when trying to give squarks
and sleptons enough mass (to be heavy enough to have remained
undetected thus far), and gauginos any mass at all. So it is generally
assumed that the soft terms arise radiatively. That is, there exists a
``hidden sector'' of hitherto unhypothesised particles in which
supersymmetry is broken. These do not couple to the ``visible sector''
of the \mbox{MSSM} particles and sparticles, but do share some interactions
with it. Supersymmetry breaking is then communicated to the visible
sector by loop corrections to the masses involving the gauge particles
of the shared interactions. Additionally, if these interactions are
flavour-blind they can offer a natural explanation for the suppression
of CP violation and FCNC's.
\subsection{The mass eigenstates}
The breaking of supersymmetry opens the possibility that the
supersymmetry eigenstates do not correspond to the mass eigenstates,
and therefore the possibility of mixing between sparticle states with
the same quantum numbers. Thus in an analogous way to the mixing of
the electroweak eigenstates following electroweak symmetry breaking,
the gaugino and higgsino supersymmetry eigenstates can mix following
supersymmetry breaking. The photino, zino, and neutral Higgsinos mix
to form the ``neutralinos'' of which there are four,
\neutralino$^0_{1,2,3,4}$. The wino and charged Higgsinos mix to form
the ``charginos'' of which there are two, \neutralino$^\pm_{1,2}$. The
subscripts in each case indicate the mass order of the states, \ie\
\[
m_{\neut^0_4} > m_{\neut^0_3} > m_{\neut^0_2} > m_{\neut^0_1} \ .
\]
\par
In general sfermions with the same quantum numbers can also mix, but
unrestricted mixing leads to the problems with FCNC's and CP violation
already discussed. Thus any acceptable model for the origin of SUSY
breaking must predict that mixing between sfermions is kept
slight. Most predict that it is negligible for the first two families,
but possibly significant for the third. Thus the two mass eigenstates
of the stau (and sbottom and stop) are differentiated by subscript
numbers indicating which is the more massive, rather than by
specifying a direct one-to-one correspondence with the Standard Model
partner's spin-states. The relationships between the Standard Model
particles, the SUSY eigenstates and the mass eigenstates are
summarised in Table~\ref{sparticles}.

\renewcommand{\arraystretch}{1.3}
\setlength{\ljlen}{\tabcolsep}
\setlength{\tabcolsep}{3pt}
\begin{table}[tpb]
\begin{tabular}{|lcllc|} \hline
Particle & Spin  & \multicolumn{2}{l}{Sparticle}   & Spin \\
&&&&\\ \hline
&&&&\\
Charged leptons \$l$ (\$e$, \$\mu$, \$\tau$)     & \$\frac{1}{2}$  & \multicolumn{2}{l}{Charged sleptons \ss{l}{R,L} (\ss{e}{R,L}, \ss{\mu}{R,L}, \ss{\tau}{1,2})} & \$0$ \\
Neutrinos \$\nu$ (\$\nu_e$, \$\nu_\mu$, \$\nu_\tau$) & \$\frac{1}{2}$  & \multicolumn{2}{l}{Sneutrinos \super{\nu} (\ss{\nu}{e}, \ss{\nu}{\mu}, \ss{\nu}{\tau})} & \$0$ \\
Quarks \$q$ (\$u$, \$d$, \$s$, ...) & \$\frac{1}{2}$   & \multicolumn{2}{l}{Squarks \ss{q}{R,L} (\ss{u}{R,L}, \ss{d}{R,L}, \ss{s}{R,L}, ...)} & \$0$ \\
Photon \$\gamma$  & \$1$  & \multicolumn{2}{l}{Photino \super{\gamma}} & \$\frac{1}{2}$\\
\$Z$ & \$1$ & \multicolumn{2}{l}{Zino \super{Z}} & \$\frac{1}{2}$\\
Neutral Higgs \$H$, \$h$, \$A$ & \$0$ & Higgsinos \ss{H}{1}, \ss{H}{2} & \renewcommand{\arraystretch}{0.7}\raisebox{3.3ex}[-3.3ex]{\$\left\}\begin{array}{c}\mbox{}\\\mbox{Neutralinos \raisebox{0.5ex}{\$\chi_{1,2,3,4}^0$}} \\\mbox{}\end{array}\right.$}\renewcommand{\arraystretch}{1} & \$\frac{1}{2}$\\
\$W^\pm$ & \$1$ & \multicolumn{2}{l}{Wino \sss{W}{}{\pm}} & \$\frac{1}{2}$\\
Charged Higgs \$H^{\pm}$ & \$0$ & Higgsinos \sss{H}{}{\pm} & \renewcommand{\arraystretch}{0.5}\raisebox{1.5ex}[-1.5ex]{\$\left\}\begin{array}{c}\mbox{}\\\mbox{\nt Charginos \raisebox{0.5ex}{$\chi^\pm_{1,2}$}} \\\mbox{}\end{array}\right.$}\renewcommand{\arraystretch}{1} & \$\frac{1}{2}$\\ 
Gluon \$g$ & \$1$ & \multicolumn{2}{l}{Gluino \super{g}} & \$\frac{1}{2}$ \\
Graviton \$G$ & \$2$  & \multicolumn{2}{l}{Gravitino \super{G}} & \$\frac{3}{2}$ \\ \hline
\end{tabular}
\caption{Summary of particle-sparticle correspondence}
\label{sparticles}
\end{table}
\setlength{\tabcolsep}{\ljlen}
\par
Of most importance in collider physics are of course the lightest of
each type of sparticle. Thus, for the sake of brevity, terms like
$\underline{\mbox{the}}$ neutralino, $\underline{\mbox{the}}$
selectron and $\underline{\mbox{the}}$ stau will be used in this
thesis, where $\underline{\mbox{the}}$ is short for `the lightest'.

\subsection{The gravitino}
The supersymmetry discussed thus far has been global
supersymmetry. When we ask it to be a local symmetry, we find the
requirement of invariance under local SUSY transformations implies
invariance under local coordinate transformations, and this is the
underlying assumption of general relativity. Thus local SUSY
incorporates gravity, and is called ``supergravity''.
\par
The breaking of supersymmetry implies the existence of a spin-1/2
fermion, the ``goldstino''. Via the super-Higgs mechanism this is
swallowed by the spin-$\frac{3}{2}$ superpartner of the spin-2
graviton, the gravitino, giving it longitudinal degrees of freedom and
thus mass. Without for the moment speculating as to the exact
mechanism of SUSY-breaking, if we assume that its ultimate source is a
VEV, $F$, in the hidden sector then by dimensional analysis we expect
that the gravitino mass will be of the order of
\[
m_\frac{3}{2} \sim \frac{F}{M_P}\ , 
\]
where ${M_P}=2.4 \times 10^{18}$\GeV/c$^2$ is the reduced Planck
mass. This is because the gravitino must lose its mass in the limits
of vanishing SUSY-breaking ($F\rightarrow0$) and/or vanishing
gravitation (${M_P} \rightarrow \infty$). The exact relationship is
\[
m_\frac{3}{2} = \frac{F}{\sqrt{3}M_P}\ .
\]
\par
There are two candidates for the interactions that mediate SUSY
breaking to the MSSM -- gravity and the ordinary gauge interactions.
The phenomenological importance of the dependence of the gravitino
mass upon $F$ will become clear in the following discussion of these
two mechanisms.
\subsection{Gravity mediated supersymmetry breaking}
\iforig{}{\enlargethispage{\baselineskip}}
The historically favoured mechanism for the breaking of supersymmetry
is via gravity \cite{supergravity}\cite{susyintro}. In this scenario
SUSY-breaking is transmitted to the visible sector by gravitational
interactions, or more generally, by interactions of gravitational
strength. Then by dimensional analysis the soft terms should be of the
order of
\[
m_{soft}\sim\frac{F}{M_P} \ ,
\]
suggesting that, for $m_{soft}$ of the order of 100\GeV, \rootF\
should be of the order of $10^{10}$-$10^{11}$\GeV, giving a gravitino
mass comparable to $m_{soft}$. Thus the LSP is almost certainly not
the gravitino. Since, under R-parity conservation, the universe will
have a relic density of LSP's left over from the Big Bang, the LSP
almost certainly cannot be charged without having already been
discovered or producing a universe incompatible with current
observations. Thus the LSP is most often assumed to be the neutralino,
and sometimes the sneutrino. The neutralino is favoured because most
models predict it to be the LSP, and because there are problems with
creating sneutrino-LSP models which predict a relic LSP density high
enough for it to form a realistic dark-matter
\iforig
{candidate (they annihilate easily via the $Z$ since their coupling to it 
is full strength, whereas the lightest neutralino, which most models
predict to be mainly bino, does not; see \cite{dark} for a
discussion).}
{candidate \cite{dark}. This is because sneutrinos annihilate easily via the $Z$
since their coupling to it is full strength, whereas the lightest
neutralino, which most models predict to be mainly bino (the
supersymmetric partner of the $B$, see Page~\pageref{Bdef}), does not.}


\subsection{\protect\iforig{Gauge mediated supersymmetry breaking}
{Gauge Mediated Supersymmetry Breaking (GMSB)}}
\label{gmsb}
Under gauge-mediated supersymmetry breaking (\cite{gmsb1} and, for
reviews, \cite{gmsb2}\cite{gmsb3}) it is the normal gauge interactions
that communicate the breaking to the visible sector. The breaking is
communicated from the hidden high-energy sector down to the visible
sector by a set of chiral supermultiplets which couple to the breaking
sector and also indirectly through gauge interactions to the visible
sector of the MSSM. Now if the mass scale of supersymmetry breaking is
denoted $F$, and the mass scale of the messenger particles is denoted
\Mmess, then by dimensional analysis
\[
m_{soft}\sim\frac{F}{\Mmess} \ ,
\]
and so for $m_{soft}$ of the correct order of magnitude \rootF\ can be
very much lower; as low as $\sim 10^4\GeV$ since there is no reason
why \Mmess\ cannot be as low as possible without contradicting any
current observations. Thus the gravitino is almost certainly the LSP
under GMSB and all supersymmetric decay chains will end in the
production of a gravitino (which forms a good dark matter
candidate).
\par
Simple GMSB models can be described by six parameters :
\par\vspace{0.5cm}
\renewcommand{\arraystretch}{1.2}
\setlength{\ljlen}{\tabcolsep}
\setlength{\tabcolsep}{3pt}
\centerline{
\begin{tabular}{ccp{12cm}}\hline
 \rootF &--& the scale of supersymmetry breaking in the messenger sector \\
 \Mmess &--& the messenger mass scale \\
 \Nfive&--& the number of complete SU(5) multiplets that make up the messenger sector \\
 \Lam &--& the universal mass scale of the SUSY particles \\
 \tanb &--& the ratio of the VEVs of the two Higgs doublets \\
 \signmu &--& $(\mu/|\mu|)$ where $\mu$ is the higgsino mass parameter \\\hline
\end{tabular}
}
\par\vspace{0.5cm}\setlength{\tabcolsep}{\ljlen}
The messenger sector is assumed to consist of a number of complete
SU(5) multiplets because then it will not interfere with the running
of the coupling constants to the extent that the MSSM prediction of
grand-unification is scuppered. But unification can still be lost for
a sufficiently large \Nfive, so it is normally assumed to be no more
than around five. The most important of these parameters is \Lam,
since it is principally responsible for the magnitude of all the
gaugino, slepton and squark masses.
\par
The gaugino mass parameters at the messenger scale are given from the
one-loop diagrams as
\[
m_a = \frac{\alpha_a}{4\pi}N_5\Lambda \ ,
\]
and the scalar masses from the two-loop diagrams (they do not receive
corrections at one-loop order) as
\[
m^2_{scalar} = 2N_5\Lambda^2 \left[ C_3\left(\frac{\alpha_3}{4\pi}\right)^2 + C_2\left(\frac{\alpha_2}{4\pi}\right)^2+\frac{3}{5}\left(\frac{Y}{2}\right)^2\left(\frac{\alpha_1}{4\pi}\right)^2 \right] ,
\]
where $C_3=\frac{4}{3}$ for colour triplets, 0 for colour singlets;
$C_2 = \frac{3 }{4}$ for weak doublets and zero for weak singlets; and
Y is hypercharge ($Q=T_3+\frac{1}{2}Y$). Thus the hierarchy, $\alpha_3
> \alpha_2 > \alpha_1$, means that strongly interacting sparticles are
more massive than weakly interacting ones.  Since the breaking
mechanism is flavour blind, all sparticles with the same electroweak
quantum numbers
 are automatically degenerate in mass.  Radiative
corrections can in general cause mixing between the left- and
right-handed sfermions, but the amount of mixing depends on the
respective Yukawa couplings and so they can be neglected for at least
the first two generations. Thus gauge-mediation offers a natural
suppression of FCNC's. Mixing in the third generation increases with
\tanb\ and causes the lightest stau to become lighter than the 
selectron and smuon. In evolving these masses down to the weak scale
using the renormalisation group equations, the masses also gain a
logarithmic dependence on \Mmess. 
\par
A principal difference between models then, and of great importance to
the manifestation of SUSY in detectors, will not be the identity of
the LSP, but of the NLSP\iforig{}{ (the \textit{next-to-}lightest
supersymmetric particle)}. This must be either the neutralino, the
right-handed sleptons collectively, the lightest stau in the case of
large mixing, or a combination of these. The analysis described in
this thesis assumes either the stau, or the three sleptons
collectively.
\par
Since the gravitino inherits the non-gravitational interactions of the
goldstino it absorbs, decay rates to it can be high enough for it to
play a role in collider physics. The lifetime of a slepton NLSP, \tl,
depends on the gravitino mass and is given by
\[
\begin{split}
c\tl\ 
&=\ 48\pi\hbar c\ \frac{M_P^2 m_{\grav}^2}{\ms^5}\ =\ \frac{48\pi\hbar c}{3}\ \frac{\rootF^4}{\ms^2}\\[\topsep]
&=\ \left(\frac{\ms}{100\GeV}\right)^{-5}\left(\frac{m_{\grav}}{10^{-7}\GeV}\right)^2 \times 17 \cm \\[\topsep]
&=\ \left(\frac{\ms}{100\GeV}\right)^{-5}\left(\frac{\rootF}{10^{6}\GeV}\right)^4 \times 99 \cm \ .
\end{split}
\]
So depending on the gravitino and slepton masses the slepton lifetime
can be such that the decay-length is anywhere between so short it is
unobservable in a detector to so long it will escape a detector before
decay. An almost identical equation holds for the neutralino if it is
the NLSP, and so the same applies.

\subsection{GMSB search signatures at LEP}
This section outlines the various ways that GMSB could be detected at
LEP. It is basically a summary of \cite{phenom}, in which the
phenomenology of a general class of GMSB models is considered.
\par
Search strategies must be geared towards hypotheses for the identity
of the NLSP. Because the gaugino, squark and slepton masses are
roughly proportional to their gauge charges only the lightest
neutralino and sleptons form plausible NLSP candidates. Since the
gaugino masses are proportional to \Nfive, while the squark and
slepton masses go only as $\sqrt{\Nfive}$, the neutralino and the
sleptons are favoured as NLSP's for low and high \Nfive\ respectively.
\par
In the neutralino-NLSP scenario, GMSB could be discovered through the
direct production of neutralinos and their subsequent decay to a
photon plus gravitino. Neutralinos can be pair-produced either through
$e^+e^-$ annihilation via the $Z$, or via t-channel selectron exchange
(see Figure~\ref{chiprod}). The t-channel is generally the more
dominant because of the relatively large coupling. The detector
signature of such events is a question of the neutralino
decay-length. In the case of very short decay-length, the signature is
two acoplanar photons (\ie\ the two photon directions and the beam
axis will not lie in a common plane) and missing energy (\ie\ the
energy of the event will be less than the LEP centre of mass
energy). The missing energy is that carried away by the gravitinos --
being neutral and colourless they will be invisible to the
detectors. If the decay-length is of the order of a typical detector
size (\ie\ $\sim1\m$) then the signature is still acoplanar photons
with missing energy, but now it can be possible to detect the fact
that the photons are not originating from the primary interaction
point (the IP) which will greatly enhance the discovery potential
because backgrounds (which should involve photons from the IP) will be
easier to remove. Should the decay-length be too long however, the
neutralinos will escape the detector before decay leaving no evidence
that any event actually occurred, becoming an invisible channel with
no discovery potential at LEP.
\begin{figure}[tbhp] 
\epsfig{figure=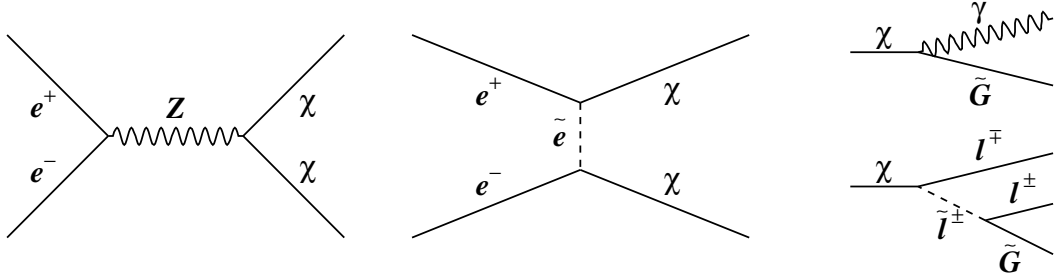,width=\textwidth}
\caption{Neutralino production and decay topologies. \ladd{Note that `$\chi$' refers to $\chi^0_1$ -- an abbreviation that will hold throughout the rest of this document. The two diagrams on the left show production in the s and t channels. The two on the right show decay in the neutralino-NLSP scenario (upper) and slepton-NLSP scenario (lower).}}
\label{chiprod}
\end{figure}
\par
If one or all sleptons are light enough, they can contribute to
neutralino production under the neutralino-NLSP scenario through
cascade decay. Sleptons can be pair-produced through annihilation via
the $\gamma$ or $Z$, and selectrons only can also be produced through
the t-channel exchange of a neutralino (see Figure~\ref{sleprod}),
although the s and t channels generally interfere destructively in
this case. They can then each decay to their respective lepton plus a
neutralino (which will be favoured over decay directly to a gravitino
in the neutralino\iforig{ }{-}NLSP scenario). The signature here then
will be two acoplanar photons and missing energy, plus two (typically
low momentum, or `soft') leptons from the IP. However the slepton
cross-section under the neutralino-NLSP scenario is expected to be
much smaller than that of the neutralino, and so this is not of great
importance.
\begin{figure}[tbhp] 
\epsfig{figure=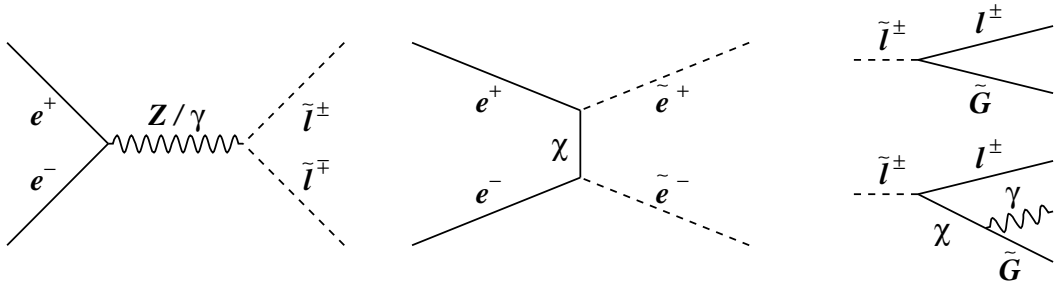,width=\textwidth}
\caption{Slepton production and decay topologies. \ladd{The two diagrams on the left show production in the s and t channels (only selectrons can be produced in the t-channel). The two on the right show decay in the slepton-NLSP scenario (upper) and neutralino-NLSP scenario (lower).}}
\label{sleprod}
\end{figure}
\par
There are two slepton-NLSP scenarios, depending on the amount of stau
mixing. In the case of little mixing the selectron, smuon and stau are
degenerate in mass (and so also in lifetime). In this case they act
collectively as the NLSP, and this is called the slepton co-NLSP
scenario. For large mixing (high \tanb), the lightest stau becomes
significantly lighter than the selectron and smuon and can act as the
sole NLSP. This is called the stau-NLSP scenario. Direct slepton
pair-production is then going to be the most obvious process to look
for in both these cases, with each slepton decaying to its respective
lepton and a gravitino. If the decay-length is short this will happen
at the IP and so the signature will be two acoplanar leptons plus
missing energy (note that the same signature is important in searches
for gravity mediated SUSY where the sleptons decay instead to stable
neutralinos). Again, if the decay-length is longer, the fact that the
leptons are not originating from the IP can be detected, enhancing the
signature in the same way as for the neutralinos. But now since the
slepton is charged its track through the detector can be visible, and
so therefore can the slepton decay vertex, enhancing the
distinctiveness of the signal even further. If the decay-length is so
long that the slepton escapes the detector before decay, then the
sleptons will be the only particles seen in the detector and the full
centre of mass energy will be observed. These events will look similar
to $e^+e^-\rightarrow\mu^+\mu^-$ events, but a detector with
sufficient single-particle mass resolution (through calculation of
$\sqrt{E^2-p^2}$ and/or measurement of the particle's rate of
ionisation) will be able to distinguish them.
\par
If the neutralino is light enough, it can contribute to slepton
production in the same way that the reverse can be true in the
neutralino-NLSP scenario. But now the contribution through cascade
decay in this way can be comparable to, and often greater than, direct
pair production. Thus in the slepton (co or stau) NLSP scenario,
slepton production though the cascade decay of neutralinos forms a
good discovery channel. In this case the signature is similar to that
from direct slepton production, but now with two additional softer
leptons from the IP. Also, the two harder leptons need not have the
same charge now since the decays of each neutralino will be
independent.
\par
It is also possible that the stau, or all three sleptons, are
sufficiently degenerate with the neutralino such as to form a
slepton-neutralino co-NLSP scenario. In this case events from both
neutralino and slepton pair production will contribute simultaneously,
and cascade decays will be highly suppressed.

\subsection{Search statuses and limits}
This section briefly summarises the searches that had been performed,
and mass limits that had been obtained, after the LEP data taking
period of 1998 during which approximately $170\pb^{-1}$ of data were
taken by each experiment at a centre of mass energy of 189\GeV. This
marks what was essentially the starting point for the analysis
described in this thesis. The current limits will be given in
Section~\ref{currentlims}.
\par
ALEPH performed searches for neutralino pair production under the
neutralino-NLSP scenario for both (effectively) zero and observable
neutralino decay-length \cite{photon-missE}\cite{gmsb189}; and for
slepton pair production under the slepton co-NLSP and stau-NLSP
scenarios for zero, observable, and very large decay-lengths
\cite{slep161}\cite{slep184}\cite{sleplong}; and for the cascade 
production of sleptons from the pair production of neutralinos under
the slepton co-NLSP and stau-NLSP scenarios for zero decay-length
\cite{gmsb189}. All these searches were brought up to date with the 
189\GeV\ data and had their results collectively interpreted in
\cite{gmsb189}.
\par
In the neutralino-NLSP scenario a lower neutralino mass limit of
91\GeV\ was obtained for zero decay-length, falling to 55\GeV\ for
$c\tau_{\chi}<100\m$. In the stau-NLSP scenario a lower stau mass
limit of 67\GeV\ was obtained for any lifetime. In the slepton co-NLSP
scenario a lower limit on the common slepton mass of 84\GeV\ was
obtained for any lifetime (under the assumption that selectron
production proceeds only via the s-channel). In the case that the
neutralino mass is less than 87\GeV\ and the neutralino-stau mass
difference is greater than the tau mass and that the stau lifetime is
negligible, the search for slepton production by cascade decay from
neutralinos allowed the limit on the stau mass in the stau-NLSP
scenario to be increased to 84\GeV\ (under the assumption that the
neutralino is mainly bino and that its mass is $\frac{5}{6}$ that of
the selectron).
\par
The results from all these searches were interpreted in terms of the
(\rootF, \Mmess, \Nfive, \Lam, \tanb, \signmu) parameter space
according to the GMSB model described in
\cite{themodel}. A scan over the parameter space used the results to 
determine the regions excluded. It provided a lower limit of 45\GeV\
on the NLSP mass for any NLSP lifetime under any scenario, a lower
limit of 9\TeV\ on \Lam, and of $2\times10^{-2}$\eV\ on the gravitino
mass.
\par
Similar work was performed by DELPHI and OPAL in \cite{delphigmsb} and
\cite{opalgmsb} respectively, but the former does not consider the
neutralino-NLSP scenario, and the latter deals only with zero NLSP
decay-length (although a slepton-with-lifetime search is detailed in
\cite{opallife}).

\chapter{The ALEPH detector at LEP}
\label{detector}
\section{The LEP collider}
`LEP' is the Large Electron-Positron collider at CERN. It is a
circular device of diameter 8,486\m\ lying in a tunnel under the
Swiss-French border near Geneva. During operation, electrons and
positrons are accelerated through an evacuated beam-pipe in opposite
directions by RF cavities. Their roughly circular orbits within the
machine are created by bending the particle beams with dipole
magnets. This inevitably leads to energy loss through synchrotron
radiation, which is replaced by the cavities. This is the reason for
the large size of LEP as the amount of energy lost in this way is
inversely proportional to the accelerator's radius. The
counter-propagating particles are allowed to collide with equal and
opposite momenta at four points, each surrounded by a detector whose
purpose is to observe and record new particles produced by the
collisions. Each detector constitutes a separate experiment, these are
ALEPH, OPAL, DELPHI and L3. For general papers on these detectors see
\cite{aleph},\cite{opal},\cite{delphi} and \cite{l3} respectively.
\par
The volume at the centre of each detector in which the beams overlap
is known as the `beam-spot' or `luminous region', and it is within this
that collisions occur and from where (primary) particles
originate. Since the beams consist of discrete bunches of particles,
the size of the luminous region is defined by the size of the
bunches. These span 200\um\ horizontally and 8\um\ vertically in
the plane perpendicular to the beam direction, and 1\cm\ parallel to the
beam direction.
\par
In the first stage of its operation, known as LEP1, LEP ran such that
the centre-of-mass energy of its colliding particles was equal to the
mass of the Z boson (\mbox{$\approx91\GeV$}), allowing it to fulfil
its first goal of performing detailed studies of the Z. Subsequently
it was upgraded and its energy increased through LEP1.5 to LEP2,
passing the threshold for, and allowing the study of, W boson pair
production and reaching in excess of 200\GeV.

\section{An overview of ALEPH}
ALEPH is a general purpose detector which allows the study of almost
all types of $e^+e^-$ interactions. It surrounds the interaction point
with near $4\pi$ solid angle coverage, as shown in the cut-away
diagram in Figure~\ref{aleph}. Its main bulk consists of several
different subdetectors forming coaxial jackets about a 5.3\cm-radius
beam-pipe, centred on the interaction point (IP).

\begin{figure}[tbh]
\centerline{\epsfig{figure=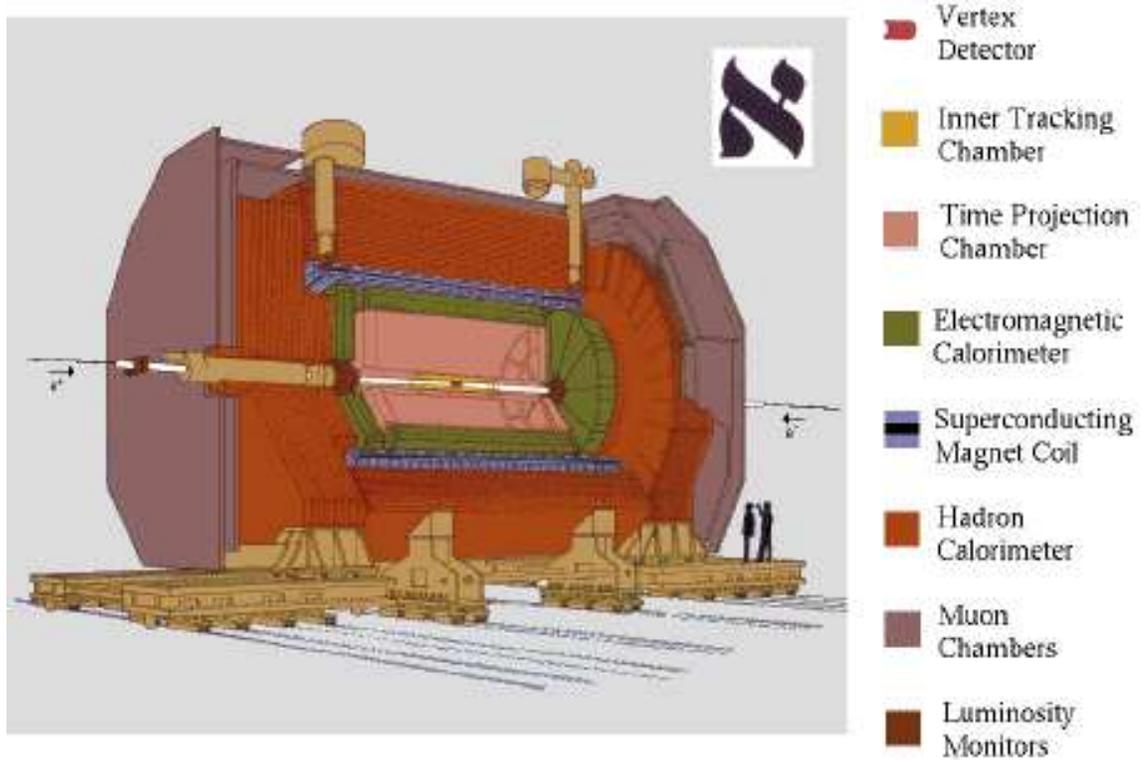,height=10cm}}
\caption{The ALEPH detector.}
\label{aleph}
\end{figure}

\par
Particles produced at the IP and travelling outwards through the
detector first encounter a series of three, low-density tracking
subdetectors designed primarily to give information on the
trajectories of charged particles. Since the tracking volume lies
within a strong magnetic field, the curvature of a charged particle's
path through the field yields its momentum. They then reach a
high-density calorimeter arrangement designed to bring them to rest
and, in doing so, make a measurement of their energy. Any particles
that manage to penetrate this, pass finally through the muon chambers
designed to tag muons leaving the detector. The magnetic field is
produced by a superconducting coil located between the electromagnetic
and hadron calorimeters, and runs parallel to the beam-axis with a
strength of 1.5\,T.
\par
The Cartesian ($x$,\,$y$,\,$z$) coordinate system used to describe
positions within ALEPH is defined as follows. The origin is the
geometrical centre of the detector, also the nominal IP. The
\mbox{$z$-axis} runs parallel to the beam-pipe in the e$^-$ beam
direction. The $x$-axis points towards the centre of LEP (to the right
as seen by an incoming positron), while the $y$-axis points vertically
upwards\footnote{This is only approximate. The slight tilt of the LEP
ring with respect to the horizontal means that there is a small angle
between the $y$-axis and the vertical.}. A cylindrical
($r$,\,$\phi$,\,$z$) coordinate system is also used, in which the
origin and $z$-axis coincide with those of the Cartesian system. The
planes $\phi=0^\circ$ and $\phi=90^\circ$ contain the $x$-axis and the
$y$-axis respectively, and $r=\sqrt{x^2+y^2}$. In addition,
it\iforig{}{ is} often useful to refer to the angle made between 
\iforig{a point or line}{the line that joins a point to the origin}
and the $z$-axis, and this is denoted by $\theta$.
\par
A detailed description of the performance of the ALEPH detector is
available in \cite{performance}. It should be noted that ALEPH has
been undergoing constant modification over the years it has been
operating, from minor changes too slight to be of note, to major
changes like the replacement of the vertex detector in 1995. This
analysis however, is only concerned with the data-taking period, 1997
onwards. Only the state of the detector in this period is described
here, and when this state differs from previous years this will not
necessarily be noted. With this in mind, there now follows a detailed
description of the main components of ALEPH.

\section{The Vertex DETector (VDET)}
\label{vdet}

The VDET is the innermost subdetector of ALEPH. It is a silicon-strip
detector which provides high-accuracy positional information on the
paths of charged particles close to the IP both in the $r\phi$ plane
and the $z$ direction. This allows accurate determination of their
trajectories in this region and thus gives information on the point of
origin of a charged particle, whether that be from the IP or not. This
is of course most useful for the study of B-physics, allowing the
accurate reconstruction of the decay vertices of B-mesons close to the
IP.
\par
The active part of the VDET is made up of 24 `faces' arranged in two,
approximately cylindrical layers. The inner layer consists of 9 faces
and is at the lowest possible radius allowed by the presence of the
beam-pipe ($\sim6.3\cm$), the outer consists of the remaining 15 and
is at the greatest possible radius allowed by the presence of the
Inner Tracking Chamber \mbox{($\sim10.3\cm$) - thus} maximising the
lever arm. The faces overlap by $\sim0.2\mm$ in $r\phi$ to ensure
complete coverage.
\par
The active units of the VDET are the `wafers', these are double-sided
silicon strip detectors of size 52.6\mm\ $\times$ 65.4\mm\ $\times$
0.3\mm. Essentially junction diodes operated in reverse bias, each is
an n-type substrate with 1021 p+ readout strips on the junction
($r\phi$) side and 640 orthogonal readout n+ strips on the ohmic ($z$)
side. The passage of a charged particle through the wafer generates
ionisation which is picked up as signal by nearby strips. Clusters of
signals are combined using a `centre-of-gravity' algorithm to produce
VDET hits.
\par

Three wafers and a support for the readout circuits all glued together
form a `module', two modules mounted with their readouts at opposite
ends on two beams form a face, active length 400\mm, total length
500\mm. One beam is made of Kevlar epoxy and serves to electrically
insulate the z side of the modules, the other is made of Carbon fibre
epoxy and provides the necessary mechanical strength of the face. The
faces are then mounted at each end on carbon fibre support flanges,
which are joined together by a carbon fibre cylinder which sits
between the two active layers and consists of two thin skins spaced
20\mm\ apart by a corrugated web. The whole support structure is water
cooled. The readout cables and water pipes are supported by a second
pair of 20\um\ thick aluminium flanges, joined by two 200\um\
thick carbon fibre cylinders which serve both to support the flanges
and as inner and outer protective skins for the
VDET. Figure~\ref{vdet1} shows a cross-section of the completed
structure in the rz plane and the face geometry as seen in the r$\phi$
plane, Figure~\ref{vdet2} shows details of a face as seen from above
and below and a three dimensional view of the fully mounted VDET.

\begin{figure}[tbh]
\centerline{\resizebox{12cm}{!}{
\rotatebox{-90}{\includegraphics*[130pt,70pt][460pt,775pt]{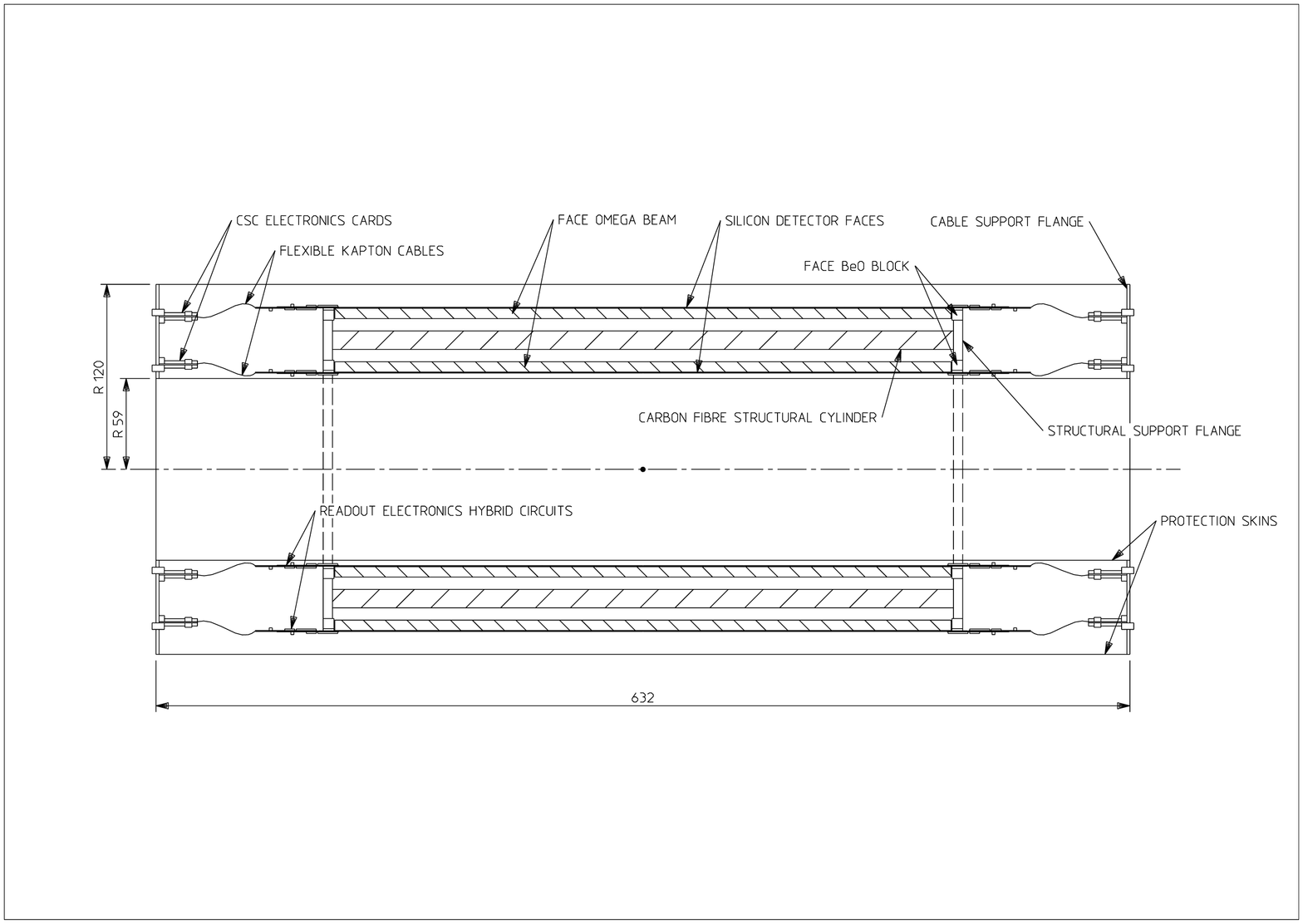}}
\rotatebox{-90}{\includegraphics[8pt,300pt][300pt,530pt]{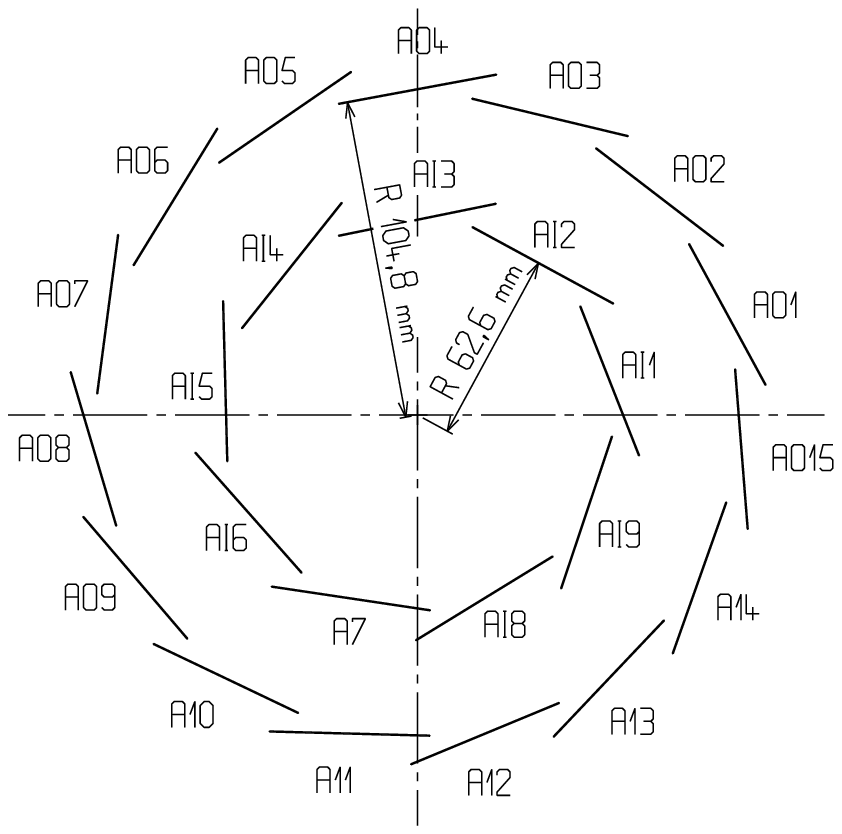}}
}}
\caption{Cross-sections of the VDET showing structural composition in the $rz$ plane \ladd{(left)} and face geometry in the $r\phi$ plane \ladd{(right)}.}
\label{vdet1}
\end{figure}

\begin{figure}[tbh]
\centerline{\resizebox{14cm}{!}{
\rotatebox{180}{\includegraphics*[30pt,68pt][550pt,750pt]{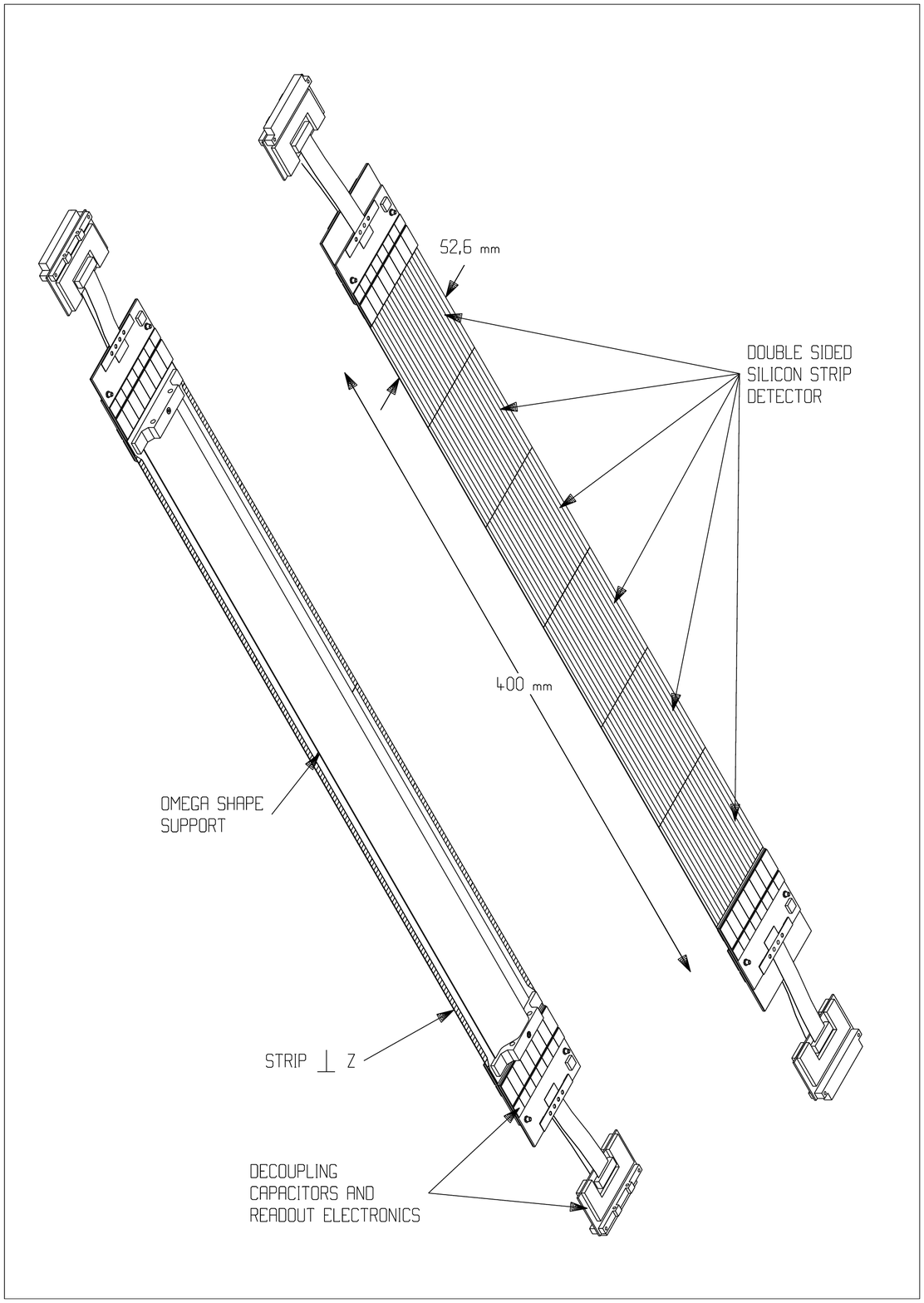}}
\rotatebox{180}{\includegraphics*[30pt,68pt][550pt,750pt]{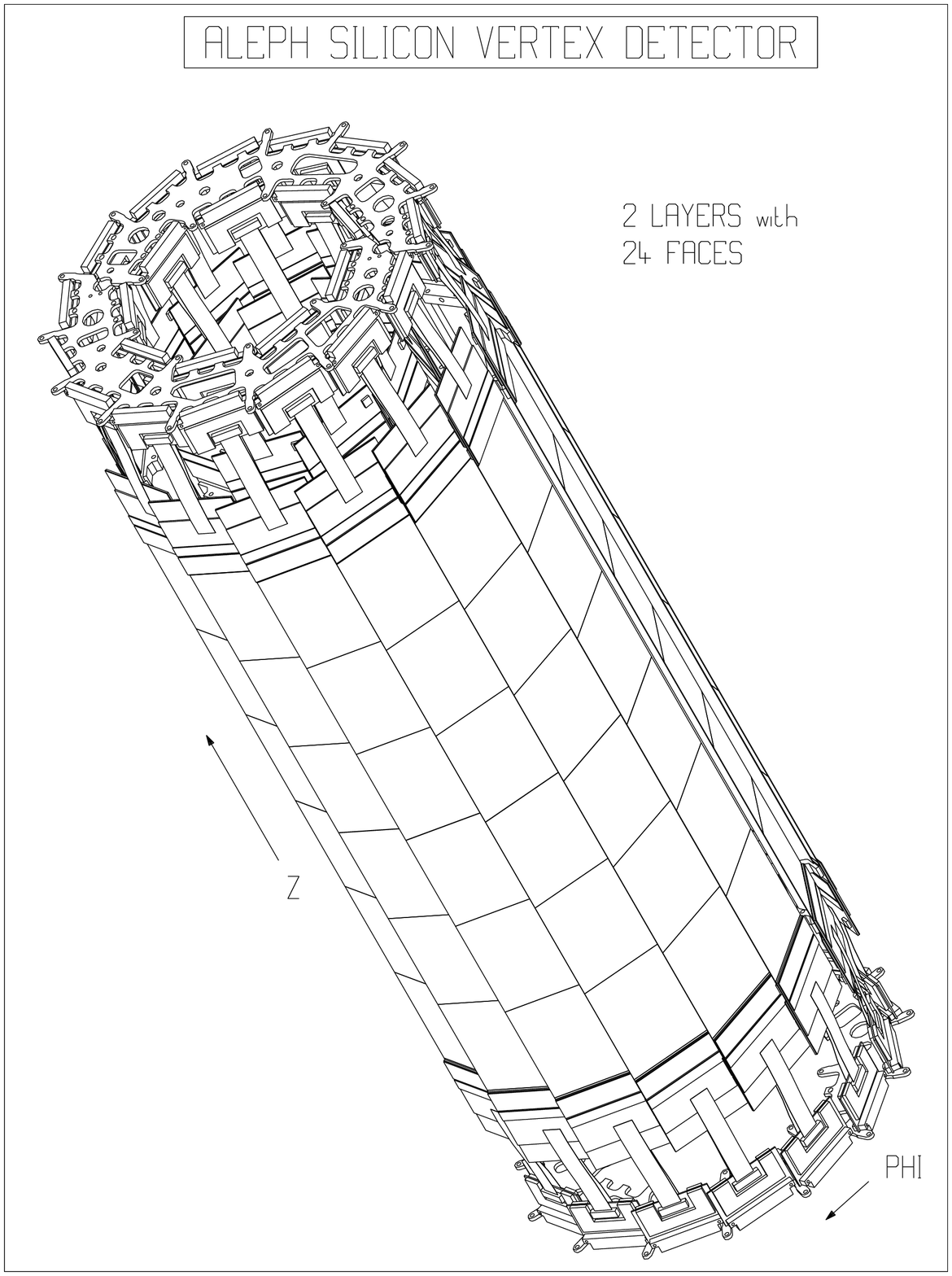}}
}}
\caption{A VDET face \ladd{(left)} and the fully mounted VDET \ladd{(right)}.}
\label{vdet2}
\end{figure}

\par
A charged particle passing through a face well inside its active
region stands a 99\% chance of leaving a reconstructed hit in both the
$r\phi$ and z views. The chance of that hit being assigned to its
reconstructed track in one view when it has been assigned in the
other, giving a three-dimensional hit for the track, is 93\% for the
$z$ view and 95\% for the $r\phi$ view. The spatial hit resolution for
primary tracks at $\theta=90^\circ$ is 10\um\ in the $r\phi$ view,
with no significant change for decreasing $\theta$, and 15\um\ in the
$z$ view, worsening to 50\um\ by
$\iforig{\cos(\theta)}{|\cos(\theta)|}=0.85$. The 40\cm\ active length
of the VDET (virtually) guarantees at least one hit up to
$\iforig{\cos(\theta)}{|\cos(\theta)|}=0.95$.

\section{The Inner Tracking Chamber (ITC)}

The ITC is a cylindrical wire drift-chamber providing tracking
information in the region $26\cm>r>16\cm$ and $|z|<100$\cm. It has a
fast readout allowing it to be used in the Level-1 trigger decision
and provides the only tracking information used in that decision. For
a charged particle passing through its full active thickness it can
provide eight hits with a resolution of $\sim150$\um\ in $r\phi$ but
only $\sim5\cm$ in $z$.
\par
The wires are strung parallel to the $z$-axis between two 25\mm\ thick
aluminium end-plates. The end-plates themselves are connected by two
carbon fibre tubes which form the cylindrical surfaces of the ITC. The
outer tube bears the tension of the wires, and has an inner radius
57\cm, a thickness of 2\mm, and has a 25\um\ layer of aluminium foil
on both surfaces to screen the chamber from RF interference and
improve the uniformity of its field. The inner tube, which provides
support to the VDET, is 600\um\ thick with a 50\um\ thick layer of
aluminium foil on its outer surface for the same reason. The high
voltage is supplied by distribution boxes mounted on electrical
end-flanges 20\cm\ beyond the end-plates. These end-flanges plus the
cylinders form a hermetically sealed volume which contains a gas
mixture of four parts argon to one part carbon~dioxide at atmospheric
pressure.
\par
There are four different types of wire. The sense wires, responsible
for attracting the ionisation produced by passing charged particles,
are held at a positive potential of $\sim2$\,kV. Each is surrounded by
five field wires and a calibration wire all held at ground potential,
forming a hexagonal `drift cell' (see Figure \ref{itc1}). Between each
pair of drift cell layers is a layer of guard wires, these support
circular hoops of aluminium wire whose purpose is to limit the damage
that would be caused should any wire break. There are 96 sense wires
per layer in the first four layers, and 144 per layer in the outer
four layers.

\begin{figure}[tb]
\centerline{\resizebox{14cm}{!}{
\epsfig{figure=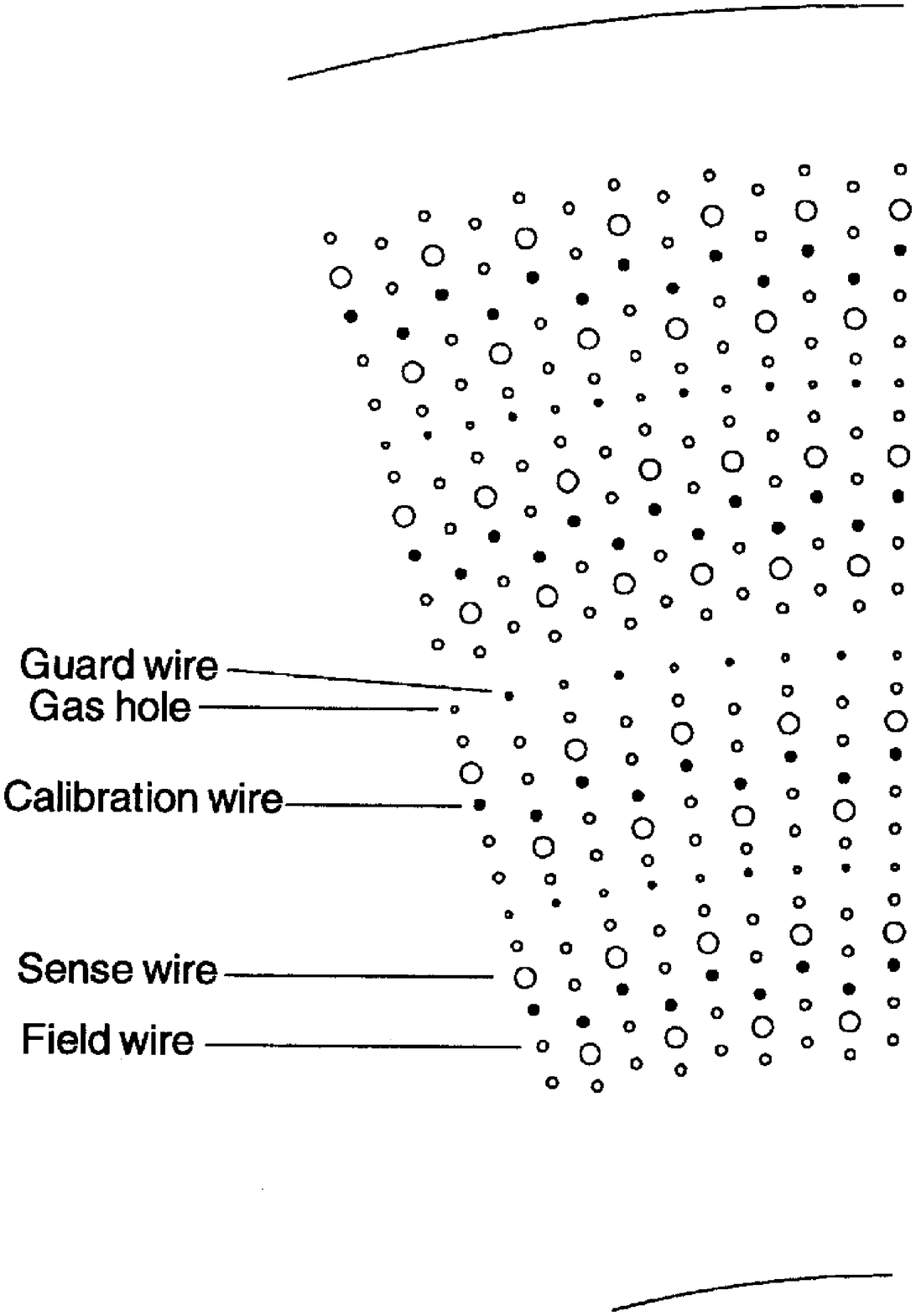,width=10cm}
\hspace{2cm}
\raisebox{12cm}{\rotatebox{-90}{\epsfig{figure=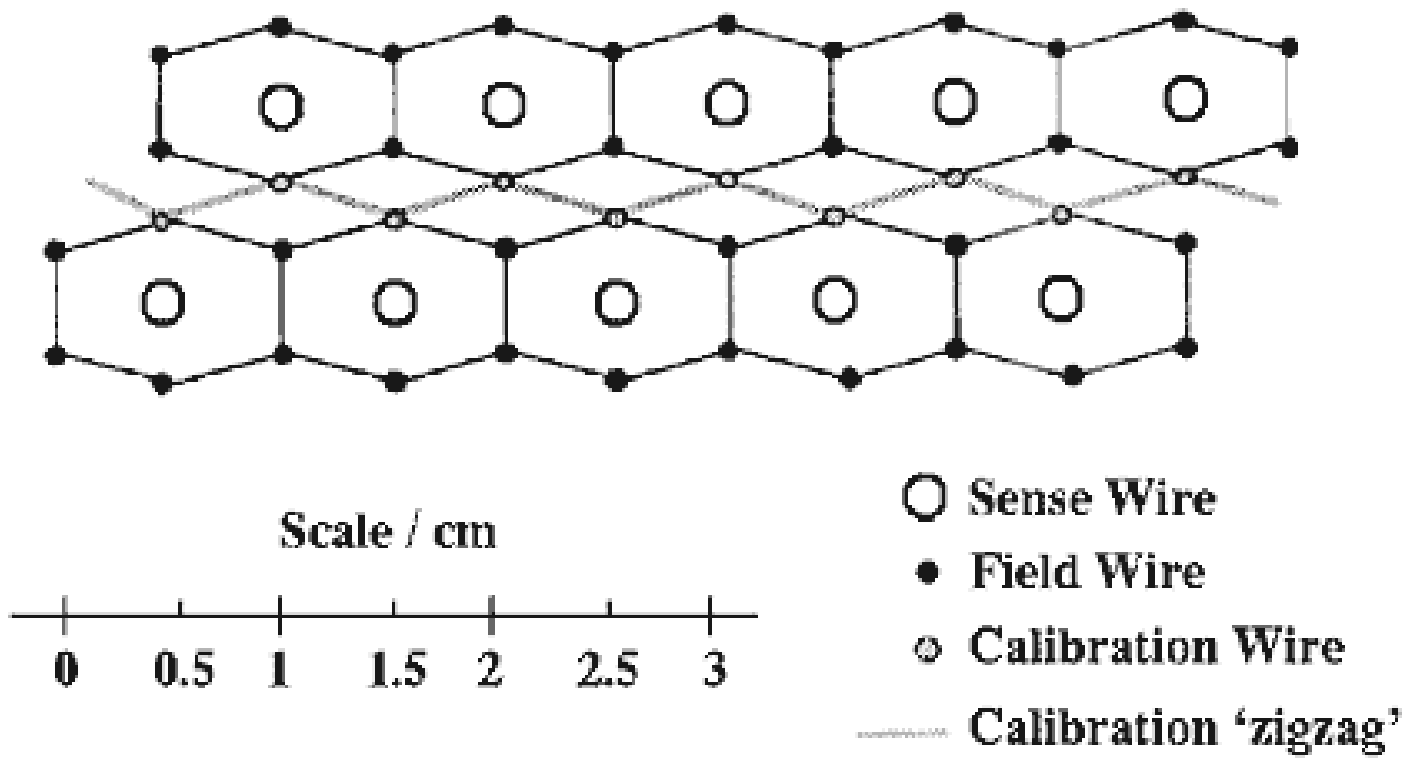}}}
}}
\caption{A detail of an ITC end-plate \ladd{(left)} and the drift cell geometry \ladd{(right)}.}
\label{itc1}
\end{figure}

\par
An ITC hit is created when a charged particle passes through a drift
cell. Its passage ionises the gas in its immediate vicinity causing a
pulse of negative charge to drift to the sense wire. The subsequent
current pulse is detected at both ends. The distance of the particle
from the wire is obtained by converting the drift time into a drift
distance using a parameterisation of the non-linear relationship
between the two. ITC hits have an inherent $r\phi$ ambiguity due to
the azimuthal symmetry of the drift cells, the result of which is that
each hit can represent one of two points on either side of the
relevant sense wire. This ambiguity can be removed in practice since
drift cells of neighbouring layers are offset from each other by half
a cell width, thus only the correct set of possible hit points will
line up to form a valid track. The $z$ coordinate of a hit is obtained
from the difference in the time of arrival of the pulse at
each\iforig{ of the}{} end of the sense wire. Note that no more than
one hit can be assigned to each wire in a given event.

\section{The Time Projection Chamber (TPC)}
\label{tpc}
The TPC is ALEPH's main tracking chamber. It accurately measures the
trajectories of charged particles by providing three-dimensional hits
at 21 separate radii in the range $170\cm>r>40\cm$ and
$|z|<220$\cm. It also measures the spatial rate of energy loss due to
ionisation (\dEdx) of a particle, which enhances particle
identification by complementing information from the
calorimeters. Structurally, the TPC consists of three main elements -
the field cage (two cylinders, inner and outer), two circular
end-plates and eight `feet' (four attached to each end-plate, they
transmit the weight of the whole structure, plus that of the ITC, to
the magnet cryostat). A diagram is shown in Figure~\ref{tpc1}\iforig{}{.}
\begin{figure}[tbh]
\centerline{\resizebox{10cm}{!}{
\epsfig{figure=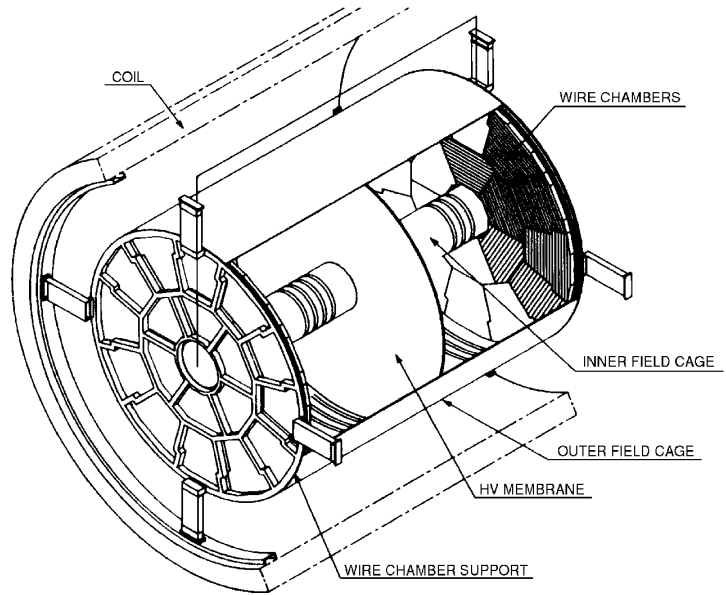,angle=-90}
}}
\label{tpc1}
\caption{TPC overall view.}
\end{figure}
\par
The cylindrical field cage is aligned with the $z$-axis and has an inner
radius of 31\cm\ and an outer radius of 180\cm. This, together with the
end-plates, forms the gas-tight volume of the TPC which contains a
mixture of argon (91\%) and methane (9\%) held at slightly above
atmospheric pressure. The volume is vertically bisected by a circular
mylar membrane coated in conducting graphite paint and held at a high
negative voltage (typically $-27$\,kV). The end-plates are held near ground
while electrodes along the inner and outer field cages are held at
potentials such that the resulting electric field (of
$\sim$115\,Vcm$^{-1}$) is uniform and aligned with the $z$-axis.
\par
A charged particle passing through the volume ionises the gas along
its path, and the electric field causes the resulting electron cloud -
an image of the particle's trajectory - to drift towards the nearest
end-plate, while its lateral diffusion is limited by the parallel
magnetic field.
\par

\begin{figure}[tb]
\centerline{\resizebox{10cm}{!}{
\includegraphics*[4.5cm,14cm][17.2cm,26cm]{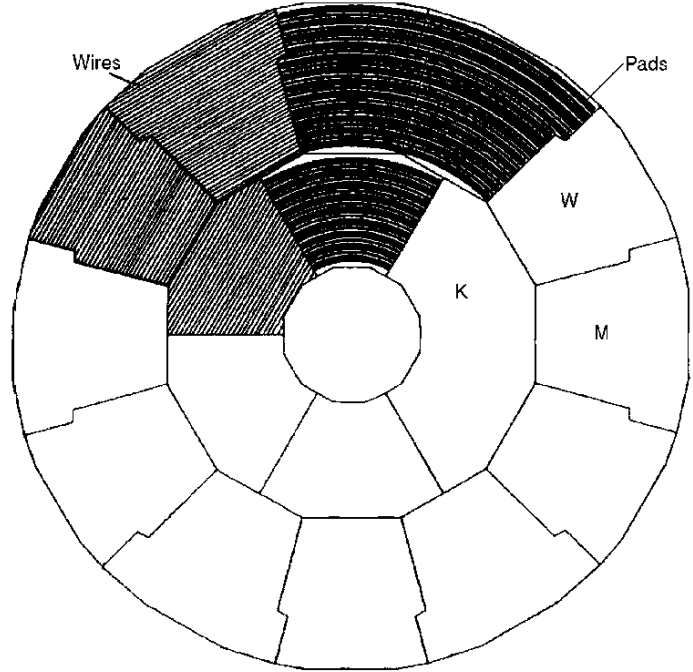}
}}
\caption{Overall geometry of a TPC end-plate showing sectors, wires and the 21 concentric pad rows.}
\label{tpc2}
\end{figure}
The end-plates are divided into `sectors' which are the active
detecting regions. Each is a proportional wire chamber over a series
of concentric `pad' rows. The pads, 6.2\mm\ $\times$ 30\mm\
($\delta$r$\phi$ $\times$ $\delta$r), provide the three-dimensional
hit coordinates, while the wire chambers provide the \dEdx\
information. A diagram of an end-plate showing the sector, wire and
pad geometry can be seen in Figure~\ref{tpc2}. There are three
different sector types (M, W and K). Significant dead regions, 24\mm\
wide, exist between neighbouring sectors' radial boundaries such that
any portion of a particle's path in r$\phi$ that lies over a dead
region will not produce hits. As such, the relative geometry of the
three sector types was chosen such that the dead regions zigzag,
limiting the number of hits a particle can lose in this way.
\par
\begin{figure}[tb]
\centerline{\resizebox{10cm}{!}{
\epsfig{figure=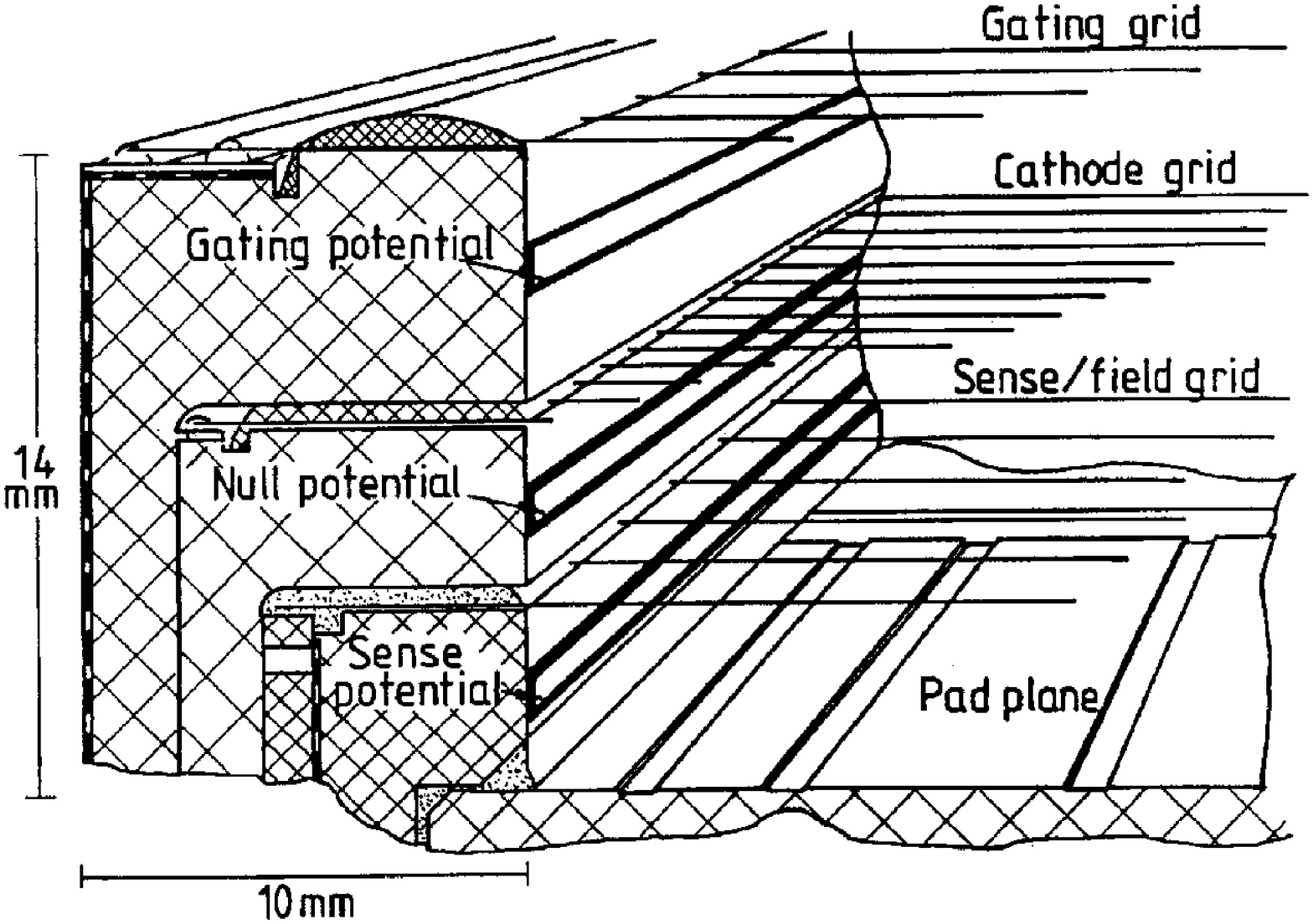,angle=-90}
}}
\caption{Schematic diagram of a sector edge showing wires and pads. \ladd{The strips on the left are held at potentials such as to reduce field distortions at sector boundaries.}}
\label{tpc3}
\end{figure}
Ionisation from a particle's path arriving at the end-plate encounters
first a layer of wires called the gating grid (see
Figure~\ref{tpc3}). If the Level-1 trigger has reached a `yes'
decision this will be held at a potential such that\iforig{}{ it} is
transparent to the passage of ionisation\footnote{\ie\ the gate will
be open. If there has been no such decision a potential difference of
200V will exist between neighbouring wires making the grid opaque,
(gate closed). This is to stop positive ions generated in the
avalanches described later from drifting back into the main TPC volume
and causing distortions in the electric field. A Level-1 `yes'
decision holds the gate open for 45\us, the maximum drift-time of the
TPC.}. Next it will encounter the cathode grid which is held at null
potential to provide a shield between the uniform field of the main
TPC volume and the field produced by the third and final layer. This
is the sense grid and consists of alternating sense and field wires,
the sense wires being held at null potential while the field wires are
held at a high positive potential. As a result, incoming electrons
avalanche towards the sense wires. The time of arrival and profiles of
the resulting charge pulses are measured both by the sense wires and,
through capacitive coupling, by the nearby pads.
\par
The integrated charge of each wire pulse is proportional to the \dEdx\
of the particle multiplied by the length of the particle's path that
projects onto the wire. They are recorded and are used to calculate
the \dEdx\ of each track once reconstruction has taken place. Clusters
of pad pulses form the ($r$,\,$\phi$,\,$z$) TPC hits. Pad pulses which
have multiple peaks in their time-profile are split, allowing more
than \iforig{once}{one} pulse per pad and thus ultimately, hits that
overlap in $r\phi$. The radius of a hit comes simply from the radius
of the relevant pad row (clusters do not transcend rows). The $\phi$
value comes from a combination of the $\phi$'s of the relevant pulses,
although the exact nature of the combination depends on the number of
pulses involved. The $z$ value is calculated from the cluster time
using the drift velocity, the cluster time being a charge-weighted
average of the time-of-arrival estimates of the pulses. The drift
velocity is known from laser calibration.
\par
A single-hit resolution of 173\um\ in $r\phi$ and 740\um\ in $z$ has
been measured from leptonic $Z^0$ decays (which involve few, high
momentum tracks leading to clear, accurately reconstructed
events). Average resolution decreases with particle momentum and dip
angle (as such, two-photon events have about the worst resolution).
\par
The fractional momentum resolution, which is proportional to momentum,
was measured from ideal $Z^0\rightarrow\mu^+\mu^-$
events\footnote{`Ideal' means the angle between the two tracks is
\iforig{less than 0.3$^\circ$}{greater than 179.7$^\circ$} and the
total ECAL energy unassociated with the muons is less than 100
MeV.}. For tracks reconstructed purely from TPC hits it was found to
be
\[ \frac{\Delta p}{p^2}=1.2\times 10^{-3} \GeV^{-1}. \]
With the benefit of ITC hits this drops to
\[ \frac{\Delta p}{p^2}=0.8\times 10^{-3} \GeV^{-1}, \]
and with VDET hits, to
\[ \frac{\Delta p}{p^2}=0.6\times 10^{-3} \GeV^{-1}. \]

\section{Material density within the tracking volume}
Clearly, from the point of creation of a particle to its entrance
into the calorimeters, any interaction with the material of the
detector is undesirable since it will alter the particle's 4-vector
and so corrupt its measurement. Such interactions can also create new
particles which may be confused with those resulting from the physics
at the interaction point. This would be especially problematic close
to the beam axis where information about a particle's momentum, or
even existence before an interaction would be minimal or
non-existent. An ideal tracking device would therefore be transparent
to the particles passing through it, \ie\ of zero density. Obviously
this is not achievable in the real world where the amount,
distribution and density of material comprising a subdetector can only
be minimised within the constraints of the design criteria of that
subdetector, both as a measuring device of sufficient accuracy and as
a physical structure of sufficient strength.
\par
The principal high-density components that exist within the ALEPH
tracking volume are summarised in Table~\ref{mat1}, and it is these
that are responsible for the vast majority of material interactions at
non-negligible polar angles. Most of this material is necessary to
isolate the various gas systems that exist within ALEPH, \ie\ the
virtual vacuum of the beam-pipe, the argon-carbon~dioxide mixture of
the ITC and the argon-methane mixture of the TPC. The total number of
radiation lengths of material to be penetrated by a straight particle
from the interaction point before reaching the first VDET layer and
first TPC pad are shown in Figures~\ref{mat2} and~\ref{mat3}
respectively as a function of polar angle. The former is important
since the first VDET layer is the first active part of the detector
that a particle will typically encounter. Any interaction that happens
at a lower radius happens `in the dark', corrupting the relevant
particle's 4-vector before it has a chance even to be detected. Also,
any new particles that are created as a result of the interaction will
be more difficult to reject since they will not be lacking any
hits. The amount of material before the first TPC pad is of interest
since there is no more solid matter between here and the outer edge of
the tracking volume. Figure~\ref{mat3} then serves as a summary of the
problematic material listed in Table~\ref{mat1} as a function of polar
angle, and shows that a high-energy photon emitted from the IP with
$\theta=90^\circ$ stands a 6\% chance of pair-converting before
entering the TPC. Note that in general, as the polar angle decreases,
the thickness of material to be penetrated increases, due to the
decreasing angle of incidence into cylindrical components of the
detector. Peaks and lumps are due to circular end-plates and
non-cylindrical support structures.

\begin{table}
\renewcommand{\arraystretch}{0.9}
\scriptsize
\centerline{
\begin{tabular}{|c|c|c|c|c|c|} \hline
& & & & & \\
Component& Purpose& Material& Geometry& Thickness & Radiation\\
& & \tiny (or principal& & (mm)	& lengths \\ 
& & \tiny materials if many)& & & (\% of X$_0$) \\
& & & & & \\ \hline
& & & & & \\
Beam-pipe 	& Holds LEP vacuum \& 	& Be 	& Cylinder, 	& 1.1 	& 0.3 	\\
		& supports VDET 	& 	& $r=5.3\cm$	& 	& 	\\
		& 			& 	& 		& 	& 	\\ 
VDET		& Active subdetector 	& Si,  	& Complex, see	& N/A	& 1.5	\\
		& + support structure	& carbon fibre& sect.\ref{vdet}&&	\\
		& 			& \& Araldite& 	 	&	&	\\
		&			& 	& 		& 	& 	\\
ITC inner wall 	& Contains ITC gas \& 	& Carbon& Cylinder,	& 0.6 	& 0.3	\\
		& supports VDET 	& fibre	& $r=12.8\cm$	& 	& 	\\
		& 			& 	& 		& 	& 	\\
ITC end-plates 	& Contain ITC gas \& 	& Al 	& Holed discs,	& 25 	& 28\iforig{.}{} 	\\
		& transmit wire tension & 	& $30\cm>r>12.8$\cm,& 	& 	\\
		& to outer wall 	& 	& $z=\pm100$\cm	&	& 	\\
		&			& 	& 		& 	& 	\\
ITC outer wall 	& Contains ITC gas \& 	& Carbon& Cylinder, 	& 2 	& 1 	\\
		& counters wire tension	& fibre	& $r=28.5\cm$	& 	& 	\\
		& 			& 	& 		& 	& 	\\
TPC inner wall & Contains TPC gas  	& Al, mylar& Cylinder, & 10.5 & 2.3 	\\
		& \& forms inner 	& \& Nomex& $r=31\cm$	& & 	\\ 
		& electric field cage		& honeycomb& 		& 	& 	\\
		&	 		& 	& 		& 	& 	\\ \hline
\end{tabular}
}
\protect\caption{A summary of the principal high-density components within the tracking volume. \ladd{Numbers quoted in the geometry column give the inner surface of the component.}}
\label{mat1}
\end{table}

\setlength{\xa}{14cm}
\begin{figure}[tbp]
\centerline{
\epsfig{figure=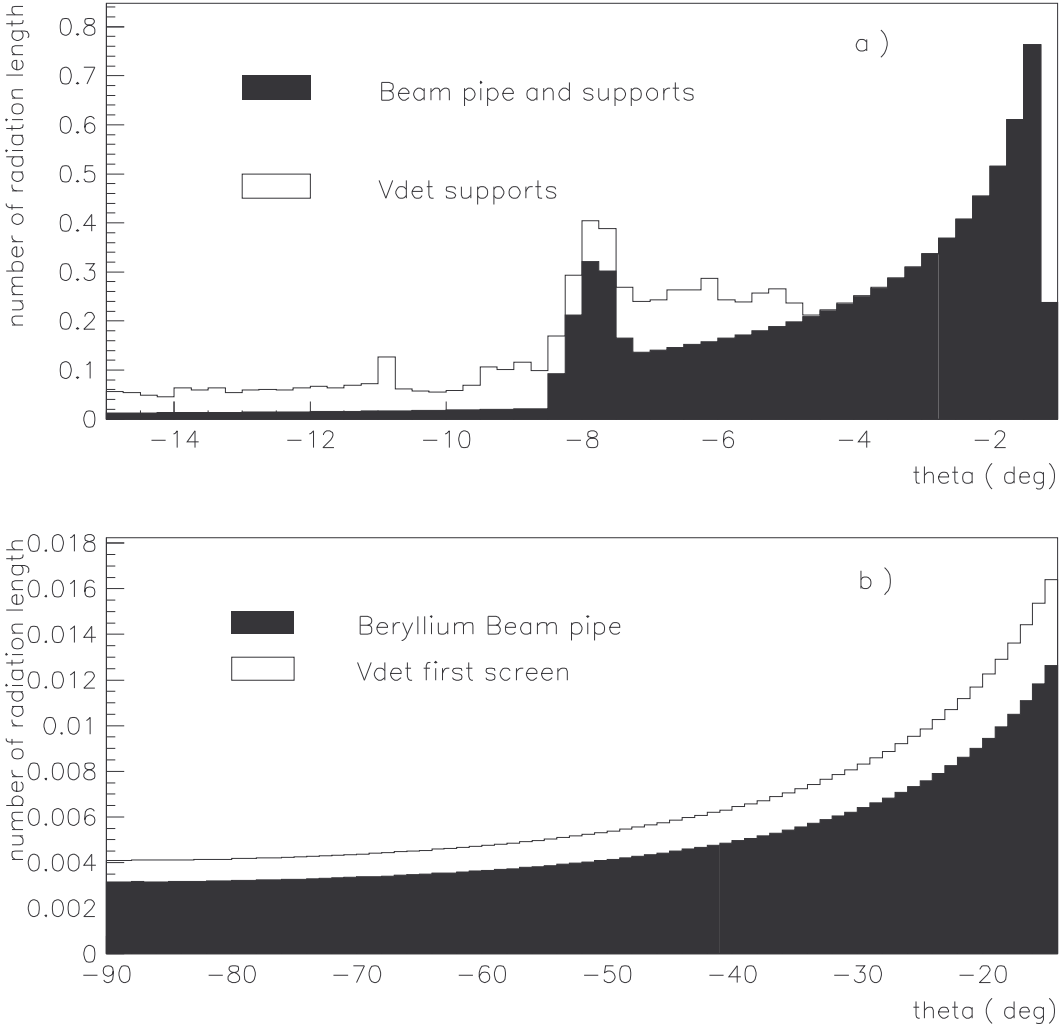,width=\xa}
}
\caption{Material before the first VDET layer as a function of polar angle, $\theta$.}
\label{mat2}
\end{figure}
\begin{figure}[tbp]
\centerline{
\epsfig{figure=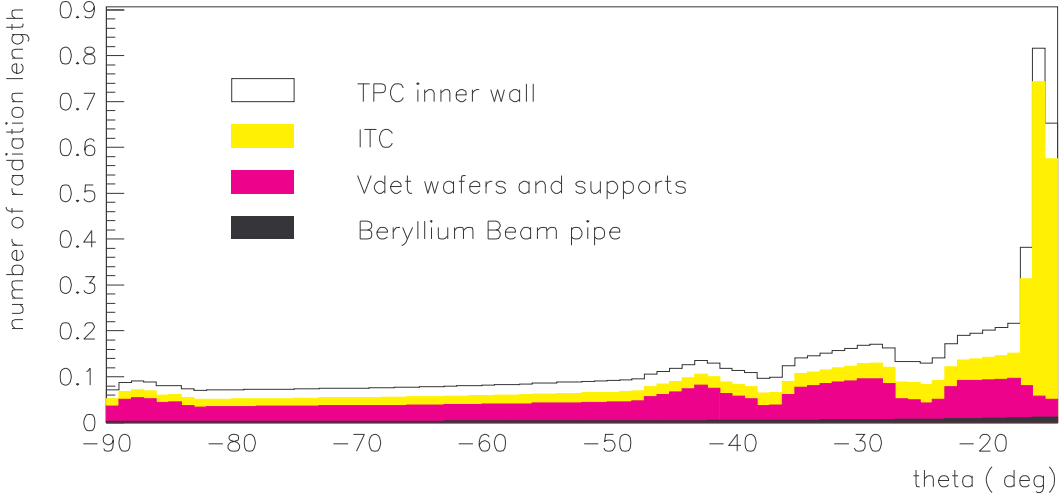,width=\xa}
}
\caption{Material before the first pad of the TPC as a function of polar \mbox{angle, $\theta$.}}
\label{mat3}
\end{figure}

\section{The Electromagnetic CALorimeter (ECAL)}
The `ECAL' is a lead and wire chamber sampling calorimeter with a
thickness of 22 radiation lengths. It performs a multi-stage energy
measurement of electromagnetic particles. Opaque to electrons and
photons, it can yield their total energy. Hadrons and muons typically
penetrate its full thickness. In measuring both the transverse and
longitudinal shower profiles created by particles it also provides
information on particle identity.
\par
Again, it is comprised of a barrel plus two endcaps aligned with the
$z$-axis. Each is segmented into twelve modules in $\phi$. A module
(there is little difference whether it be in the barrel or an endcap)
is a hermetically sealed unit containing 45 `layers' in an 80\%-xenon
to 20\%-carbon dioxide gas mixture held at slightly above atmospheric
pressure. A layer consists of a lead sheet, an aluminium extrusion
containing an anode wire plane, and a plane of cathode pads separated
and insulated from the wires by a graphite-coated mylar layer (see
Figure~\ref{ecal1}). An incoming photon or charged particle will
generate an electromagnetic shower upon collision with the lead
sheet. Electrons freed by the resulting ionisation of the gas are
avalanched towards the anode wires where the resulting charge pulse
capacitively induces a signal on nearby cathode pads (each measuring
$30\mm\times30\mm$). Thus the pads measure the position of the
e.m. showers whilst the wires, which have a fast readout, can be used
in the trigger. The 45 layers of a module are split into three
`stacks'. The first stack contains the first 10 layers and is 4
radiation lengths thick, the second contains the next 23 and is 9
radiation lengths thick, the third contains the final 12 and is also 9
radiation lengths thick since its lead sheets are double the thickness
(4\mm\ as opposed to 2\mm). Cathode pads from consecutive layers are
grouped and connected internally into `towers', which are skewed such
as to point to the IP. Energy deposits from all pads belonging to a
particular stack within a tower are summed in readout. This provides
three energy readings per tower, the magnitudes of which (absolute and
relative) depend on the incoming particle ID and energy.

\begin{figure}[tb]
\centerline{
\epsfig{figure=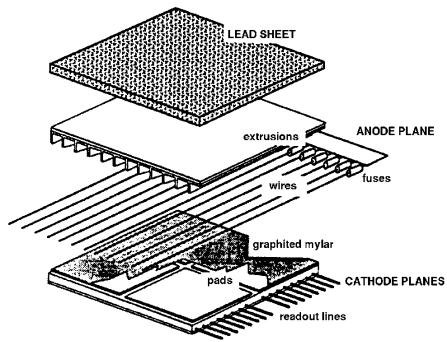,height=6cm,angle=-90}
}
\caption{Portion of a separated ECAL layer.}
\label{ecal1}
\end{figure}

\par
In excess of 73,000 towers provide the high granularity required for
good $e\!-\!\pi$ separation in jets. Solid angle coverage is $3.9\pi$
with dead regions due to module boundaries representing 2\% of the
barrel surface and 6\% of the endcap surface. Endcap and barrel
modules are offset from each other by 15$^\circ$ to prevent
coincidence of boundaries. Calibration over the full energy range is
performed using Bhabha, $\tau$$\tau$ and $\gamma$$\gamma$ events, and
also independently of any events using radioactive sources mounted
close to the gas inlets and injected periodically into the gas
system. The fractional energy resolution, $\Delta$E/E, for electrons
and photons is
\iforig{(18\(/\sqrt{E}\))\%}{$(18\GeV^\frac{1}{2}/\sqrt{E})$\%}. The
spatial resolution is
\iforig{(6.8\(/\sqrt{E}\))\mm}{$(6.8\mm\GeV^\frac{1}{2}/\sqrt{E}$)}.

\section{The Hadron CALorimeter (HCAL)}
The HCAL forms the final barrier to particles travelling through the
detector, and is designed to stop hadrons and provide a measure of
their energy. Muons typically penetrate its full thickness, and their
departure from the detector is tagged by muon chambers\iforig{ at the}{}
surrounding the main body of the HCAL. A particle at normal incidence
encounters 1.2\m\ of iron, equivalent to 7.2 interaction lengths for
hadrons. The HCAL also serves as the main support structure of ALEPH,
and as the return path for the magnetic field.
\par
It consists of a barrel of twelve modules and two endcaps of six
modules each. A module consists of 22 consecutive layers of 5\cm-thick
iron and streamer tubes, and is typically $\sim7\m$ long depending on
position. Each has a final layer of 10\cm-thick iron. The tubes
consist of 8\cm-wide `comb' profile PVC extrusions contained in
plastic boxes. Nine `teeth' form eight cells, each containing an anode
wire running the length of the tube in a gas mixture of 22.5\% Ar to
47.5\% CO$_2$ to 30\% isobutane. A photograph is shown in
Figure~\ref{hcal}. The internal surfaces of the cells are coated in
graphite paint. Showers from collisions in the iron layers create
ionisation in the \iforig{tubes and localised electron
avalanches}{tubes, and the resulting electrons avalanche} towards the
anode wires. Signals are induced on cathodes on both sides of the
tube. The upper (open) side supports copper pads, the lower supports
aluminium strips which run the length of the tube. Like the ECAL, the
pads are grouped into towers pointing to the IP. There are just under
2,700 towers in all, and the angular ($\Delta\phi\times\Delta\theta$)
size they present to the IP ranges from $3.7^\circ\times3.0^\circ$ at
$\theta=90^\circ$, to $15.0^\circ\times2.5^\circ$ at
$\theta=6^\circ$. The aluminium strips are 4\mm\ wide. Their output is
binary, indicating whether or not a tube has been fired at least
once. They provide a detailed two-dimensional view of shower
propagation within the detector which is important for muon
identification.

\begin{figure}[tb]
\centerline{\mbox{
\includegraphics[width=7cm]{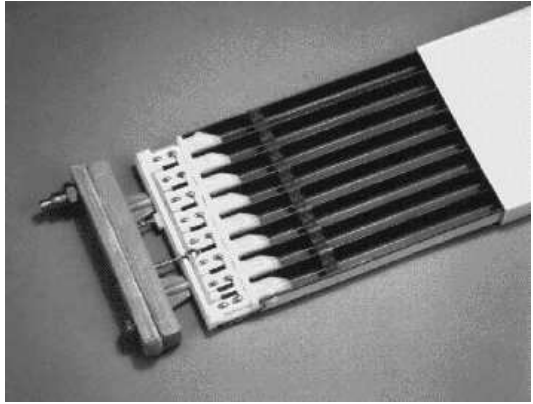}
}}
\caption{Photograph of a streamer tube.}
\label{hcal}
\end{figure}

\par
Surrounding the iron structure are two double layers of tubes, 50\cm\
apart around the barrel, 40\cm\ apart at the endcaps. These are the
muon chambers. Here the pad layers are replaced by additional strip
layers which run orthogonally to the tubes, enabling three-dimensional
hits to be obtained for charged particles leaving the
detector. Together with information from the main bulk of the detector
this provides powerful muon identification.
\par
The energy resolution of the HCAL for pions is
\iforig{$\Delta E/E=84\%/\sqrt{E}$}{$\Delta E/E=(84\GeV^\frac{1}{2}/\sqrt{E})$\%}.

\section{The Luminosity CALorimeter (LCAL)}
The LCAL is a lead and wire chamber sampling calorimeter which
provides the main luminosity measurement for ALEPH. An accurate
measurement of the luminosity is essential in order to obtain reaction
cross-sections from event-rates. The measurement is obtained from the
event-rate of the QED t-channel Bhabha process, which is well
understood. The differential cross-section is proportional to
$\theta$$^{-4}$, and receives corrections due to interference from the
s-channel process,
e$^+$e$^-$$\rightarrow$Z/$\gamma$$\rightarrow$e$^+$e$^-$, which
decrease with $\theta$. Thus in order to obtain the high event-rates
needed for an accurate measurement of the luminosity, and to make
insignificant the s-channel corrections\footnote{necessary since they
depend on the properties of the Z which were not accurately known at
the start of LEP.}, the LCAL sits at very low $\theta$.
\par
The LCAL consists of two annuli about the beam-pipe, 25 radiation
lengths thick, beginning at $z=\pm263$\cm. Each covers the range
44\,mr to 160\,mr and is formed by two semi-annular modules with 38
sampling layers each. The modules are almost identical to ECAL
modules, with layers consisting of lead and proportional wire chambers
(see Figure~\ref{ecal1}). Like the ECAL, the layers are grouped into
three stacks, and pads are grouped into towers which point to the
IP. There are 384 towers per module.
\par
Bhabha events are tagged by requiring back-to-back hits in both units
above a threshold energy. The integrated luminosity is obtained by
dividing the number of events seen by the theoretical cross-section
multiplied by the experimental efficiency. An accuracy of the order of
0.5\% is obtained.

\section{The trigger system.}
Necessary to the operation of ALEPH is a system that will judge when
an event of note (e$^+$e$^-$ annihilation, Bhabha or two-photon) has
taken place and will subsequently initiate the readout of the
subdetectors. This is the trigger system. It allows the TPC to exist
in its insensitive state (`gate closed', see Section~\ref{tpc}) for
much of the running time without loss of efficiency. It also reduces
the amount of uninteresting data written to tape, and the dead-time of
the detector that results from readout.
\par
The trigger decision is split into three levels, each based on
increasingly more complex information. The Level-1 trigger reaches a
conclusion within 5\us, less than the bunch crossing time. It
requires a good charged track in the ITC and/or information from the
calorimeters indicative of a particle deposit. Should the decision be
`yes', the TPC gate is held open and the Level-2 decision is processed
within 50\us. This is essentially a recalculation of the Level-1
decision based now on three-dimensional information from the TPC. In
the case of a `yes-decision' readout of the whole detector is
initiated, otherwise data-acquisition is halted and reset.
\par
During LEP1 the Level-1 trigger performed much better than was
anticipated, making Levels-2 and 3 relatively unimportant as they
vetoed only a small fraction of events. Since running at LEP2 energies
however, noisier beam conditions have led to a large increase in the
number of unwanted background `events'. As a result the importance of
Level-2 has risen greatly, now reducing the event-rate by in excess
of 50\%. Level-3 is an offline trigger, remains unimportant and is not
described here.

\section{Event reconstruction}
\label{recon}
Event reconstruction refers to the processing that is performed on the
raw data output from the detector in order to correlate the
information and produce final data that is more representative of what
actually \iforig{ocurred}{occurred} in the event. Most important from
the point of view of the analysis described in this thesis, is the
track reconstruction. The tracks are the reconstructed helical
trajectories of the charged particles and are formed by connecting the
hits produced in the tracking chambers. The reconstruction algorithm
starts in the TPC, linking nearby hits to form track
segments. Segments which are compatible with a single helical
trajectory with an axis parallel to the $z$-axis are connected to make
tracks. At least four TPC hits are required for a track. These are
then extrapolated down into the ITC and VDET where compatible hits are
added. If there are ITC hits left unassigned to a track then the
algorithm attempts in a similar way to form these into tracks (which
will then not have assigned TPC hits). These tracks are required to
have a minimum of four ITC hits. The tracks are parameterised by a set
of five helix parameters. These are: the inverse radius of a track's
circular projection in the $xy$ plane, 1/$R$; the tangent of its `dip
angle' (equal to the ratio of its longitudinal and transverse momentum
components), tan($\lambda$); its distance of closest approach to the
$z$-axis,
\dzero; and the ratio of its $y$ and $x$ momentum components and its
$z$ coordinate at that point, \phizero\ and \zzero\ respectively. A
final track fit is performed using Kalman filter techniques
\cite{kalman}.
\par
Once an event is reconstructed, the energy flow algorithm is run. The
purpose of this is principally to create associations between charged
tracks and calorimeter deposits, enabling the identification of
particles and an improvement in the overall energy resolution, and is
described in detail in \cite{eflow} and
\cite{performance}. Unfortunately energy flow was not written with
charged particle decay in the tracking volume in mind. It disregards
tracks with a \dzero\ greater than 2\cm\ or a $|\zzero|$ greater than
10\cm, and so is of limited use in this analysis.

\chapter{The Search}
\label{search}

\section{Introduction}
\label{introintro}
This thesis details a search for the supersymmetric process by which
(lightest) neutralinos are pair-produced in $e^+e^-$ collisions, each
then independently and promptly decaying to a slepton plus
corresponding lepton, the sleptons then travelling a measurable
distance in the detector before themselves decaying to a lepton plus
gravitino.
\[e^+\ e^- \rightarrow \chi\ \chi \rightarrow \tilde{l}l\ \tilde{l}l \rightarrow \tilde{G}ll\ \tilde{G}ll\]
The final state is thus four leptons (two from the IP, two from kinks
or with large \dzero\footnote{See Section~\ref{recon}.}) plus missing
energy (since the gravitinos will not be detected). Each neutralino
decays independently giving six distinct channels, \selsel, \smusmu,
\staustau, \selsmu,
\selstau\ and \smustau. An example event is shown in Figure~\ref{event}. 
This process can contribute to the exclusion of the GMSB parameter
space (or potential for GMSB discovery) since, for a significant
portion of the kinematically allowed space, the cross-section for the
neutralino is greater than that for the slepton, and so when
$m_{\chi}>m_{\tilde{l}}$, slepton production via neutralino decay can
dominate over direct production.
\begin{figure}
\centerline{\epsfig{figure=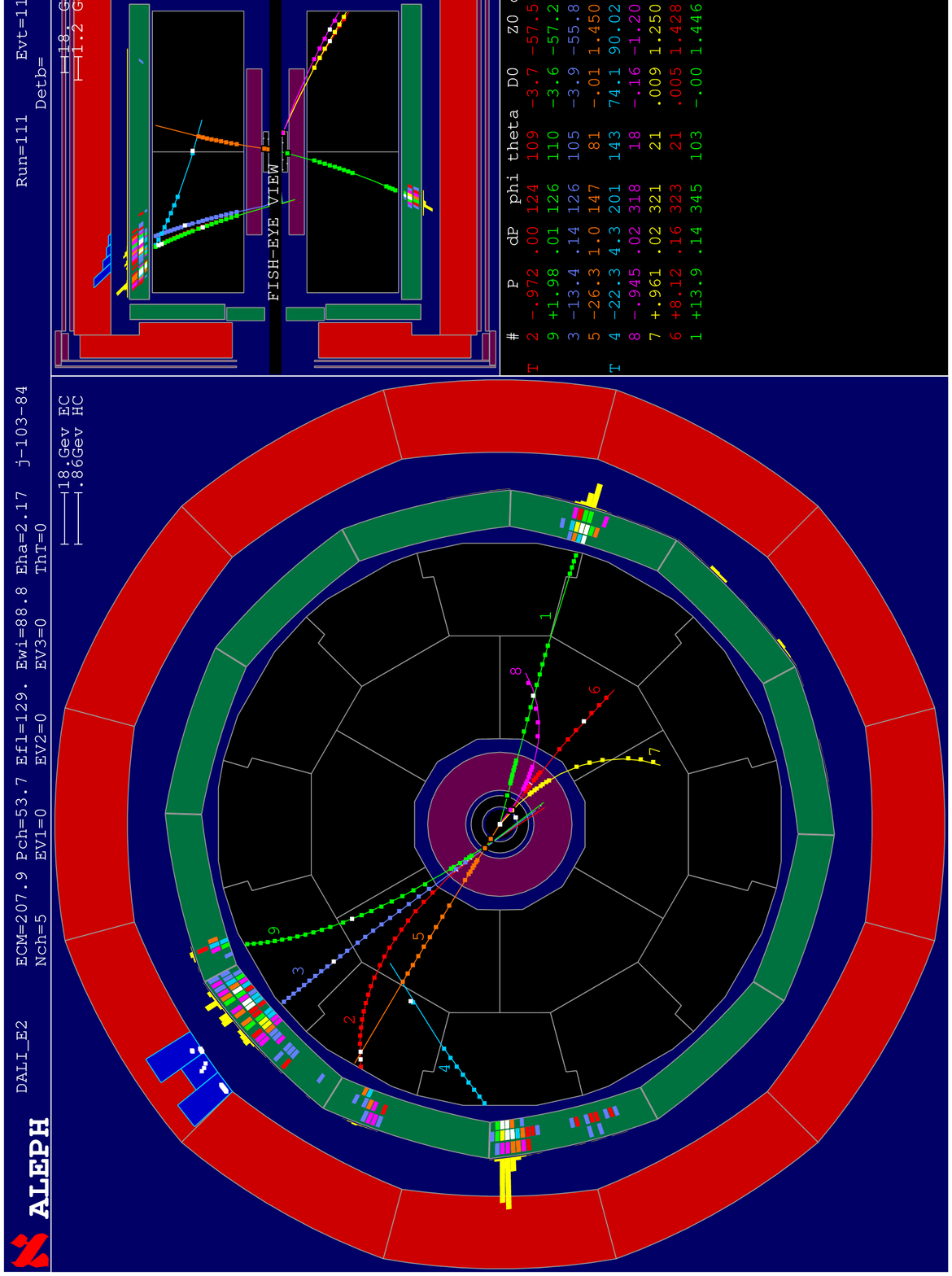,width=0.96\textwidth}}
\caption{An example Monte Carlo \selstau\ event at 208\GeV. \ladd{The detector is shown in the `fish-eye' view to magnify the tracking volume relative to the calorimeters. The \stau\ decayed after travelling a few centimeters, the \selectron\ after travelling $\sim1\m$. Both $\tau$'s have undergone three-prong decay.}}
\label{event}
\end{figure}
\par
The ultimate goal of the search was to discover evidence for
supersymmetry. In the absence of any such evidence, it was to
formulate model-independent cross-section limits for the process and
to exclude as much of the kinematically allowed parameter space as
possible under a minimal GMSB model.
\par
The search was performed using a total of 628\pb$^{-1}$ of LEP2 data
taken over a range of centre-of-mass energies from 189 to $208\GeV$,
recorded between 1998 and 2000 (see Table~\ref{lums}). The data was
stored on `DST's. These contain all the information about the
reconstructed ALEPH events (unlike the usual format of MINIs), but
events must fulfil some basic criteria to be written to DST from `POT'
(essentially the same but containing all events). An event must have
\begin{itemize}
\item more than $3\GeV$ of charged energy coming from within 5\cm\ in \dzero\ and 20\cm\ in \zzero
\item OR 2 or more charged tracks with \dzero\ and \zzero\ within this range
\item OR total ECAL wire energy greater than 15\GeV\ and ECAL $|t_0|<500$\ns\ ($t_0$ is the difference between the time of the event as recorded by the ECAL wires and the time of the nearest LEP bunch crossing: events due to conventional $e^+e^-$ collision typically have $|t_0|<100$\ns)
\item OR 1 or more photon candidates and no charged tracks.
\end{itemize}
\par

A method was required to distinguish events from the process being
searched for (`signal' events) from events due to known Standard Model
processes (`background' events). The technique used is called Monte
Carlo analysis, and requires the simulation of both the signal and the
background. The simulated events are called Monte Carlo data. The
method for distinguishing signal from background can then be developed
using the Monte Carlo data, and subsequently applied to the real
data. The results can be compared to the results that would be
expected in the absence of the signal as given by the background Monte
Carlo, and the degree of evidence for the signal process can be
quantified.

\begin{table}
\centerline{
\renewcommand{\arraystretch}{1}
\begin{tabular}{|c|c|r@{.}l|}
\hline
Energy (GeV) & Year & \multicolumn{2}{|c|}{Integrated Luminosity (pb$^{-1}$)}\\
\hline
188.6 & 1998 & \hspace{1.7cm} 174&2 \\
191.6 & 1999 & 29&0  \\
195.5 & 1999 & 79&9  \\
199.5 & 1999 & 86&3  \\
201.6 & 1999 & 41&9  \\
199.8 & 2000 & 0&8   \\
201.8 & 2000 & 0&7   \\
202.7 & 2000 & 1&8   \\
203.8 & 2000 & 8&3   \\
205.0 & 2000 & 71&6  \\
206.3 & 2000 & 65&5  \\
206.6 & 2000 & 60&8  \\
208.0 & 2000 & 7&3   \\
\hline\multicolumn{2}{|c}{Total  =} & 628&0 \\
\hline
\end{tabular}
}
\protect\caption{Integrated luminosities recorded by ALEPH at each centre-of-mass energy achieved by LEP between 1998 and 2000. \ladd{Year 2000 energies formed a continuum and have been placed in 1\GeV\ bins about integer values. The energy shown is the luminosity weighted average.}}
\label{lums}
\end{table}

\section{Monte Carlo simulation}
\subsection{Signal}
\label{signalmc}
Signal Monte Carlo data was produced using SUSYGEN 2.2 \cite{susygen}
before subsequent processing by GALEPH \cite{galeph} (the ALEPH
detector simulation program based on GEANT \cite{geant},\cite{geant3})
and JULIA \cite{julia} (the ALEPH event reconstruction program used to
reconstruct both Monte Carlo and real data events). The sleptons were
set stable at generator level, then tracked and decayed by
GALEPH\footnote{Some modification to GALEPH was needed for the
tracking of sleptons. This was performed by Clemens Mannert.}.
\par
The characteristics of a signal event as observed in the detector
depend only on the neutralino mass, and the slepton masses and
lifetimes. GMSB dictates that either the sleptons are degenerate in
mass, and therefore share the role of the NLSP (the co-NLSP scenario),
or that the stau is lighter and so is the sole NLSP (the stau-NLSP
scenario). In the first case the mass degeneracy also leads to
lifetime degeneracy, and so all three sleptons are described by a
single mass and a single lifetime. In the second case the mass
difference between the stau and the selectron/smuon becomes an extra
free parameter. But it is of little importance since if the stau is
significantly lighter than the selectron/smuon then the neutralino
will decay almost exclusively to it, and so no channel other than the
\staustau\ is of importance and the mass of the selectron/smuon is
irrelevant.
Thus although GMSB parameter space is six-dimensional, the working
parameter space for this search can be reduced to three dimensions,
($m_\chi$, $m_{\tilde{l}}$, $\tau_{\tilde{l}}$), where $l$ can be $e$,
$\mu$ or $\tau$ in the case of degeneracy, and $\tau$ otherwise. These
are of course not the only relevant parameters. The neutralino
production cross-section and branching ratios are of critical
importance to the success of the search since they define the event
rates for each channel. But for the purposes of signal Monte Carlo
generation and the formulation of the selections they are not
important.
\par
For the purposes of Monte Carlo generation it is necessary to identify the
approximate region of parameter space in which the selections are
likely to be sensitive. In particular, slepton lifetimes should be
chosen such that a good mapping of the variation of the selection
efficiency with slepton lifetime is obtained over the region of
interest. Thus it is natural to convert slepton lifetime to slepton
decay length, since this is the principal feature of the signal
events, and thus will have a far more direct relationship to the
selection efficiencies. The natural working parameter space is then
($m_\chi$,$m_{\tilde{l}}$, $d_{\tilde{l}}$), where $d_{\tilde{l}}$ is
the slepton decay length in the detector.
At the time of Monte Carlo
generation, sensible values can be chosen for the slepton decay length
based on the detector size and geometry, and then translated into
slepton lifetimes ensuring a good efficiency mapping. The slepton
decay length is a function of $m_\chi$, $m_{\tilde{l}}$, $\sqrt{s}$ as
well as $\tau_{\tilde{l}}$. In general this function is not simple
because the ionisation caused by the slepton as it passes through the
material of the detector will cause it to slow down, and so the decay
length is also a `function' of the detector set-up. However, it is not
necessary that the translation from decay length to lifetime be exact,
and the following approximation is used:
\begin{gather*}
\tau_{\tilde{l}}=\frac{d_{\tilde{l}}}{\gamma\beta c} \\
\parbox{\textwidth}{\flushleft where,} \\
\gamma\beta=P_{lab}/m_{\tilde{l}} \ , \\ 
P_{lab}=\sqrt{E_{rest}^2-m_{\tilde{l}}^2+\left(E_{rest}\ p_{\chi}/m_{\chi} \right)^2} \ , \\
E_{rest}=(m_\chi^2+m_{\tilde{l}}^2-m_l^2)/2 m_\chi \ , \\
p_{\chi}=\sqrt{E_{beam}^2-m_{\chi}^2} \ .
\end{gather*}

Here $\gamma$ and $\beta$ pertain to the slepton in the lab frame,
$P_{lab}$ is the (initial) momentum of the slepton in the lab frame if
it is emitted at 90$^\circ$ to the lab direction in the neutralino
rest-frame, $E_{rest}$ is the energy of the slepton in the neutralino
rest-frame, $p_{\chi}$ is the momentum of the neutralino in the lab
frame, $E_{beam}$ is the LEP beam energy and $m_l$ is the mass of the
relevant lepton. In the case of a mixed channel (\ie\ the two sleptons
have different flavours) the lifetime is calculated with both lepton
masses and the average is taken.
\par
Initial work showed that for a given decay length, the selection
efficiencies were only weak functions of the neutralino and slepton
masses. Thus Monte Carlo data was generated to focus on the variation
of selection efficiency with decay-length more than on the
masses. Eight decay-lengths were chosen, and seven points in
($m_\chi$, $m_{\tilde{l}}$) space were generated for each of them,
apart from the two highest decay lengths for which an extra two points
in mass space were used (since some initially surprising results were
obtained for long decay lengths which will be described in
Section~\ref{effs}), making a total of sixty points. This was done for
each channel at both the lowest and the highest LEP centre-of-mass
energies under consideration: 189\GeV\ and 208\GeV. An exception was
the
\selsel\ channel at 189\GeV\ which was used as a test case, and so 21
mass points were generated per decay length. 500 events were generated
at each point. The slepton masses ranged from 67\GeV\ to 102\GeV\, and
the neutralino masses from 69\GeV\ to 103\GeV\, with never less than a
1\GeV\ mass difference between them. The decay lengths ranged from
2\mm\ to 20\m. The exact points at which Monte Carlo data was
generated for each channel at each energy are given in
Appendix~\ref{mc}. Slepton mass degeneracy was always assumed in the
generation of the mixed channels (\selsmu,
\selstau, \smustau).
\par
All the signal Monte Carlo data was run through the POT $\rightarrow$
DST selection described in Section~\ref{introintro}, and events
failing were appropriately tagged and not included in any signal
efficiencies.

\subsection{Background}
Due to the nature of the signal a large array of Standard Model
processes form significant backgrounds to the search. Only processes
which are very easily cut against, or which have a very low
cross-section can be disregarded as sources of background. A list of
the backgrounds analysed, together with their cross-sections, the
number of events that were analysed, and the generators used for their
production, is given in Table~\ref{bgmc}. At 189\GeV\ (at which
174\pb$^{-1}$ of data was collected) roughly 20 times the number
expected in data were analysed for each process. At 208\GeV\ (at which
only 7.3\pb$^{-1}$ of data was collected) roughly 200 times the number
expected in data were analysed for each process. These factors were
not achieved for the two-photon to tau and up/down quark processes
however, since limited Monte Carlo samples were available.
\begin{table}[hptb]
\renewcommand{\arraystretch}{1.4}
\centerline{
\begin{tabular}{|c|>{$}c<{$}|c|c|c|c|c|c|}
\hline
\multicolumn{2}{|c|}{Process} & \multicolumn{2}{c|}{At 189\GeV} & \multicolumn{2}{c|}{At 208\GeV} & Generator \\ \cline{3-6}
\multicolumn{2}{|c|}{}& $\sigma$ (pb) & \multicolumn{1}{c|}{No. analysed} & $\sigma$ (pb) & No. analysed & \\
\hline
\multirow{5}{2ex}{\rotatebox{90}{Annihilation}} & qq  		& 99.4	& 340,000	& 78.8	& 120,000 	& \footnotesize KORALZ 4.2\\
						& \tau\tau 	& 8.21	& 30,000 	& 6.61	& 10,000	& \footnotesize KORALZ 4.2\\
						& WW     	& 16.6	& 60,000 	& 17.5	& 26,000	& \footnotesize KORALW 1.21 \\
						& ZZ     	& 2.76	&  10,000	& 2.79	& 4,000	 	& \footnotesize PYTHIA 5.7 \\
						& Zee    	& 99.1	& 350,000	& 98.9	& 144,000	& \footnotesize PYTHIA 5.7 \\
\cline{1-1}
\multirow{5}{2ex}[2ex]{\rotatebox{90}{Two-photon}}& \gamma\gamma\rightarrow\tau  & 431	& 800,000& 461	& 600,000 & \footnotesize PHOT02 \\
						& \gamma\gamma\rightarrow ud   & 487	& 400,000& 487	& 250,000 & \footnotesize PHOT02 \\
						& \gamma\gamma\rightarrow c    & 93.2	& 160,000& 94.4	& 50,000  & \footnotesize PHOT02 \\
						& \gamma\gamma\rightarrow s    & 23.9	& 40,000 & 23.7	& 17,000  & \footnotesize PHOT02 \\
\hline
\end{tabular}
}
\caption{The Standard Model background processes that were analysed as backgrounds to the search. \ladd{$\sigma$ is the production cross-section. \protect\iforig{}{The final states of annhilation processes are created via the annihilation of the electron and positron, whereas the final states of two-photon processes are created via the collision of two photons which originated from the electron and positron. For information about KORALZ, KORALW, PYTHIA and PHOT02 see \cite{koralz}, \cite{koralw}, \cite{pythia} and \cite{phot02} respectively.}}}
\label{bgmc}
\end{table}
\par
The generated events were processed by GALEPH and JULIA in the same
way as for signal\footnote{This was performed by other ALEPH members,
generating Monte Carlo data for general ALEPH use.}. Kinematic cuts
were applied in the generation of the two-photon Monte Carlo data. For
all the stated two-photon processes, events were required to have a
final state invariant mass greater than $2.5\GeV$. For the two-photon to
tau process, final states were required to have at least $0.15\GeV$ of
transverse momentum. For the two-photon to quark processes a scattering
angle of at least 5\,mr was required for the electron or positron.
\par
The samples were split into two equal halves, one to act as a guide in
formulating the selections, the other to give the final background
expectation once the selections were finalised (so as to minimise bias
in the background expectation). Exceptions were the two-photon to quark
processes where the high cross-section and limited Monte Carlo samples
meant that the same sample had to be used for both.

\chapter{Event Analysis}
\label{analysis}
\section{Introduction}
The technique for the selection of candidate signal events is a
two-step process. Firstly, using the ALPHA \cite{alpha} package which
facilitates access to the full event information, preliminary data
candidates are identified using some simple preselection. For those
events satisfying the preselection, variables are calculated that are
designed to be sensitive to the differences between signal and
background. These are written to HBOOK \cite{hbook} ntuples
(essentially tables of data in which rows correspond to successive
events and columns correspond to variables) forming a greatly reduced
data set. The second stage is the application of a selection,
consisting of a set of cuts on the variable values, to determine the
final signal candidates. This is the technique used by most searches
for physics beyond the Standard Model at LEP.
\par
Many, if not all, of the variables should pertain to the event as a
whole, such as the number of charged tracks, total energy and
invariant mass. In order to calculate such `global' variables the
relationships between the various reconstructed detector objects such
as tracks and calorimeter deposits should be known. This is the
purpose of the energy flow algorithm. But energy flow has only limited
abilities to make sense of processes that are happening away from the
interaction point. It ignores tracks with a \dzero\ greater than 2\cm\ or a
\zzero\ greater than 10\cm, and since JULIA does not reconstruct the
slepton decay vertex, it does not recognise the relationship between
the reconstructed slepton track (if there is one) and the resulting
lepton track. Thus energy flow will only yield correct results if the
slepton decay length is short enough for the resulting lepton to
satisfy the cuts on \dzero\ and \zzero, or long enough for the slepton to reach
the calorimeters before decay; and so it is of limited use in this
analysis.
\par
Even if energy flow did treat charged particle decay in the tracking
volume correctly, the signal would still not be well-defined by global
variables in the case of the \staustau\ channel (and the \selstau\ and
\smustau\ channels to a lesser extent) because of the larger number of 
invisible particles (up to eight neutrinos on top of the two
gravitinos) and the larger range in the charged track multiplicity
(from multi-prong tau decays). The selection then relies primarily on
the presence of high-\dzero\ tracks: that being the primary feature of the
signal. Several other processes can generate high-\dzero\ tracks though,
such as nuclear interactions, ECAL splash-backs, photon conversions,
multiple scattering and cosmic rays. The frequency with which these
processes occur means that their rejection efficiency must be close to
100\% if they are not to swamp any signal.
\par
After the event preselection is described in Section~\ref{presel},
this chapter goes on to detail the procedures that are used to
discriminate between high-\dzero\ tracks due to the signal process, and
those due to background. Sections~\ref{splashbacks} and \ref{NIs}
describe the identification of ECAL splash-backs and nuclear
interactions respectively. Then Section~\ref{fromslep} describes the
procedure that is used to confirm (or reject) that a high-\dzero\ track is
compatible with the hypothesis of originating from slepton decay. The
final cut-based selections are described in
Chapter~\ref{selections}. In order to facilitate the descriptions of
the procedures described in this chapter a terminology is used which
is explained below and in Figure~\ref{jargon}. The figure also
introduces some points of interest concerning tracks reconstructed
from particles that have not been produced at the IP. There are three
terms which are important to understand:
\par
\textbf{The post(pre)-\dzero\ trajectory of a particle.} This is the section of the particle's trajectory after it has passed (before it has reached) its point of closest approach to the beam-axis.
\par
\textbf{A post(pre)-\dzero\ track.} This is a track that has been formed by hits created on the post(pre)-\dzero\ portion of a particle's trajectory. Both this and the term above refer to truths about the event that are not known a priori from the reconstructed information.
\par
\textbf{The post(pre)-\dzero\ section of a track.} This is the section of a track which is post(pre)-\dzero\ according to the track direction (assigned by JULIA, and which may be incorrect). The track will only have assigned hits on its post-\dzero\ section. If it is a pre-\dzero\ track then its pre-\dzero\ section will probably follow closely the post-\dzero\ section of a post-\dzero\ track (see Figure~\ref{jargon}).
\par
Although this terminology might seem confusing, it is an unfortunate
reflection of the subtleties involved in the relationship between the
reconstructed and `truth' information of tracks not originating from
the IP.
\par
Many of the procedures \iforig{involving}{involve} the fitting of
tracks to vertices, either to test whether a group of tracks form a
new vertex, or whether they fit to an existing vertex. This is
performed using the ALEPH YTOP package described in \cite{ytop} and
\cite{lutz}. It provides a $\chi^2$ for each fit performed, which is
normalised to the number of degrees of freedom.

\begin{figure}[htpb]
\setp{108mm}{70mm}{100mm}{1.4}
\centerline{
\framebox{\resizebox{10.3cm}{!}{\includegraphics*[\xa,\ya][\xb,\yb]{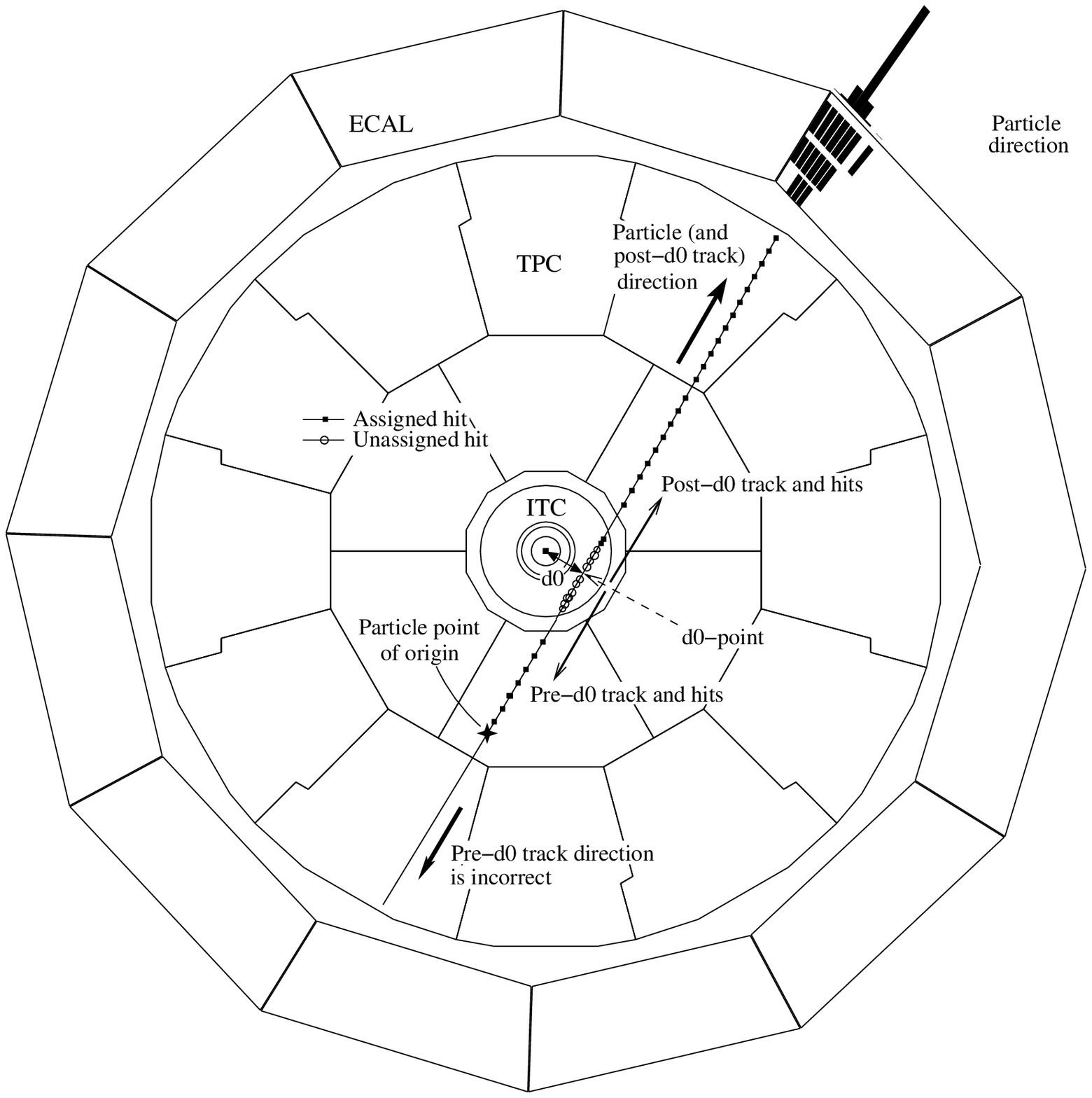}}}
}
\caption{\lscap{The reconstructed track(s) of a particle that has been produced in the TPC, moving towards the beam-axis, through the ITC, then back out into the TPC and on to the ECAL. The \dzero-point is the point at which the track is closest to the beam-axis. Before it reaches the \dzero-point it is travelling along its pre-\dzero\ trajectory, afterwards, along its post-\dzero\ trajectory. Since the idea that particles are produced at the IP is a paradigm within JULIA, it assumes that the particles responsible for all tracks were moving away from the beam-axis and did not originate significantly before their \dzero-point. For this reason it will not form tracks from hits that would lie on both sides of the \dzero-point; and so the particle trajectory shown here is reconstructed as two tracks: one pre-\dzero\ track reconstructed from the pre-\dzero\ hits, and one post-\dzero\ track reconstructed from the post-\dzero\ hits. JULIA incorrectly assigns the directions of pre-\dzero\ tracks as being radially outwards, and so both their momentum 3-vector and charge have the wrong sign. Tracks that have no TPC hits will be referred to as ITC tracks, tracks that do have TPC hits will be referred to as TPC tracks and may or may not have ITC hits. Either type of track may or may not have VDET hits.}{An example of a high-\dzero\ track with labelling to indicate the meaning of certain terms.}}
\label{jargon}
\end{figure}
\section{Preselection}
\label{presel}
In order to speed up analysis, events were first subjected to some
very basic preselection aimed at fast rejection of events that were
clearly not signal candidates. The event was rejected if
\begin{itemize}
\item The scalar sum of the momenta of all tracks was less than \mbox{$4\times(E_{LEP}/188.6)\GeV$} (to reduce the number of two-photon events)
\item The total number of tracks was less than 3 or greater than 30 (to reduce both two-photon and $q\overline{q}$ events)
\item The total number of tracks was less than 4 and the two highest momentum tracks had momenta greater than half the beam energy and the angle between them was greater than $179^\circ$ (to reduce $e^+e^-$ and $\mu^+\mu^-$ events)
\item The total number of energy flow charged tracks was greater than 11 (to further reduce $q\overline{q}$ events)
\item The energy flow event energy was less than \mbox{$7\times(E_{LEP}/188.6)\GeV$} and the energy flow event momentum was less than \mbox{$1\times(E_{LEP}/188.6)\GeV$} (to further reduce two-photon events)
\end{itemize}

\section{Rejection of ECAL splash-backs}
\label{splashbacks}
The electromagnetic shower created when a charged particle enters the
electromagnetic calorimeter typically propagates in the direction of
the incoming particle. Some of the electrons and positrons in the
shower can however, scatter back into the TPC where they can be
reconstructed as tracks. This is known as an `ECAL splash-back'.
\par
\begin{figure}[htb]
\centerline{
\resizebox{10cm}{!}{\includegraphics*[45mm,80mm][150mm,170mm]{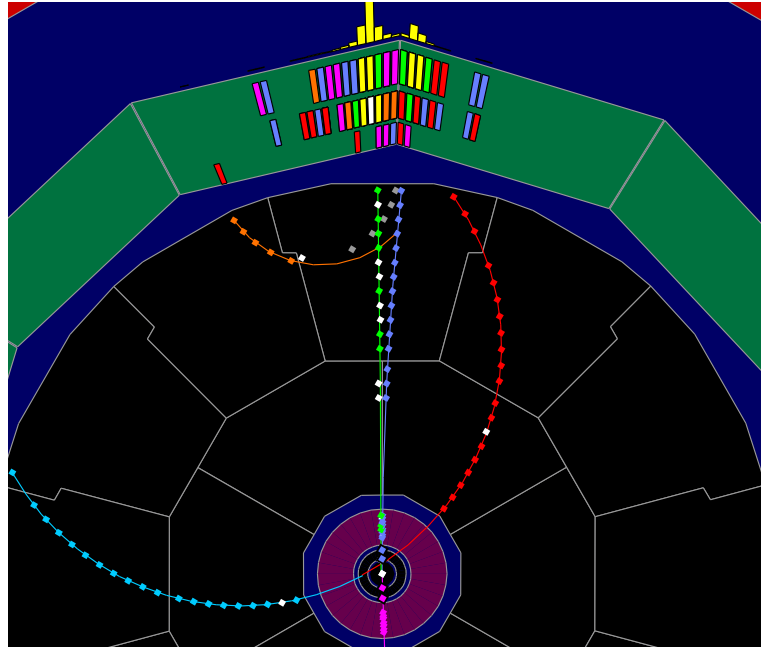}}
}
\caption{An example of an ECAL splash-back in a Monte Carlo $\tau\tau$ event at $189\GeV$. \ladd{Two electrons have escaped the ECAL to be reconstructed as high \dzero\ tracks in the TPC, one arcs through the ITC and passes out the other side to form a third track.}}
\label{splashbackpic}
\end{figure}
They are characterised by a high-\dzero\ track `entering' the
\iforig{TPC}{ECAL} (in fact the particle responsible is leaving but
JULIA assumes all tracks move radially away from the beam-axis: the
track is pre-\dzero) close to another track that has higher momentum,
with both forming a common vertex in the region of the ECAL (see
Figure~\ref{splashbackpic} for an example). The only circumstance
under which this is likely to occur in a signal event is when the
slepton decays inside the ECAL and the resulting lepton is ejected
back into the TPC. Since this analysis is only intended to be
sensitive to slepton decay in the tracking volume then the possibility
of rejecting signal events by confusing slepton decays with ECAL
splash-backs is not regarded a problem.
\par
The track that results from the splash-back will be referred to as the
output track, the track causing the splash-back will be referred to as
the input track. The conditions that must be satisfied by both a
candidate output and input track for the output track to be tagged as
a splash-back are as follows:\\
\begin{minipage}{\linewidth}
\begin{itemize}
\item The output track must have TPC hits and momentum, $p_{output}<\\5\times (E_{LEP}/188.6)\GeV$ and \dzero$>0.5\cm$ or \zzero$>2\cm$.
\item The input track must also have TPC hits and have $p_{input}>0.5\GeV$ and $p_{input}>p_{output}$.
\item The points at which the output and input tracks leave the TPC volume must be within 50\cm\ of each other.
\item The tracks must form a common vertex in the region
\begin{tabbing}
The tracks must for\= \kill
\> $165<r<230\cm$ and $|z|<300\cm$, \\
or \> $200<|z|<300\cm$ and $r<230\cm$
\end{tabbing}
(these roughly correspond to the vicinity of the ECAL barrel and endcaps respectively).
\end{itemize}
\end{minipage}
\section{Rejection of nuclear interactions}
\label{NIs}
Hadrons interacting with nuclei in the structure of the detector can
cause hadronic showers leading to multiple high-\dzero\ tracks. These are
characterised by many tracks originating from a common vertex which is
displaced from the IP (see Figure~\ref{NIpic} for an example). Like
splash-backs these have distinct characteristics that can be
tagged. Care must be taken though to avoid confusing a
\stautomultiprongtau decay vertex with a nuclear interaction.
\par
\begin{figure}[htb]
\centerline{
\resizebox{10cm}{!}{\includegraphics*[55mm,110mm][160mm,205mm]{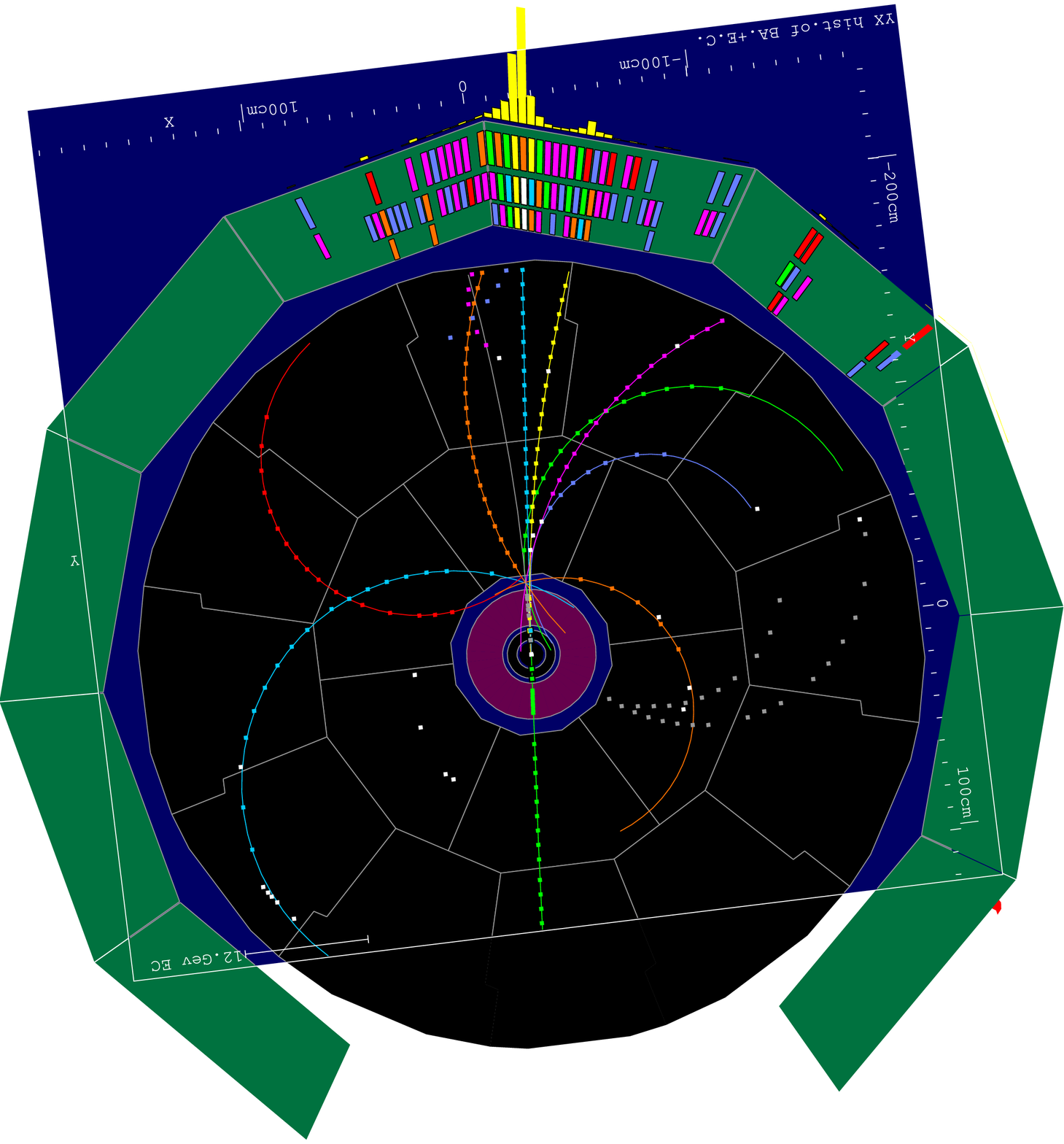}}
}
\caption{An example of a nuclear interaction in a Monte Carlo $\tau\tau$ event at $189\GeV$. \ladd{A pion from the decay of a tau has interacted with a nucleus in the inner TPC wall, giving a shower of high-\dzero\ tracks, mainly protons.}}
\label{NIpic}
\end{figure}
Clearly an algorithm that can find multiple tracks originating from a
common vertex is required. Initially this was done by testing all
track pairs for compatibility with a common vertex, and then
attempting to merge these 2-track vertices. This was very CPU
intensive however, and a method which used the fitting routine less
often was sought. The procedure settled upon relied on starting with a
single high-\dzero\ track, and trying to add further tracks to it to build
up a multi-track vertex. Once all the tracks that could be added had
been (which may be none at all), and if there were one or more other
high-\dzero\ tracks in the event which had not been included in the vertex,
then one of these was used in the same way as the first track to try
to build a new vertex. Unlike the first method tried, this method
required making a choice as to the order in which the tracks should be
selected so as to give the best chance of all tracks from a given
vertex actually being included. Using tracks in order of increasing
fractional momentum error was found to be successful. The exact
procedure used is shown in the flow diagram of Figure~\ref{NIflow}. In
this diagram `most reliable' means smallest fractional momentum
error. A track passes `basic cuts' if it has not been tagged as an
ECAL splash-back, has TPC hits, has a momentum smaller than half the
beam energy and greater than $0.1\GeV$, and a \dzero$>0.15\cm$ or a
\zzero$>3\cm$. A track is `hit-compatible' with a vertex if at least all
bar one of its hits lie on one side of the vertex, so that it is
compatible with the hypothesis that it was produced at that vertex.
\par
\begin{figure}[htbp]
\centerline{\iforig{\epsfig{angle=-90,file=flow_original.eps}}{\epsfig{file=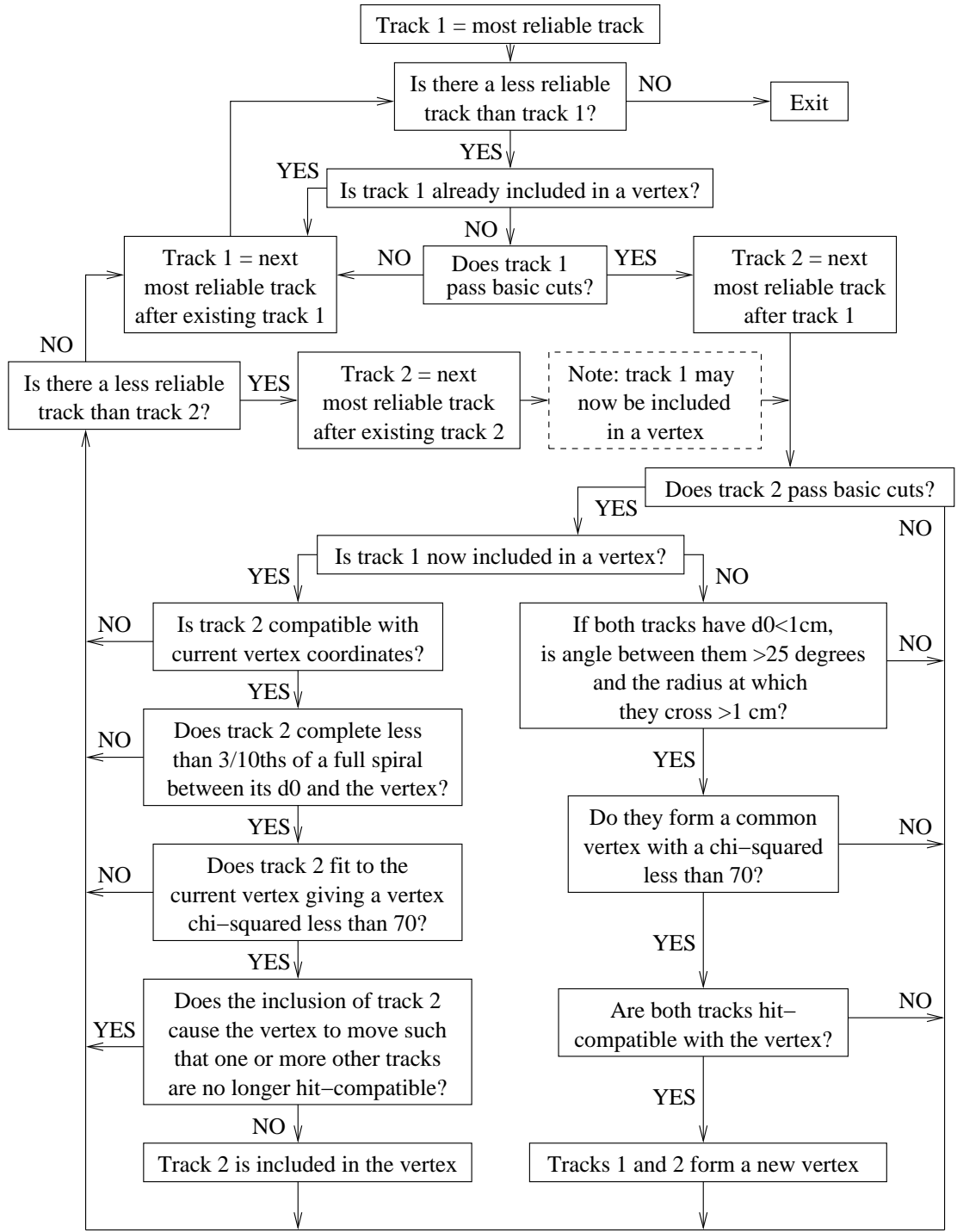,width=\textwidth}}}
\label{NIflow}
\caption{Flow diagram showing the procedure used to find and tag nuclear interaction vertices.}

\end{figure}
Once one multi-track vertex has been found with this procedure, then
if each new vertex after the first is less than $3\cm$ from, or shares
more than 50\% of its tracks with, a previously found vertex then an
attempt is made to merge the two. If this results in a new vertex with
a $\chi^2$ (given by YTOP) less than 70 then the merge is considered
successful, and the two vertices are replaced with the single merged
vertex.
\par
Once the vertices have been finalised, certain properties are
calculated for each, such as the summed momentum and charge of the
constituent tracks. Care was taken however, to account for a
potentially problematic feature of the reconstruction algorithm. Since
JULIA always assumes that particles move radially away from the beam
axis, a particle that is produced far enough from the beam axis and
moving back towards it can result in two reconstructed tracks (a pre-\dzero\
and a post-\dzero), as is shown in Figure~\ref{jargon}. In this case the
track corresponding to the portion of the particle trajectory before
it reaches its point of closest approach to the beam axis is assigned
the wrong direction. Thus its momentum 3-vector and charge have the
wrong sign, and this will corrupt the overall properties of any
multi-track vertex in which it is included. Also, since the total
number of tracks in the vertex is used to classify it, then a single
particle giving two tracks would corrupt this as well. To correct this,
any track whose TPC hits lie at a smaller radius than the vertex has
its charge and momentum 3-vector reversed. Then the \dzero, \zzero\ and
directions of these tracks are compared with any tracks which are part
of the same vertex and for which the vertex lies on the pre-\dzero\
portion. If the tracks differ in \dzero\ and \zzero\ by less than 30\% of their
respective average, and the angle between them at their \dzero-points is
greater than $170^\circ$, then both are assumed to have originated
from the same particle trajectory and the one with fewer TPC hits is
removed from the vertex. As an example, if both the tracks shown in
Figure~\ref{jargon} were found to originate from a nuclear
interaction, then the pre-\dzero\ track would be removed.
\par
Each vertex is then placed into one of 3 categories based on its
properties. A vertex is labelled as a \stautomultiprongtau vertex
if it satisfies all of the following conditions --
\begin{itemize}
\item its charge is $\pm1$
\item its largest two-track angle at the vertex is less than $20^\circ$
\item its momentum is $>4\times(E_{LEP}/188.6)\GeV$ and $<55\times(E_{LEP}/188.6)\GeV$
\item its kink-angle is between $5^\circ$ and $170^\circ$
\item its kink-angle is $>15^\circ$ or its $|\cos(\theta)|$ is less than 0.6
\item its number of tracks is less than 6
\item its distance from the IP is greater than $0.3\cm$
\item and its distance from the IP is greater than $1\cm$ or its $\chi^2$ is less than 3.
\end{itemize}
The charge of the vertex is the sum of the charges of all the
constituent tracks. The momentum is the vector sum of the momentum
3-vectors of all the constituent tracks. The kink-angle is the angle
between the vertex momentum 3-vector and the straight line from the IP
to the vertex. The $|\cos(\theta)|$ of the vertex refers to the vertex
position, \ie\ $\iforig{|z_{vertex}/r_{vertex}|}{|z_{vertex}/\sqrt{r_{vertex}^2+z_{vertex}^2}\,|}$.
\par
Although this is a reasonably stringent set of cuts, background events
can \mbox{still} have vertices labelled as \stautomultiprongtau since a
charged particle emitting a bremsstrahlung photon which quickly
undergoes pair-conversion can create something very similar.
\par
If a vertex fails this selection then it is classified as a nuclear
interaction vertex if --
\begin{itemize}
\item its radius is greater than $4.8\cm$ (a nuclear-interaction is highly unlikely inside the beam-pipe, so the minimum radius is set at the beam-pipe radius of 5.3\cm, minus 5\mm\ leeway)
\item its largest two-track angle at the vertex is greater than $20^\circ$
\item and its number of tracks is greater than 2.
\end{itemize}
If it fails this also then it is classified as `unknown', and the
constituent tracks are still considered valid.

\section{Confirmation of slepton-decay hypothesis}
\label{fromslep}
Although this analysis attempts to reject high-\dzero\ tracks from known
sources such as those discussed in Sections~\ref{splashbacks} and
\ref{NIs}, and these attempts are largely successful, they are not sufficient. The number of residual high-\dzero\ tracks is still too many. Hence a procedure was developed, not to test whether a high-\dzero\ track comes from a known background process, but to test whether it is compatible with coming from the signal process, \ie\ the decay of a heavy charged particle. The hits of a candidate track are examined to identify the region of its length over which the responsible particle could have originated, and this is then used to calculate what can be expected as the detector response to the parent slepton. In the absence of such a response the track can be rejected. For instance, if a track's hits indicate that the responsible particle must have been produced far out in the TPC, the parent slepton should have its own reconstructed track which forms a good kink-vertex with the high-\dzero\ track. In the absence of such a 2-track vertex the high-\dzero\ track can be rejected. This section describes the exact procedure in detail. It should be noted that only tracks that have TPC hits, a momentum greater than $1\GeV$, a \dzero\ greater than $0.1\mm$ and a $\chi^2_{BS}$ (the $\chi^2$ of the track-fit to the beam-spot per degree of freedom) greater than 5 are considered as candidate high-\dzero\ tracks.

\subsection{A particle's point of origin}
\label{origin}
While a reconstructed track is a reliable indicator of a particle's
trajectory, it does not tell us exactly where that particle originated
or where it stopped/decayed. Nominally, the point of origin is
considered to be the \dzero-point. This is fine for tracks originating
from the IP, since the distance scale associated with the uncertainty
on the particle's exact point of origin will be at most the size of
the beam spot, which will be several orders of magnitude lower than
the track's radius of curvature. Thus the track is effectively
straight over the relevant range, and the uncertainty on its point of
origin does not translate into significant uncertainty on its initial
momentum 3-vector. For a track not originating from the IP no such
simple assumption can be made, and only analysis of its hits can yield
information as to its point of origin. This will take the form of a
range of the track length over which the particle could have
\iforig{orginated}{originated}.
\par
The following procedure for finding this range assumes that the track
in question is post-\dzero\ (\ie\ reconstructed from hits created as the
particle moves away from the beam axis), and so the track direction
and charge are correctly assigned by \mbox{JULIA}. Although the
following procedure is not sensible for a pre-\dzero\ track, it will be
explained later that such a track is in fact dealt with in the correct
way.
\par
The range over which the particle could have been produced is between
the \iforig{highest}{lowest}-radius point at which there is evidence
that the particle was following the track trajectory, and the
\iforig{lowest}{highest}-radius point at which there is evidence that 
it was not. To find evidence that the particle was not following the
track trajectory, a scan is performed backwards along the track
length, starting from the point at which it achieves its highest
radius in the TPC, looking for missing hits. Nominally, a hit is
expected at every TPC pad radius that the track crosses, every ITC
wire radius that the track crosses while it is inside the ITC active
volume, and every intersection that the track makes with a VDET
wafer. There are several legitimate reasons why a track can be missing
a hit however. The TPC sector boundaries form dead regions of the
order of 1\cm\ wide, and in the ITC certain wires are not
active. Also, even if the particle has produced a hit it may not be
assigned to the respective track since the reconstruction algorithms
can simply fail to assign the hits correctly, or may only have one hit
to assign between two tracks (if they are closer than the minimum
two-track resolution). Thus if the track fulfils any of the conditions
given in Table~\ref{misshit} at a point where a hit is expected but
not observed, the lack of a hit is not considered abnormal.
\begin{table}
\renewcommand{\arraystretch}{1.3}
\centerline{
\begin{tabular}{|c|>{\scriptsize$\bullet$\hspace*{-2.5mm}}cp{12cm}|} \hline
\mr{4}{5ex}[-2ex]{TPC}& & The track is $<$\,1.5\cm\ in $r\phi$ from a sector boundary \\
		& & The track is $<$\,1.5\cm\ in $r\phi$ and $<$\,4\cm\,(10\cm) in $z$ from another TPC (ITC) track on the same side (on either side) of the central membrane \\
		& & The track is $<$\,2\cm\ in $r\phi$ and $<$\,4\cm\ in $z$ from an unassigned hit \\
		& & The radius of the point in question is $<$\,3\mm\ greater than the \newline track's \dzero\ \\ \hline
\mr{4}{5ex}[-1ex]{ITC}& & The nearest wire has a hit \\
		& & The nearest wire is dead \\
		& & The track is $<$\,1\mm\ in $r\phi$ from the cell edge and the next-nearest wire has a hit\\
		& & The point in question is at a radius $<$\,0.1\mm\ above the track's \dzero\ \\ \hline
\mr{2}{6ex}[0ex]{VDET}& & There is a hit within 0.5$^\circ$ in $\phi$ \\
		& & The track is within one sigma of the wafer edge \\ \hline
\end{tabular}
}
\caption{Conditions under which the lack of a hit is not considered abnormal in each tracking subdetector.}
\label{misshit}
\end{table}
\par
If a point is encountered at which there is no hit and at which none
of the conditions of Table~\ref{misshit} are satisfied, then it is
assumed that the responsible particle was not following the track
trajectory at this point. Then if there are no hits at lower radii,
the range for the particle point of origin is set between this point
and the lowest-radius assigned hit. If there are one or more hits at lower
radii, then the track information appears to be inconsistent with a
genuine particle trajectory. In this case, if the point under
consideration is in the VDET the track is rejected as bad. If it is in
the ITC then the track is spared for now and the scan continues, but
if another such point is encountered the track is rejected. If the
point is in the TPC, then the location of the bad point is merely
recorded. These bad TPC points will be referred to again later.
\par
If the scan reaches the track's \dzero-point without the track being
rejected or the particle range-of-origin being determined then the
scan continues along the pre-\dzero\ section of the track, as far as the
highest radius that the pre-\dzero\ section of the track reaches in the TPC
if necessary. On the pre-\dzero\ section the track will not have any
assigned hits and so Table~\ref{misshit} governs completely whether
the track is considered good at each point. If the scan reaches the
end of the pre-\dzero\ section without conclusion, then the track is
rejected (since it is probably a cosmic muon that has passed right
through the detector).

\subsection{Where is the slepton?}
\label{mother}
Once a range for the particle's point of origin has been calculated,
it is possible to translate this into a range of possible trajectories
for the parent slepton, and to make the requirement that the event
information supports one of these trajectories and thus the hypothesis
that the candidate track originated from slepton decay. This section
describes the procedure that is applied to do this.
\par
There are four distinct scenarios for the reconstructed detector
response to the slepton depending on how far the slepton travels
before decay and in what direction. If the slepton achieves a high
enough radius before decay it can produce a reconstructed TPC track,
which should form a well-reconstructed vertex with the lepton
track. In the following discussion this will be referred to as a
TPC-vertex. Failing this, it can produce an ITC track which will form
a vertex with the lepton track that is not so well reconstructed in
$z$ due to the low $z$-resolution of the ITC. This will be referred to
as an ITC-vertex. Failing this, if the slepton at least passes through
some active tracking components before decay, then it should leave
some hits which point to the lepton track. These will be referred to
as pointing-hits. Finally there is the case that the slepton decays
before creating any hits, and so the only sign of its existence is the
\dzero\ of the lepton track. Examples of each case are shown in
Figure~\ref{mother2to5}. The range for the point of origin of the
particle responsible for the candidate high-\dzero\ track determines which
of these scenarios are possible, and which are not.
\setlength{\ljlen}{\topsep}
\setlength{\topsep}{0mm}
\setlength{\ljlenb}{7cm}
\renewcommand{\arraystretch}{0.5}
\begin{figure}[htbp]
\centerline{
\begin{tabular}{>{\PBS\centering}p{\ljlenb}>{\PBS\centering}p{\ljlenb}}
\setp{100mm}{131mm}{65mm}{1.2}
\resizebox{\ljlenb}{!}{\rotatebox{0}{\includegraphics*[\xa,\ya][\xb,\yb]{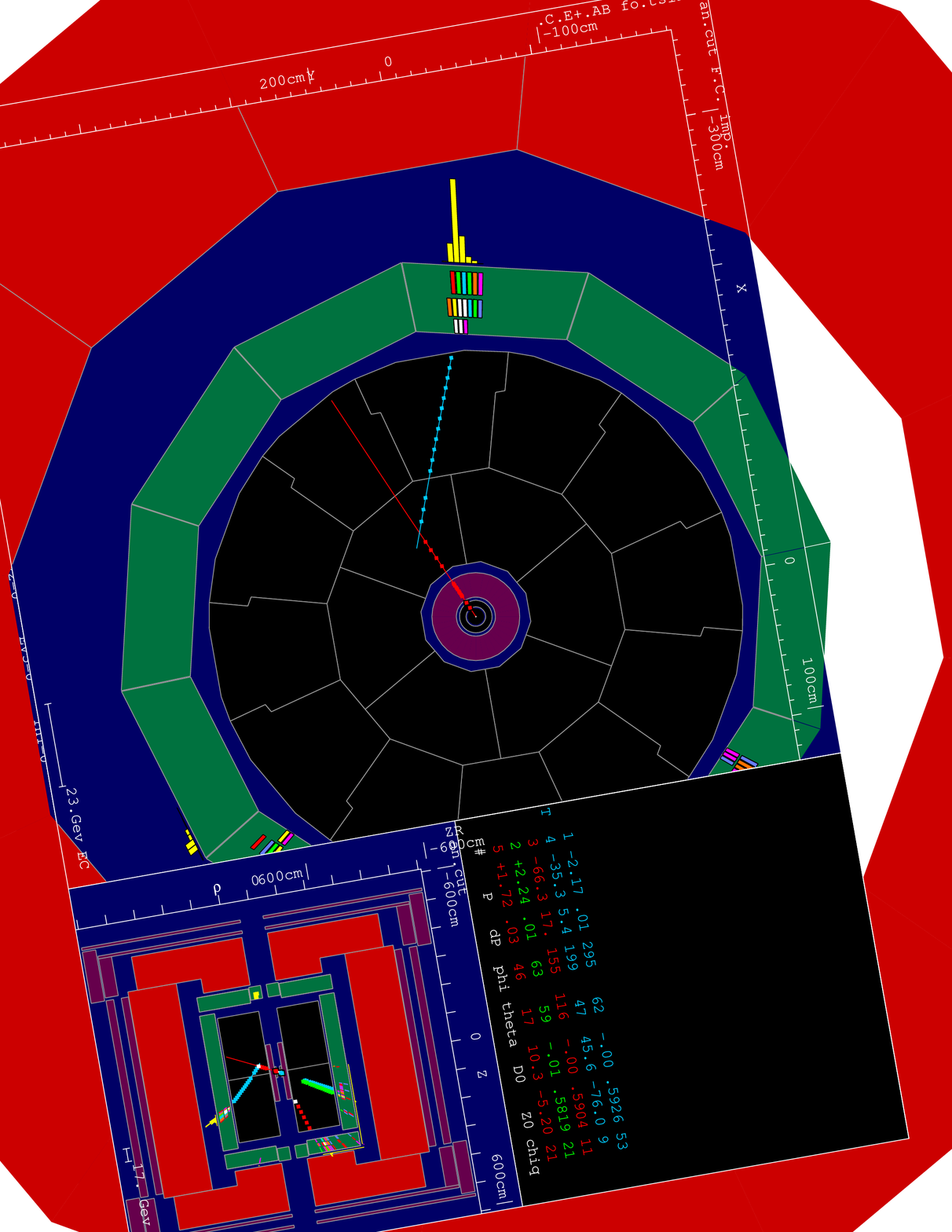}}} & 
\setp{148mm}{132mm}{55mm}{1.2}
\resizebox{\ljlenb}{!}{\rotatebox{0}{\includegraphics*[\xa,\ya][\xb,\yb]{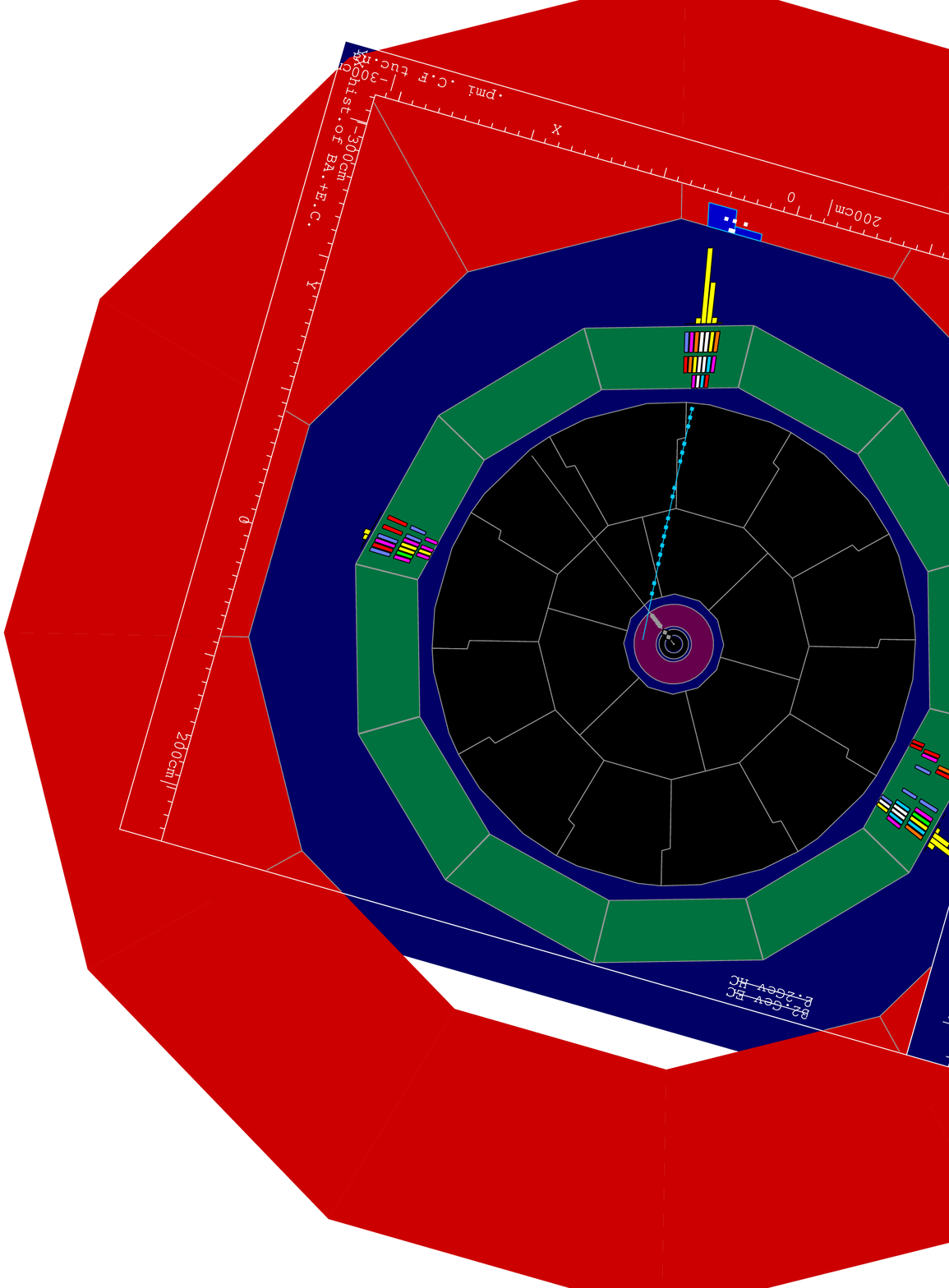}}} \\
a). & b). \\
& \\
\setp{100mm}{130mm}{75mm}{1.2}
\resizebox{\ljlenb}{!}{\rotatebox{0}{\includegraphics*[\xa,\ya][\xb,\yb]{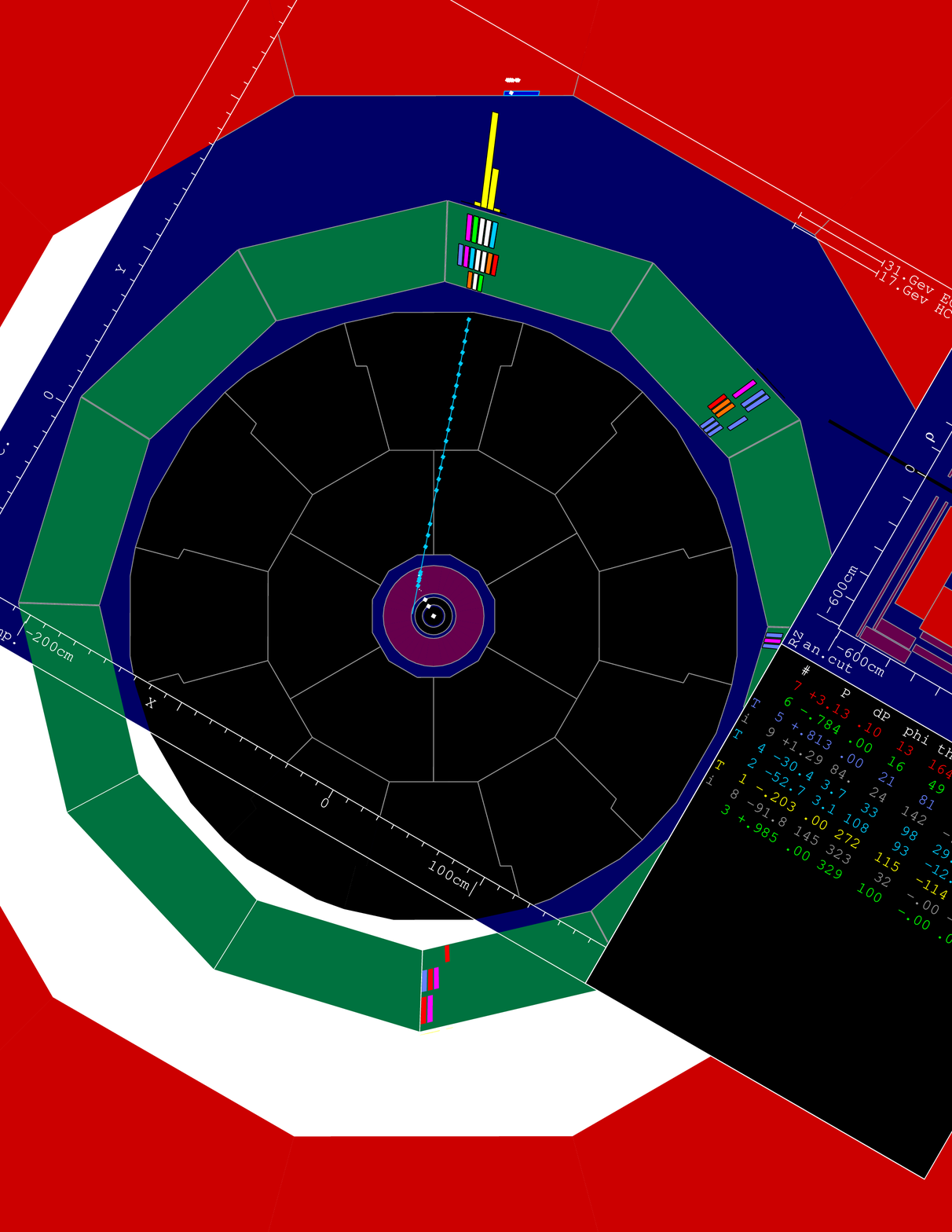}}} &
\setp{105mm}{55mm}{65mm}{1.2}
\resizebox{\ljlenb}{!}{\rotatebox{0}{\includegraphics*[\xa,\ya][\xb,\yb]{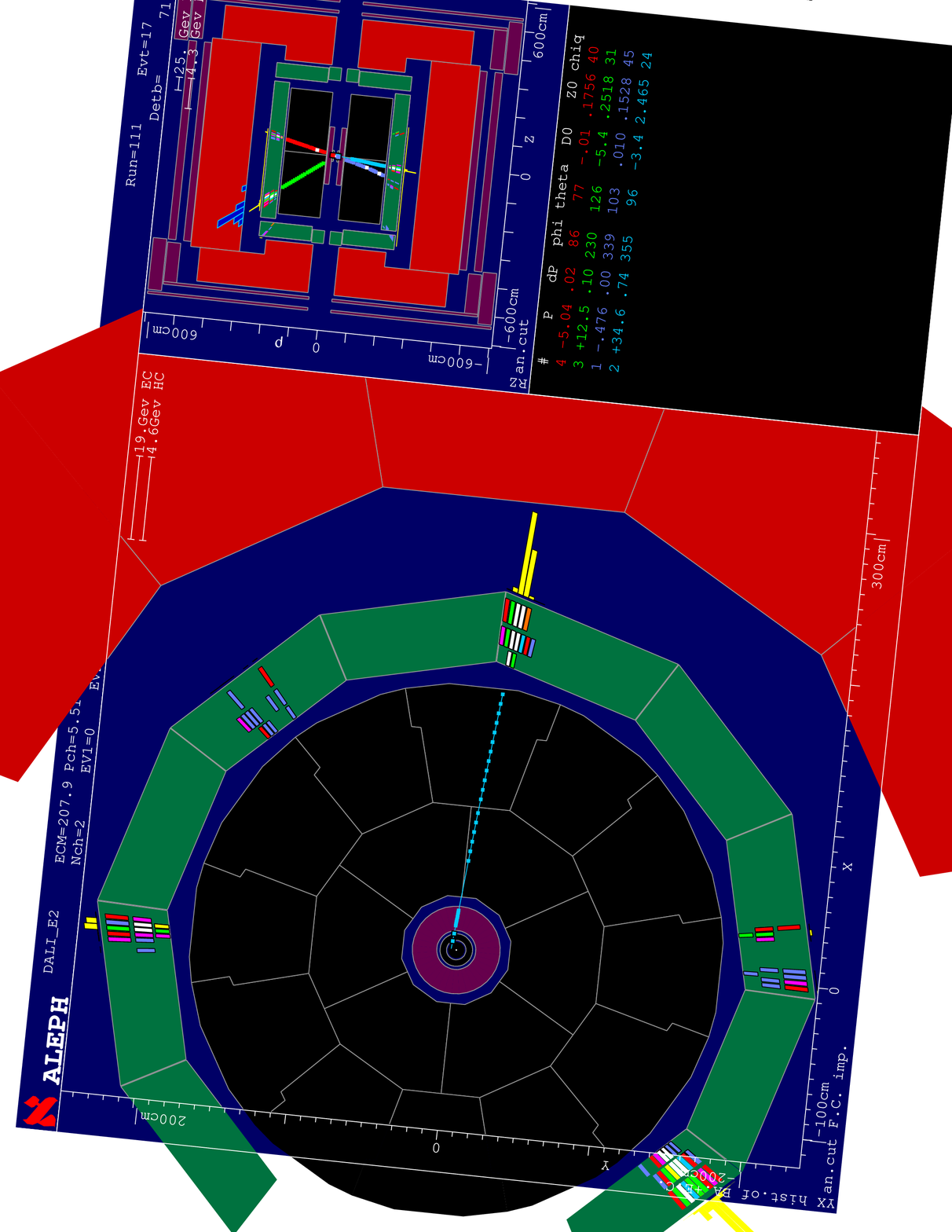}}} \\
c). & d). \\
\end{tabular}
}
\caption{Examples of the typical reconstructed detector response to the slepton and lepton for decreasing slepton decay lengths. Shown are \ladd{(a)} a TPC-vertex, \ladd{(b)} an ITC-vertex, \ladd{(c)} pointing-hits and \ladd{(d)} a simple high-\dzero\ track from signal Monte Carlo data. \ladd{All tracks and hits other than those of the slepton-lepton pair shown have been removed for clarity.}}
\label{mother2to5}
\end{figure}
\setlength{\topsep}{\ljlen}
\par
The relationship between the point of origin range and the possible
scenarios is as follows. If, for at least part of the range,
\begin{itemize}
\item the track radius\iforig{}{ (\ie\ the distance of the track from the $z$-axis)} is above that of the 4th TPC pad row, then a TPC-vertex is possible.
\item the track radius is above that of the 4th ITC wire layer and the track is either inside the ITC or the straight line from the track to the IP enters the ITC at a radius greater than this (and thus the slepton, under the assumption that its trajectory was straight, crossed at least 4 ITC wires), then an ITC-vertex is possible.
\item the straight line from the track to the IP passes through some tracking components, then pointing-hits are possible.
\item the straight line from the track to the IP passes through no tracking components, then no slepton response at all is possible.
\end{itemize}
These possibilities are not mutually exclusive, they could all be
open. However, of those that are open, it may be possible to exclude
some based on likelihood. For instance, if the track radius is
significantly above that of the 4th TPC pad row for the entire range,
then anything other than a TPC-vertex is unlikely, and a TPC-vertex
can be demanded. Thus if any of these conditions, with $3\cm$ added to
the radius necessary for a TPC-vertex and $3\mm$ added to the radius
necessary for an ITC-vertex, are true over the entire range, then the
first of the respective possibilities becomes a demand. If the demand
is not satisfied, the track is rejected. If it is found that no single
response can be demanded in this way, then the track is tested for the
possible slepton responses in the order TPC-vertex, ITC-vertex,
pointing-hits. If the track does not satisfy any of the possible
responses (bearing in mind that if the case that the slepton did not
produce any hits has been identified as a possibility, then no
response is possible), then it is rejected.
\par
The procedure involved in looking for TPC-vertices, ITC-vertices and
pointing-hits will now be described.
\subsubsection{TPC-vertices}
The first step of testing a track for a TPC-vertex is to identify
candidates for the slepton track. These are required to have TPC hits
(by definition), a \dzero\ less than $2\cm$ and a \zzero\ less than
$2.5\cm$, and must not have been identified as tracks from a
nuclear-interaction or an ECAL splash-back. Its \dEdx\ (if \dEdx\
information is available for the track) must be no less than 10
standard deviations below that expected for a singly-charged particle
of mass $60\GeV$ (the approximate GMSB slepton mass limit), and the
track should show evidence that the particle responsible ceased to
follow the track trajectory before leaving the TPC (\ie\ be missing at
least one hit from its end without observable reason -- see
Table~\ref{misshit}). If there are no candidate slepton-tracks, there
can be no TPC-vertices.
\par
The high-\dzero\ track is tested for a good TPC-vertex with each
candidate slepton track. They must form a common vertex which lies on
the post-\dzero\ section of the slepton track and has a $\chi^2$ (from
YTOP) less than 30. The $z$-coordinates of the two tracks at the vertex
must differ by less than $7\cm$, the angle between them at the vertex
must be greater than $10^\circ$, and the last assigned hit of the
slepton track must not be at a radius more than $1\cm$ above that of
the vertex.
\par
The checks made on the high-\dzero\ track's hits are more complex,
since there are three possibilities for their locations. If the
decay-angle (defined here as the angle between the slepton and lepton
directions at the point of decay in the $r\phi$ plane) is less than
90$^\circ$ then all the high-\dzero\ track's hits lie at radii greater
than the vertex radius, and both the track directions are correct and
so the decay-angle will be calculated correctly (see
Figure~\ref{mother2to5}.a). If the decay-angle is greater than
90$^\circ$ and the high-\dzero\ track is post-\dzero, then again the
track directions are correct and the decay-angle will be calculated
correctly, but now the hits can lie both above and below the vertex
radius so no constraints can be made on their positions (see
Figure~\ref{tpcver}). However, if the decay-angle is greater than
90$^\circ$ and the high-\dzero\ track is pre-\dzero\ (so all its hits
lie at a radius smaller than that of the vertex), then the decay-angle
will be calculated incorrectly to be less than 90$^\circ$. So if the
calculated decay-angle is less than 90$^\circ$, all the hits are
required to lie \textit{either} at radii above that of the vertex
minus 1\cm, \textit{or} at radii below that of the vertex plus
1\cm. If the calculated decay angle is greater than 90$^\circ$ then no
constraints are made on the positions of the hits. Thus if a
high-\dzero\ track is found to have a TPC-vertex it can be tagged as
post-\dzero\ or pre-\dzero\ based on the positions of its hits
relative to the vertex.
\begin{figure}[btp]
\setr{130mm}{142mm}{150mm}{0.22}
\resizebox{\textwidth}{!}{\includegraphics*[\xa,\ya][\xb,\yb]{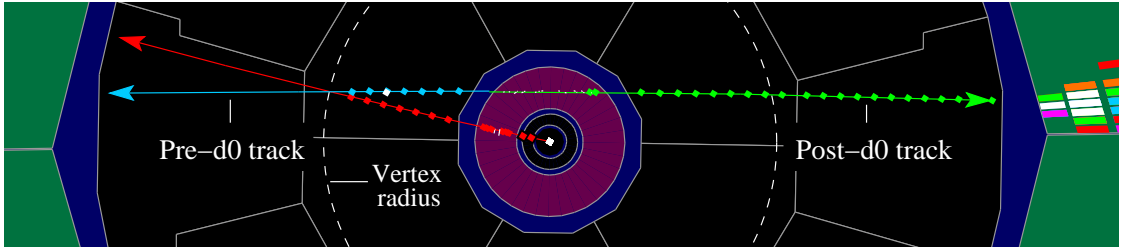}}
\caption{An example of a TPC-vertex with a decay angle greater than 90$^\circ$. \ladd{The lepton has formed a pre-\dzero\ and a post-\dzero\ track. The arrows show the track directions as assigned by JULIA.}}
\label{tpcver}
\end{figure}
\par
At this stage, the number of bad TPC-points for the track (see
Section~\ref{origin}) can be fixed. For pre-\dzero\ tracks it is set as the
number lying at least $6.5\cm$ (normal pad-row spacing plus $1\mm$)
below the radius of the vertex. For post-\dzero\ tracks it is the number
lying above the radius. The number of bad TPC points forms a track
variable which can be used as an indicator of the reliability of the
track, and is referred to in the selections described in the next
chapter.

\subsubsection{ITC-vertices}
As with TPC-vertices, the first step towards finding ITC-vertices is
to identify candidate slepton tracks. These are required to have no
TPC hits (by definition), a \dzero\ less than $1\cm$ (no constraint on \zzero\
due to the low ITC $z$-resolution) and a momentum greater than $1\GeV$.
\par
The high-\dzero\ track is then tested for a good ITC-vertex with each
candidate slepton track. Because of the low $z$-resolution of the ITC
a full vertex fit is not performed. It is merely required either that
the tracks' circular $r\phi$ projections cross in the $r\phi$ plane,
in which case the crossing point closest to the beam axis is taken as
the two-dimensional vertex, or that their distance of closest approach
in $r\phi$ is less than $5\mm$, in which case the vertex position is
the point of closest approach.
\par
The vertex radius is required to be no more than $5\mm$ below that of
the 4th ITC wire layer. The upper limit on the radius is determined by
a scan along the ITC track outwards into the TPC, looking for evidence
that the particle responsible was not following the track
trajectory. At successive pad-row radii a set of conditions similar to
those summarised in the TPC section of Table~\ref{misshit} are
evaluated, with the $z$ requirements relaxed\footnote{The separation
in $z$ for \iforig{proxmity}{proximity} to another track to be
considered a valid reason for missing a hit is set at 10\cm\ for both
ITC and TPC tracks, and no $z$ information at all is used in judging
the proximity of unassigned hits.}. The upper limit on the radius is
then set at the radius of the second point at which the track is
missing a hit without observable reason plus $5\mm$, unless the first
and second points are consecutive in which case the first is used. The
first point is in general not used since the track errors in the TPC
are much greater in the absence of assigned TPC hits, and so some
leeway is appropriate.
\par
The ITC track is required \iforig{to not}{not to} have any assigned
hits on wire-layers above the layer that is itself just above the
vertex. The constraints on the TPC hits of the high-\dzero\ track are
the same as for TPC-vertices, and so if it passes it is known whether
it is post-\dzero\ or pre-\dzero, and its number of bad TPC points is
fixed in the same way. If it is found to be post-\dzero\ with a
decay-angle less than 90$^\circ$ then it is further required to have
no ITC hits below the wire layer that itself is just below the vertex,
and no VDET hits at all.
\par
While TPC-vertices are very rare in background events, ITC-vertices
are less so because of the weaker constraints in the $z$ direction, and
because multiple scattering in the ITC outer wall and TPC inner wall
can cause the ITC and TPC hits of a particle to be reconstructed as
separate tracks, giving vertices very similar to those in signal. Some
rejection of these is attained by the requirement that the decay-angle
(corrected in the case of pre-\dzero\ tracks) is greater than
15$^\circ$.

\subsubsection{Pointing-hits}
The object here is to find a line of hits compatible with marking the
slepton's trajectory from the IP to the high-\dzero\ track. Although the
slepton's trajectory is helical, it should have sufficient momentum
that it can be considered straight over the distances involved. This
is done to keep matters simple, but a small curvature of the slepton's
path can be accommodated in the leeway used in the following
procedure.
\par
The two endpoints of the particle's point-of-origin range are used to
define the $\phi$-range in which the line of hits should lie. If the
track lies at sufficiently high $|\cos(\theta)|$ over the whole range,
then the straight line from the IP to the track at any point in the
range will not cross any tracking components and it is therefore
possible that the slepton did not produce any hits. The track is
accepted without further requirements in this case. If this is not the
case, the procedure continues to look for pointing-hits using
hypothesised slepton trajectories.
\par
Firstly it is determined if a slepton hit is required in the first
VDET layer. This is the case if the track radius is above the minimum
VDET layer-1 radius (VDET layers are not cylindrical and so do not sit
at a single radius), and the track $|\cos(\theta)|$ is below the
maximum VDET layer-1 $|\cos(\theta)|$ for the entire range. In this
case, unassigned layer-1 VDET hits are sought in the relevant $\phi$
range. If none are found the track is rejected. If one or more are
found, they dictate the possible trajectories for the parent
slepton. These are straight lines from the IP with $\phi$'s equal to
those of the VDET hits, and $\theta$'s such that they intersect the
track. If they also intersect the second VDET layer, an unassigned hit
is required here also, with a $\phi$ within $1.2^\circ$ of that of the
first VDET hit (\ie\ no more than $\sim2\mm$ from the slepton
trajectory given by the first hit). Trajectories for which a second
VDET hit is expected but not observed are rejected. If a slepton hit
in the VDET is only required for part of the track's point-of-origin
range, then VDET hits in the relevant $\phi$ range are still sought,
but the track is not rejected if none are found. For the part of the
range where slepton hits are not required (which may be all of the
range), hypothesis slepton trajectories are taken from the $\phi$'s of
the ITC wires in the first layer that lie in the relevant $\phi$
range. For each trajectory which crosses at least one ITC wire-layer,
ITC information is required to support the hypothesis that the slepton
followed this trajectory. This is tested by applying the conditions
summarised in the ITC section of Table~\ref{misshit} at each
wire-layer radius. If a slepton trajectory is found not to be valid in
this way, it is rejected. If the remaining number of slepton
trajectories is non-zero, the track is considered to have
pointing-hits.

\subsubsection{A note on pre-\dzero\ tracks}
It was mentioned in Section~\ref{origin} that the procedure for
calculating the point-of-origin range assumes the track is post-\dzero,
and is nonsensical for pre-\dzero\ tracks. However, the ordering in
which the procedures are applied in the computer code is not exactly
as is implied in this chapter. In particular, the procedure for
finding TPC-vertices precludes the possibility that a TPC-vertex could
be found which is incompatible with the relevant track's point of
origin range. So the first thing that is actually done is that a
TPC-vertex is sought. If one is found then the track is accepted
without the particle point-of-origin range ever being
calculated. While this makes sense from a pragmatic viewpoint since it
speeds up the analysis, it also solves the problem of pre-\dzero\
tracks. Since a pre-\dzero\ TPC track can only be formed if the
particle point-of-origin is at a radius greater than that of the 4th
TPC pad row (the particle will need to create at least 4 TPC hits
before the \dzero-point), it is reasonably safe to say that the
slepton must also have crossed 4 TPC pad rows, and so also have a
reconstructed TPC track, and so the track should have a
TPC-vertex. The unconditional acceptance of tracks with TPC-vertices
therefore means that pre-\dzero\ tracks are treated correctly.

\subsection{The track variable, \mother}
All tracks which are tested for compatibility with the hypothesis that
they originated from slepton decay using the procedures described in
this chapter, are assigned an integer value to indicate the level and
type of success. This value is the track variable, \mother.
\par
An \mother\ of 0 indicates that the track is not compatible with the
hypothesis, and such tracks are rejected as good high-\dzero\ tracks. A
track has an \mother\ of 5 if it has a TPC-vertex, an \mother\ of 4 if
it has an ITC-vertex, an \mother\ of 3 if it has pointing-hits, and an
\mother\ of 1 or 2 if the track is compatible with the hypothesis that 
the slepton did not produce any hits. A value of 2 is
given in the case that the track has sufficiently small \dzero\ that it is
possible that the slepton decayed before reaching a high enough radius
to produce a hit. A value of 1 is assigned when the slepton did reach
a high enough radius, but had a $|\cos(\theta)|$ high enough that it
still managed not to pass through any tracking components before
decay. This distinction is drawn since an \mother\ of 1 is the most
unlikely of the values under the signal hypothesis, but this relative
probability is not mirrored in the background, where detector effects,
cosmic muons and beam-gas events mean that an \mother\ of 1 is not so
improbable.
\par
Examples of tracks with \mother's from 2 to 5 are shown in
Figure~\ref{mother2to5}, and an example of a track with an \mother\ of
1 is shown in Figure~\ref{mother1}.

\begin{figure}[htbp]
\setlength{\ljlenb}{7cm}
\centerline{
\renewcommand{\arraystretch}{1}
\begin{tabular}{>{\PBS\centering}p{\ljlenb}>{\PBS\centering}p{\ljlenb}}
\setp{114mm}{117mm}{50mm}{1}
\resizebox{\ljlenb}{!}{\includegraphics*[\xa,\ya][\xb,\yb]{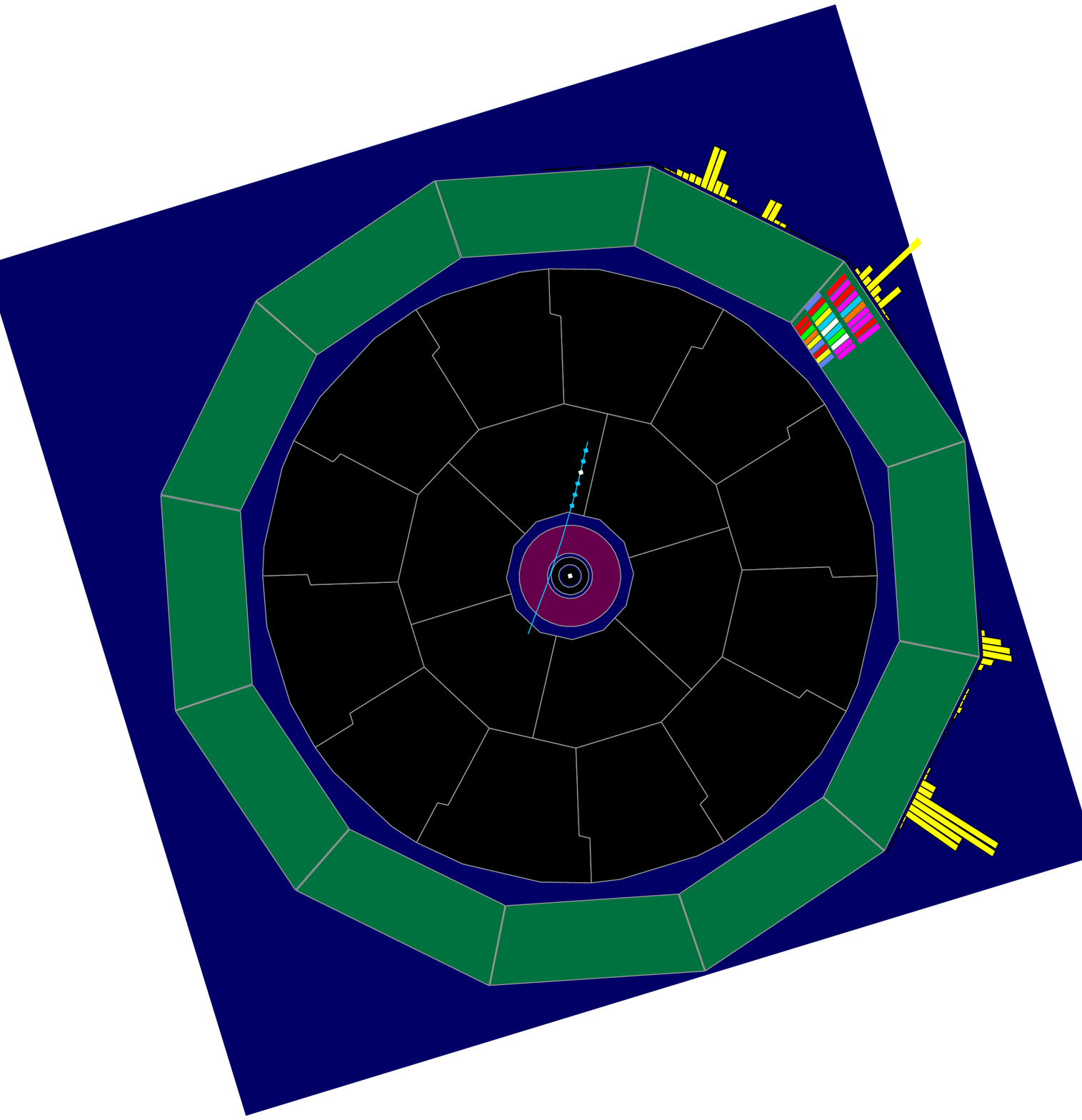}} &
\setp{74mm}{117mm}{80mm}{1}
\resizebox{\ljlenb}{!}{\includegraphics*[\xa,\ya][\xb,\yb]{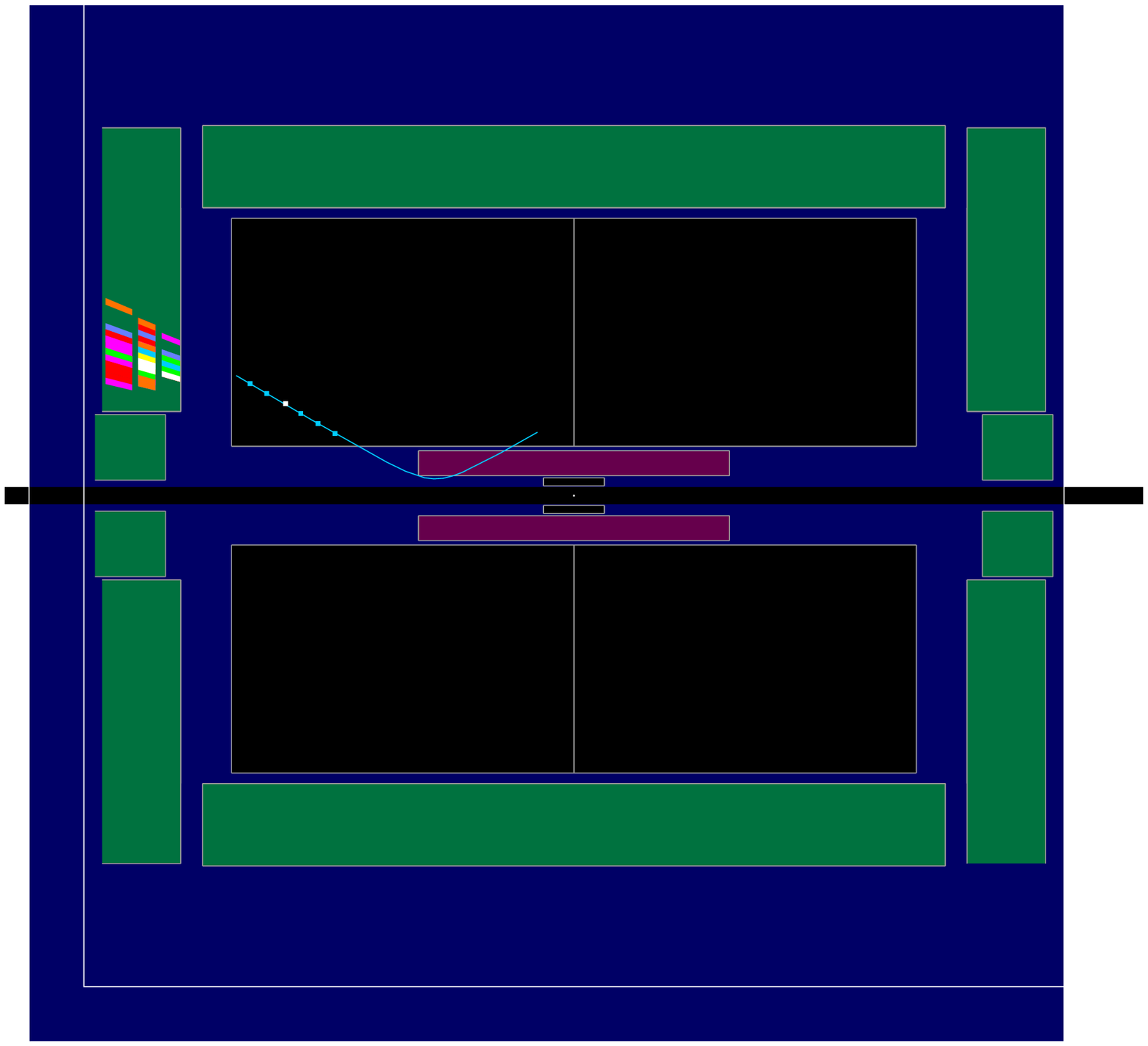}} \\
a). & b). \\
& \\
\end{tabular}
}
\caption{An example of a track with $\mother=1$ from signal Monte Carlo\ladd{, viewed in (a) the $r\phi$ plane and (b) the $\rho z$ plane ($\rho$ is equal in magnitude to $r$, but with a sign that makes it negative for a certain half of the $\phi$ range which is chosen to make the event as clear as possible). Although the \dzero\ of the track shows that the slepton reached a radius above at least one VDET layer before decaying, it did so at such a high $|\cos(\theta)|$ that it did not produce hits.}}
\label{mother1}
\end{figure}

\chapter{Selections}
\label{selections}

\newcommand{\andd}{\HS \HS}
\newcommand{\oandd}{\HS \textbf{and} \HS}
\newcommand{\orr}{\HS \textbf{or} \HS}

\newcommand{\gtt}{\!>\!}
\newcommand{\ltt}{\!<\!}
\newcommand{\nchnoef}{N_{ch}}
\newcommand{\nch}{N_{ch}'}
\newcommand{\nelefnsecel}{N_{e}}
\newcommand{\eprbbest}{\overrightarrow{P_{e}}}
\newcommand{\nchtpc}{N_{TPC}}
\newcommand{\motherbest}{\overrightarrow{\mother}}
\newcommand{\angmaxptx}{c_{2ch}}	
\newcommand{\ndz}{N_{good}}	
\newcommand{\sbtwo}{S_{2\beta}'}
\newcommand{\acopjet}{\Phi_{aco}'}
\newcommand{\ncheftpc}{N_{TPC}'}
\newcommand{\etot}{E_{tot}'}
\newcommand{\ebeam}{E_{beam}}
\newcommand{\mtot}{m_{tot}'}
\newcommand{\acol}{\alpha'}
\newcommand{\hhzmninety}{\alpha_{hemi-z}'}
\newcommand{\pt}{p_t'}
\newcommand{\pz}{p_z'}
\newcommand{\pleadbest}{\overrightarrow{p}}
\newcommand{\ytwothree}{y_{23}'}
\newcommand{\cmiss}{c_{miss}'}
\newcommand{\cthru}{c_{thrust}'}
\newcommand{\cthleadbest}{\overrightarrow{c}}
\newcommand{\etwelve}{E_{12}'}
\newcommand{\ncompsixty}{N_{60\GeV}}
\newcommand{\nmuefnsecmu}{N_\mu}
\newcommand{\pnoefvec}{p_{tot}^\pm}
\newcommand{\pnoefscl}{E_{tot}^\pm}
\newlength{\rse}
\setlength{\rse}{1.8ex}

This chapter describes the sets of cuts that are made on the variables
calculated from the full event information that form the final
discrimination between signal and background. There are two sets of
six selections. The six selections of a given set are aimed at
selecting with high efficiency events corresponding to each of the six
possible final states (\selsel, \smusmu, \staustau,
\selsmu, \selstau\ and \smustau), while minimising the acceptance for 
background events. The difference between the two sets is only that
the three most critical cuts are tuned under different conditions,
such that one set gives optimal sensitivity in the case of low slepton
decay-length, and the other in the case of high slepton
decay-length. As such the two sets of six selections will be called
the low and high decay-length optimisations. Both sets are applied
independently to the real data and the Monte Carlo data, and so there
are two sets of results described in Chapter~\ref{results}. More
details on the ``hows and whys'' of optimisation will be given in
Section~\ref{opt}. The exact differences between the optimisations
\iforig{is}{are} given in Section~\ref{trackcuts}.
\par
Each selection is split into track and global cuts. Track cuts are
cuts applied to variables that pertain to a given track within an
event and are used to select good high-\dzero\ tracks. Global cuts are cuts
on variables that pertain to the event as a whole, such as the total
energy or the number of charged tracks. The global cuts are quite
loose in that they cause little (typically $<5\%$) signal
rejection. The track cuts are more stringent and so the existence of
one or more good high-\dzero\ tracks is the most important property of a
signal event. The relative importance of track and global cuts and a
rationale for the general selection technique is discussed in
Section~\ref{additional}.

\section{Cuts common to all selections}
\subsubsection{Track cuts}
\label{trackcuts}
All selections require at least one good high-\dzero\ track ($\ndz>0$ in
the notation of Section~\ref{additional}). This section describes the
track cuts that define what is meant by `good'.
\par
The momentum, \dzero\ and $\chi^2_{BS}$ (the $\chi^2$ of the track fit to
the beam spot per degree of freedom) are the three variables whose
associated cuts are tuned, and so are required to be greater than
values which depend both on the channel and on the optimisation in
question. Table~\ref{cutvalues} shows these values. A track is allowed
to fail the \dzero\ cut if its \zzero\ is $>8\cm$. No other (global or track)
cut depends on the optimisation, and so Table~\ref{cutvalues}
completely summarises the difference between the low and high
decay-length optimisations.
\begin{table}
\centerline{
\renewcommand{\arraystretch}{1}
\begin{tabular}{|c|l|c|c|c|c|c|c|} \hline
\multicolumn{2}{|c|}{} &\selsel&\smusmu&\staustau&\selsmu&\selstau&\smustau \\ \hline
& low $d_{\tilde{l}}$  & 19.0 & 20.0 & 4.7 & 14.9 & 10.0 & 8.1 \\
\raisebox{1.5ex}[-1.5ex]{$p$ (GeV)}
& high $d_{\tilde{l}}$ & 19.8 & 18.6 & 6.1 & 20.0 & 8.5 & 4.3 \\ \hline
& low $d_{\tilde{l}}$  & 0.17 & 0.50 & 0.13 & 0.09 & 0.24 & 0.5 \\
\raisebox{1.5ex}[-1.5ex]{\dzero\ (cm)}
& high $d_{\tilde{l}}$ & 0.50 & 0.50 & 0.50 & 0.50 & 0.49 & 0.48 \\\hline
& low $d_{\tilde{l}}$  & 700 & 550 & 230 & 180 & 210 & 700 \\ 
\raisebox{1.5ex}[-1.5ex]{$\chi^2_{BS}$} 
& high $d_{\tilde{l}}$ & 680 & 690 & 700 & 700 & 590 & 610 \\ \hline
\end{tabular}
}
\caption{The cuts on track momentum ($p$), \dzero\ and $\chi^2_{BS}$ for each channel under each optimisation. \ladd{A track must have parameters greater than these values to be considered as a good high-\dzero\ track in the corresponding selection.}}
\label{cutvalues}
\end{table}

\par
The \mother\ of a track is required to be greater than 1, and it must
have at least 4 assigned TPC hits and less than 3 bad TPC points
(defined as a point where the track is missing a TPC hit for no
observable reason -- see Sections~\ref{origin} and \ref{mother}).
\par
For the purpose of eliminating photon conversions the track has a
variable called $r_{point}$. This is the radius at which the track's
circular $r\phi$ projection is pointing directly away from the
\iforig{$z$-axis}{origin}. If the track has more than one hit at a radius below this
then $r_{point}$ is set to zero. Tracks originating from photon
conversions should be pointing away from the \iforig{$z$-axis}{origin}
at the point where the conversion occurred, and should have no hits at
lower radii. Thus tracks with $185\cm>r_{point}>4.5\cm$ are
rejected. This is almost never satisfied by signal tracks, and is a
very effective cut against photon conversion tracks since it does not
rely on both the electron and the positron from the conversion being
reconstructed.
\par
To further reduce the background from nuclear interactions the
distance of closest approach in three dimensions of each track to its
nearest tagged nuclear interaction vertex (if there is one)
\iforig{are recorded and form}{is recorded and forms} the variable
$d_{NI}$. Tracks with $d_{NI}<5\cm$ are rejected. If the track is part
of a multi-track vertex that failed both the nuclear interaction and
\stautomultiprongtau criteria, then it is rejected if the $\chi^2$ of
the vertex is less than 100, and its maximum two-track angle is
greater than 20$^\circ$ or its kink angle is less than 20$^\circ$. The
latter two conditions reduce the probability of rejecting vertices
that are due to slepton decay but which have not been identified as
such.
\par
To reduce the background from cosmic rays comparisons between the
candidate tracks of an event are made. If two tracks do not have a
reconstructed slepton-lepton vertex with a kink angle in $r\phi$
greater than 90$^\circ$ at a radius greater than the 4th TPC pad row,
and the difference between their \dzero's is less than 10\% of their
average \dzero, then they are both rejected.
\par
Tracks with $\mother= 5$ are required to have a slepton-lepton vertex
radius less than that of the 20th TPC hit pad and a slepton momentum
greater than $2\GeV$ (to further reduce background from ECAL
splash-backs). \iforig{Plus}{Also,} the track-fit $\chi^2$ (per degree of freedom)
must be less than 4 (this is the fit of the helical track to the hits,
not of the two tracks to the TPC-vertex).
\par
Tighter cuts are made on tracks with $\mother< 5$. They are required
to have at least one VDET hit with $r\phi$ information or \dzero$>2\cm$ or
\zzero$>10\cm$. The track-fit $\chi^2$ is required to be less than 2.5. To
reduce the cosmic ray background the momentum is required to be less
then $150\GeV$. To reduce the background from low angle $\gamma\gamma$
events the $|\cos(\theta)|$ of the track is required to be less than
0.94, and if $\mother=4$ the $|\cos(\theta)|$ of the slepton-lepton
vertex is also required to be less than 0.94. \iforig{A cut is also
made on the minimum possible $\slep\rightarrow l $ decay angle in 2
and 3 dimensions, $\alpha_{min2}$ and $\alpha_{min3}$.}{Cuts are also
made on the minimum possible $\slep\rightarrow l$ decay angle in two
dimensions (the $r\phi$ projection), $\alpha_{min2}$, and in three
dimensions, $\alpha_{min3}$.} These are defined as the angle between
the track and the line joining the track to the IP at the track's
innermost hit location. It is required that $\alpha_{min2}> 1.5^\circ$
and $\alpha_{min3}> 5^\circ$.

\subsubsection{Global cuts}
\newcommand{\gt}{\,$>$\,}\newcommand{\lt}{\,$<$\,}
A certain degree of cosmic elimination is obtained by requiring that
the event is not `class 3'. Classes are groups of basic cuts that an
event is tested against after reconstruction. There are 25 in total
and all events satisfy at least one of them. They are intended to
speed up analyses by providing fast rejection of events that
\iforig{an}{a} user wishes to ignore. Class 3 is intended to flag
cosmic rays passing through the VDET. It requires exactly two tracks
with \dzero\lt10\cm,
\zzero\lt50\cm, $|\cos(\theta)|<0.95$, p\gt10\GeV\ and with $>$\,9 TPC
hits. It then requires that at least one of these must have a
\dzero\gt1\cm\ or a \zzero\gt2\cm, and that their directions differ by
$>179^\circ$ in $\phi$, $<1^\circ$ in $\theta$, and their \dzero's differ
by $<0.5\cm$.
\par
Further rejection of cosmic events is obtained by requiring
$t_0$\lt100\ns, and that either the \mother\ of the best track is
greater than 4 \textit{or} that the total charged momentum from the IP
is greater than $1\GeV$ \textit{or} the cosine of the angle between the
two highest momentum tracks that have TPC hits is $>-0.999$. If the
\mother\ of the best track is less than 4 the additional
constraint that the cosine of the angle between the two highest momentum
tracks (regardless of hits) is also $>-0.999$.
\par
Beam-gas events are rejected by requiring that the charged momentum
from the IP be non-zero or that the ratio of transverse to
longitudinal momentum from the IP be $>0.4$ (beam gas events are low
angle and not centred in the detector).

\section{Additional cuts for individual selections}
\label{additional}
In addition to the track cuts described in Section~\ref{trackcuts}, tracks are
also subject to an additional cut if they are to pass the \selsel,
\smusmu\ or \selsmu\ selections. For the \selsel\ selection only tracks 
with a non-zero probability of being an electron are accepted. The
probability is calculated during reconstruction using calorimeter
information. It is equal to zero if the track passes some very loose
muon identification criteria. For the \smusmu\ selection only tracks
with a muon probability (calculated during reconstruction using
calorimeter and muon chamber information) greater than 0.2 are
accepted. For the \selsmu\ selection a track must pass either of these
conditions. No such cuts can be used in the channels involving staus
since the high-\dzero\ tracks can be electrons, muons or pions.
\par
Events are also subject to additional selection-dependent global cuts.
The global variables that are used are described in
Table~\ref{vars}. Primed variables are calculated only using tracks
and calorimeter deposits selected by the energy flow algorithm. For
charged tracks this puts a cut of $2\cm$ on \dzero\ and $10\cm$ on
\zzero\ since it is only intended for physics at the IP. The phrase
`from the IP' in Table~\ref{vars} can be translated as `defined by
energy flow'. Variables with an $\overrightarrow{\mathrm{arrow}}$
above them pertain to the `best' track in the event. This is defined
as the track with the highest value of \mother. If there is more than
one track with a \mother\ of 5 (the maximum) then the best is the one
with the highest momentum. All energies and momenta are in units of
\GeV, all angles are in degrees.
\par
Several variables refer to `hemispheres' or `jets'. The event
hemispheres are formed by finding the thrust axis of the event. This
is the unit vector, $\mathbf{n}$, that minimises
$\sum\mathbf{p}.\mathbf{n}$, where the sum is over all energy flow
objects (charged tracks, photons and neutral hadrons) and $\mathbf{p}$
refers to the momentum 3-vector of a given object. Objects are then
sorted into the hemispheres depending on which side of the axis they
are closest to. The direction of a hemisphere is the direction of the
sum of the 3-vectors of all its constituent objects. The jets are
formed using the commonly used DURHAM algorithm\iforig{}{ \cite{durham}}.
The $y$-cut is chosen such that the event has exactly two jets, unless
the event has only one object in which case jets are not defined. The
direction of a jet is the direction of the sum of the 3-vectors of all
its constituent objects.
\setlength{\ljlen}{\topsep}
\setlength{\topsep}{0mm}
\vspace{5mm}
\renewcommand{\arraystretch}{1.1}
\setlength{\ljlenb}{\parskip}
\setlength{\parskip}{0mm}
\begin{spacing}{1}
\tablehead{\hline $Variable$ & Description \\ \hline}
\tabletail{\hline \multicolumn{2}{r}{\small continued on next page \hspace*{1cm}}\\ \hline}
\tablelasttail{\hline \label{vars}}
\bottomcaption{The variables used in the selections and their definitions.}
\begin{supertabular*}{\textwidth}{>$c<$p{5.1in}}
\nchnoef     & Total number of reconstructed charged tracks \\
\nch         & Number of reconstructed charged tracks from the IP \\
\nelefnsecel & Number of electrons \\
\eprbbest    & Probability that the best track is an electron \\
\motherbest  & The \mother\ of the best track \\
\angmaxptx   & The cosine of the angle between the two highest momentum tracks \\
\ndz         & The number of tracks passing the track cuts (must be $\ge1$) \\
\sbtwo       & $\sqrt{1-0.5(\beta_1^2+\beta_2^2)}$; where $\beta_1$ and 
$\beta_2$ are the boosts of the two hemispheres (one of which can be zero if 
the hemisphere contains no energy from the IP) \\
\acopjet     & The acoplanarity of the jets (the angle between their $r\phi$ 
projections; defaults to 999 if the event cannot be forced into 2 jets) \\
\nchtpc      & The number of tracks that have TPC hits (tracks that have been 
tagged as originating from a non-SUSY secondary vertex are not included) \\
\ncheftpc &  The number of tracks from the IP that have TPC hits \\
\hhzmninety  &  The minimum angle between the plane of the 
hemisphere directions and the z-axis; defaults to 999 if one or more
hemisphere contains no energy from the IP\\
\etot        & The total energy from the IP \\
\pleadbest   & The momentum of the best track \\
\pt          & The transverse momentum from the IP \\
\pz          & The longitudinal momentum from the IP (unsigned) \\
\ytwothree   & The DURHAM algorithm $y$ cut for which the event goes from 2 to 3 jets \\
\cmiss       &  The $|\cos(\theta)|$ of the momentum missing from the IP \\
\ebeam       &  The LEP beam energy (constantly updated for real data by the LEP Energy Working group using 15 minute chunks of data\iforig{}{, see \cite{LEPnrg}}) \\
\cthru       & The $|\cos(\theta)|$ of the thrust axis \\
\cthleadbest & The $|\cos(\theta)|$ of the best track \\
\etwelve     & The energy from the IP within 12 degrees of the beam axis (for Monte Carlo data this has \iforig{a}{an} extra amount added taken randomly from a distribution determined from random triggers in order to simulate beam noise) \\
\ncompsixty & The number of tracks with \dzero$<0.5\cm$ and \zzero$<3\cm$ and a \dEdx\ no less than 10 standard deviations below that which would be expected from a singly charged particle with a mass of $60\GeV$ (the approximate GMSB slepton mass limit) \\
\mtot      & The invariant mass at the IP \\
\acol      & The \iforig{acolinearity}{acollinearity} of\iforig{}{ the} hemispheres (the angle between them; defaults to 999 if one or more hemisphere contains no energy from the IP) \\
\nmuefnsecmu & The number of muons \\
\pnoefscl  & The total charged energy (tracks that have been tagged as originating from a non-SUSY secondary vertex are not included) \\
\pnoefvec  & The total charged momentum (tracks that have been tagged as originating from a non-SUSY secondary vertex are not included) \\
\end{supertabular*}
\end{spacing}
\setlength{\parskip}{\ljlenb}\vspace{\intextsep}
\par
It may seem best to develop global cuts which, like the track cuts,
tightly define the signal properties and so exclude background by the
fact that it is non-signal-like. For example,\iforig{}{ it may seem
sensible} to require in the
\selsel\ selection that $4\le\nchnoef\le6$ (since there should be four
lepton tracks with possibly two slepton tracks) and
$\nelefnsecel=4$. But such simple cuts, while intuitive, cause
significant losses in the signal efficiency. The reason for this is
twofold. Firstly it is a matter of how large a loss of efficiency is
considered significant. Because of the observable slepton lifetime the
signal is extremely distinctive (see Figure~\ref{event}), and so it
should be possible to gain close to 100\% efficiency with little
background as long as the decay-length is of the right size. Thus
while a cut which reduces the efficiency by (for example) 5\%
\iforig{maybe}{may be} considered perfectly acceptable in most SUSY
searches, here it is considered to have a significant detrimental
impact and alternative cuts would be sought. The other reason is that
the signal is simply not well defined by global variables. The slepton
lifetime makes the usually indispensable tool of energy flow of very
limited value, and interferes with the particle identification since
the lepton tracks can be foreshortened and can cross the TPC wires and
enter the calorimeters at unusually high angles to the radial
direction. To go back to the earlier example, the cut
$4\le\nchnoef\le6$ causes an average loss of efficiency in excess of
10\% over all the \selsel\ Monte Carlo samples generated at 208\GeV,
and just the cut $\nelefnsecel\ge2$ causes more than a 10\% loss in
nearly half the samples. Thus to obtain the high efficiencies that
must be achievable given the nature of the signal, the cuts on global
variables must be set loose.
\par
The strategy then was to focus the global cuts more on the properties
of the background rather than that of the signal, and to ensure that
they had virtually no effect on the signal efficiency for the whole
(\mc, \ms, \dl) signal parameter space. Of course, in the case of slepton 
decay-length comparable to the \dzero-resolution of the detector, some
cuts will have to have a severe impact on the efficiency. But these
cuts should be those related to the one, well-defined and principal
feature of the signal -- the slepton decay-length, and so should be
track cuts. Also, if the number of cuts with the potential to harm the
efficiency are kept low, they can all be tuned by an optimisation
process and the analysis as a whole will then be close to
optimal. These are the cuts on track momentum, \dzero\ and $\chi^2_{BS}$,
which have already been described.
\par
So the most important cuts in the analysis (those cuts which reject
the large majority of the background) are the track cuts, and the
global cuts serve only as a means of \iforig{futher}{further}
discrimination in the case that an event has at least one good
high-\dzero\ track.
\par
The additional global cuts are given in tabular form in
Sections~\ref{selsel} to \ref{smustau}. The third column in each table
indicates which Standard Model background the cut is intended to
remove. If a cut is quite general in that it is effective against many
backgrounds, then a `-' is used. The cuts are grouped into numbered
sets, where each set is intended to discriminate against the stated
background. Ideally, every cut would have \iforig{their}{its}
associated plots of the distributions of the associated variable for
signal and for background presented here, but given the number of cuts
this is impractical, and so plots will only\iforig{}{ be}
shown for a few selected cuts.

\subsection{Additional cuts for the \selsel\ selection}
\label{selsel}
The cuts are given in Table~\ref{tselsel}. The event is required to
pass cut 1, and is only required to pass cut 2 if $\ndz=1$ and
$\motherbest=2$ (\ie\ minimally signal-like -- only one track passing
the cuts with no evidence of the parent slepton). Cut 2 is an
anti-$\tau\tau$ cut that discriminates against the high energy and
(frequently) back-to-back nature of $\tau\tau$ events. It, or a slight
variant of it, is used in every selection, and the distributions of
the associated variables for the \selsel\ channel and $\tau\tau$
background are shown in Figure~\ref{tautau} on
page~\pageref{tautau}. $\tau\tau$ is a major background due to the
$\tau$ lifetime which gives it a decay length of $\sim 5\mm$ at
$\ebeam \sim 100\GeV$, and the fact that the multi-prong decays are
narrow which can lead to reconstruction difficulties as hits can be
incorrectly assigned amongst the tracks from a single decay.
\setlength{\ljlenc}{\tabcolsep}
\setlength{\tabcolsep}{4pt}
\renewcommand{\arraystretch}{1}
\begin{table}[H]
\begin{center}
\begin{tabular}{|c|>{\centering \HS}p{5in}<{\HS \hspace{0mm}}|c|} \hline
1 & $20\gtt \nchnoef\gtt2$ \andd \$\nch\ltt7$ \andd
\lcurl$\nelefnsecel\gtt0$ \orr $\eprbbest\gtt0.1999$ \orr $\motherbest\gtt4$\rcurl
& $qq$ \\ \hline
2 & $0.99\gtt\angmaxptx\gtt-0.999$ \andd \lcurl$\sbtwo\gtt0.1$ \orr $\acopjet\ltt174$\rcurl & $\tau\tau$ \\
\hline
\end{tabular}
\end{center}
\caption{Additional cuts for the \selsel\ selection}
\label{tselsel}
\end{table}
\begin{figure}[h]
\epsfig{figure=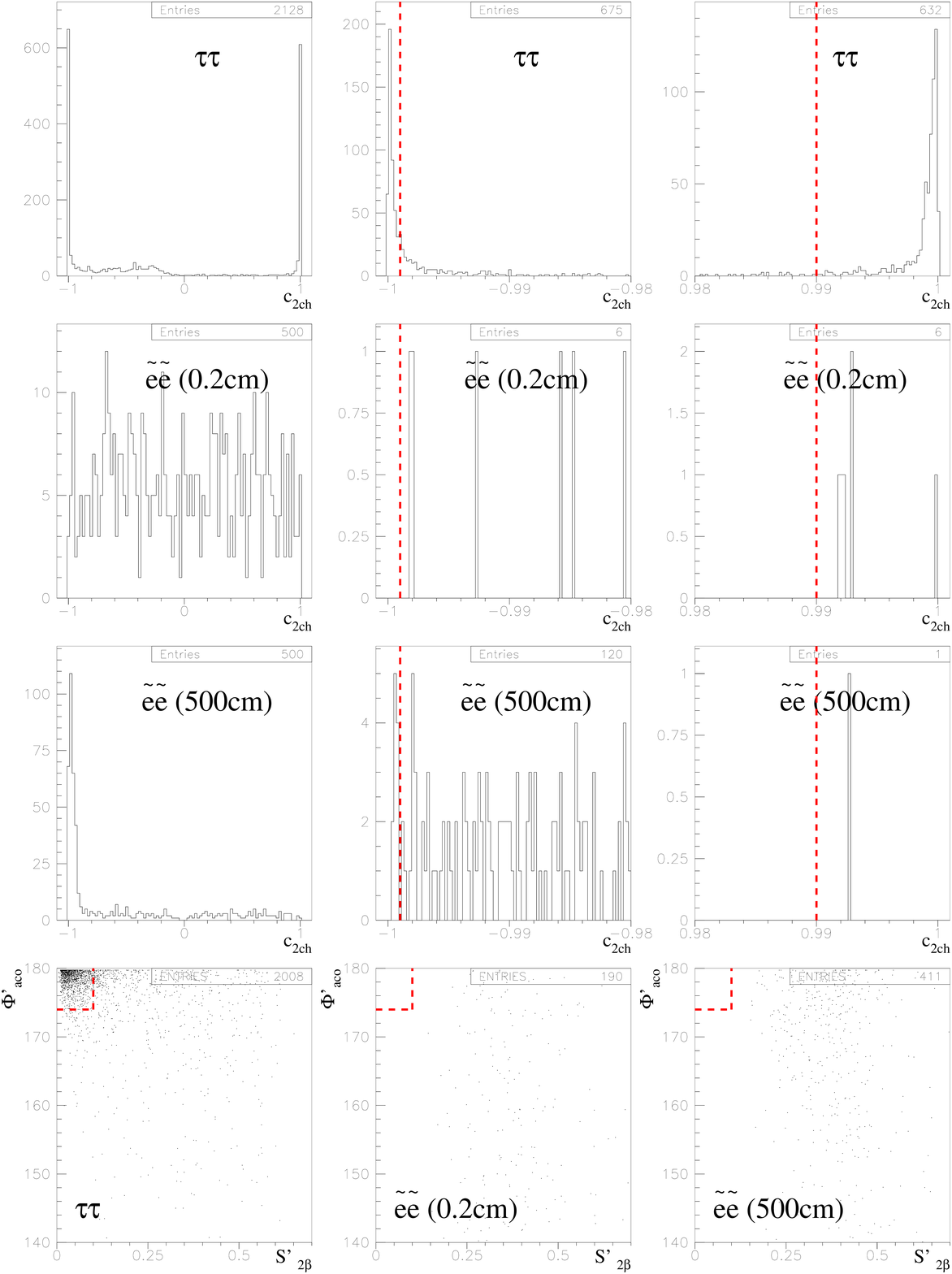,width=\textwidth}
\caption{The distributions of $\angmaxptx$, $\sbtwo$ and $\acopjet$ for $\tau\tau$ and \selsel\ Monte Carlo events at 189\GeV. \ladd{Top left is the full distribution of $\angmaxptx$ for $\tau\tau$ events. The two other plots on the same row show sub-ranges of the full distribution and the cut positions used in the anti-$\tau\tau$ cuts. The second and third rows show the same for \selsel\ events with decay-lengths of 0.2 and 500\cm\ respectively. The \protect\iforig{third}{fourth} row shows scatter plots of $\acopjet$ versus $\sbtwo$ for the same samples. The rectangle in the top left of each plot shows the anti-$\tau\tau$ cut on these variables (events inside the rectangle fail the cut).}}
\label{tautau}
\end{figure}

\subsection{Additional cuts for the \smusmu\ selection}
The cuts are given in Table~\ref{tsmusmu}. The event is required to
pass cut 1. As with the \selsel\ selection it is only required to pass
cut 2 if $\ndz=1$ and $\motherbest=2$.
\begin{table}[H]
\begin{center}
\begin{tabular}{|c|>{\centering \HS}p{5in}<{\HS \hspace{0mm}}|c|} \hline
1 & $20\gtt\nchnoef\gtt2$ \andd $\nch\ltt7$ & - \\ \hline
2 & $0.99\gtt\angmaxptx\gtt-0.999$ \andd \lcurl$\sbtwo\gtt0.1$ \orr $\acopjet\ltt174$\rcurl & $\tau\tau$ \\
\hline
\end{tabular}
\end{center}
\caption{Additional cuts for the \smusmu\ selection}
\label{tsmusmu}
\end{table}

\subsection{Additional cuts for the \staustau\ selection}
The \staustau\ selection is, as might be expected, the most complex of
the six. The high-\dzero\ tracks can be electrons, muons or pions and so no
additional cuts can be made on the track properties. There are also
difficulties with the global properties of the events. Whereas in a
channel without staus there are only two invisible particles, in a
\staustau\ event the neutrinos from tau decay mean that there can\iforig{}{ be}
up to ten, leading to a much greater spread in the visible energy, and
increased background from $\gamma\gamma$ events. The multi-prong
decays also push up the multiplicity of the events which make $qq$
events more difficult to remove.
\par
The cuts are given in Table~\ref{tstaustau}. The event is required to
\textit{fail} cut 1. This is because cut 1 is intended to pick out
$\gamma\gamma$ events from \staustau\ events rather than the other way
around. If $\motherbest=5$ then this is the only cut the event has to
pass, although this is very rare in background events. If
$\motherbest<5$ then the event also has to pass cuts 2 and 3 which are
simple cuts on energy, mass, multiplicity and
\iforig{acolinearity}{acollinearity}. If there is no
\stautomultiprongtau candidate in the event it is also required to
pass cuts 4 and 5, and to \textit{fail} cut 6 (another
pro-$\gamma\gamma$ cut). Under the hypothesis that the event is
$\gamma\gamma$ and that only the electron/positron travelling in the
direction of the missing momentum was deflected, then
$\pt/(\ebeam+\pz)$ is the $\tan(\theta)$ of that electron/positron
after deflection. Finally if $\motherbest<5$ and there is no
\stautomultiprongtau candidate and $\hhzmninety<6$ and $\ndz=1$ and
$\motherbest\le3$ then the event is also required to pass cut 7, which
is a variant of the anti-$\tau\tau$ cut used in the previous two
selections.
\par
Cut 4 is the most convoluted of the cuts. This is because it uses
variables whose distributions for signal and background overlap
significantly, but since $\nchtpc$ and $\hhzmninety$ are not
correlated with each other or with $\cthru$ and $\cmiss$ (although
$\cthru$ and $\cmiss$ are correlated), requiring events to pass only
one of the cuts on these variables means the overall cut has
significant impact on $qq$ events (approximately 90\% fail) while
leaving signal quite untouched (approximately 1\% loss in
efficiency). Figure~\ref{qqcut} on page~\pageref{qqcut} shows the
distributions of these variables and the cut values.
\begin{table}[H]
\begin{center}
\begin{tabular}{|c|>{\centering \HS}p{5.2in}<{\HS \hspace{0mm}}|c|} \hline
 & $\etot\ltt40$ \andd $\pt\ltt8$ \andd $\pz\ltt20$ \andd $\ndz=1$ \andd $\pleadbest\ltt7$ & \\ \cline{2-2}
\raisebox{\rse}{1} & $(\ytwothree/0.1)^2+(1-\cmiss)^2+(\etot/2\ebeam)^2\ltt0.4$ & \raisebox{\rse}{$\gamma\gamma$} \\ \hline
2 & $14\gtt\nchtpc\gtt2$ \andd $\ncheftpc\ltt11$ & - \\ \hline
3 & $(140\ebeam/94.3)\gtt\etot\gtt6$ \andd $\mtot\gtt7.7$ \andd $\acol\ltt178$ & $qq$ \\ \hline
4 & $\etwelve\ltt25$ \andd \lcurl$\nchtpc\ltt8$ \orr $\hhzmninety\gtt6$ \orr $(\cthru\ltt0.8$ \oandd $\cmiss\ltt0.8)$\rcurl & $qq$ \\ \hline
5 & $\pleadbest\gtt5$ \orr $\cmiss\ltt0.95$& - \\ \hline
 & $\pt/(\ebeam+\pz)\ltt0.05$ \andd $\ytwothree\ltt0.03$ & \\ \cline{2-2}
\raisebox{\rse}{6} &$\cmiss\gtt0.8$ \andd $\cthru\gtt0.8$ \andd $\ndz=1$ \andd $\cthleadbest\gtt0.8$ & \raisebox{\rse}{$\gamma\gamma$} \\ \hline
7 & \lcurl$0.99\gtt\angmaxptx\gtt-0.999$ \orr $\motherbest\gtt2$\rcurl \andd \lcurl$\sbtwo\gtt0.2$ \orr $\acopjet\ltt174$\rcurl & $\tau\tau$ \\ \hline
\end{tabular}
\end{center}
\caption{Additional cuts for the \staustau\ selection}
\label{tstaustau}
\end{table}
\begin{figure}
\epsfig{figure=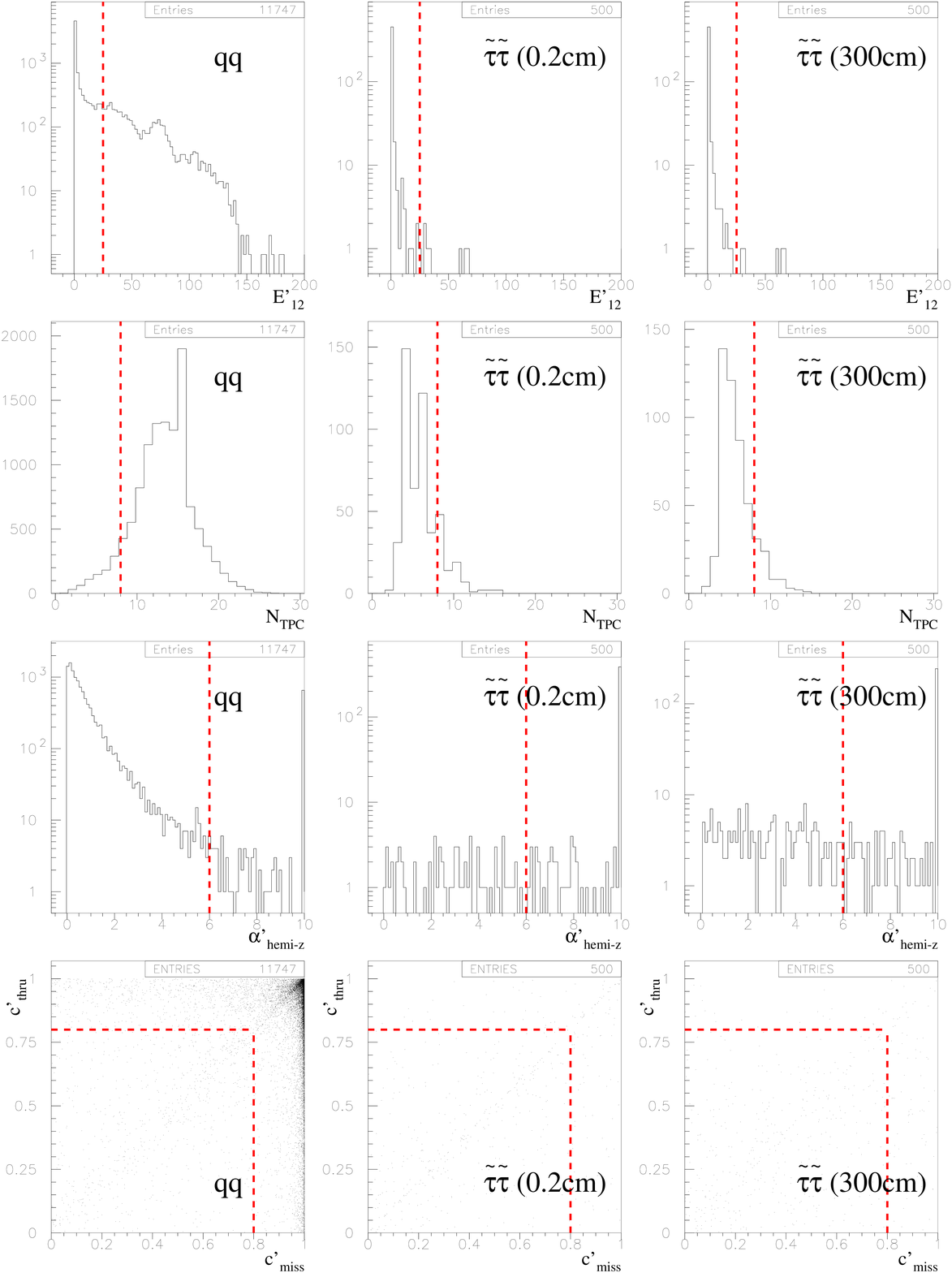,width=\textwidth}
\caption{The distributions of $\etwelve$, \protect\iforig{$\hhzmninety$, $\ncheftpc$}{$\nchtpc$, $\hhzmninety$} and of $\cthru$ versus $\cmiss$ for $qq$ and \staustau\ Monte Carlo events at 189\GeV. \ladd{These are the variables used in cut 4 of the \staustau\ selection. In the distributions of $\hhzmninety$, values in excess of 10 have been placed in the last bin to retain the overall normalisation. The plots on the left are for $qq$ events, the centre plots for \staustau\ events with a decay-length of 0.2\cm, and the right plots for \staustau\ events with a decay-length of 500\cm. The cut values used in the \staustau\ selection are shown as lines in the top three rows of histograms, and as rectangles in the bottom row (events outside the rectangle fail the cut).}}
\label{qqcut}
\end{figure}

\subsection{Additional cuts for the \selsmu\ selection}
The cuts are given in Table~\ref{tselsmu}. The event must pass cut 1
and, if $\ndz=1$ and $\motherbest=2$, cut 2 also.
\begin{table}[H]
\begin{center}
\begin{tabular}{|c|>{\centering \HS}p{5in}<{\HS \hspace{0mm}}|c|} \hline
1 & $20\gtt\nchnoef \gtt2$ \andd $\nch \ltt7$ & $qq$ \\ \hline
2 & $0.99\gtt\angmaxptx\gtt-0.999$ \andd \lcurl$\sbtwo \gtt0.1$ \orr $\acopjet \ltt174$ \orr $\hhzmninety\gtt6$\rcurl & $\tau\tau$ \\
\hline
\end{tabular}
\end{center}
\caption{Additional cuts for the \selsmu\ selection}
\label{tselsmu}
\end{table}

\subsection{Additional cuts for the \selstau\ selection}
The cuts are given in Table~\ref{tselstau}. The event is required to
pass cuts 1 and 2. Cut 3 is only applied if $\motherbest\le3$ since it
can be damaging to signal efficiency when the slepton decay length is
long. In particular, an upper limit on $\etot$ is not applied if it
looks like there is a fully reconstructed slepton in the detector
($\ncompsixty>0$ and $\nchtpc<11$) since the slepton's energy will be
included by energy flow and so there \iforig{maybe}{may be} no missing
energy. Cut 4 is the standard $\tau\tau$ cut and is only applied if
$\ndz=1$ and $\motherbest=2$.
\begin{table}[H]
\begin{center}
\begin{tabular}{|c|>{\centering \HS}p{5in}<{\HS \hspace{0mm}}|c|} \hline
1 & $\nch \ltt9$ \andd $\nchnoef \gtt2$ \andd $\ncheftpc \ltt10$ \andd $\nchtpc \ltt12$ & $qq$ \\ \hline
2 & $\pt/(\ebeam+\pz) \gtt0.02$ \andd \lcurl$\mtot \gtt20$ \orr $\pt/(\ebeam+\pz) \gtt0.1$\rcurl & $\gamma\gamma$ \\ \hline
  & $\etot\gtt6$ \andd $\mtot \gtt7.7$ \andd $\acol \ltt178$ \andd $\cmiss \ltt0.999$ & \\ \cline{2-2}
\raisebox{\rse}{3} & $\etot\ltt140\ebeam/94.3$ \orr \lcurl$\ncompsixty \gtt0$ \oandd $\nchtpc\ltt11\rcurl$ & \raisebox{\rse}{-} \\ \hline
4 & $0.99\gtt\angmaxptx\gtt-0.999$ \andd \lcurl$\sbtwo \gtt0.1$ \orr $\acopjet \ltt174$\rcurl & $\tau\tau$ \\ \hline
\end{tabular}
\end{center}
\caption{Additional cuts for the \selstau\ selection}
\label{tselstau}
\end{table}

\subsection{Additional cuts for the \smustau\ selection}
\label{smustau}
The cuts are given in Table~\ref{tsmustau}. The event is required to
pass cuts 1 and 2. Cuts 3 and 4 are the same as in the \selstau\
selection and are only applied under the same circumstances. Cut 5 is
an additional anti-$\gamma\gamma$ cut and is only applied if $\ndz=1$.
\begin{table}[H]
\begin{center}
\begin{tabular}{|c|>{\centering \HS}p{5in}<{\HS \hspace{0mm}}|c|} \hline
1 & $\nch \ltt9$ \andd $\nchnoef \gtt2$ \andd $\ncheftpc \ltt10$ \andd $\nchtpc \ltt12$ & $qq$ \\ \hline
2 & $\nmuefnsecmu \gtt0$ \orr $\pleadbest \gtt10$ & $\gamma\gamma$ \\ \hline
  & $\etot \gtt6$ \andd $\mtot \gtt7.7$ \andd $\acol \ltt178$ \andd $\cmiss \ltt0.999$ & \\ \cline{2-2}
\raisebox{\rse}{3} & $\etot \ltt140\ebeam/94.3$ \orr \lcurl$\ncompsixty \gtt0$ \oandd $\nchtpc \ltt11$\rcurl  & \raisebox{\rse}{$\tau\tau$} \\ \hline
4 & $0.99\gtt\angmaxptx\gtt-0.999$ \andd \lcurl$\sbtwo\gtt0.1$ \orr $\acopjet\ltt174$\rcurl & $\tau\tau$ \\ \hline
5 & $\pnoefscl \gtt10$ \andd \lcurl$\cthru\ltt0.8$ \orr $\pnoefscl \gtt30$ \orr $\pnoefvec \gtt25$\rcurl & $\gamma\gamma$ \\ \hline
\end{tabular}
\end{center}
\caption{Additional cuts for the \smustau\ selection}
\label{tsmustau}
\end{table}

\setlength{\topsep}{\ljlen}
\setlength{\tabcolsep}{\ljlenc}

\section{A note on the exclusivity of the selections}
It should be noted that the selections are not mutually
exclusive. That is to say, a single event can pass more than one of
them. In fact, there is no combination of selections which an event,
in principle, cannot satisfy. This is because the channels all give
very similar final states, and so reliably disentangling one from
another is practically impossible.
\par
But exclusivity is very desirable amongst the selections of a search
with more than one channel, since the results of the selections are
then statistically independent, making any subsequent combination far
simpler. Deliberate efforts could be made to ensure the selections are
exclusive, but the method would have to be quite arbitrary, and would
have a similarly arbitrary effect on the discovery/exclusion power of
the search.
\par
It would of course be possible simply to have a single selection aimed
at selecting signal events from any channel. But in order to encompass
all six channels without reproducing the same level of complexity of
six individual selections, an all-inclusive selection would
necessarily have to have a higher level of background acceptance.
\par
The final states of the channels of the signal process fall mid-way
between being different enough to allow exclusive selections, and
being similar enough to allow a single selection, without harming the
final result. Individual non-exclusive selections have been used then,
and the complexity that follows is dealt with in the next chapter.

\section{Optimisation}
\label{opt}
As has already been stated, the cuts on track momentum, \dzero\ and
$\chi^2_{BS}$ are the most critical of all the selection cuts. They
cannot be set loose to encompass all the signal for any decay-length
since they are very important in rejecting background, but if set too
tight they will have a severe detrimental impact on the signal
efficiency. They are therefore tuned to obtain the optimal cut values.
\par
The definition of the optimal cut value is that which gives the lowest
average expected 95\% confidence level upper limit on the signal
cross-section, $\overline{\sigma}_{95}$\iforig{. That is, }{, that is}
the lowest average 95\% confidence level cross-section limit that
would be obtained over a large number of experiments in which the
outcomes were as expected under the background-only hypothesis, as
dictated by background Monte Carlo studies (the real data is not
used). The methods for calculating the confidence level and
$\overline{\sigma}_{95}$ will be given in Sections~\ref{cl} and
\ref{xslim} respectively.
\par
Ideally $\overline{\sigma}_{95}$ should be obtained as a function of
all the cut values simultaneously.  But since there are eighteen of
these (three variables $\times$ six channels), and the method for
calculating just a single cross-section limit is quite complex and
CPU-intensive, this is impractical. Thus $\overline{\sigma}_{95}$ is
obtained as a function of each cut value individually, first in the
selection order \selsel, \smusmu, \staustau, \selsmu, \selstau,
\smustau\ for the cut on momentum, then the same for the cut on
$\chi_{BS}^2$, and then for the cut on \dzero. At each stage the cut
value that gave the smallest $\overline{\sigma}_{95}$ was identified
and the cut set there. Cuts waiting to be tuned were set at sensible
initial values.
\par
A $\overline{\sigma}_{95}$ can only be calculated though, for a given
signal hypothesis. That is, a given point in the (\mc, \ms, \dl)
parameter space and a given set of values for the
$\chi\rightarrow\sel,\smu,\Stau$ branching ratios
($B_{\sel,\smu,\Stau}$). Eight such signal hypotheses were used,
corresponding to all permutations of high and low decay-length, high
and low neutralino mass, and either slepton co-NLSP
($B_{\sel,\smu,\Stau}=\frac{1}{3}$) or stau-NLSP ($B_{\sel,\smu}=0$,
$B_{\Stau}=1$) branching ratios. The high and low decay-lengths used
were 80\cm\ and 0.2\cm, and the high and low neutralino masses were
95\GeV\ and 88\GeV.
\par
It was found however, that good sensitivity could be obtained under
all the signal hypotheses using just the two optimisations obtained
from the low decay-length, low neutralino mass, slepton co-NLSP
combination, and the high-decay length, high neutralino mass,
stau-NLSP combination. Since it was found that the optimised cut
values were most strongly influenced by the choice of slepton
decay-length, these optimisations were dubbed the low and high
decay-length optimisations respectively, and the other optimisations
were not used.

\chapter{Results and interpretation}
\label{results}

\newcommand{\eff}{\varepsilon}
\newcommand{\peff}{\epsilon}

\section{Outcome}
The number of background events expected to pass at least one
selection, calculated from applying the selections to the background
Monte Carlo samples, is shown in Table~\ref{result1} for both
optimisations (see previous chapter) along with the number of events
observed in real data. The data is consistent with the background
expectation, and so there is no evidence of supersymmetry. The numbers
are broken down into seven bins by LEP energy -- a re-binning with
respect to Table~\ref{lums}. Since Monte Carlo background studies were
only performed at energies of 189 and 208\GeV, the expected numbers of
background events for intermediate energies were gained by
re-normalising the values at these two energies to the correct
luminosity, and then linearly interpolating between them.
\newlength{\achar}
\settowidth{\achar}{X}
\par\newcommand{\err}[1]{\mbox{\scriptsize\,$\pm$\,{#1}}}
\begin{table}[hptb]
\centerline{
\renewcommand{\arraystretch}{1.2}
\iforig{ 
\begin{tabular}{|c|c|m|c|m|c|} \hline
Energy & \Lum         & \multicolumn{2}{c|}{Low \dl\ optimisation} & \multicolumn{2}{c|}{High \dl\ optimisation} \\ \cline{3-6}  
(GeV)  & ($\pb^{-1}$) & $No. expected$ & No. observed & $No. expected$ & No. observed \\ 
\hline
188.6 & 174.2 & 1.4\err{0.4} & 0 & 0.5\err{0.3} & 1 \\ 
191.6 & 28.9  & 0.2\err{0.1} & 2 & 0.1\err{0.0} & 1 \\ 
195.5 & 79.9  & 0.7\err{0.2} & 1 & 0.2\err{0.1} & 1 \\ 
199.5 & 87.1  & 0.7\err{0.2} & 0 & 0.2\err{0.1} & 0 \\ 
201.6 & 44.4  & 0.4\err{0.1} & 0 & 0.1\err{0.1} & 0 \\ 
205.0 & 79.9  & 0.7\err{0.2} & 0 & 0.2\err{0.1} & 0 \\ 
206.7 & 133.7 & 1.3\err{0.5} & 2 & 0.3\err{0.2} & 1 \\
\hline
\multicolumn{2}{|c|}{Total}    & 5.25 & 5 & 1.51 & 4 \\ \hline
\end{tabular}
}{        
\begin{tabular}{|c|c|>{\hspace*{2\achar}}r<{\,\hspace*{-\tabcolsep}}>{\hspace*{-\tabcolsep}\scriptsize$\pm$}l|c|>{\hspace*{\achar}}r<{\,\hspace*{-\tabcolsep}}>{\hspace*{-\tabcolsep}\scriptsize$\pm$}l|c|} \hline
Energy & \Lum         & \multicolumn{3}{c|}{Low \dl\ optimisation} & \multicolumn{3}{c|}{High \dl\ optimisation} \\ \cline{3-8}  
(GeV)  & ($\pb^{-1}$) & \multicolumn{2}{c|}{No. expected} & No. observed & \multicolumn{2}{c|}{No. expected} & No. observed \\ 
\hline
188.6 & 174.2 & 1.4&0.4 & 0 & 0.5&0.3  & 1 \\ 
191.6 & 28.9  & 0.2&0.06& 2 &0.08&0.04& 1 \\ 
195.5 & 79.9  & 0.7&0.2 & 1 & 0.2&0.1  & 1 \\ 
199.5 & 87.1  & 0.7&0.2 & 0 & 0.2&0.1  & 0 \\ 
201.6 & 44.4  & 0.4&0.1 & 0 & 0.1&0.06 & 0 \\ 
205.0 & 79.9  & 0.7&0.2 & 0 & 0.2&0.1  & 0 \\ 
206.7 & 133.7 & 1.2&0.5 & 2 & 0.3&0.2  & 1 \\
\hline
\multicolumn{2}{|c|}{Total}    & 5.2&1.4 & 5 & 1.5&0.8 & 4 \\ \hline
\end{tabular}
}}
\caption{A comparison of the expected number of events from background Monte Carlo studies with the number observed in data, broken down by LEP energy. \ladd{\Lum\ is the total integrated luminosity taken at each energy.\protect\iforig{}{ The quoted errors are purely statistical in origin.}}}
\label{result1}
\end{table}
Since the selections for each channel are not mutually exclusive, a
channel-by-channel breakdown of the number expected versus observed is
not appropriate. Rather, the numbers (expected and observed) are
broken down into those passing particular
\textit{sets} of selections. Since there are six selections there are
($2^6=$) 64 possible outcomes for the combination of selections that
any event passes, including the complete set (passing all selections)
and the null set (failing all selections, and therefore of no
interest). The selection sets are numbered from 0 to 63 according to
the convention
\[ \S_{0\rightarrow63}=32\,\delta_{\selsel} + 16\,\delta_{\smusmu} + 8\,\delta_{\staustau} + 4\,\delta_{\selsmu} + 2\,\delta_{\selstau} + \delta_{\smustau} \ ,\]
where $\delta_{chan}$ is 0 for events failing the respective channel's
selection, and 1 for events that pass. The expected versus observed
number of events is shown broken down by selection set in
Table~\ref{result2}. Only sets for which the expected background was
non-zero, or for which a data event was observed are listed.  No
errors are quoted on the expected numbers of events in this table
since, by breaking the numbers down by selection set, the events
passing the selections become so thinly spread that each figure is
typically based on no more than one event; and the procedure for
calculating the (asymmetric) error on a weighted sum of Poisson
variable estimators, when the estimators themselves are based on very
few events, is extremely complex. Even without errors though, it is
clear that the agreement for sets with a non-zero expected number of
background events is good. Two events under each optimisation are
observed however, with sets under which there were no events
expected. Given that, overall, the number of background Monte Carlo
events analysed was only around three times the number of events
expected in the data set however, this does not constitute an
incompatibility between real and background Monte Carlo data.



\begin{table}
\centerline{
\renewcommand{\arraystretch}{1.2}
\begin{tabular}{|c|c|c!{\vrule width 1pt}c|c|c|}
\hline
Selection & \multicolumn{2}{c!{\vrule width 1pt}}{Low $d_{\tilde{l}}$} & Selection & \multicolumn{2}{c|}{High $d_{\tilde{l}}$} \\
\cline{2-3}\cline{5-6}
set       & No. expected & No. observed    & set       & No. expected & No. observed \\
\hline
8  & 3.0 & 2 & 2  & 0.6 & 1 \\ 
2  & 0.6 & 0 & 5  & 0.4 & 0 \\ 
5  & 0.4 & 0 & 3  & 0.2 & 1 \\ 
3  & 0.2 & 0 & 11 & 0.2 & 0 \\ 
10 & 0.2 & 1 & 39 & 0.2 & 0 \\ 
12 & 0.2 & 0 & 8  & 0.0 & 1 \\ 
11 & 0.2 & 0 & 9  & 0.0 & 1 \\ \cline{4-6}
36 & 0.2 & 0 & \multicolumn{3}{c}{}\\ 
39 & 0.2 & 0 & \multicolumn{3}{c}{}\\ 
44 & 0.2 & 0 & \multicolumn{3}{c}{}\\ 
4  & 0.0 & 1 & \multicolumn{3}{c}{}\\ 
14 & 0.0 & 1 & \multicolumn{3}{c}{}\\ 
\cline{1-3}
\end{tabular}
}
\caption{A comparison of the number of expected events from background Monte Carlo studies with the number observed in data, broken down by selection set.}
\label{result2}
\end{table}

\section{Efficiencies}
\label{effs}
Figures~\ref{eff1} to \ref{eff3} show, for each channel, the
probability of a signal event passing its own channel's selection, and
the probability of it passing at least one selection no matter which,
as a function of decay length (estimated using the prescription
described in Section~\ref{signalmc}) for both the low and high decay
length optimisations. All plots are created from the Monte Carlo data
generated at \mbox{$\sqrt{s}=189\GeV$}.
\par
It can be seen that, in general, the selections other than that of the
channel's own selection add little to the total efficiency for a
channel (\ie\ if an event is not selected by its own channel's
selection, it is unlikely to be selected by another), as might be
expected.
\par
In each plot the different efficiency values at each decay length
correspond to the different points in (\mc, \ms) space at which Monte
Carlo data was generated (listed in Appendix~\ref{mc}). All the points
associated to the same (\mc, \ms) point are joined by a dotted line.
This is for guidance only, it does not represent an interpolation. The
dashed lines are formed by the straight line extrapolations from the
two lowest and two highest decay length points. More will be said
about these in Section~\ref{interp}. Note that there are many more
mass points per decay length for the \selsel\ channel since it was
used as a test case to determine how many points in mass space would
be needed for the subsequent Monte Carlo samples.
\par
The very weak explicit dependence of the efficiencies on the
neutralino and slepton mass is manifest. Most of the variation in the
efficiency between the mass points at a given decay length can be
accounted for by statistical fluctuation from a common value. In each
plot there is one obvious exception however, most clearly at high
decay length. In every plot the point with the highest efficiency at
the highest decay length is the point with the least initial
(\ie\ before slepton decay) phase space open to the
events. Specifically, for channels not involving staus it is the
\mbox{(\mc, \ms)} point
\mbox{(94\GeV, 93\GeV)}; for channels that do involve a stau it is the point
\mbox{(94\GeV, 91\GeV)}. In both cases this is only a 0.5\GeV\
difference between the beam energy and the neutralino mass, and
approximately a 1\GeV\ difference between the neutralino mass and the
mass of the slepton plus that of the (heavier) lepton. The reason for
the larger efficiency at high decay length for these points is that
their decay length is greatly overestimated. The lack of phase space
results in the slepton moving very slowly, so that its \dEdx\ is very
large. Thus the slepton is stopped by energy loss through ionisation
almost as soon as it encounters any matter, and so typically does not
penetrate the beam-pipe. Thus its decay length is effectively limited
to $\sim5\cm$. The efficiency does peak and then drop with decay
length in a not too dissimilar way to the other points, but the
drop-off at high decay length is not because the sleptons are
increasingly escaping the tracking volume before decay. It is because
in order to give such a slow-moving slepton such a large (estimated)
decay length, the slepton lifetime has to be set very large (0.2\us\
for a selectron at the stated mass point with $\dl=500\cm$). Once it
is stopped in the beam-pipe it waits out the remainder of its lifetime
before decaying. This can exceed the time period during which the ITC
is active ($\sim 1\us$), and so the resulting decay products (the
high-\dzero\ tracks) do not have ITC hits and so fail the quality
criteria described in the previous chapters. This is increasingly
likely with increasing estimated decay length and so causes the
associated drop in efficiency, but still results in better
efficiencies than the other points for the highest decay length.
\label{eff}
\setlength{\xa}{6cm}
\setlength{\xb}{4cm}
\setlength{\ljlen}{\tabcolsep}
\setlength{\tabcolsep}{0pt}
\begin{figure}
\centerline{\resizebox{\textwidth}{!}{
\renewcommand{\arraystretch}{1}
\begin{tabular}{cc}
\epsfig{file=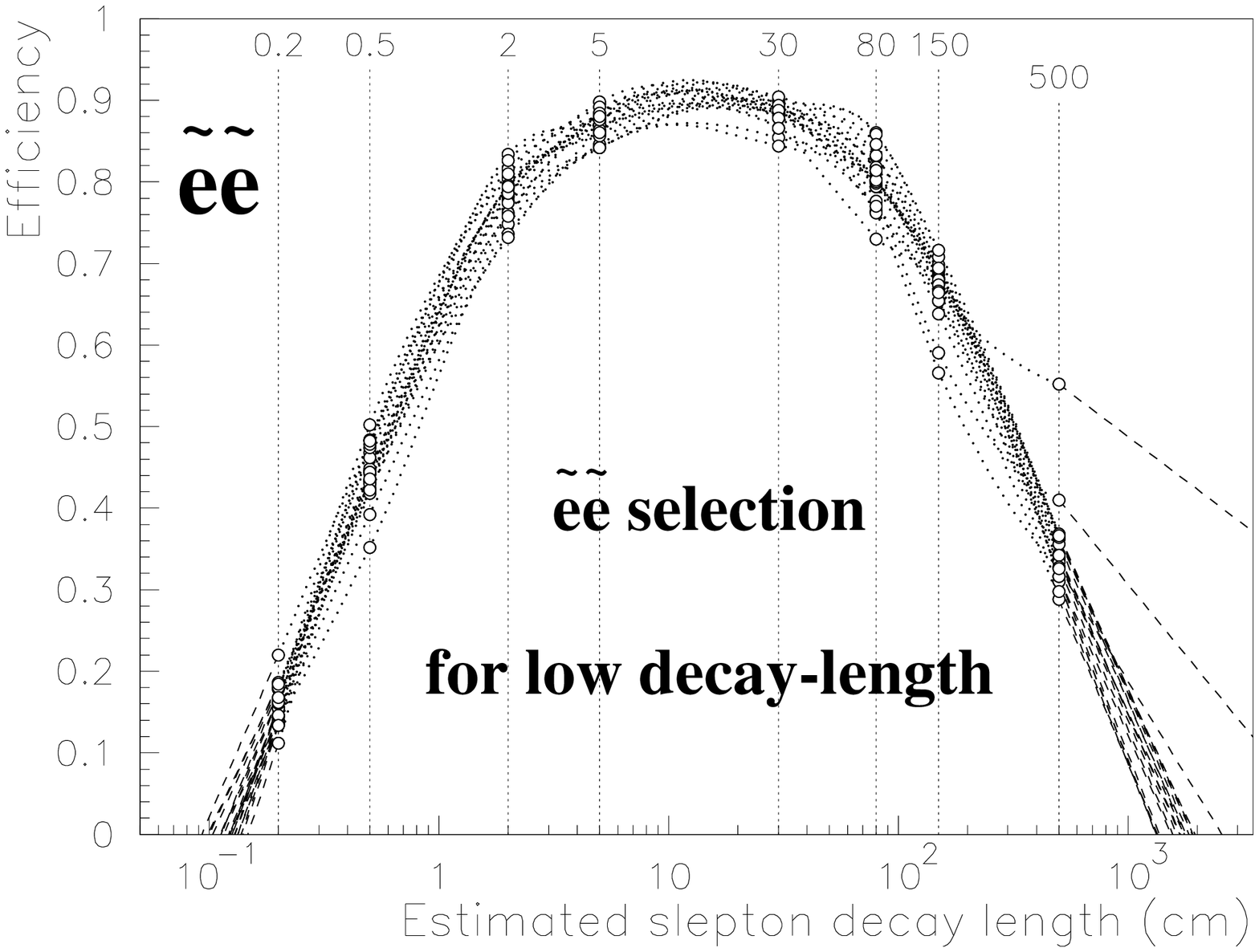,height=\xb,width=\xa} & \epsfig{file=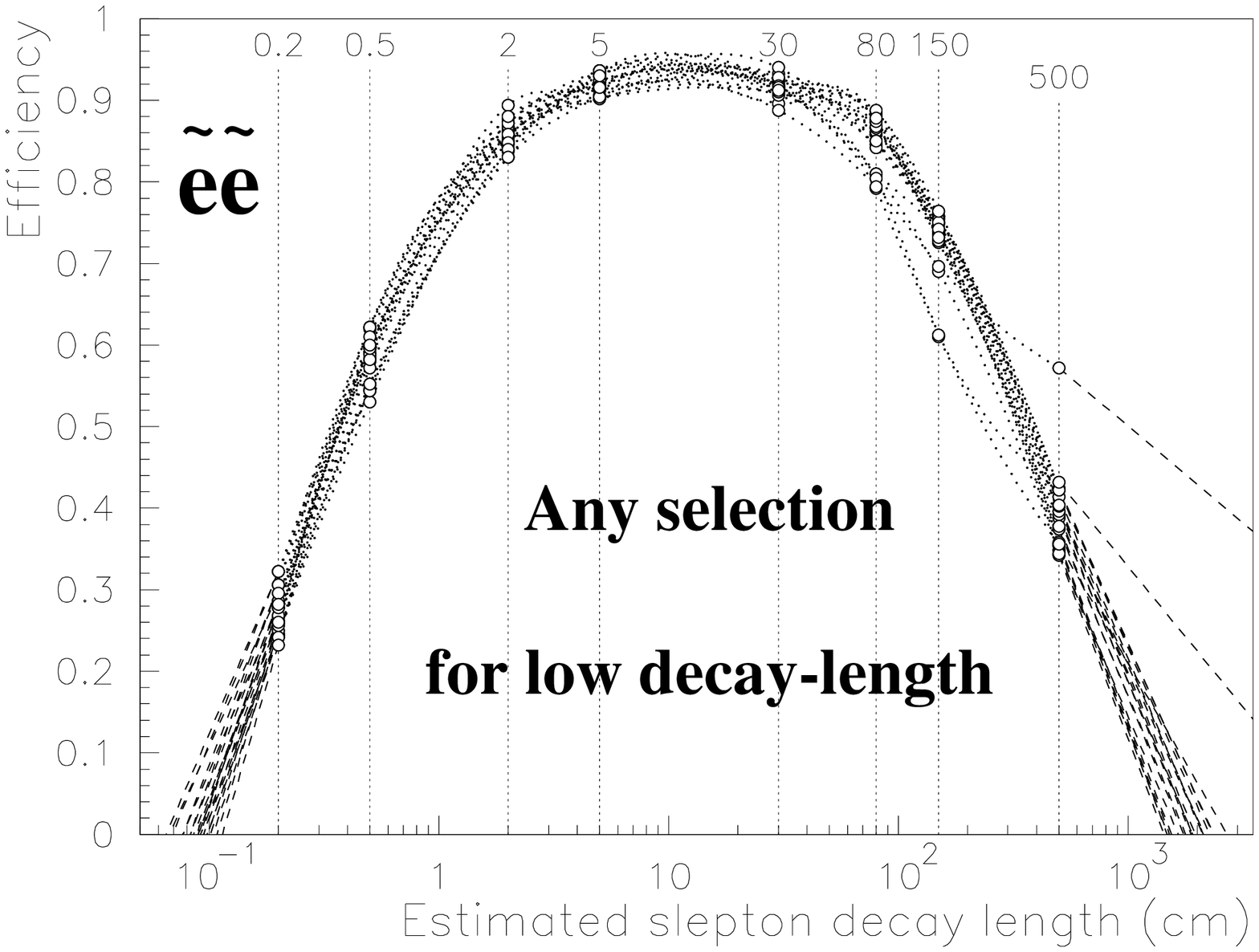,height=\xb,width=\xa} \\
\epsfig{file=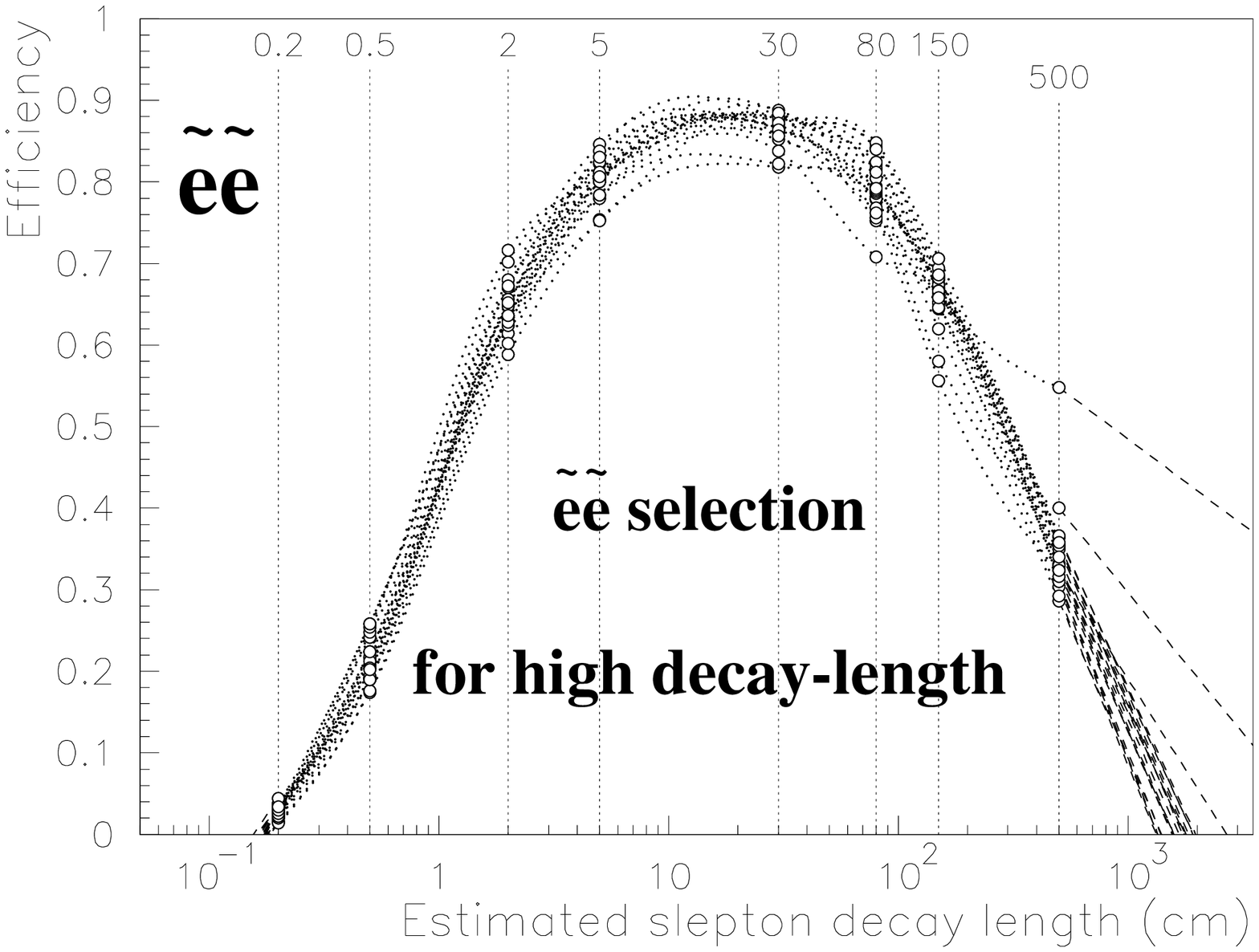,height=\xb,width=\xa} & \epsfig{file=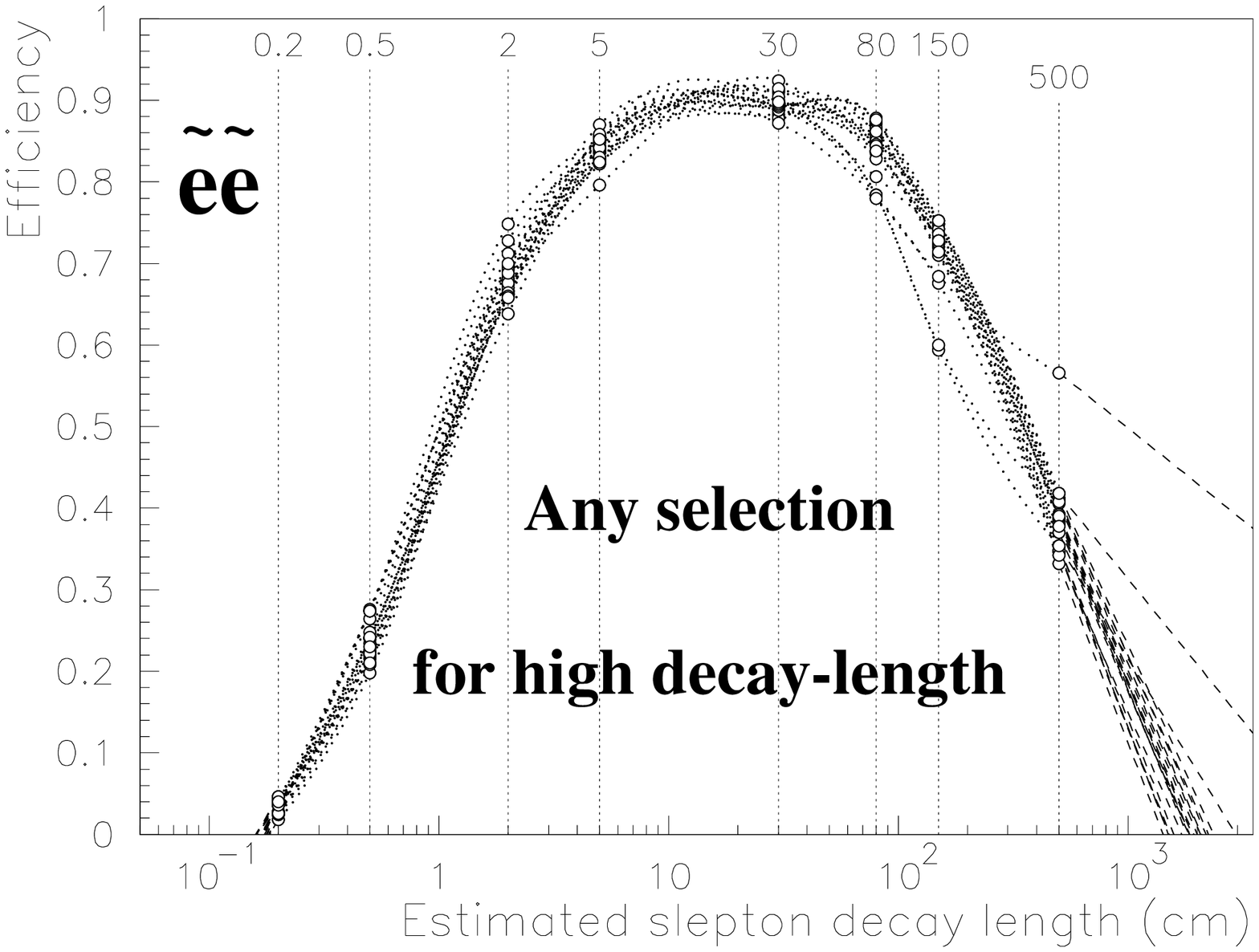,height=\xb,width=\xa} \\
\epsfig{file=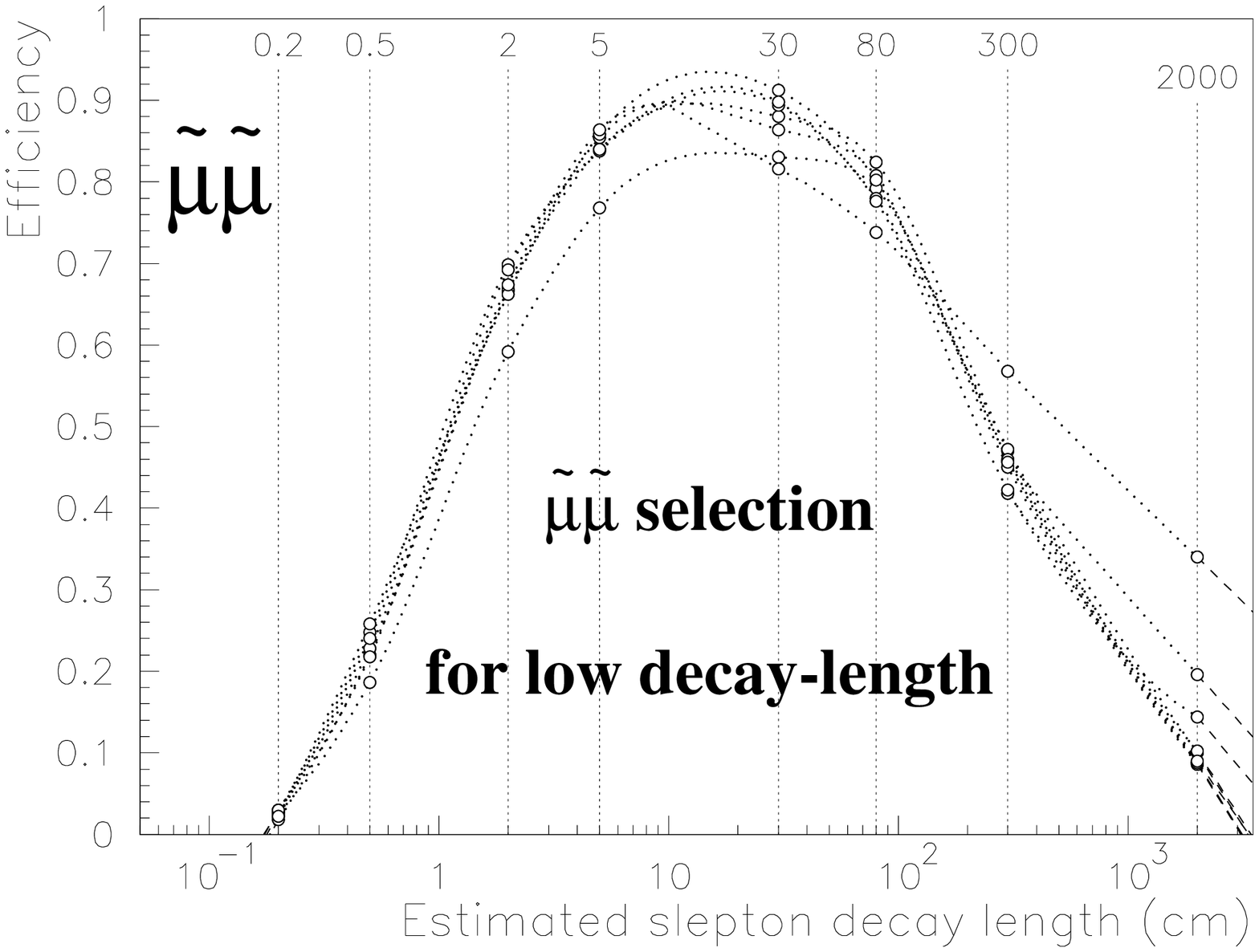,height=\xb,width=\xa} & \epsfig{file=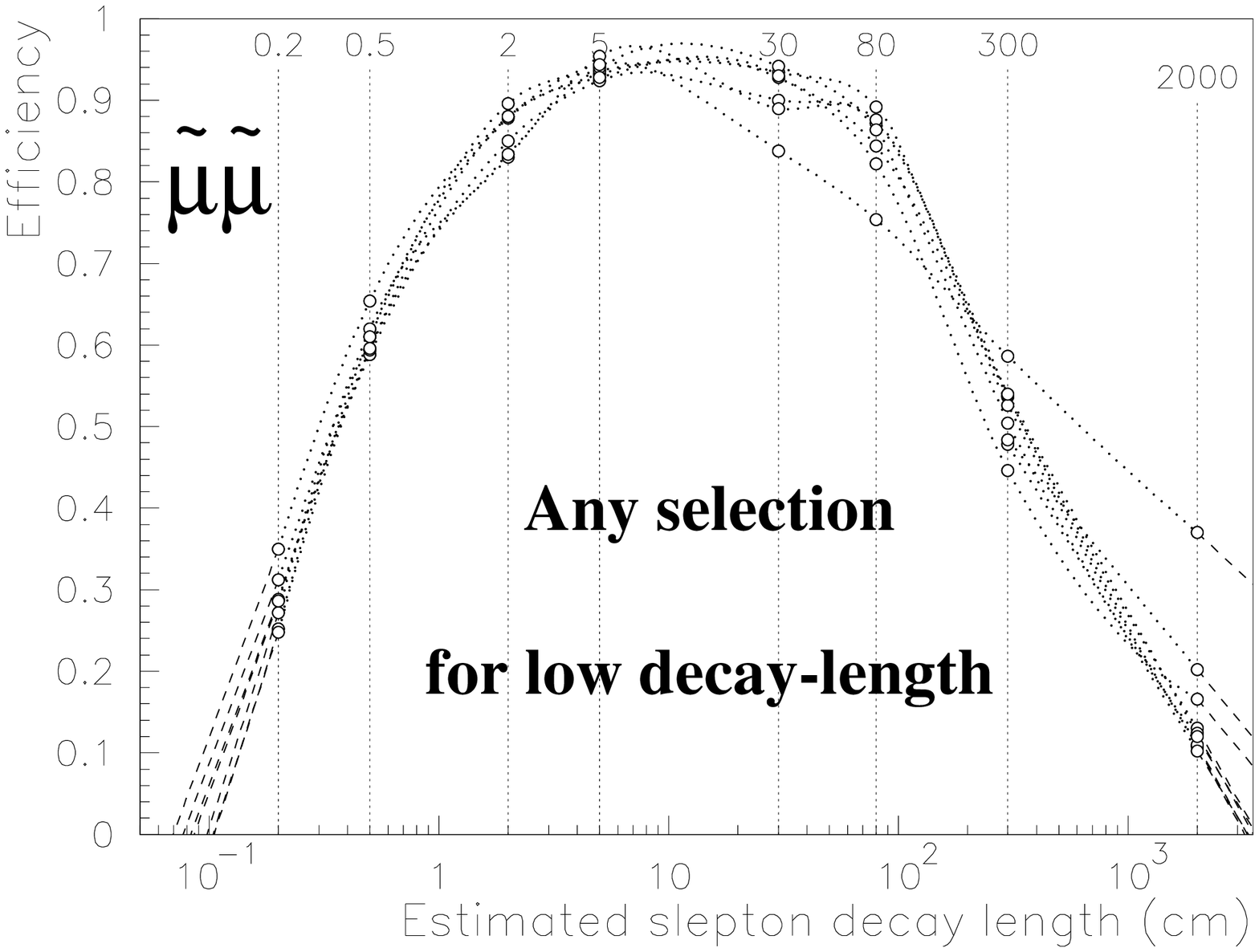,height=\xb,width=\xa} \\
\epsfig{file=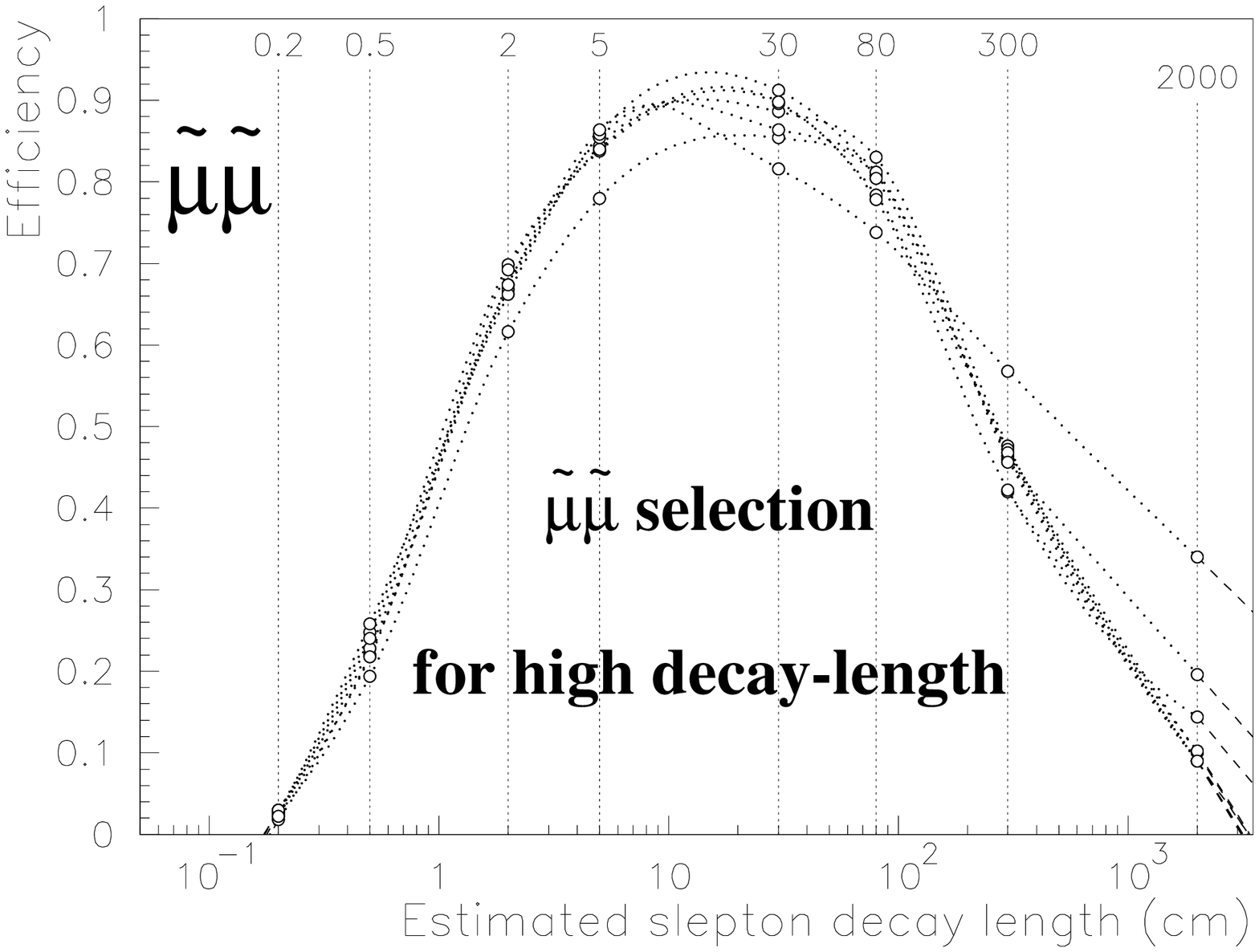,height=\xb,width=\xa} & \epsfig{file=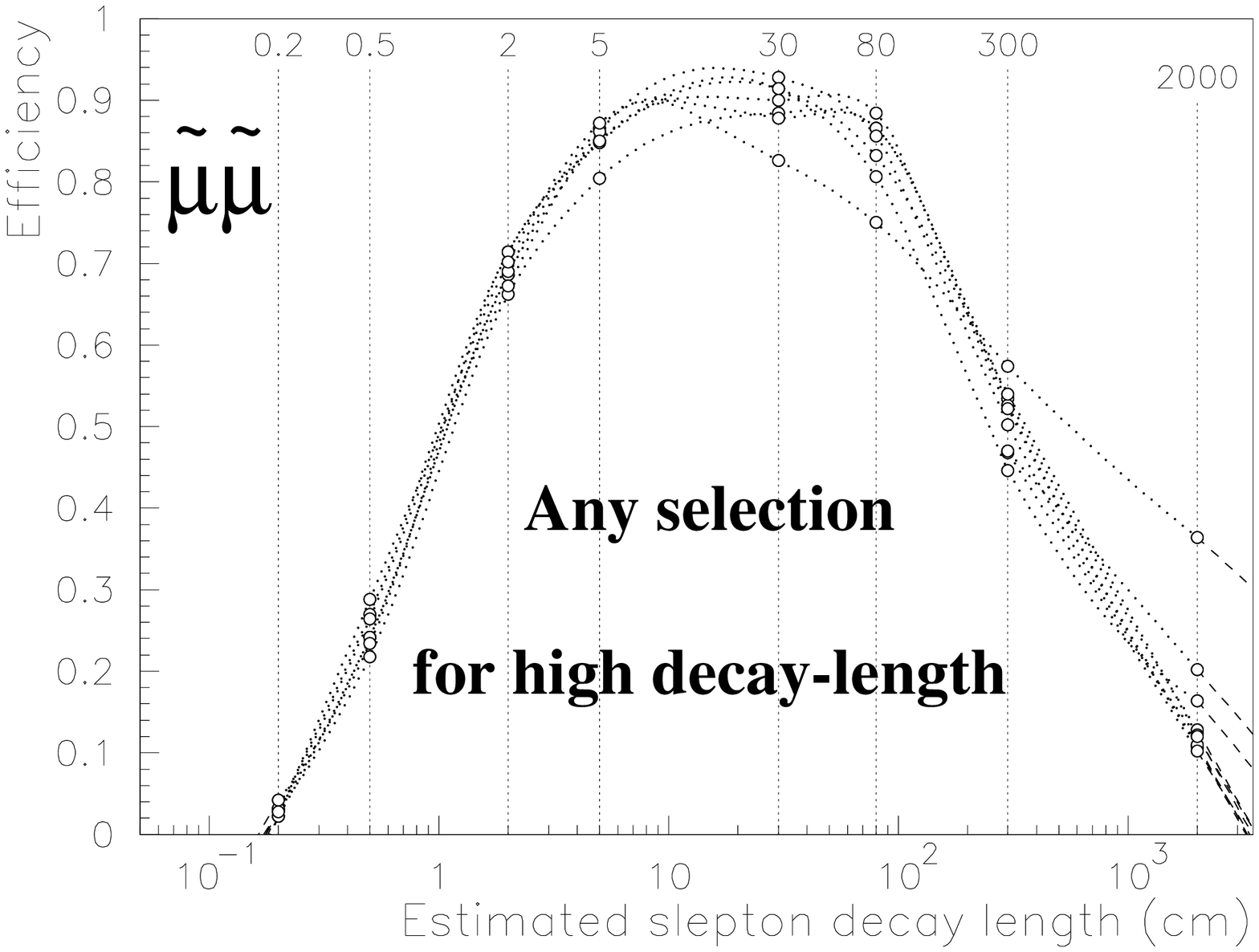,height=\xb,width=\xa} \\
\end{tabular}
}}
\caption{Efficiencies for the \selsel\ and \smusmu\ channels as a function of estimated slepton decay length. \ladd{The different curves correspond to different points in the (\mc, \ms) mass space.}}
\label{eff1}
\end{figure}
\begin{figure}
\centerline{\resizebox{\textwidth}{!}{
\renewcommand{\arraystretch}{1}
\begin{tabular}{cc}
\epsfig{file=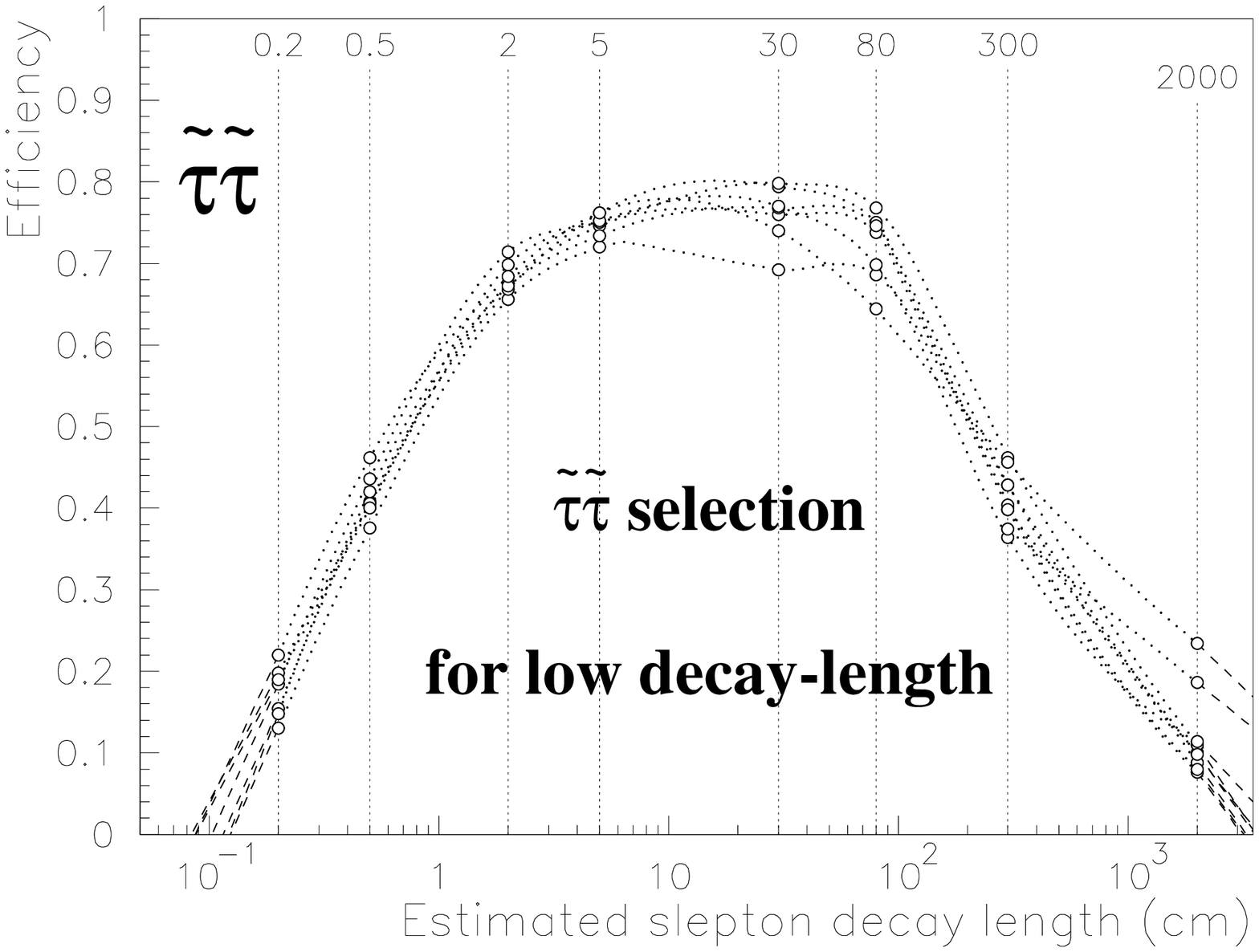,height=\xb,width=\xa} & \epsfig{file=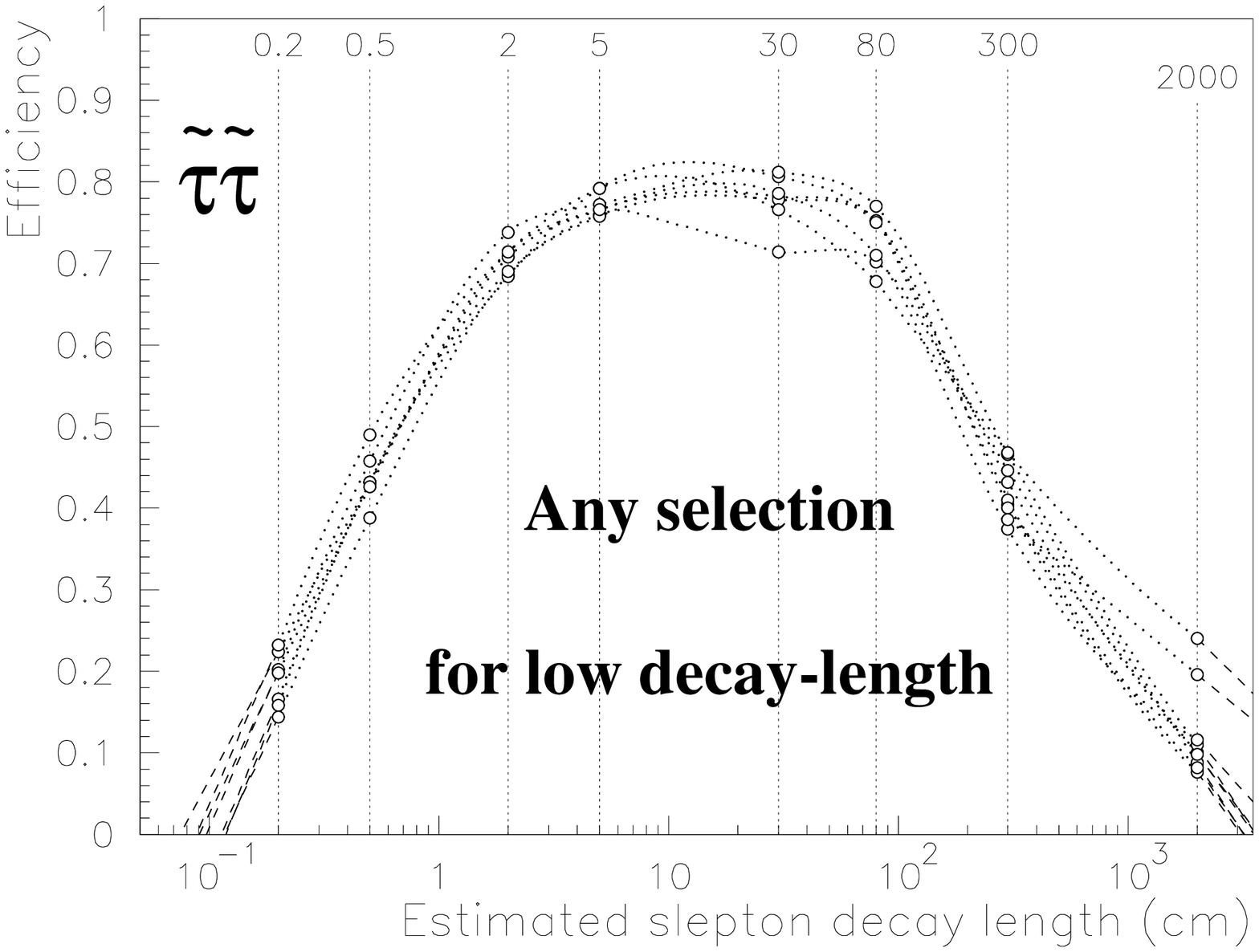,height=\xb,width=\xa} \\
\epsfig{file=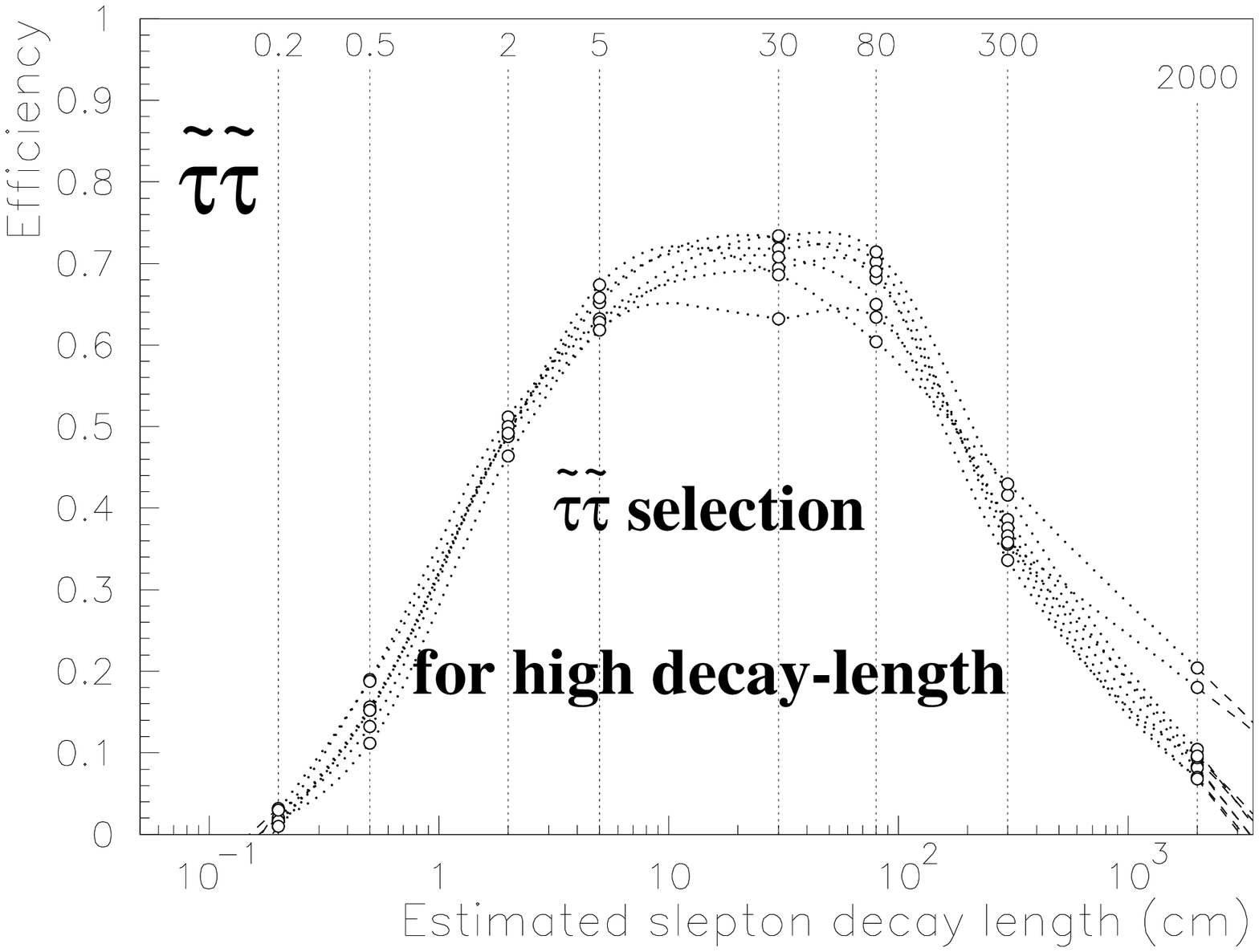,height=\xb,width=\xa} & \epsfig{file=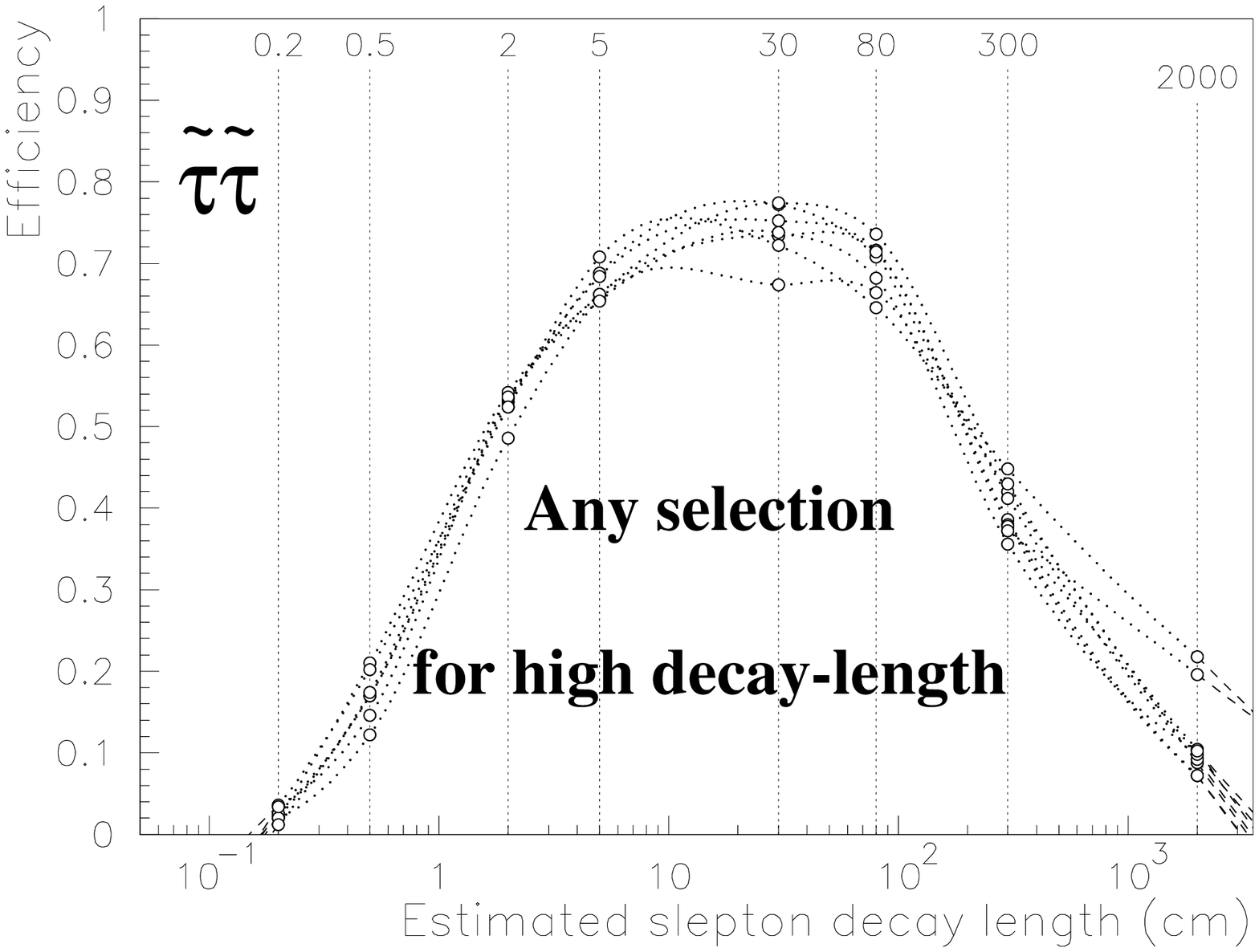,height=\xb,width=\xa} \\
\epsfig{file=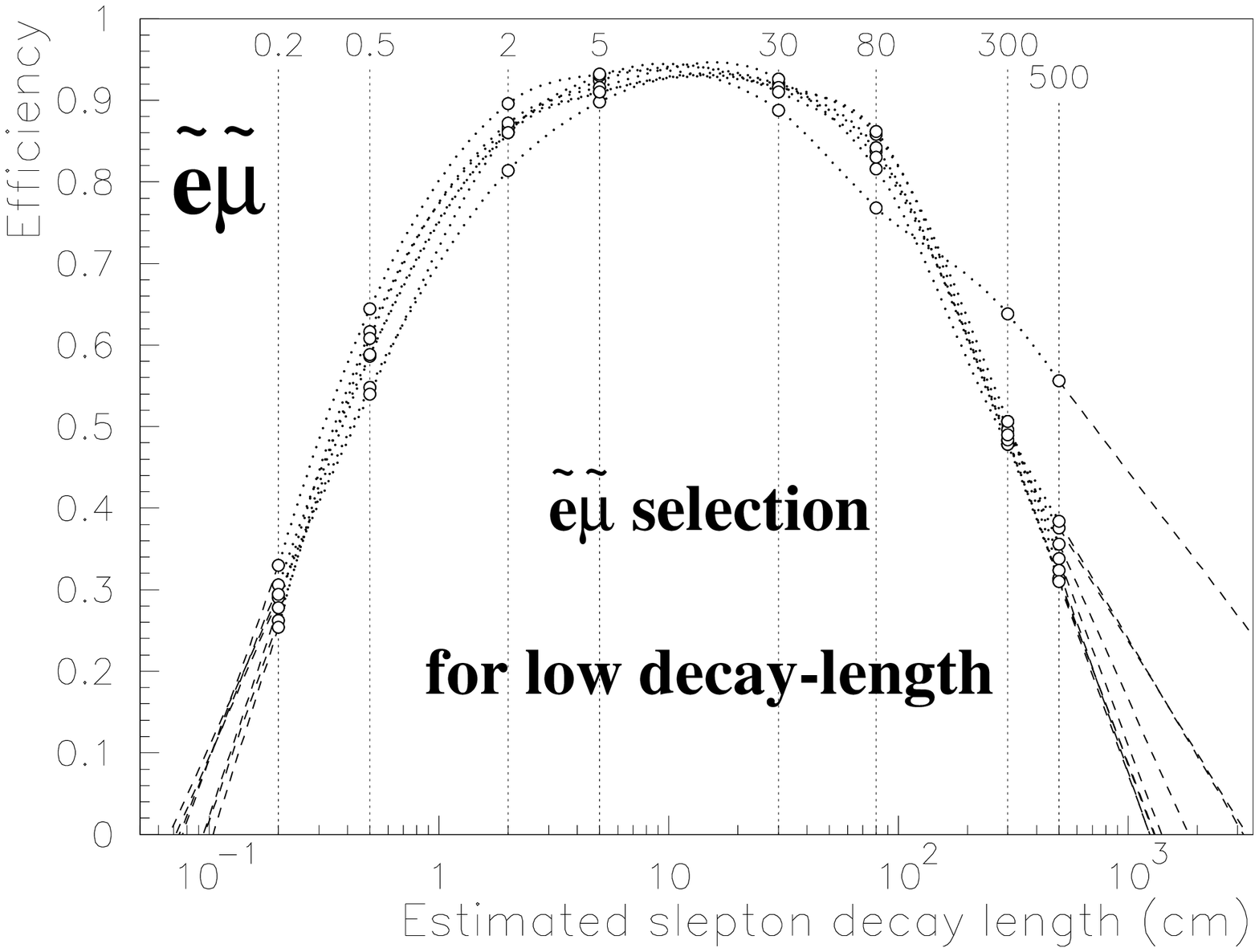,height=\xb,width=\xa} & \epsfig{file=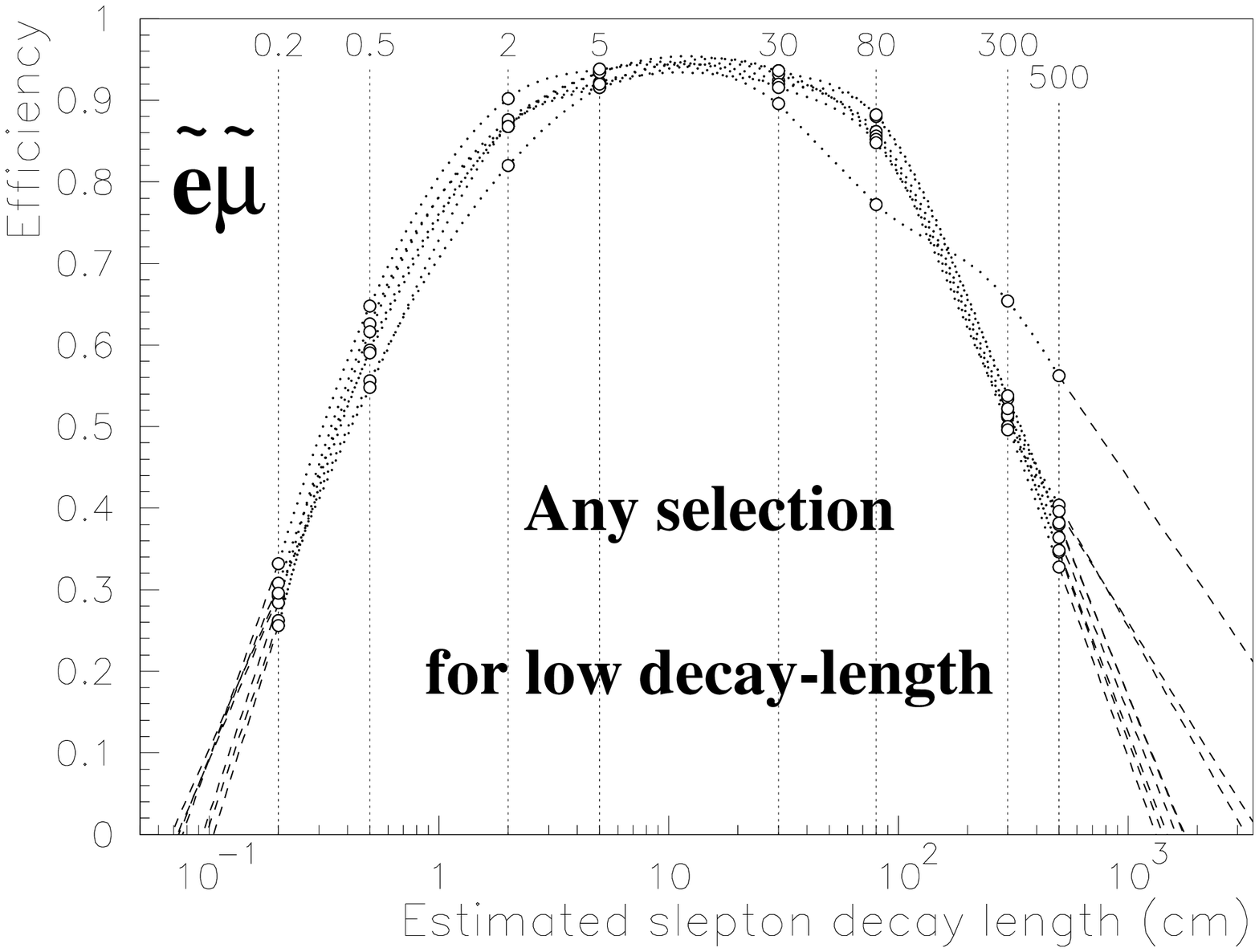,height=\xb,width=\xa} \\
\epsfig{file=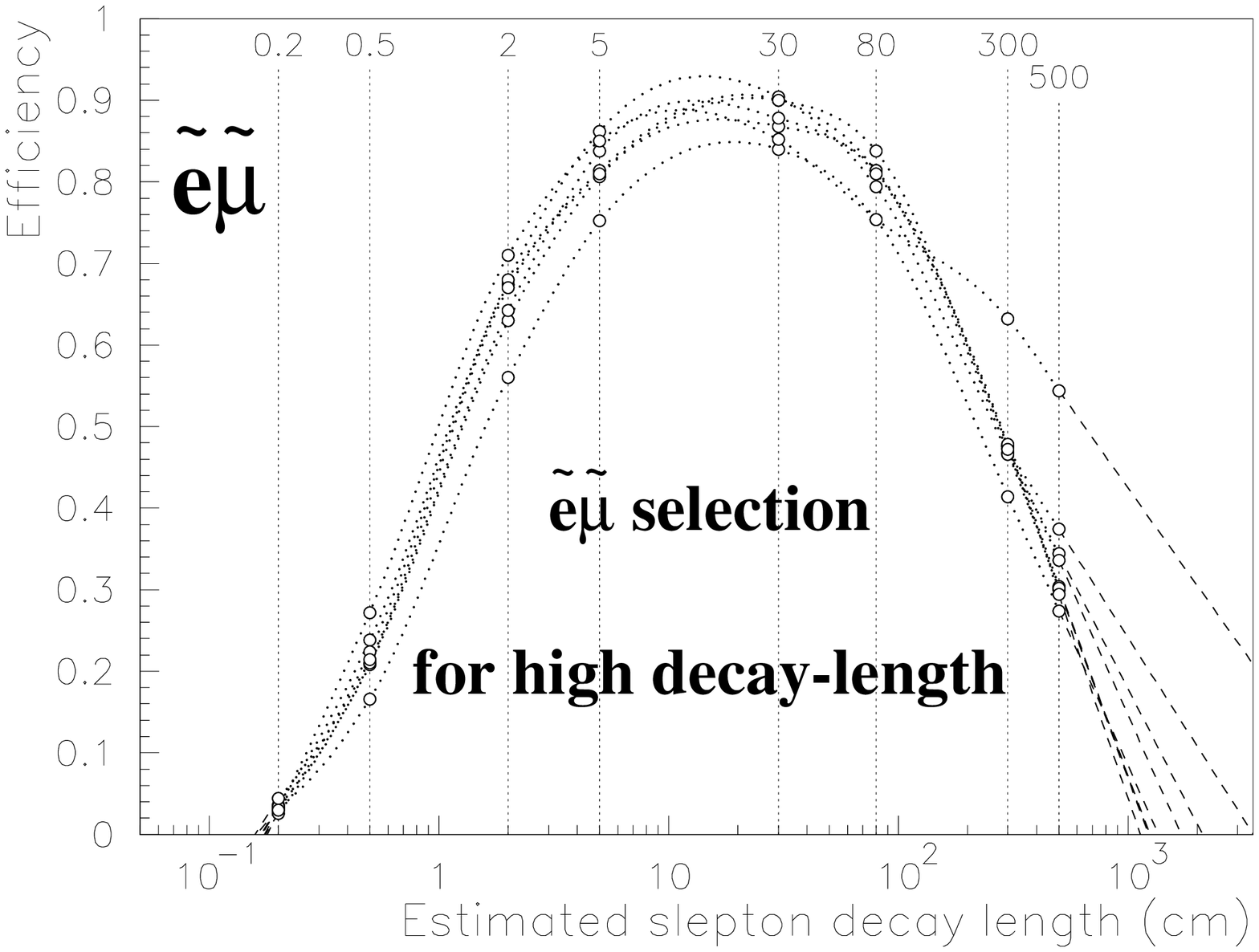,height=\xb,width=\xa} & \epsfig{file=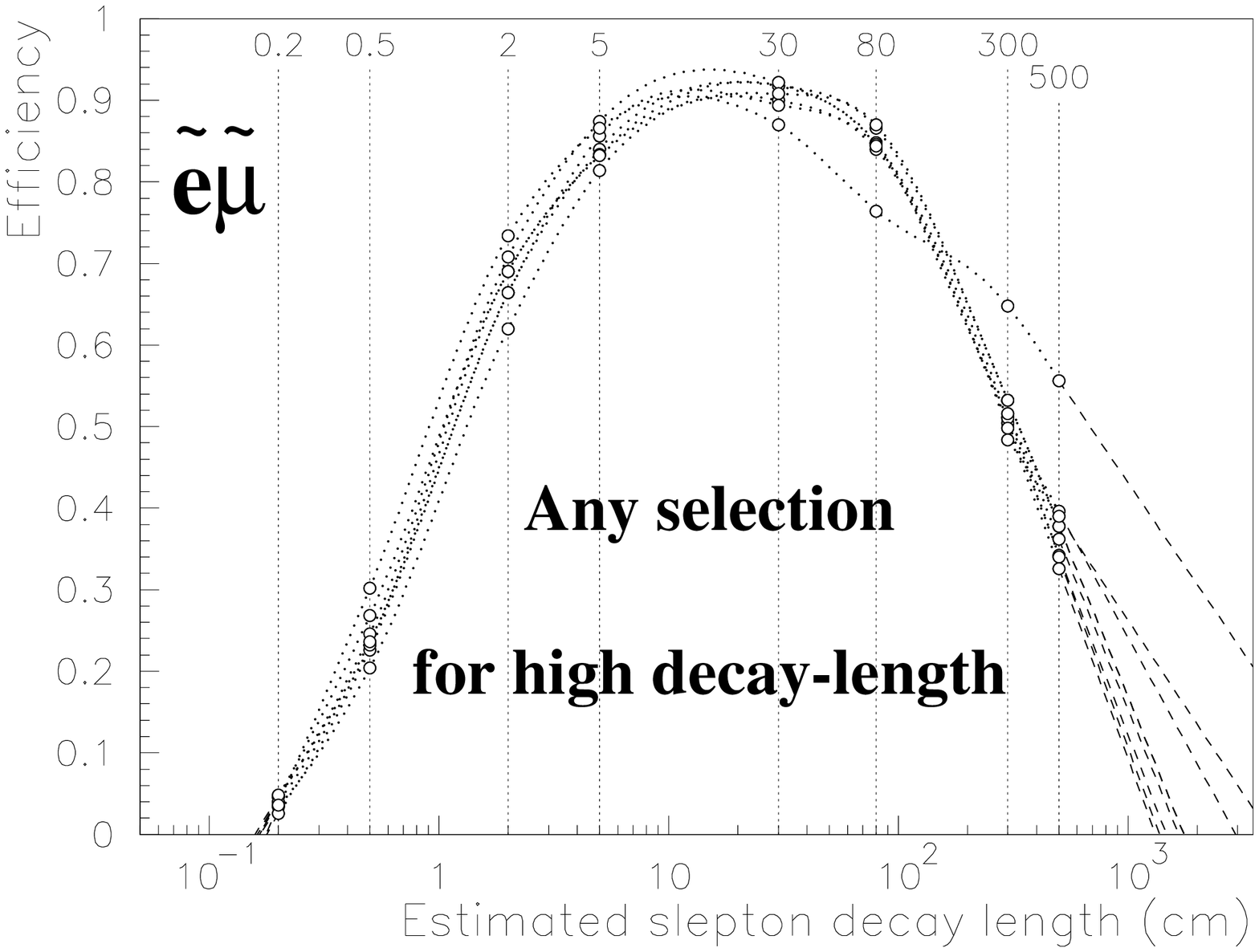,height=\xb,width=\xa} \\
\end{tabular}
}}
\caption{Efficiencies for the \staustau\ and \selsmu\ channels as a function of estimated slepton decay length. \ladd{The different curves correspond to different points in the (\mc, \ms) mass space.}}
\label{eff2}
\end{figure}
\begin{figure}
\centerline{\resizebox{\textwidth}{!}{
\renewcommand{\arraystretch}{1}
\begin{tabular}{cc}
\epsfig{file=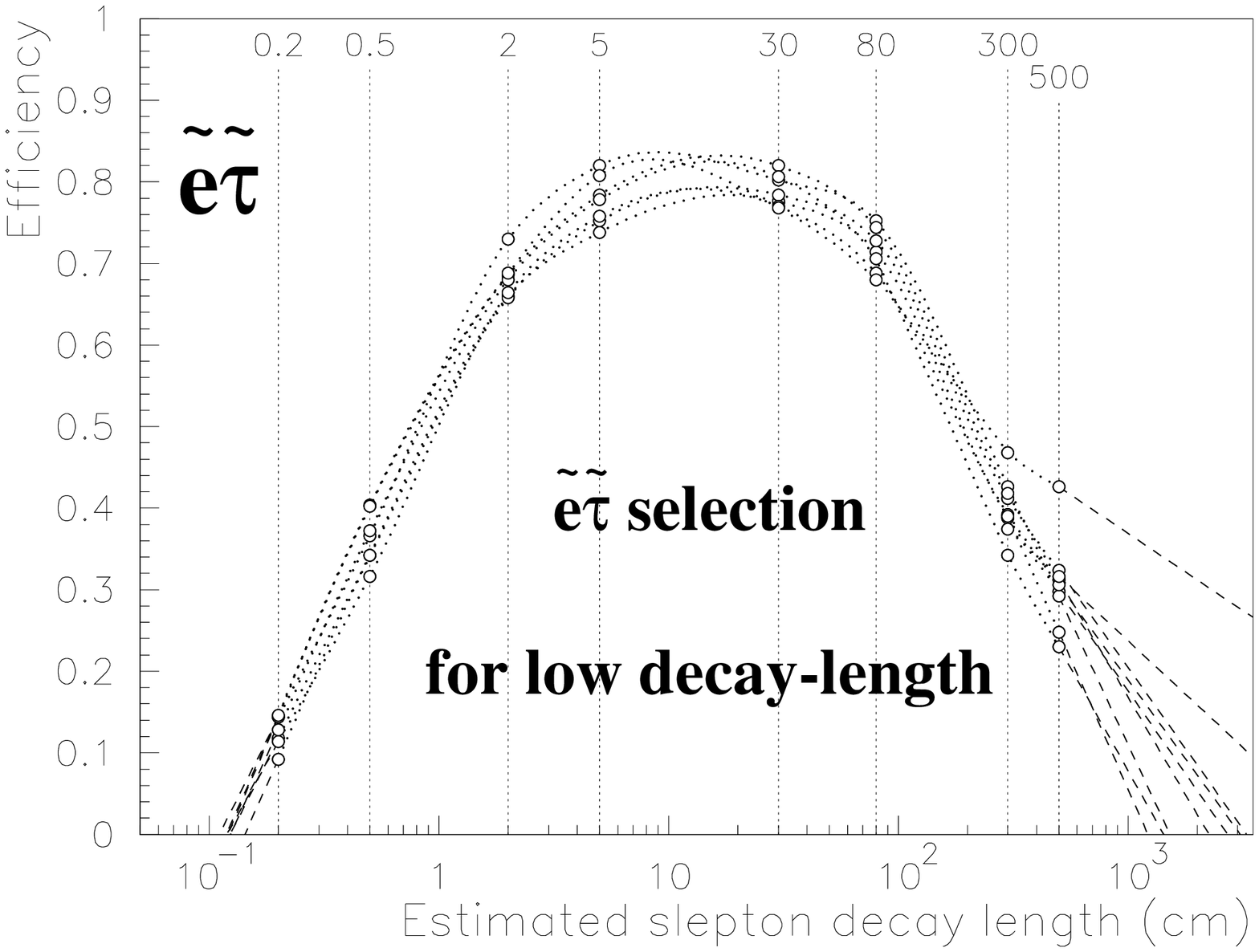,height=\xb,width=\xa} & \epsfig{file=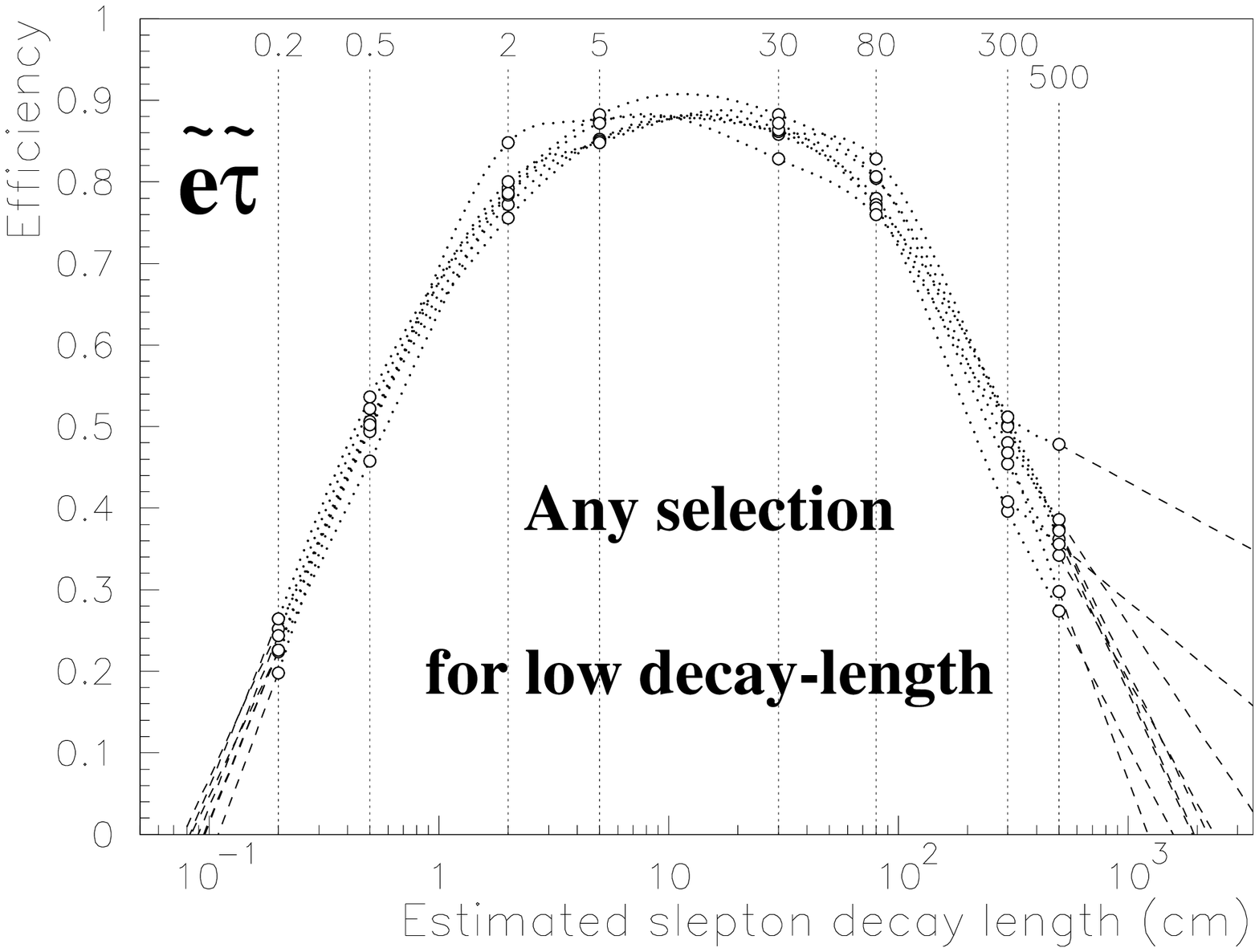,height=\xb,width=\xa} \\
\epsfig{file=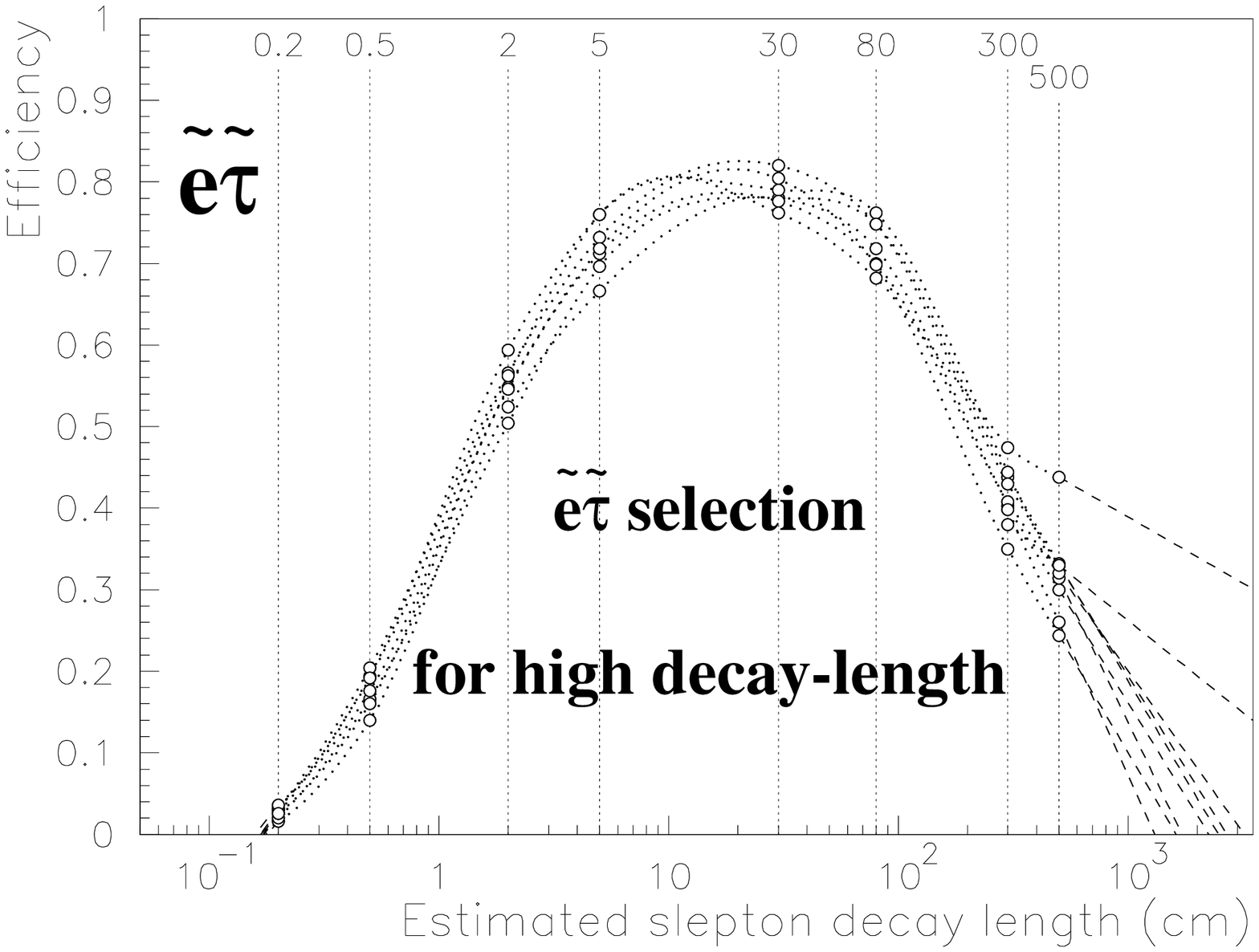,height=\xb,width=\xa} & \epsfig{file=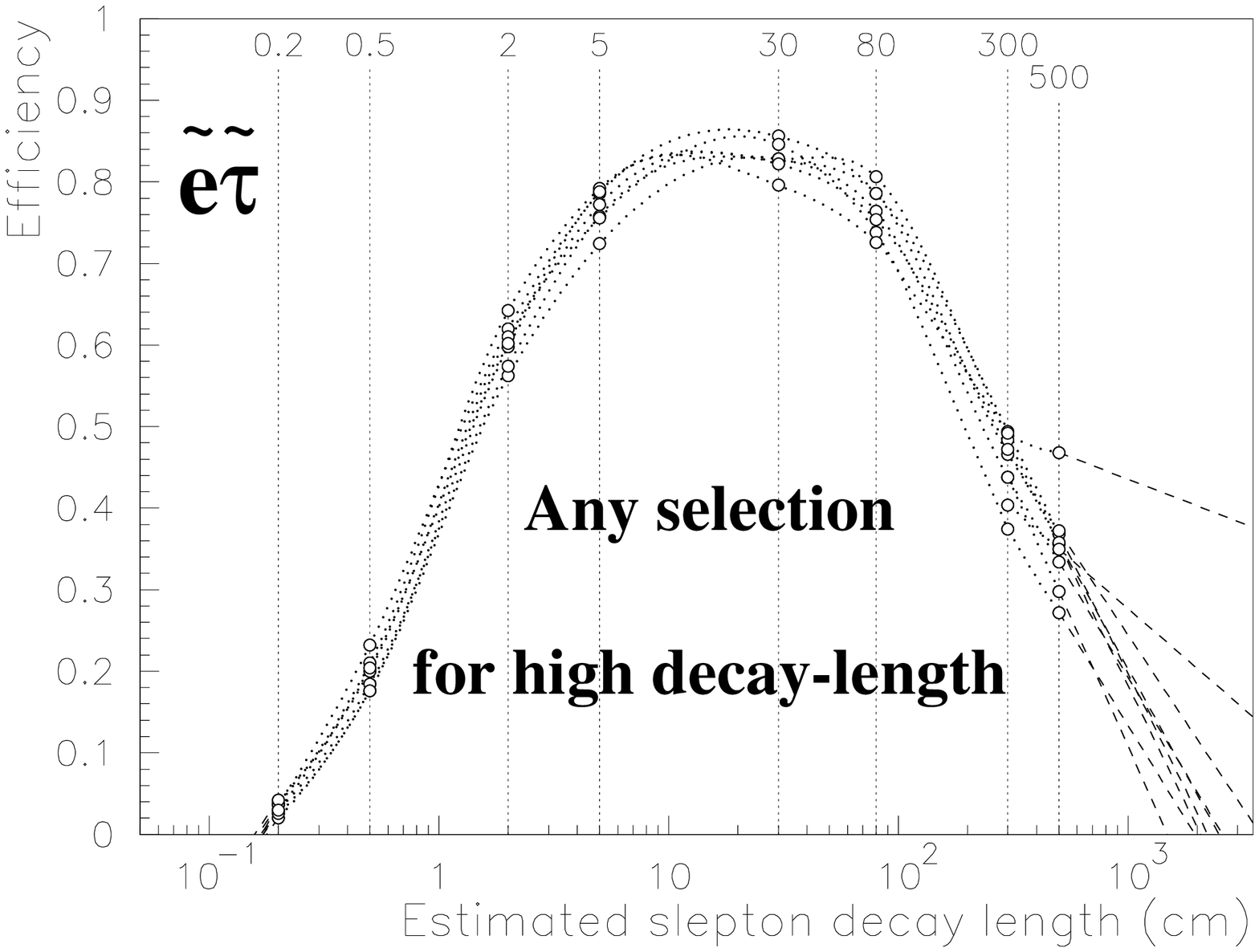,height=\xb,width=\xa} \\
\epsfig{file=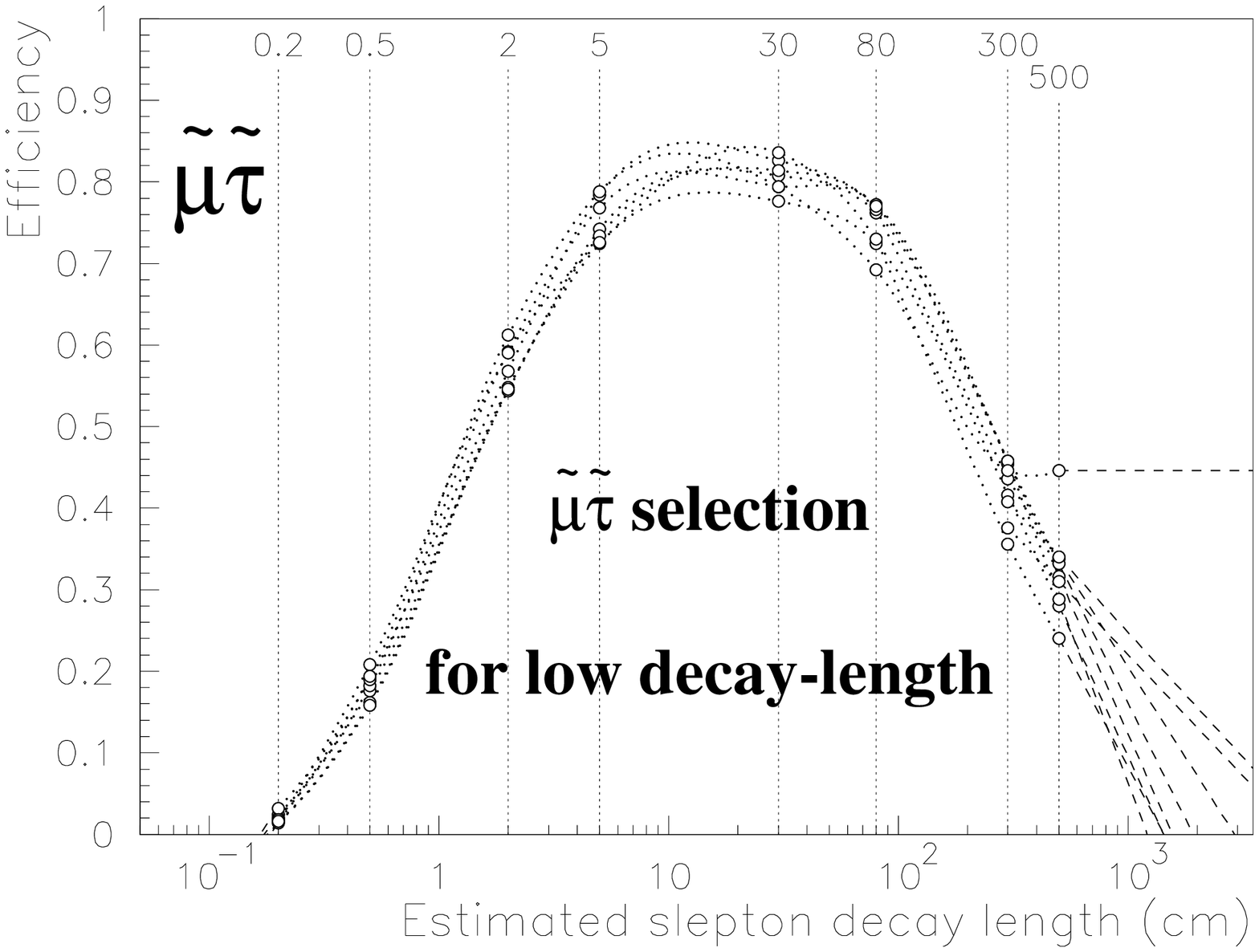,height=\xb,width=\xa} & \epsfig{file=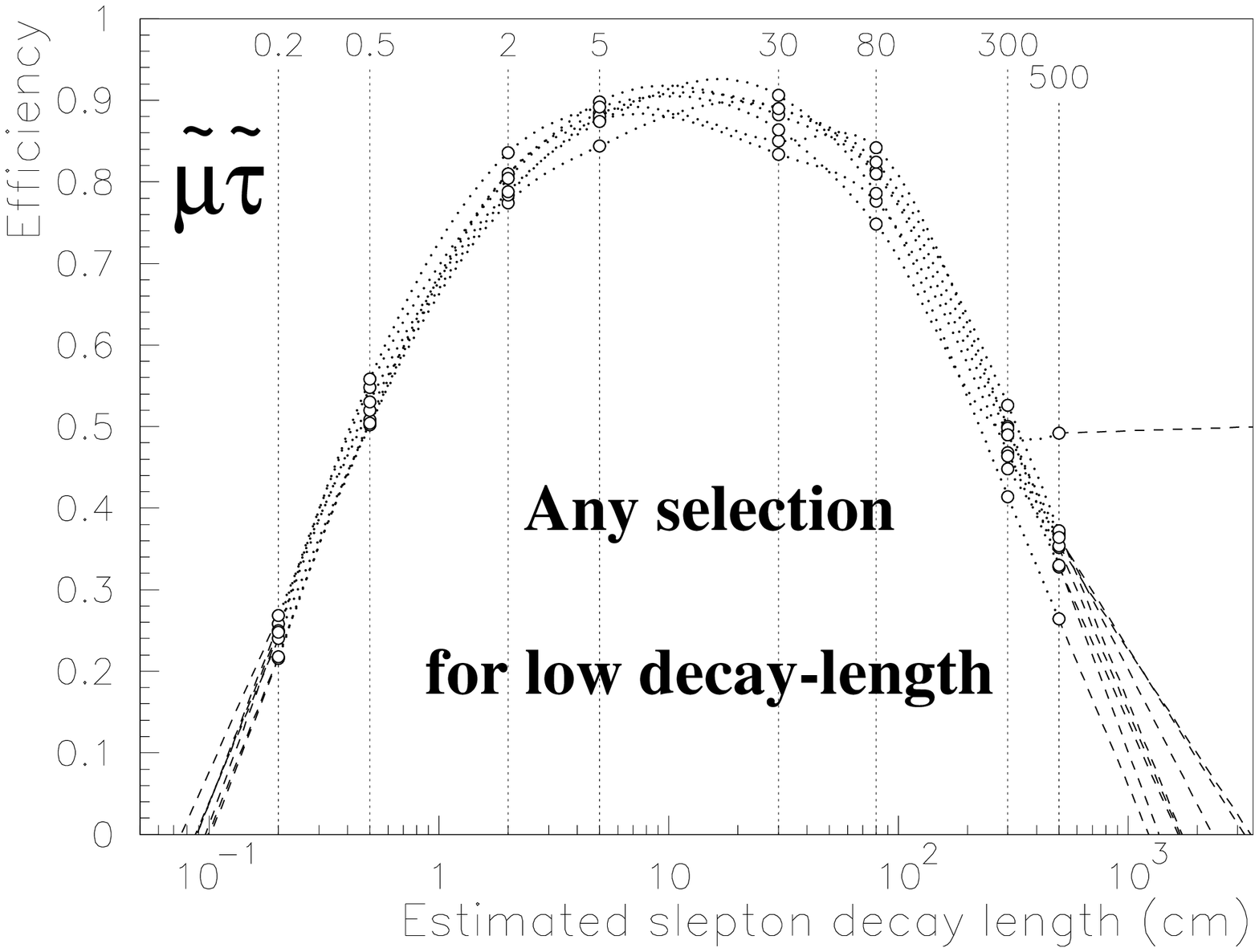,height=\xb,width=\xa} \\
\epsfig{file=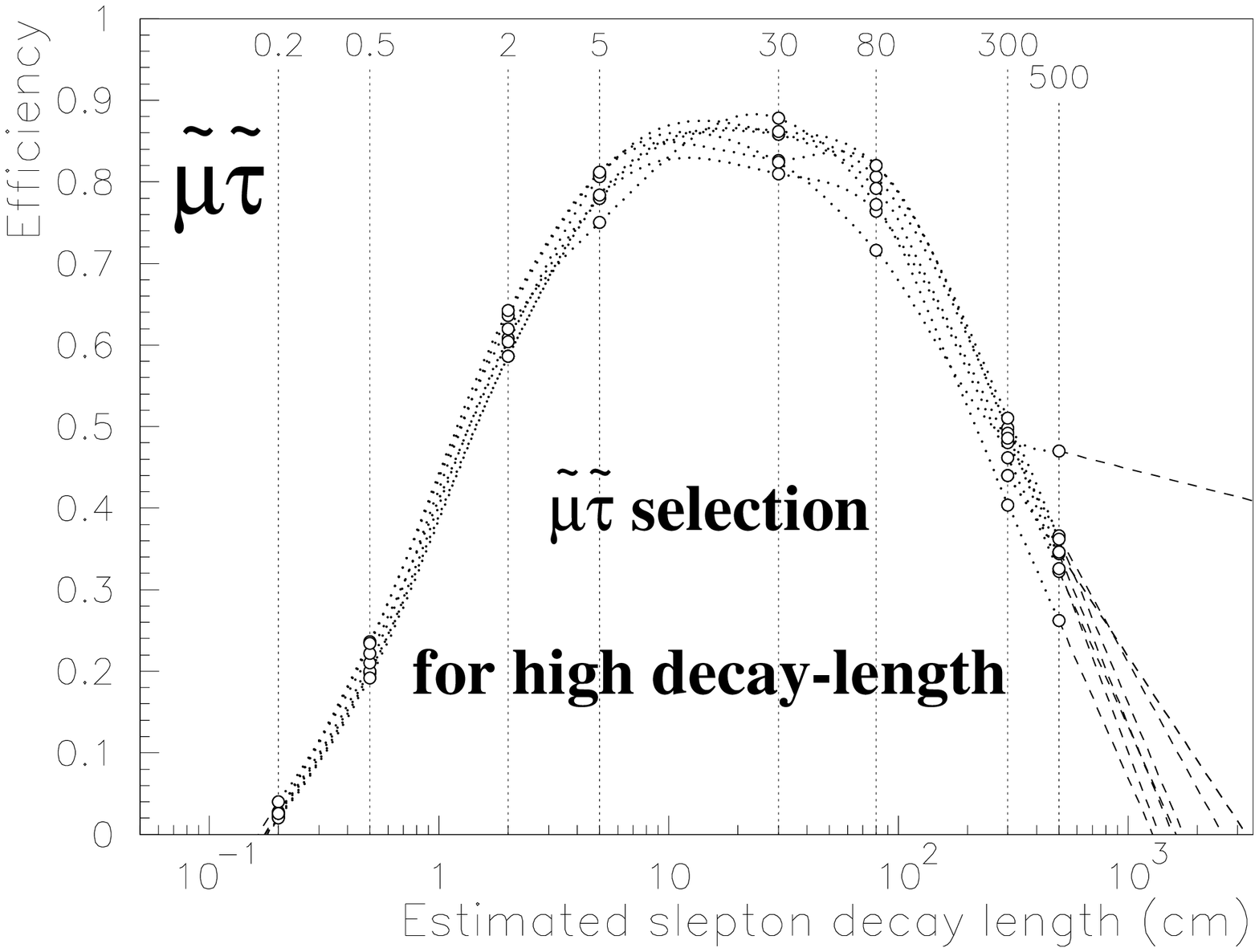,height=\xb,width=\xa} & \epsfig{file=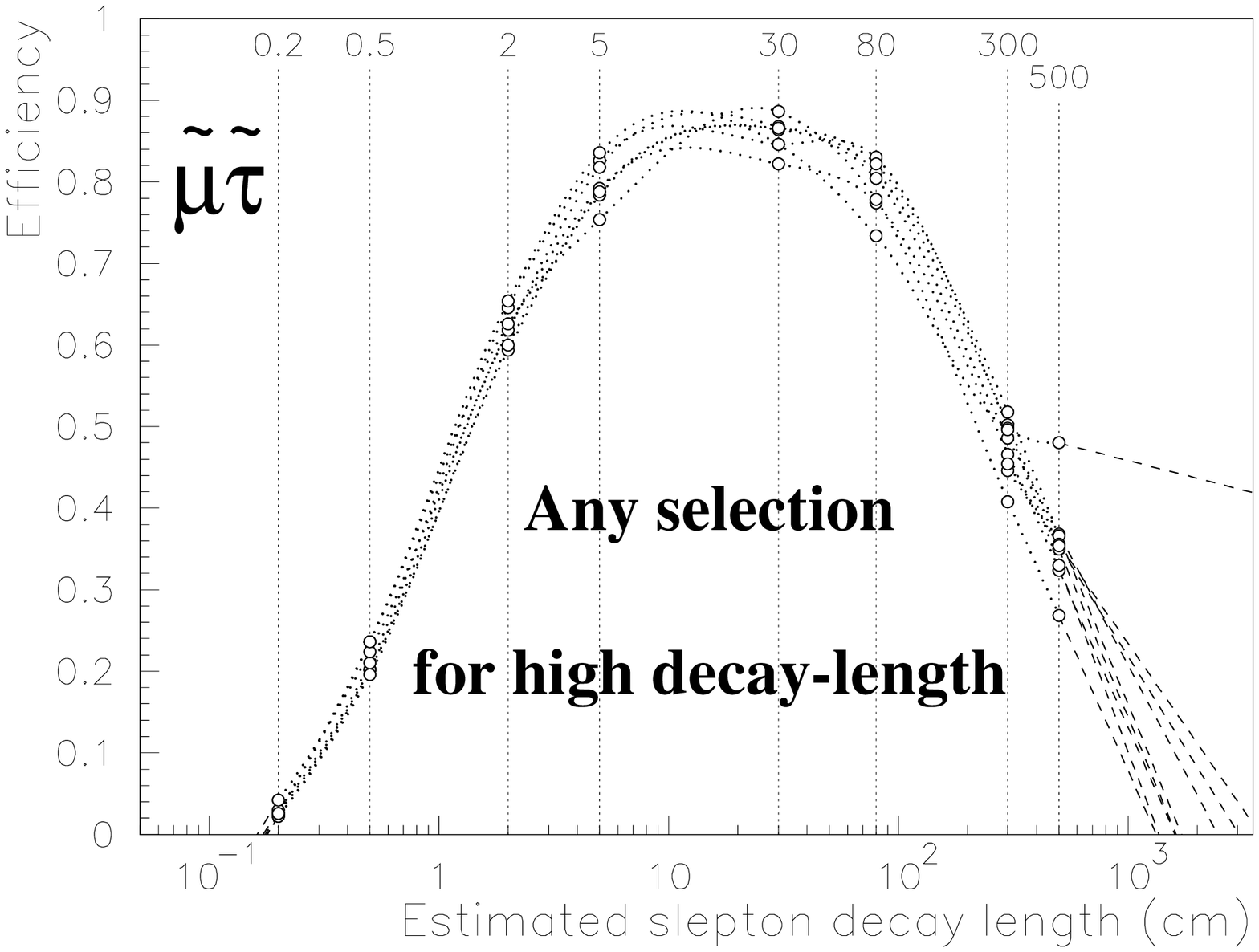,height=\xb,width=\xa} \\
\end{tabular}
}}
\caption{Efficiencies for the \selstau\ and \smustau\ channels as a function of estimated slepton decay length. \ladd{The different curves correspond to different points in the (\mc, \ms) mass space.}}
\label{eff3}
\end{figure}
\setlength{\tabcolsep}{\ljlen}
\par
The efficiencies for the $\sqrt{s}=208\GeV$ Monte Carlo data are not given
since they are virtually identical when compared with the appropriate
scaling (more will be said about this in Section~\ref{interp}). The
slow-slepton effect resulting in the higher efficiencies for high
decay lengths is not observed in the 208\GeV\ Monte Carlo data though,
since the minimum $E_{beam}-\mc$ energy difference was limited to
1\GeV. This shows that the effect is only confined to a very small
corner of the mass space.

\section{The confidence level}
\label{cl}
The ultimate goal of a search, in the absence of observing significant
evidence for the signal, is to place the most stringent possible
limits on the parameters of the underlying theory. Due to the
statistical nature of particle physics, a search can never, with 100\%
certainty, rule out the hypothesised phenomenon. Only a `confidence
level', necessarily less than 100\%, can be attached to the assertion
that the phenomenon was not occurring. The confidence level is the
probability that, if the hypothesised process were actually occurring,
and if the experiment were done again in exactly the same way, a
result more compatible with the hypothesis would be obtained. A high
confidence level then indicates an experiment that is not compatible
with the hypothesis, a low confidence level indicates one which
is. The benchmark confidence level for `exclusion' (disproof of the
hypothesis) is 95\%. At each point in the relevant parameter space the
confidence level can be computed from the slepton and neutralino
properties, and those points for which it is $\ge95\%$ are said to be
excluded. This section describes the way in which the confidence level
is constructed. However, the method of construction used for this
analysis is unusually complex for a SUSY search, and so there is first
a discussion highlighting the motivation for this complexity.
\par
The confidence level is computed from a `test statistic', $t$, which
itself is computed from the information yielded by the search. The
test statistic should be a measure of how `signal-like' (how
compatible with the search hypothesis) the search outcome is. The
higher its value, the more signal-like. Then if the expected
probability distribution function of the test statistic under the
signal hypothesis, $\rho(t)$, is known, the confidence level is given
by
\[\mbox{CL}=1-\int\limits_{t_{min}}^{t_{meas}} \raisebox{1cm}{}\!\rho(t)\ \mathrm{d}t \ , \]
where $t_{min}$ is the minimum possible value of the test statistic,
and $t_{meas}$ is the actual value from the search result. The
definition of $t$ is, formally, completely free. Clearly it should be
constructed such as to have good discriminating power between the
signal and no-signal scenarios, but a bad choice only leads to a
non-optimal confidence level (lower than the best possible). It does
not make the confidence level wrong. 
\par
For the most optimal confidence level, $\rho(t)$ should take into
account that some contribution to the value of $t$ is expected from
background as well as from signal, and so should be the density of the
test statistic under the signal-plus-background hypothesis. Any
systematic errors in the simulation of the background however, then
find their way into the confidence level. So the background
contribution to $\rho(t)$ should be left out unless the background is
well understood and its simulation is trusted. Since the background to
this search comes from unconventional processes (principally particle
interaction with the structure of the detector), the simulation of
which have not been checked with the same rigour as more conventional
sources, the background contribution to $\rho(t)$ is left out.
\par
Typically with searches, as is the case with this search, the larger
the number of observed events, the more signal-like the outcome. If
there is no additional information associated with each event to
discriminate signal from background, the only sensible choice for the
test statistic is then just the number of observed events,
$n_{obs}$. The confidence level is then given by
\[\mbox{CL}=1-\sum\limits_{n=0}^{n_{obs}}P_n=1-\sum\limits_{n=0}^{n_{obs}} \frac{e^{-n_{exp}}n_{exp}^n}{n!}\ , \]
where $P_n$ is the (Poisson) probability of obtaining $n$ events as a
search outcome, and $n_{exp}$ is the mean number of expected events
under the signal (with no background) hypothesis.
\par
In this way, a confidence level can be obtained for each channel by using 
\begin{gather*}
n_{obs}= \mbox{number of events passing channel's selection,} \\
n_{exp}=\Lum\,\sigma\,P_{chan}\,\eff_{chan} \ ,
\end{gather*}
where \Lum\ is the integrated luminosity, $\sigma$ is the neutralino
cross-section, $P_{chan}$ is the probability that the neutralinos will
decay such as to give this channel (given by the branching ratios),
and $\eff_{chan}$ is the efficiency with which events of this channel
will be selected. For statistical reasons it cannot be said that the
point is excluded if any one of these CLs is greater than 95\%. The
channel that appears most likely to exclude the point based on Monte
Carlo studies only should be chosen, and if the CL of that channel
(and that channel alone) is greater than 95\% then the relevant point
in parameter space is excluded.
%
\par
This is clearly, however, a non-optimal approach. The information from
five of the six channels is being thrown away. It should be possible,
rather than to perform CL calculations for each channel individually,
to calculate just one CL using all the data which would be greater
than any one of the single-channel CLs. This cannot be calculated in
a simple way from the single-channel CLs directly because the fact
that the associated selections are not mutually exclusive means that
the CLs are not statistically independent.
A solution would be to use
\begin{gather*}
n_{obs}= \mbox{number of events passing \textit{any} selection,} \\
n_{exp}=\Lum\sigma \sum\limits_{chan=1}^{6} P_{chan} \eff_{chan} \ .
\end{gather*}
\par
But this is still non-optimal because it treats all observed events in
an equal way. The number of expected signal events will be different
between channels, and the number of expected background events will be
different between the selections. Thus the contribution of an event to
the value of the test statistic should be in some way representative
of the discriminating power of the selection(s) it passes. In
particular, in the case that the stau is significantly lighter than
the selectron and smuon, and so effectively becomes the sole NLSP, it
can be expected \iforig{than}{that} including events that do not pass
the $\tilde{\tau}\tilde{\tau}$ selection will only worsen the result.
\par
A solution is to use the non-exclusivity of the selections as an extra
source of information. The distribution of the signal efficiency
amongst the 63 selection sets (disregarding the null set) is quite
different from the equivalent distribution of the expected
background. Thus the set of selections an event passes can be
considered a discriminating variable that gives information on how
signal-like it is. If the sets are numbered we have an integer
variable, denoted \S, defined for each event with a value between 1
and 63. The test statistic can then be defined as the ratio of the
probability of the search outcome under the signal-plus-background
hypothesis to that of just the background hypothesis (\ie\ a relative
likelihood) taking into account the known numbers of signal and
background events expected with each selection set:
\begin{equation}
t=\frac{L_{s+b}}{L_b}=\frac{\prod\limits_{\S} P(s_\S+b_\S,n_\S)}{\prod\limits_{\S} P(b_\S,n_\S)} \ . \label{single}
\end{equation}
$P(x,n)$ is the Poisson probability of observing $n$ events from an
expected mean of $x$, $s_\S$ and $b_\S$ are the mean numbers of signal
and background events expected to pass the set of selections, \S,
respectively, and $n_\S$ is the observed number of events passing
selection set \S. This test statistic now makes use of all the
available information and copes naturally with variations in the
number of expected signal events per channel and number of expected
background events per selection.
\par
A drawback is that $\rho(t)$ cannot be determined analytically now,
only by a `toy' Monte Carlo experiment. That is to say, using random
numbers, a succession of hypothetical search results has to be
generated in accordance with that which would be expected under the
signal-only hypothesis. The corresponding test statistic can be
calculated for each and a normalised histogram of these values built
up to form $\rho(t)$.
\par
The sorting of events by selection set now requires an associated
decomposition of the efficiency. Thus we refer to the `total'
efficiency, defined as the probability of a signal event passing at
least one selection, and the `partial' efficiencies, defined as the 63
probabilities of a signal event passing each respective selection
set. Then
\begin{gather*}
s_\S=\Lum\sigma \sum_{chan} P_{chan}\peff_{\S\,chan} \ , \\
\parbox{\textwidth}{and} \\
\eff_{chan}=\sum_{\S=1}^{63} \peff_{\S\,chan} \ \
\end{gather*}
where $\peff_{\S}$ denotes the partial efficiency associated with
selection set \S, and $\eff$ denotes the total efficiency.
\par
So far the discussion has not incorporated the fact that the search
detailed in this thesis spans a range of energies at which different
amounts of data were taken, and for which the signal cross-sections
and partial efficiencies will be different. This should also be built
into the confidence level. If the data is split into a number of
energy bins then the total test statistic is just the product of the
single-energy test statistics calculated for each bin. Thus the
generalisation of Eq.~\ref{single} to multiple energies is
\begin{gather}
t=\prod\limits_e t_e=\prod\limits_e\prod\limits_{\S}\frac{P(s_{\S\,e}+b_{\S\,e},n_{\S\,e})}{P(b_{\S\,e},n_{\S\,e})} \ , \label{multiple} \\
\parbox{\textwidth}{\flushleft where $e$ denotes the energy bin and} \notag \\
\notag \\
s_{\S\,e}=\Lum_e\sigma_e \sum_{chan}P_{chan}\peff_{\S\,chan\,e} \ .\notag
\end{gather}
All the terms have the same basic definitions as before, but now a
subscript $e$ means that value as defined at the centre-of-mass energy
corresponding to energy bin $e$.
\par
It is important to note that although numbers based on the background
expectation are used to calculate the confidence level, any systematic
error in these numbers can only\iforig{}{, on average,} reduce the
confidence level. This goes back to a point mentioned at the start of
this section -- that the definition of the test statistic is
completely free. A bad choice for the test statistic (or using
incorrect numbers in the calculation of it) merely yields a less than
optimal number. It is only in the calculation of the density of the
test statistic under the signal hypothesis, $\rho(t)$, that systematic
errors can cause \iforig{an}{a systematically} inflated value, and the
background expectation is not included in this.\iforig{}{\par For a
more detailed discussion on the construction of test statistics,
including the case of possible systematic errors in the background
expectation, see \cite{CLs}.}

\section{Interpolation and extrapolation}
\label{interp}

This section describes the methods of interpolation and extrapolation
employed to obtain the signal efficiencies for any given point in
(\mc, \ms, \dl, $\sqrt{s}$) space. Such methods are required for the
scans that will be described in Sections~\ref{xslim} and
\ref{parspace}. The process is performed for each channel individually
in stages. It is the partial efficiencies that are interpolated, and
the total interpolated efficiency is their sum. Interpolation is first
performed for four points in mass space to the correct mass point at
decay lengths and $\sqrt{s}$'s above and below the desired values,
then in decay length to form two points with the correct \dl, and then
in $\sqrt{s}$ to arrive at the desired point. The following describes
the procedure for a single channel.

\subsubsection{Interpolation in mass space}
For a given slepton decay length and $\sqrt{s}$, a method to gain the
efficiencies for any point in (\mc,\ms) space was required. It was
decided that a linear method was preferable for reasons of simplicity
and transparency. The simplest method for linear interpolation in two
dimensions is bilinear interpolation, but since the points in mass
space do not form a rectangular grid, bilinear interpolation would
lead to discontinuities in the (partial and total) efficiencies as a
function of the masses. To get a continuous function the mass points
have to be joined to make a tessellating set of polygons, and then an
interpolation performed over each polygon. Since there is no obvious
benefit from using polygons with a larger than minimal number of
sides, only added complexity and less transparency, triangles were
chosen as the polygons into which to split the plane. A commonly used
and natural method for dividing a plane into triangles based on a set
of points is called Delaunay triangulation (originally described in
\cite{delaunay}), and this was the method used.
\par
The Delaunay triangulation of a point-set is the collection of
triangles which satisfy an ``empty circle'' property, \ie\ the circle
drawn through the vertices of any triangle does not enclose any other
point from the set. As long as no circle can be drawn through four or
more points then the collection is unique. If this is not the case
then there are degenerate collections (\ie\ more than one collection
that satisfies the Delaunay criteria). This is the case with the sets
of points in (\mc, \ms) space considered here, which do have subsets
of four points (but never more) lying on common circles. Thus the
degeneracy necessitates a choice for the diagonal down which the
quadrilaterals defined by the co-circular points are to be split in
order to form the triangles. The decision is always taken to split
such quadrilaterals down the diagonal with the lowest average total
efficiency in order to be conservative. The triangulation is thus
performed for each channel, energy, decay length, and optimisation
separately. It should be noted that no extrapolation is done in mass
space -- the efficiency is assumed to be zero outside the convex hull
of the point-set.
\par
Figure~\ref{deltri} shows examples of interpolated surfaces created by
this method, and their associated Delaunay triangulations. Note that
the range of the vertical scale has been limited to accentuate the
variation of the surface. If viewed with a vertical scale from 0 to 1
both the surfaces appear almost flat. It is the fact that the
efficiency is in general such a weak explicit function of the masses
for a given decay length that allows the interpolation in mass space
to be performed with so few points. Exceptions exist however, at high
decay length. It can be seen from the efficiency plots of
Section~\ref{eff}, that when the slepton mass is sufficiently close to
the neutralino mass, and the neutralino mass is sufficiently close to
the beam energy, the efficiency fails to fall as quickly with high and
increasing decay length as for the other points in mass space. Thus
the efficiency takes a sudden rise in the relevant corner of the mass
space at the largest decay length. Since few points exist to map this
rise an interpolation in this region would be dubious. In order to
deal with this, the point with largest total efficiency at the maximum
decay length has its total efficiency artificially reduced to that of
the point with the second highest total efficiency, and the partial
efficiencies are scaled accordingly. This was judged preferable to the
large amount of extra Monte Carlo generation that would be needed to
map this rise in such a small corner of the parameter space.
\begin{figure}[bthp]
\centerline{\resizebox{\textwidth}{!}{
\iforig{
  \epsfig{figure=deltri2_original.eps,width=6cm}\epsfig{figure=deltri1_original.eps,width=6cm}
  }{
  \epsfig{figure=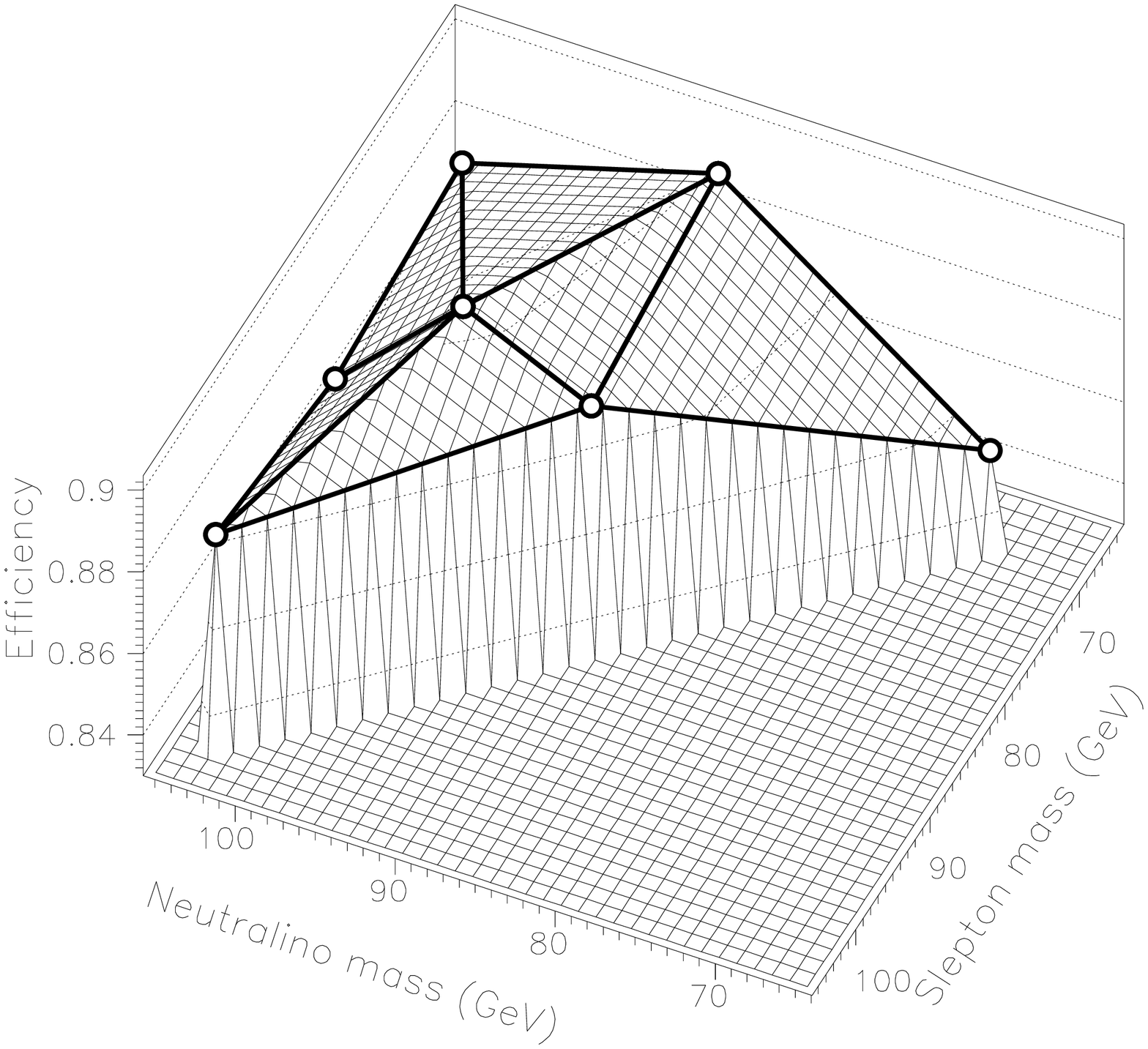,width=6cm}\epsfig{figure=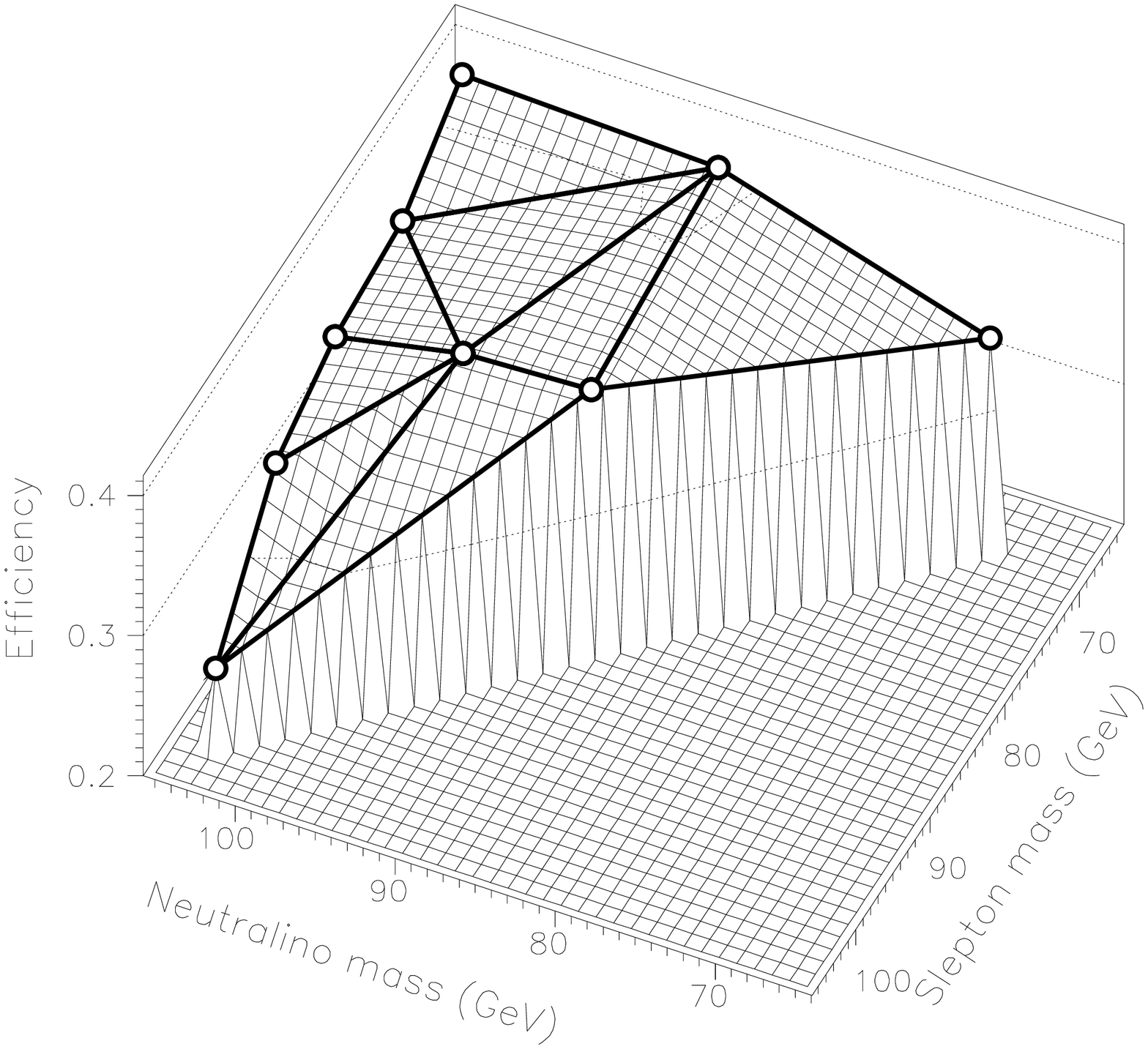,width=6cm}
  }
}}
\settowidth{\xa}{a).}
\settowidth{\xb}{b).}
\setlength{\ya}{0.25\textwidth}\addtolength{\ya}{-0.5\xa}
\setlength{\yb}{0.5\textwidth}\addtolength{\yb}{-0.5\xa}\addtolength{\yb}{-0.5\xb}
\hspace*{\ya}a).\hspace{\yb}b).\\
\caption{\lscap{The total efficiency as a function of neutralino mass 
and slepton mass for the \selsel\ channel at 208\GeV\ under the
high-\dl\ optimisation at decay lengths of a). 30\cm\ and
b). 500\cm. The highlighted points on the surface show the points at
which Monte Carlo data was generated. All other points on the surface
are interpolated from these using the Delaunay triangulation which is
shown superimposed.}{Examples of efficiencies interpolated in the
(\mc, \ms) plane using Delaunay triangulation.}}
\label{deltri}
\end{figure}
\subsubsection{Interpolation in decay length}
As can be seen from the efficiency versus decay length plots of
Section~\ref{eff}, a good method of interpolation in decay length
would simply be to join the associated points with straight lines, and
this is what is done. Although this is linear interpolation, it is
linear with respect to $\log(\dl)$ rather than \dl\ itself. Unlike the
treatment of mass space, extrapolation is used to extend the region of
sensitivity to decay lengths beyond the maximum and minimum values at
which Monte Carlo data was generated. This is performed by simply extending
the straight lines joining the two highest and two lowest points in
decay length down to the $x$-axis, as shown in the plots.
\par
Extrapolation is possible because it is a mathematical certainty that
the efficiency cannot drop completely to zero no matter how high or
low the decay length. The efficiency can only approach zero
asymptotically as the decay length falls to zero or extends to
infinity. Thus the roughly straight lines that form the left and right
sides of the efficiency `humps' must level out at more extreme
values. The extrapolations will thus give estimated efficiencies lower
than the true efficiencies, and the extension of the lifetime range
will be conservative. The beginning of this curve on the low decay
length side can be seen in the plots for the high-\dl\ optimisation.
\par
There is an obvious problem with one (\mc, \ms) point in the \smustau\
channel under the low decay length optimisation however, since the
extrapolated line at high decay length does not descend to the
$x$-axis. For one plot it is exactly horizontal, for the other it
actually increases. But this is fixed by the reduction of the
efficiency of this mass point at its highest decay length that is
performed in order to make the interpolation in the mass plane valid
(as described in the previous section). The plots of Section~\ref{eff}
are shown without this reduction. With the reduction the line drops to
zero with the other mass points.

\subsubsection{Interpolation in $\sqrt{s}$}
\newcommand{\eb}{\ensuremath{E_{beam}}}
The method of interpolation in $\sqrt{s}$ (the LEP centre of mass
energy) is again linear, and is based on an empirical observation. The
mass points chosen for the generation at 208\GeV\ were, for the most
part, simply scaled from those chosen at 189\GeV. That is to say,
while the points occupy different positions in (\mc, \ms) space, they
are at approximately the same positions in ($\frac{\mc}{\eb}$,
$\frac{\ms}{\mc}$) space (where $\eb=\sqrt{s}/2$). It was observed
that, for any given channel and decay length, the total and partial
efficiencies of a point generated at 189\GeV\ were virtually identical
to those of the point generated at 208\GeV\ which was closest in
($\frac{\mc}{\eb}$, $\frac{\ms}{\mc}$) space with the same decay
length. This simple scaling indicated that no further Monte Carlo data
was required at intermediate energies -- the efficiencies could be
obtained for any intermediate energy by an appropriate interpolation.
\par
\newcommand{\mca}{\ensuremath{\m_{\chi}}}
\newcommand{\mcb}{\ensuremath{\m_{\chi}'}}
\newcommand{\msa}{\ensuremath{\m_{\tilde{l}}}}
\newcommand{\msb}{\ensuremath{\m_{\tilde{l}}'}}
To obtain the efficiencies at a given decay length \dl\ and a given
mass point \mbox{(\mc, \ms)} at LEP energy \eb, the equivalent scaled
masses (\mcb, \msb) at 189\GeV\ and 208\GeV\ were found using the
prescription:
\begin{gather*}
\frac{\mca}{\eb}=\frac{\mcb}{\eb'} \\
\parbox{\textwidth}{\flushleft and} \\
\frac{\msa+\overline{m}_{lepton}}{\mca}=\frac{\msb+\overline{m}_{lepton}}{\mcb}
\end{gather*}
where $\eb'$ is either 189\GeV\ or 208\GeV\ to obtain the respective
equivalent masses, and $\overline{m}_{lepton}$ is the average of the
masses of the two leptons involved in the channel (e.g. $(m_\mu +
m_\tau)/2$ for the \smustau\ channel). The inclusion of the lepton
masses in the scaling protects the $\mc-\ms$ mass difference from
dropping below the mass of either of the leptons when scaling down in
energy. Then the efficiencies can be obtained at the scaled mass
points for the relevant decay length for both 189 and 208\GeV, and
then linearly interpolated to the correct energy, although the
interpolation is not very important since the values will be very
similar.
\par
A method of quantifying the agreement of the partial and total
efficiencies between 189 and 208\GeV\ was developed. For each point in
(\mc, \ms, \dl) space at which 208\GeV\ Monte Carlo data was
generated, the 189\GeV\ point closest in the ratio-space defined above
with the same decay length was found. Note that although 360 points in
total were generated at 208\GeV, only 333 ($>90\%$) are directly
comparable to 189\GeV\ points because the same decay lengths were not
always generated at both energies. For each pair two $\chi^2$'s were
calculated\iforig{. One}{, one} as a measure of the agreement of the
total efficiencies and one as a measure of the agreement of the
partial efficiencies without reference to their sum (it is thus a
measure only of the relative sizes of the partial efficiencies, with
any disagreement of the absolute sizes caused by a disagreement of the
total efficiency removed). The former is defined as:
\begin{gather*}
\chi^2=\sum_{i=1}^2 \frac{(\overline{\eff}-\eff_i)^2}{V_i} \ , \\
\parbox{\textwidth}{where} \\
\overline{\eff}=\frac{\eff_1 V_2 + \eff_2 V_1}{V_1+V_2} \ ; \hspace{1cm} \eff_i=\frac{N_i}{500} \ ;  \hspace{1cm} V_i=\frac{N_i}{500^2}\left(1-\frac{N_i}{500}\right) \ .
\end{gather*}
$\eff_i$ is a total efficiency, $N_i$ is the number of events out of
the total sample of 500 passing at least one selection, and the
subscript $i$ refers to either 189 or 208\GeV\ Monte Carlo data depending on
its value. The $\chi^2$ from the partial efficiencies was defined in a
similar way for each selection set \S, and summed over, but now since
the total number of events being considered is not fixed (events
failing all selections, $500-N_i$ of them, are ignored) the errors are
treated as Poisson as opposed to binomial:
\newcommand{\corind}[1]{\iforig{_#1}{_{\S #1}}}
\iforig{\newcommand{\fing}{\peff}}{\newcommand{\fing}{\eta}}
\begin{gather*}
\chi^2=\frac{1}{M}\sum_{\S=1}^{63}\sum_{i=1}^2 \frac{(\overline{\fing}_\S-\fing_{\S i})^2}{V\corind{i}} \ , \iforig{}{\displaybreak[0]} \\
\parbox{\textwidth}{where} \\
\overline{\fing}_\S=\frac{\fing_{\S 1} V\corind{2} + \fing_{\S 2} V\corind{1}}{V\corind{1}+V\corind{2}} \ ; \hspace{1cm} \fing_{\S i}=\iforig{\frac{N_{\S i}}{N_i}}{\frac{N_{\S i}}{N_i}\equiv\frac{\epsilon_{\S i}}{\eff_i}} \ ; \hspace{1cm} V\corind{i}=\frac{N_{\S i}}{N_i^2} \ .
\end{gather*}
$\peff_{\S i}$ is a partial efficiency corresponding to selection set
\S, $N_{\S i}$ is the number of events out of the total sample passing 
selection set \S, and \iforig{$i$ and $N_i$}{the other terms} have the
same definitions as for the previous $\chi^2$. $M$, the number of
degrees of freedom to which the $\chi^2$ is normalised, is taken as
the number of selection sets for which $N_{\S 1}+N_{\S 2}>0$, as
opposed to the total of 63. This is because many of the selection sets
have zero associated events at both energies and so do not contribute
to the $\chi^2$. If they are included in the number of degrees of
freedom the $\chi^2$ ends up artificially low.
\begin{figure}[thpb]
\setlength{\xa}{7.2cm}
\HS \epsfig{figure=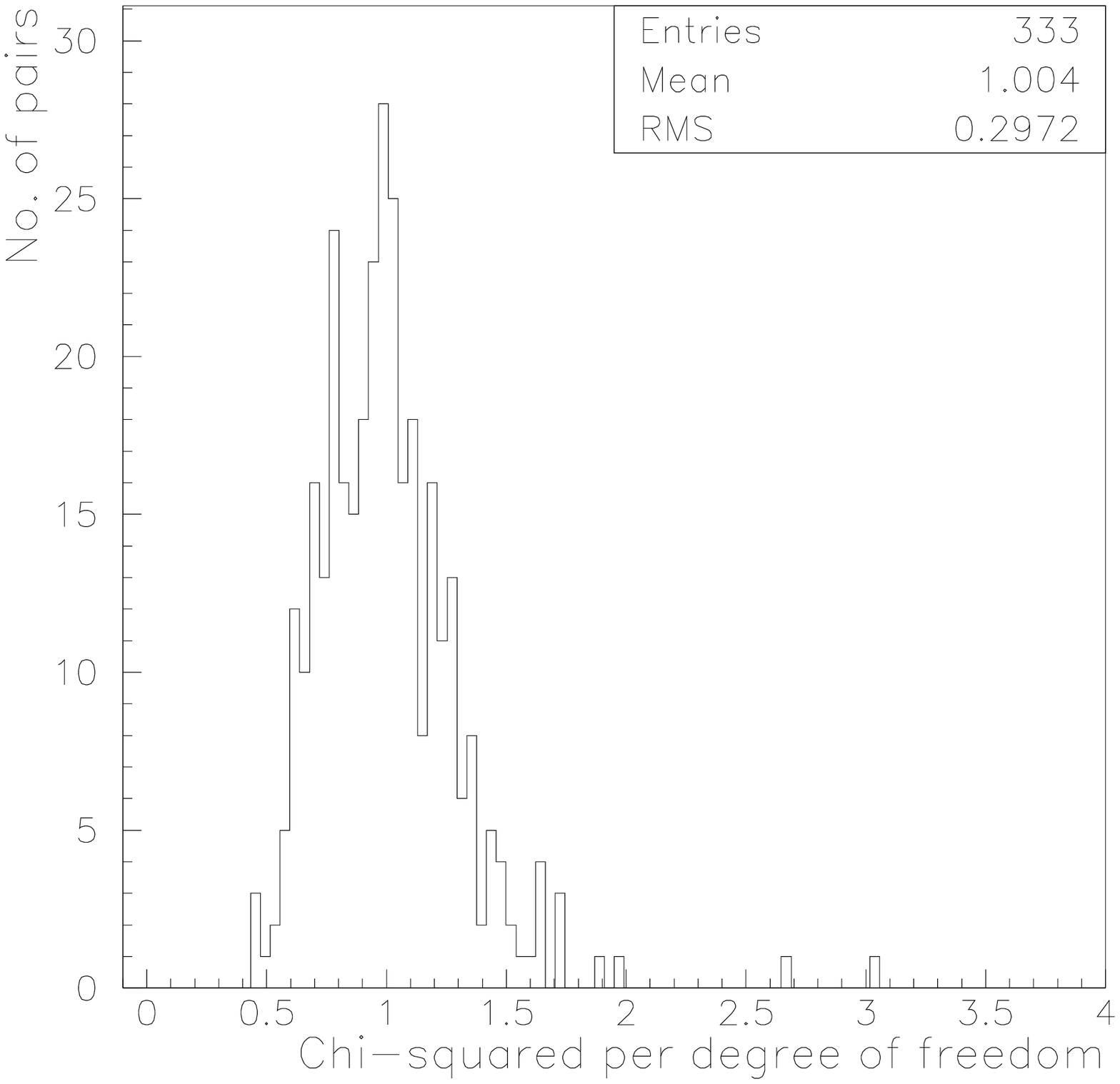,width=\xa} \HS \epsfig{figure=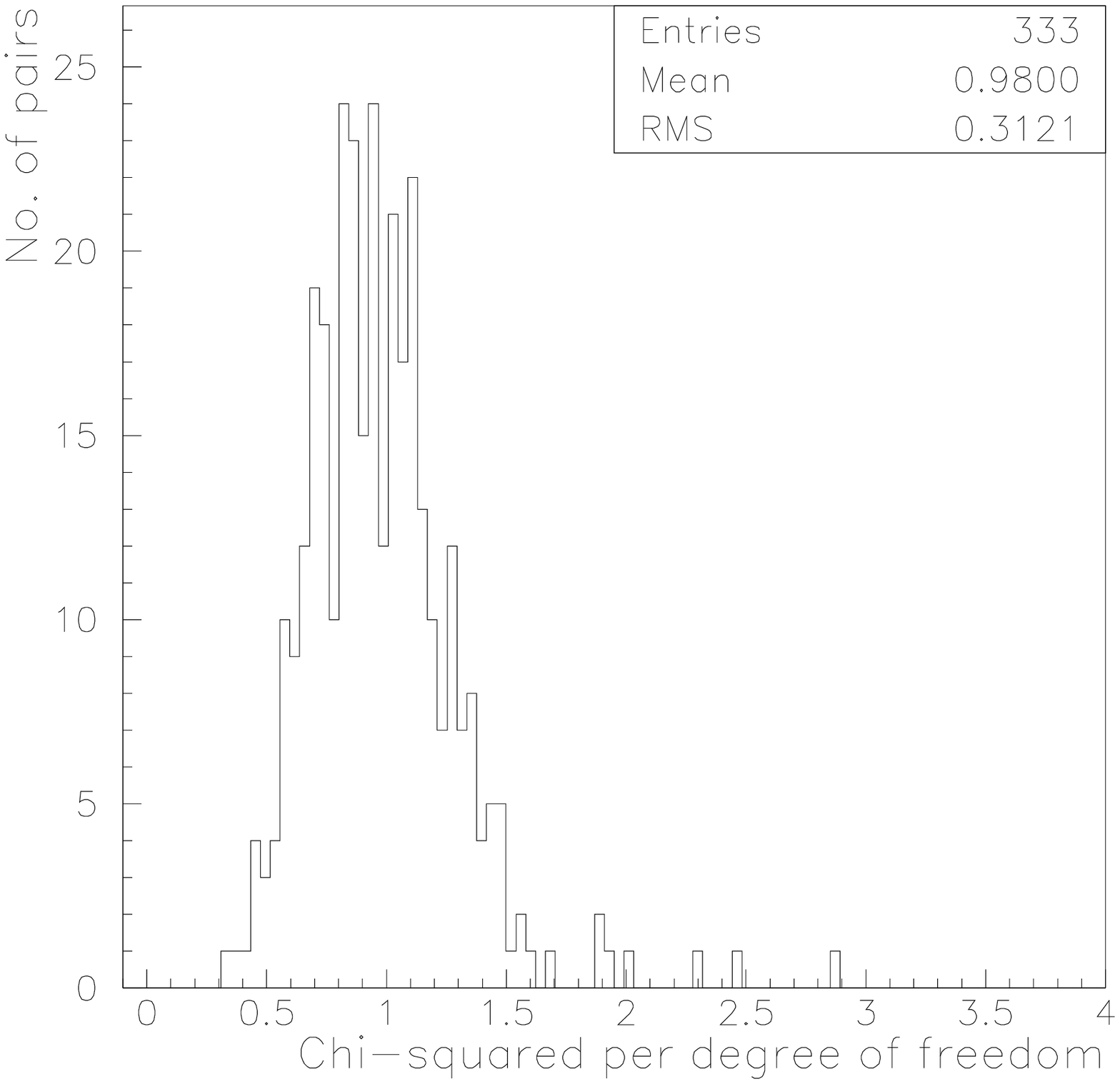,width=\xa} \HS \hspace{0mm} \\
\centerline{\HS \parbox{0.4\textwidth}{a). Comparison of partial efficiencies, low \dl\ optimisation.} \HS\HS \parbox{0.4\textwidth}{b). Comparison of partial efficiencies, high \dl\ optimisation.}\HS\hspace{0mm}} \\
\hspace*{0mm}\HS \epsfig{figure=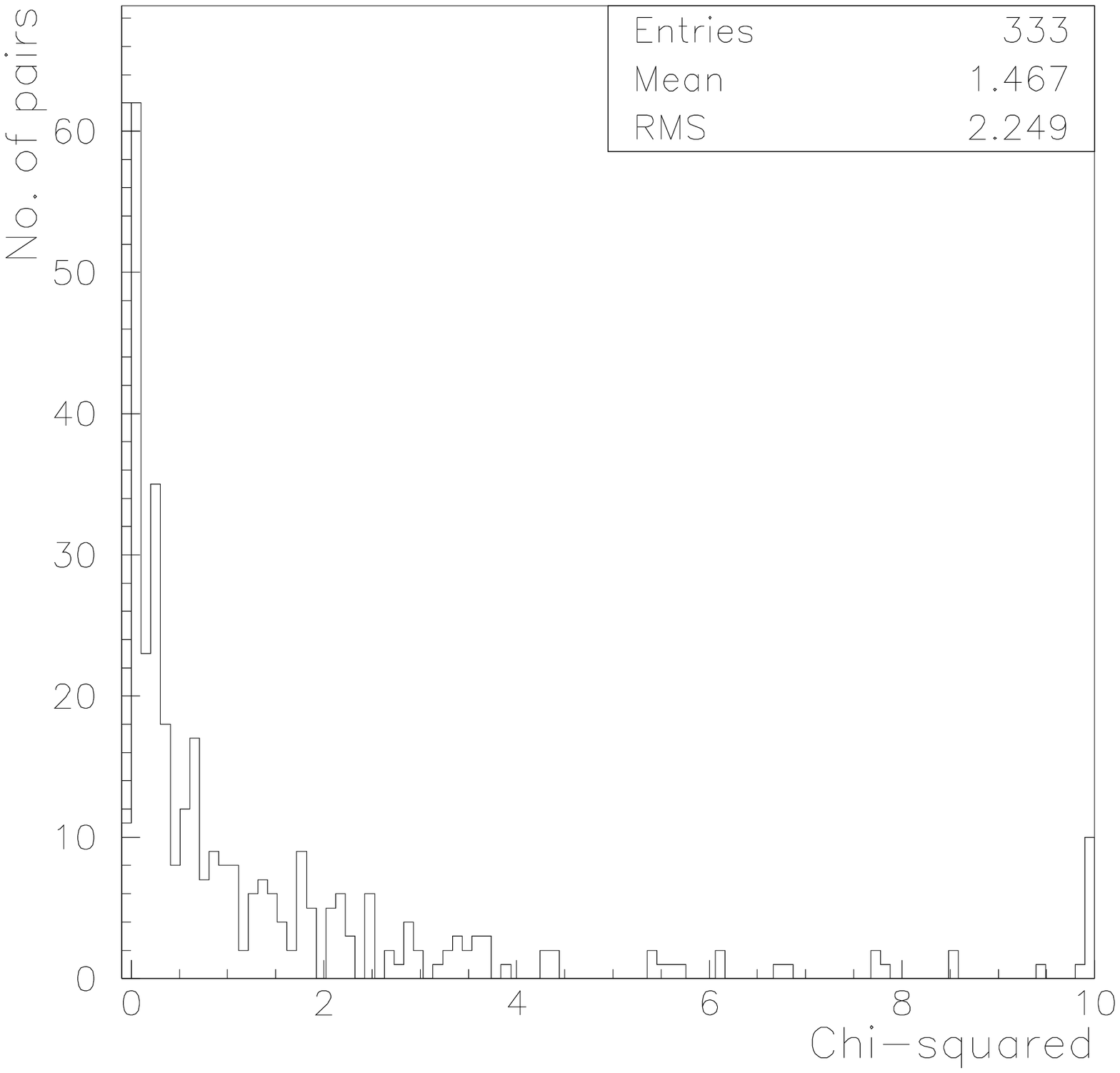,width=\xa} \HS \epsfig{figure=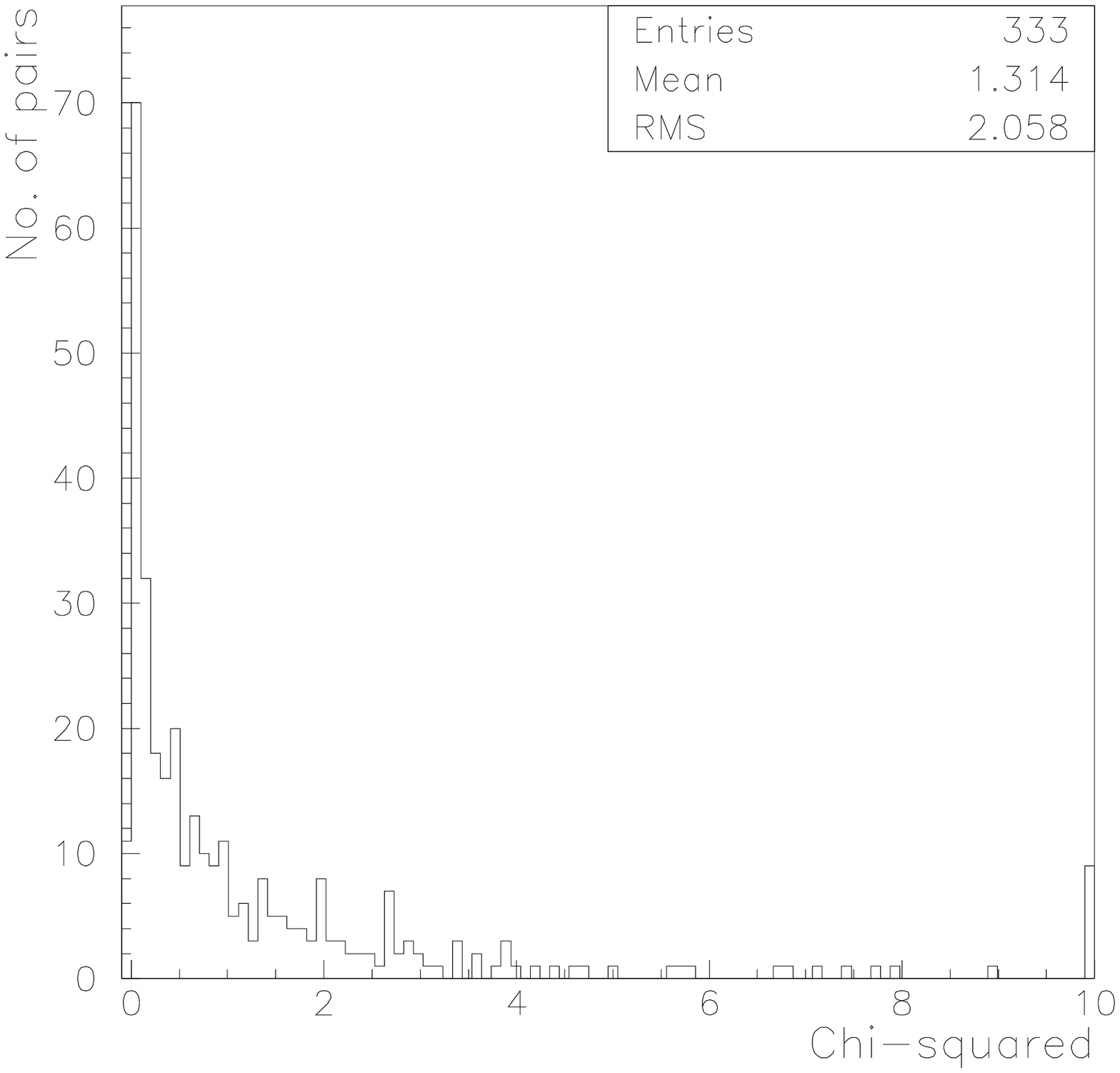,width=\xa} \HS\hspace{0mm} \\
\centerline{\HS \parbox{0.4\textwidth}{ c). Comparison of total efficiencies, low \dl\ optimisation.} \HS\HS \parbox{0.4\textwidth}{ d). Comparison of total efficiencies, high \dl\ optimisation.}\HS\hspace{0mm}}
\caption{The $\chi^2$ distributions resulting from compatibility tests between 189 and 208\GeV\ Monte Carlo data. \ladd{In the two lower plots not all pairs of points gave a $\chi^2$ in the range shown. Such pairs have been placed in the last bin. Good agreement is shown amongst all the partial efficiencies, and most of the total efficiencies.}}
\label{ecmagree}
\end{figure}
\par
Both of these $\chi^2$'s are plotted in Figure~\ref{ecmagree} for the
low and high \dl\ optimisations. Good agreement between the relative
sizes of the partial efficiencies is observed for all pairs of
points. Good agreement between the total efficiencies is observed for
most pairs of points, but some have $\chi^2$'s well in excess of 10
(seven are in excess of 100), and these have been placed in the last
bins of the respective plots. All these pairs contain the 189\GeV\ low
phase-space `problem' points described in Section~\ref{eff}. It is not
unsurprising that disagreement is observed here, since the same level
of $E_{beam}-\mc$ degeneracy is not replicated in the 208\GeV\ Monte
Carlo data, and the comparison is performed without the previously
described efficiency reduction. The difference in the total
efficiencies can thus be taken as a result of their slightly different
positions in the mass-ratio space in a region where an anomalous
effect is causing rapid change.
\par
In summary, the signal efficiencies are almost imperceptibly weak
functions of $\sqrt{s}$ for a given point in the stated mass-ratio
space and a given \dl. Thus efficiencies can be reliably gained at all
energies between 189 and 208\GeV\ by linear interpolation of the
efficiencies calculated at these two energies.



\section{Model independent cross-section limits}
\label{xslim}
The cross-section limit is defined here as the 95\% confidence level
upper limit on the lightest neutralino production cross-section at a
specific centre of mass energy for a given set of the relevant
sparticle parameters (the neutralino mass and branching ratios, and
slepton masses and lifetimes). Cross-section limits cannot be
calculated analytically. A series of trial values are chosen and the
confidence level is computed for each. Successive values are picked so
as to converge on a computed confidence level of 95\%. Once a
confidence level sufficiently close to 95\% is attained, the trial
value can be taken as the cross-section limit.
\par
Since the cross-section ($\sigma$) is a function of $\sqrt{s}$
however, and the search spans a range in $\sqrt{s}$, a different
cross-section value must be used at each. These are related by the
function, $\sigma(\sqrt{s})$, which defines the evolution of the
cross-section with centre of mass energy. But the use of a specific
model to gain $\sigma(\sqrt{s})$ will lead to limits that are not
universally valid. This problem can be circumvented if the
cross-section limit is calculated at (or above) the highest energy
considered in the analysis, $\sqrt{s}_{max}$, and the cross-sections
at lower $\sqrt{s}$ are obtained by a linear interpolation of the
cross-section at the highest energy down to zero at threshold. Then
the cross-section limits are valid as long as $\sigma=0$ and
$\mathrm{d}^2\sigma/\mathrm{d}(\sqrt{s})^2$ is negative at
$\sqrt{s}=2\m_\chi$ and there is no more than one stationary point in
$\sigma(\sqrt{s})$ for $\sqrt{s}<\sqrt{s}_{max}$ (see
Figure~\ref{evol}). Then if the real cross-section is larger than the
cross-section limit at $\sqrt{s}_{max}$, it is also larger at all
lower energies. This underestimation of the cross-section at lower
energies for a given cross-section at $\sqrt{s}=\sqrt{s}_{max}$ leads
to an underestimation in the confidence level, and thus an
overestimation of the magnitude of the cross-section at
$\sqrt{s}_{max}$ needed to give a confidence level of 95\%. Thus the
cross-section limit, quoted at $\sqrt{s}=\sqrt{s}_{max}$, will be
conservative under these basic assumptions.
\begin{figure}[htpb]
\setq{0mm}{175mm}{210mm}{0.5}
\resizebox{\textwidth}{!}{\includegraphics*[\xa,\ya][\xb,\yb]{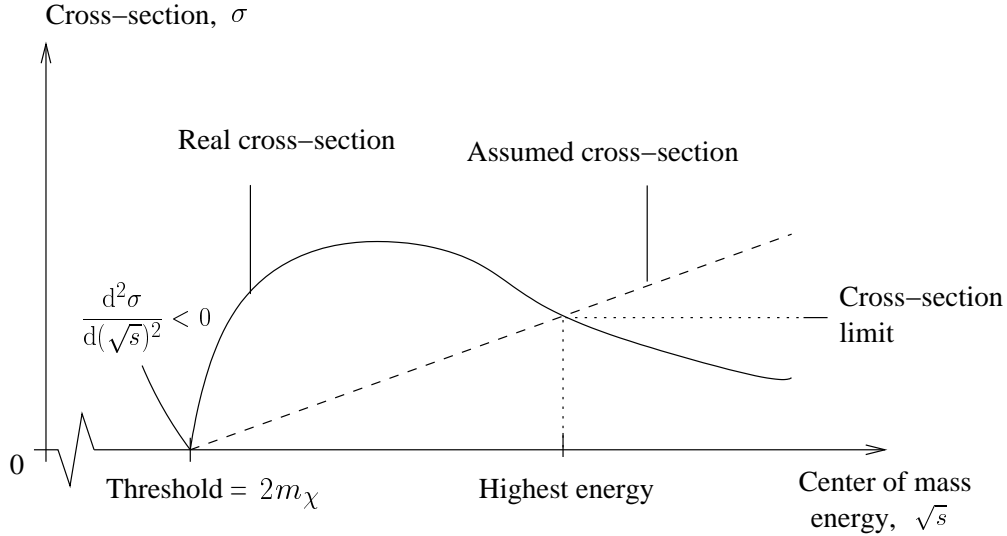}}
\caption{Schematic diagram of a hypothetical cross-section evolution with centre of mass energy and the interpolation used to evolve a cross-section at the highest energy down to lower energies. \ladd{If the real cross-section is larger at the highest energy it is also higher at all lower energies.}}
\label{evol}
\end{figure}
\par
Two scans over the (\mc, \ms, \tl) space were performed, with
neutralino masses ranging from 68 to 103\GeV, slepton masses ranging
from 67 to $(\mc-1)$\GeV\ and values of $\log_{10}(\tl)$ (where \tl\
is measured in seconds) ranging from $-11.25$ to $-6.75$. The step
size was 1\GeV\ in both the masses (\ie\ integer values were used). The
values of $\log_{10}(\tl)$ were non-equidistant. For one scan the
neutralino branching ratios to each slepton were set equal at
$\frac{1}{3}$ (\ie\ slepton co-NLSP scenario), for the other the
neutralino was assumed to decay exclusively to stau plus tau
(\ie\ stau-NLSP scenario). The energy binning for the data was as shown
in Table~\ref{result1} (\ie\ seven bins from 188.6 to 206.7\GeV). At
each point the partial efficiencies were calculated for each channel,
energy and optimisation using the interpolation and extrapolation
techniques described in Section~\ref{interp}. The total efficiencies
were reduced by one standard deviation (and the partial efficiencies
scaled accordingly) to be conservative. The background expectation
values (the $b_{\S\,e}$ of Eq.~\ref{multiple}) were linearly
interpolated from the values determined from background Monte Carlo data at
189 and 208\GeV. Note that in the case of an event being observed at
an energy and with a selection set for which $b_{\S\,e}=0$, the value
of the test statistic is infinite. This is clearly not sensible since
the finite number of Monte Carlo background events analysed means that
although no events may have been observed with a given selection set
at a given energy, it does not necessarily mean that the mean expected
number is non-zero. Thus, since approximately 10 times the number of
events expected in data were analysed at 189\GeV, and 100 times at
208\GeV, a minimum value was placed on the $b_{\S\,e}$ of
$(1-t)/10+t/100$, where $t=(\sqrt{s}_e-189)/(208-189)$ and
$\sqrt{s}_e$ is the energy of the respective bin. A trial value for
the cross-section limit at ($\sqrt{s}_{max}=$) 208\GeV\ was run down
to the individual values at the seven bin energies. This gave all the
information required to calculate a test statistic, as defined by
Eq.~\ref{multiple}, for any given experiment outcome.
\par
Since there are two optimisations, there are two confidence
levels. The confidence level chosen is the one which gives the higher
average expected value under the background-only hypothesis (with no
reference to the actual search outcomes). The exact procedure is as
follows. Firstly the two test statistics for the actual observed
outcomes under both optimisations are calculated. Then a series of 100
`toy' Monte Carlo experiment outcomes are generated under the
background-only hypothesis for each optimisation (\ie\ using random
numbers, 100 sets of hypothetical $n_{\S\,e}$ are generated in line
with the values given by the Monte Carlo background expectation), and
the test statistic is calculated for each. Then a sequence of toy
outcomes are generated in the same way but under the signal-only
hypothesis, and the test statistic is calculated for each. For each
successive pair of signal outcomes the confidence levels for the two
actual observed outcomes and the two sets of 100 toy background
outcomes are updated. The confidence level for a given actual or toy
background outcome after $N$ signal-only outcomes have been generated
is given by
\[ \mbox{CL}=1-\frac{n_{lower}}{N} \ ,\]
where $n_{lower}$ is the number of the toy signal outcomes which gave
a test statistic less than or equal to the test statistic of the
outcome under consideration. This obviously has an associated binomial
error. Once the average of the 100 confidence levels under one
optimisation becomes larger than that of the other's to a good degree
of statistical significance, the optimisation corresponding to the
lower average is dropped and the process continues with just one
optimisation.
The upper and lower limits for the value of the confidence level
corresponding to the one actual observed outcome still under
consideration are taken as the upper and lower bounds of its 99.9\%
central confidence interval: CL$_{high}$ and CL$_{low}$
respectively. These are calculated for $N=30$, 100, 300, 1,000, 3,000,
and 10,000. If CL$_{high}$ is less than 95\% then the point is
considered un-excluded and the procedure stops. If CL$_{low}$ is
greater than 95\% then the point is considered excluded and the
procedure stops. If neither of these is true the procedure continues
until the next value of $N$ at which they will be re-calculated. If
$N$ reaches 10,000 with neither of these conditions being met then it
is not known whether the confidence level is above or below 95\%, only
that it must be close. This is the requirement that must be satisfied
for the trial value of the cross-section at 208\GeV\ to be considered
the cross-section limit.
\par
Figures~\ref{xslim1} to \ref{xslim3} show the cross-section limit
($\sigma_{95}$) at $\sqrt{s}=208\GeV$ in units of nanobarns as a
function of neutralino and slepton mass for each value of
$\log_{10}(\tl)$ considered. The left and\iforig{ and}{} right plots
correspond to the slepton co-NLSP and stau-NLSP scenarios
respectively. The contours show lines of constant cross-section
limit. The $x$ (\mc) and $y$ (\ms) scales are identical (and so the
line of slepton-neutralino degeneracy is the $45^\circ$ line joining
the corners), and the cross-section limit is calculated at the
mass-bin centres. The colour shading scheme is the same for all the
plots, and the numbers on the keys to the \iforig{left}{right} of each
refer to values of $\log_{10}(\sigma_{95})$ (where $\sigma_{95}$ is in
\nb) and range from 0.924 ($\sigma_{95}=8.4\nb$) to 4.1
($\sigma_{95}=12,590\nb$). In the lowest and highest lifetime plots
for each scenario no limit exists beyond the outer contour (at
$\sigma_{95}=10,000\nb$). The trial cross-section values were required
to converge on $\sigma_{95}$ within a maximum number of iterations in
order to ensure CPU-time was kept reasonable. In a few bins this did
not occur, and so there are some spurious points where no limit is
shown (appearing as isolated black squares). These are purely random
occurrences and not the result of any physical effect.
\par
The region of maximum sensitivity moves from bottom left to top right
with increasing lifetime, since when the lifetime is short (long), low
(high) neutralino and slepton masses are preferred to increase
(reduce) the slepton decay length such that it is in the most
sensitive region. Figure~\ref{cldist} shows the confidence level
distributions for each scan. They are centred on 0.95 (95\%) with a
small spread, as expected. For comparison with the quoted limits, a
perfect search (100\% efficiency and zero observed events) would yield
a limit of
\[
\sigma_{95} = -\ln(1-0.95)\frac{(208-2\mc)}{\sum\limits_{e}\Lum_e(\sqrt{s}_e-2\mc)}
\]
where the 0.95 is the confidence level, and elements in the sum over
energy bins for which $\sqrt{s}_e<2\mc$ are ignored. The \mc\
dependence comes from the assumed linear evolution of the
cross-section with $\sqrt{s}$. This function is plotted in
Figure~\ref{xslimperf} together\iforig{}{ with} a comparison with the
obtained limit at $\log\tl=-8.75$ under the co-NLSP scenario.
\renewcommand{\arraystretch}{0.5}
\begin{figure}[H]
\begin{tabular}{c}
\resizebox{\textwidth}{!}{\epsfig{figure=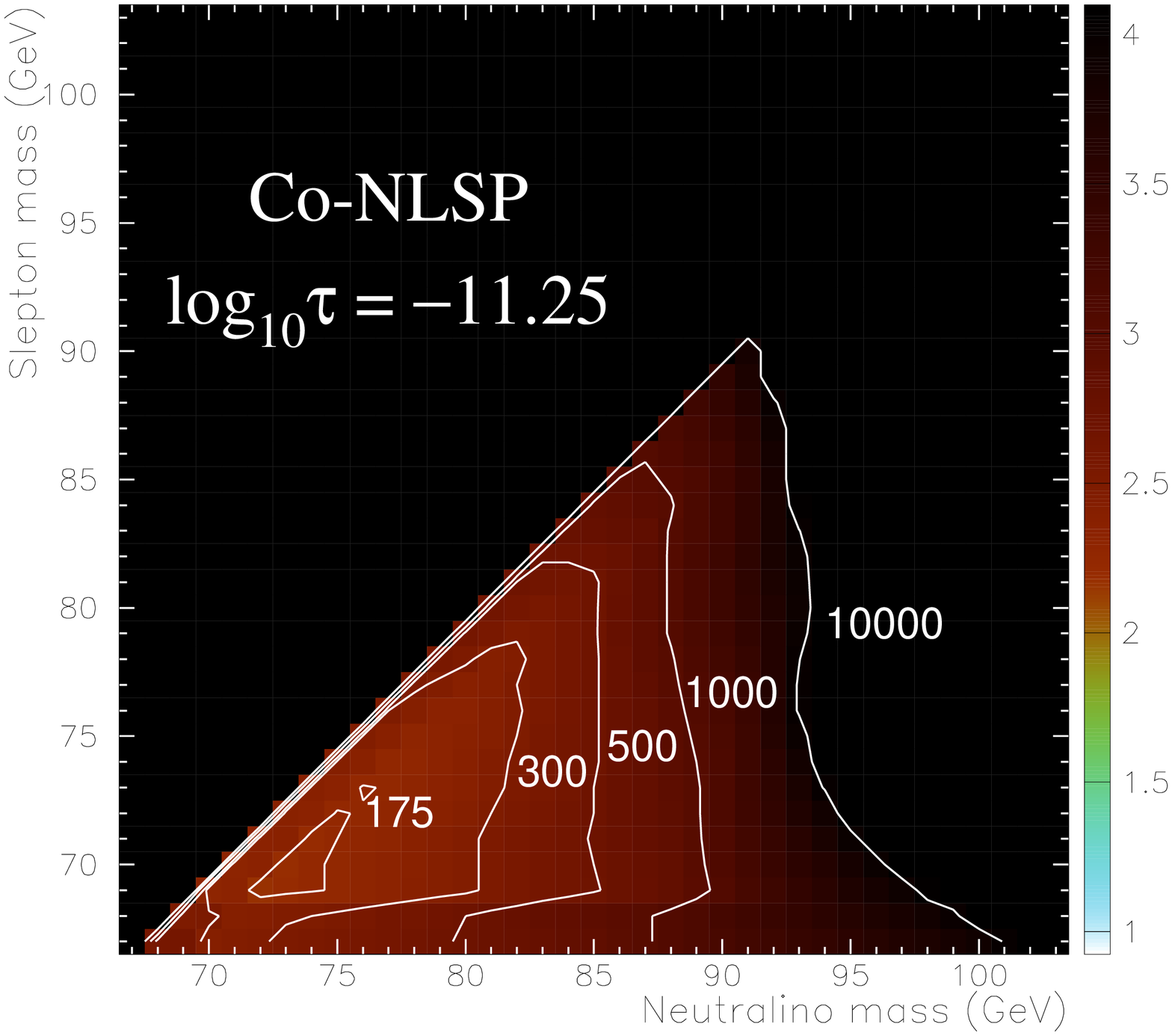}\epsfig{figure=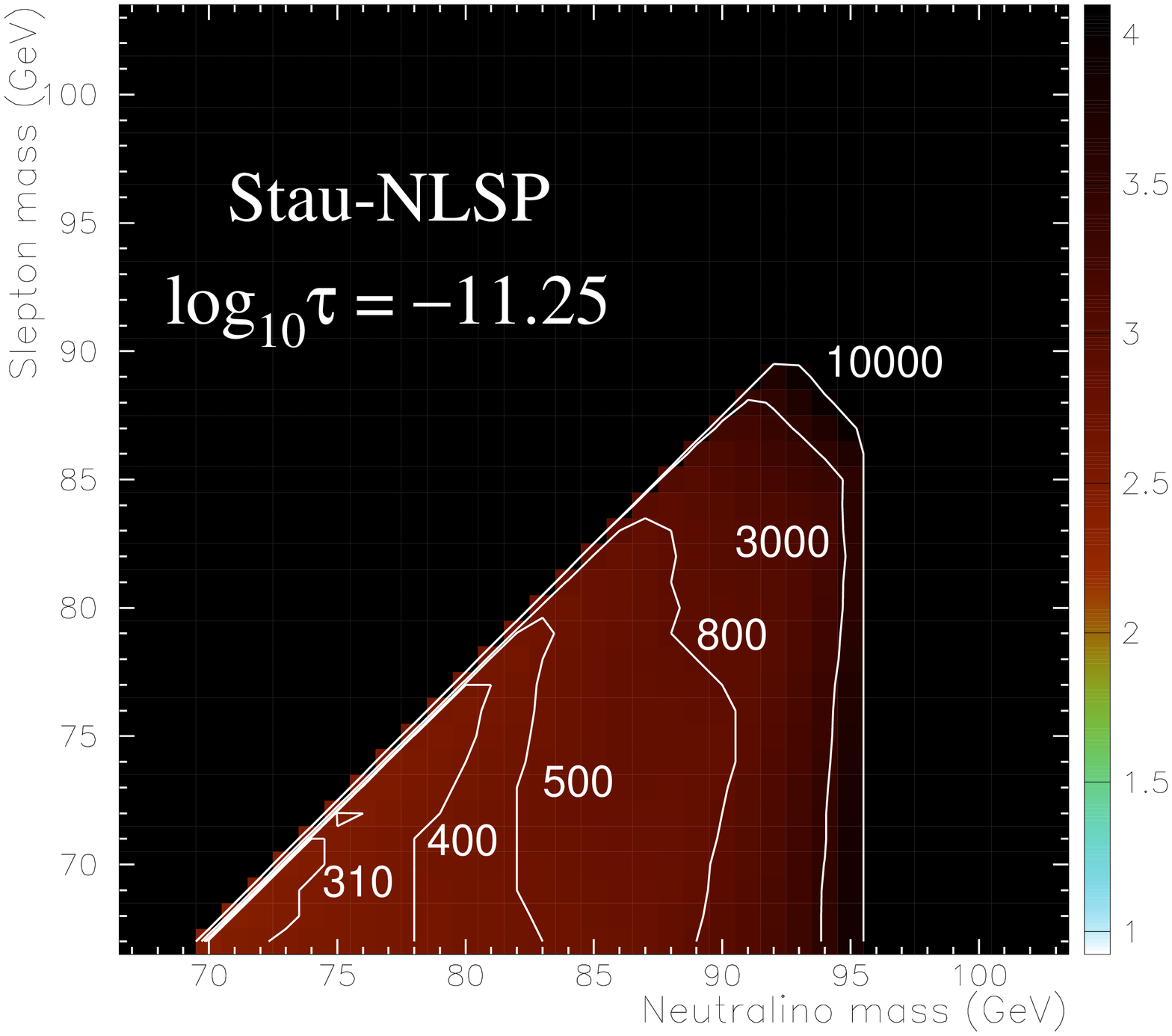}} \\
\resizebox{\textwidth}{!}{\epsfig{figure=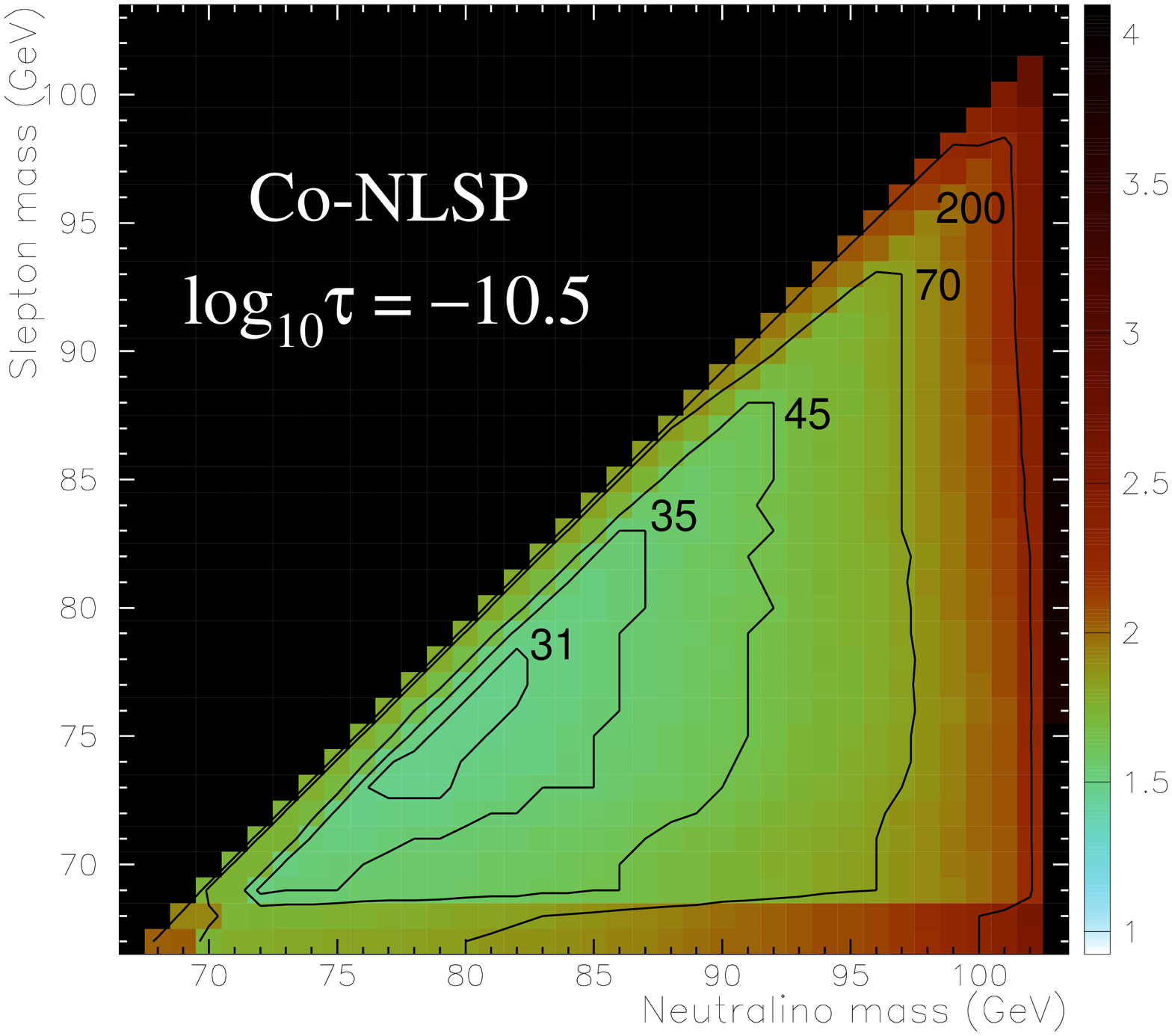}\epsfig{figure=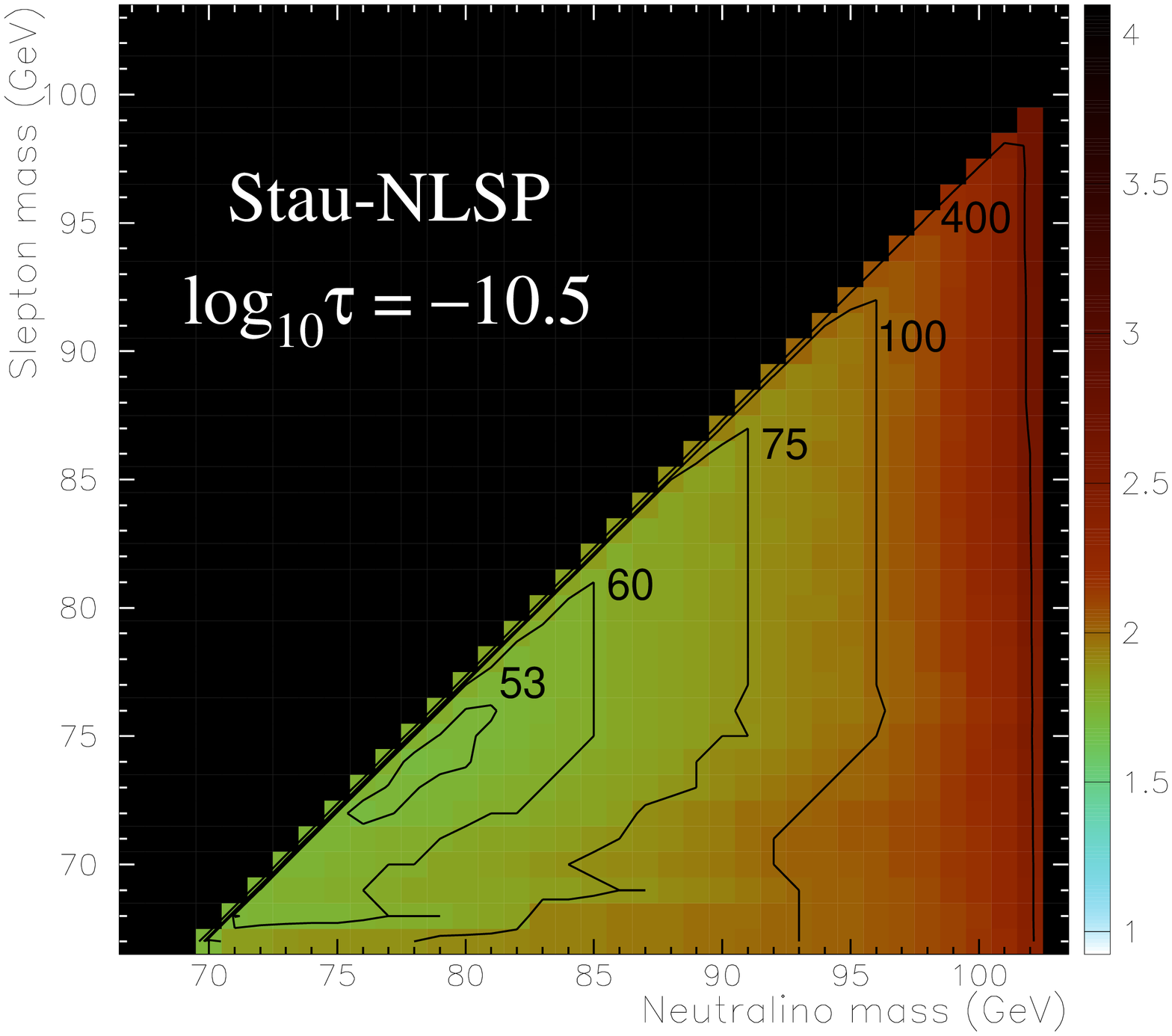}} \\
\resizebox{\textwidth}{!}{\epsfig{figure=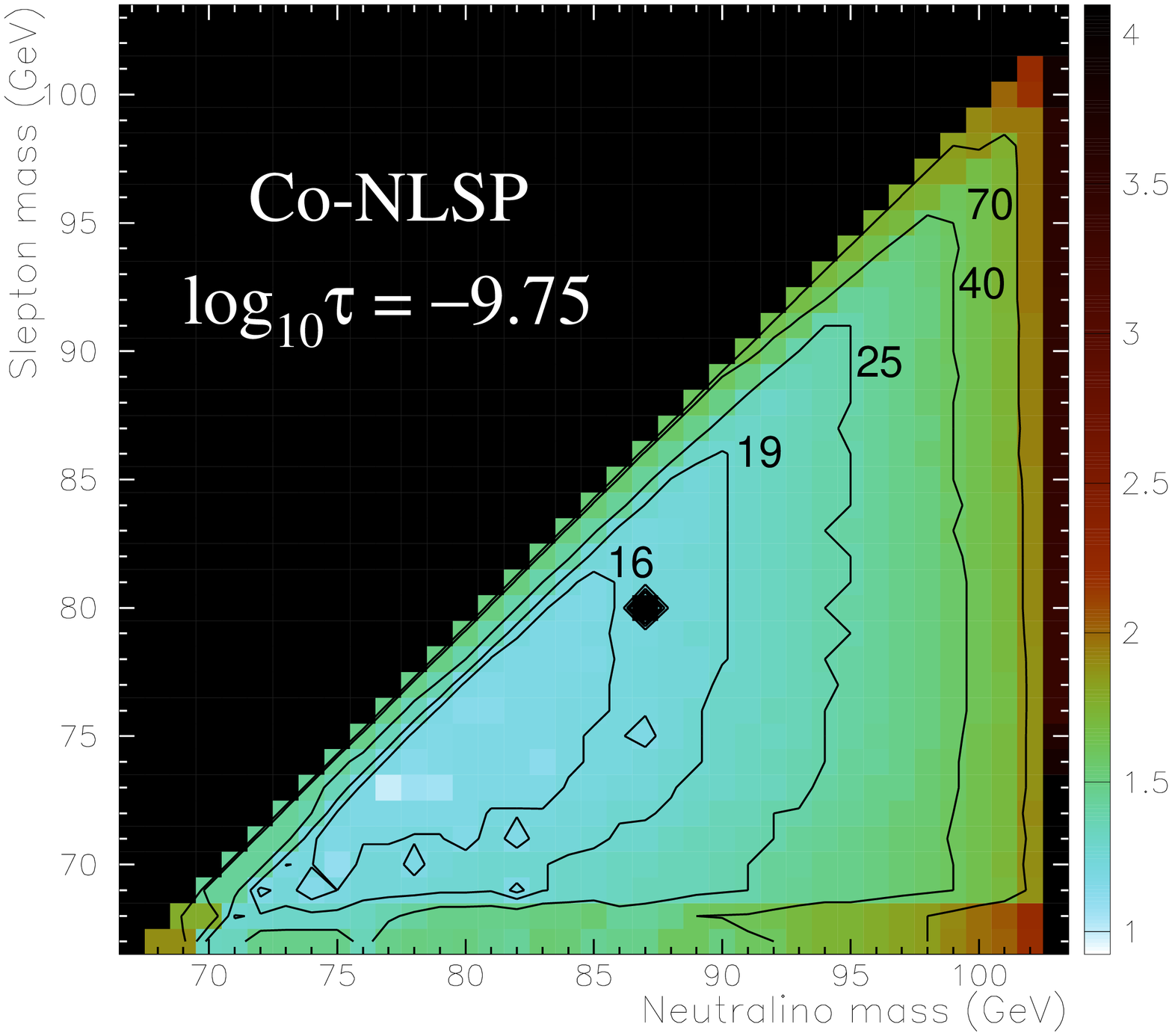}\epsfig{figure=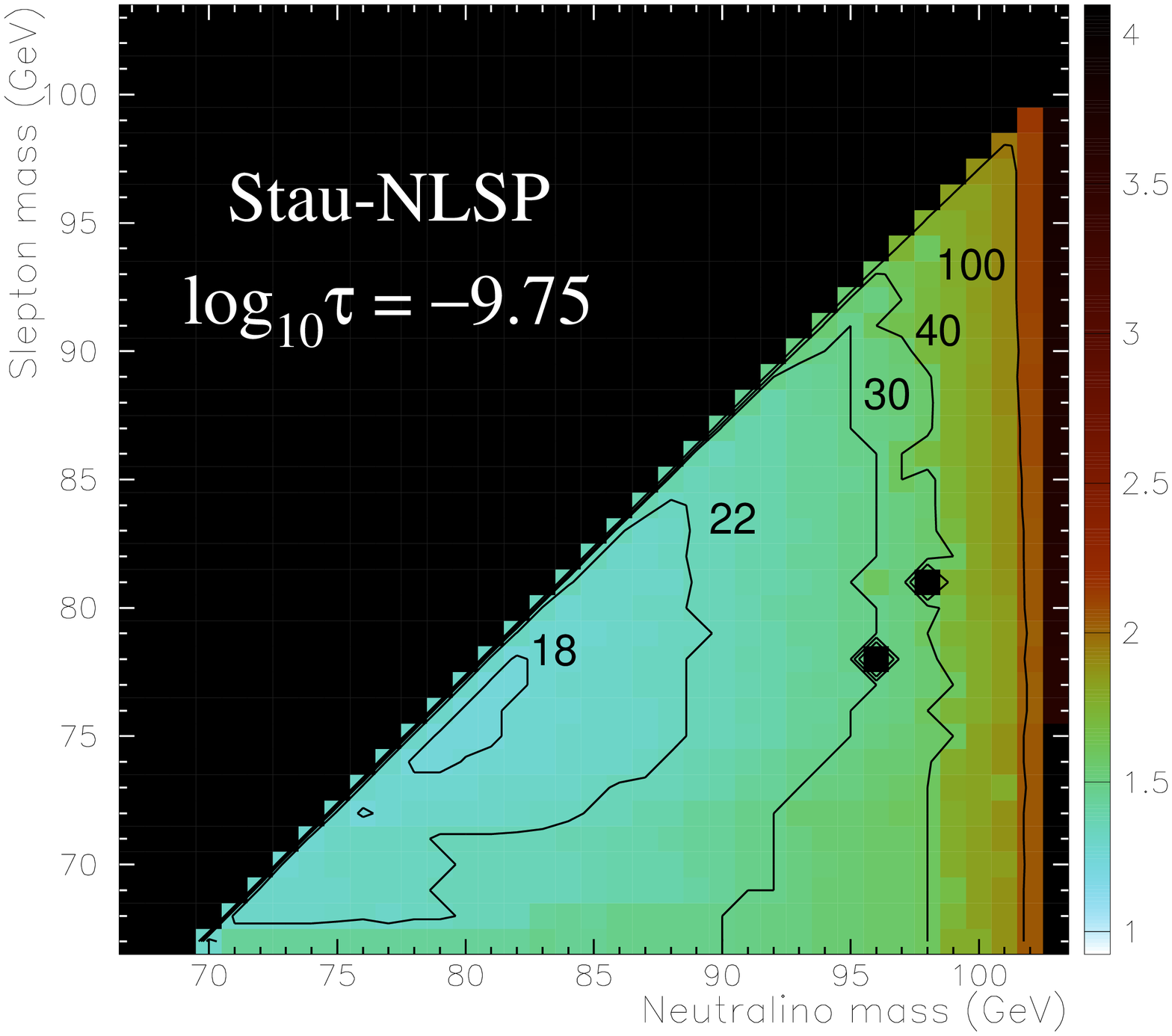}}
\end{tabular}
\caption{
\protect\iforig
{Model independent limits on the neutralino production cross-section at a LEP energy of 208\GeV\ as a function of neutralino and slepton mass. \ladd{`Co-NLSP' means the neutralino was assumed to decay to the three sleptons with equal branching ratios of $\frac{1}{3}$. `Stau-NLSP' means the neutralino was assumed to decay exclusively to the stau. The slepton lifetime is the `$\tau$' in $\log_{10}\tau$ and is measured in seconds. The limits are in units of nanobarns.}}
{Model independent limits on the neutralino production cross-section at a LEP energy of 208\GeV\ as a function of neutralino and slepton mass. \ladd{The limits are at 95\% confidence level and in units of nanobarns. `Co-NLSP' means the neutralino was assumed to decay to the three sleptons with equal branching ratios of $\frac{1}{3}$. `Stau-NLSP' means the neutralino was assumed to decay exclusively to the stau. The slepton lifetime is the `$\tau$' in $\log_{10}\tau$ and is measured in seconds.}}
}
\label{xslim1}
\end{figure}
\begin{figure}[H]
\resizebox{\textwidth}{!}{\epsfig{figure=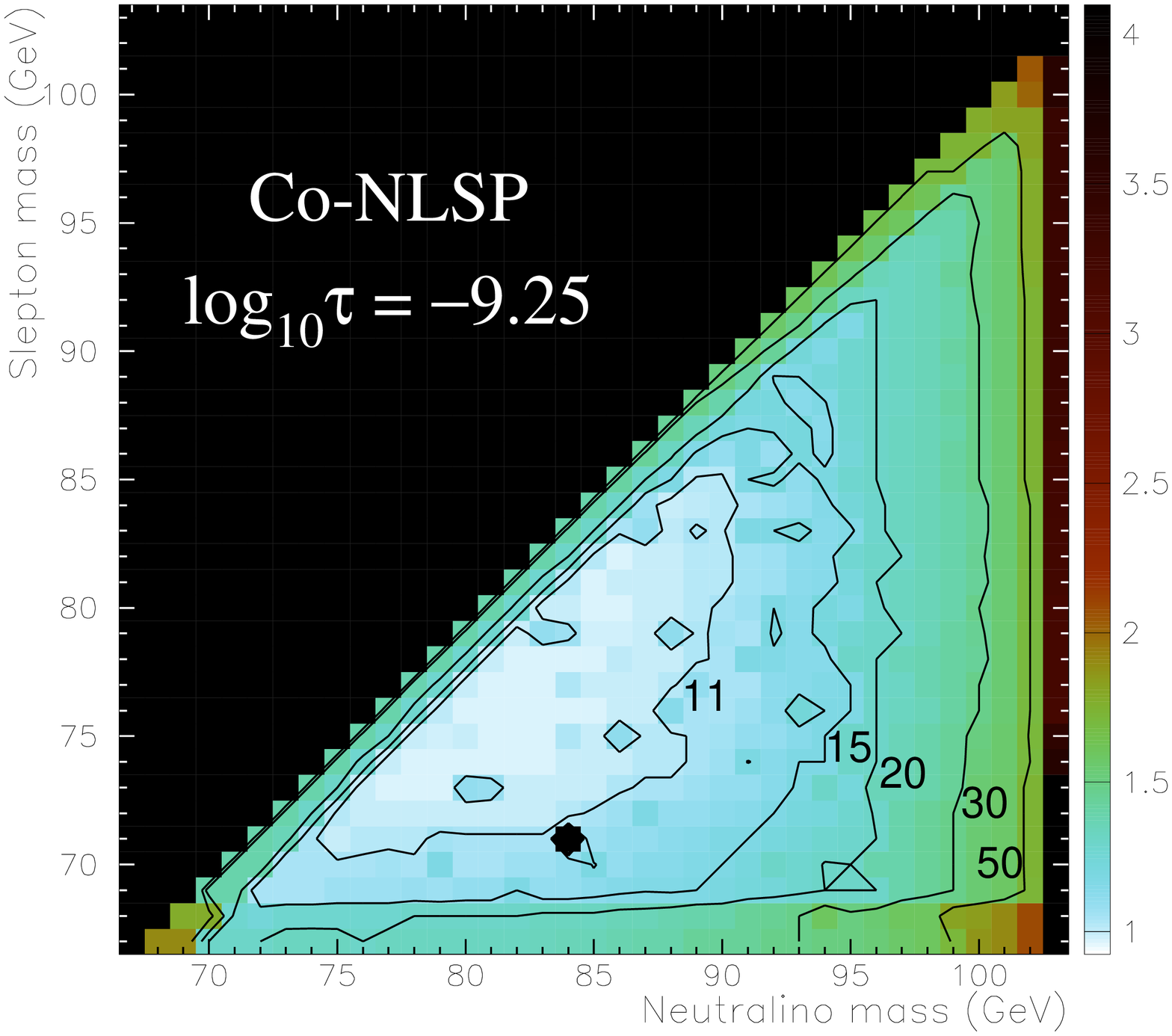}\epsfig{figure=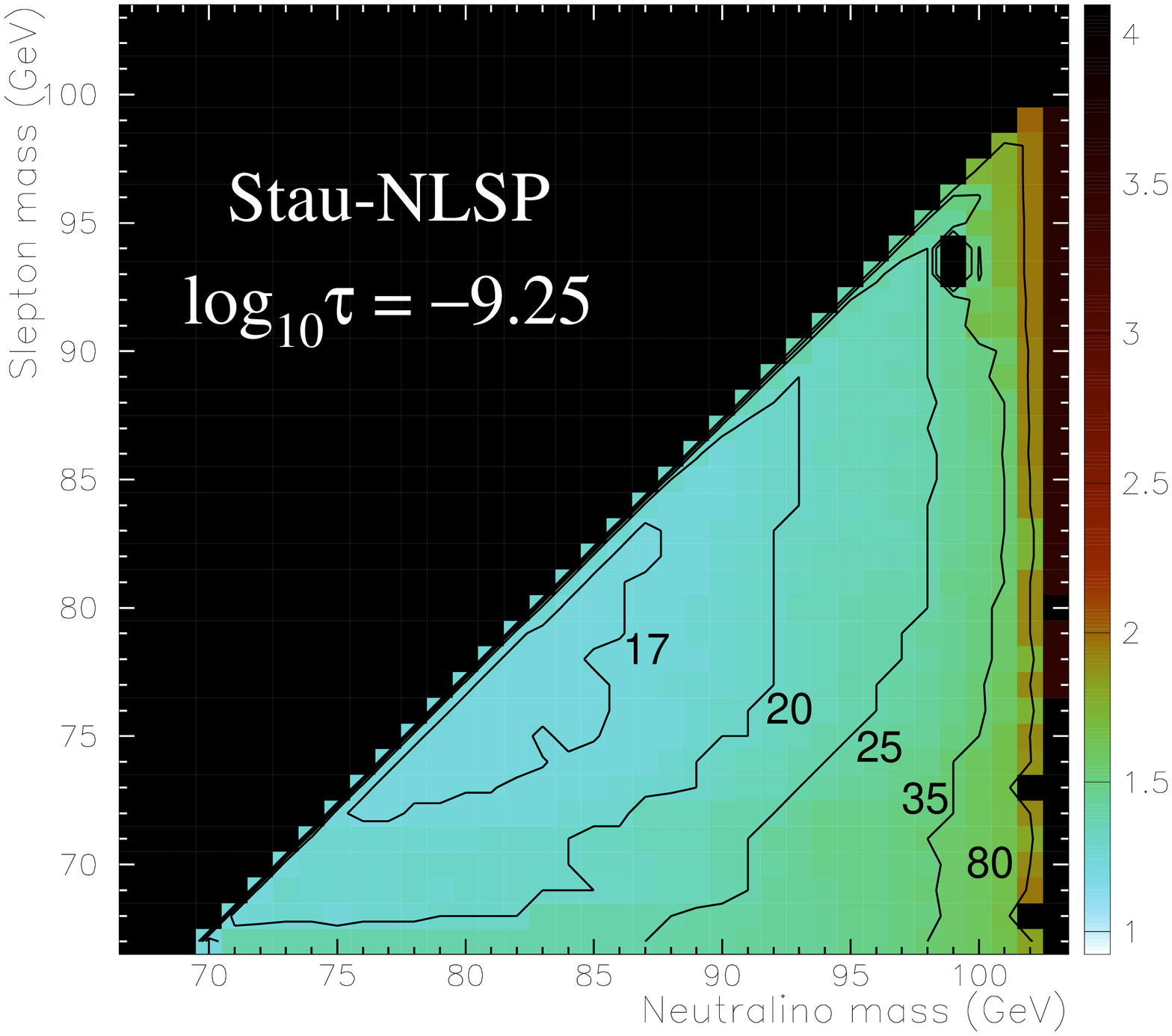}}
\resizebox{\textwidth}{!}{\epsfig{figure=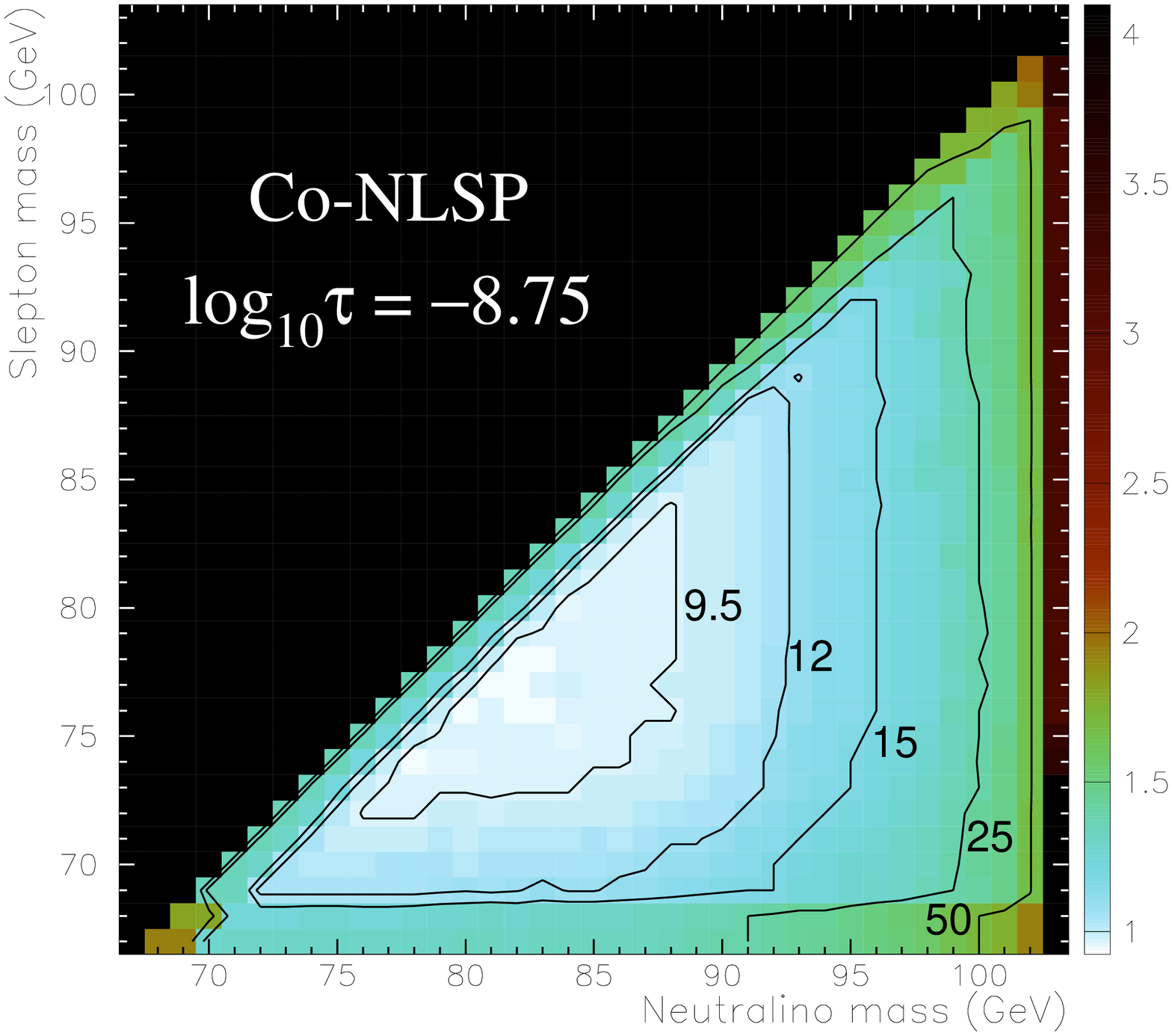}\epsfig{figure=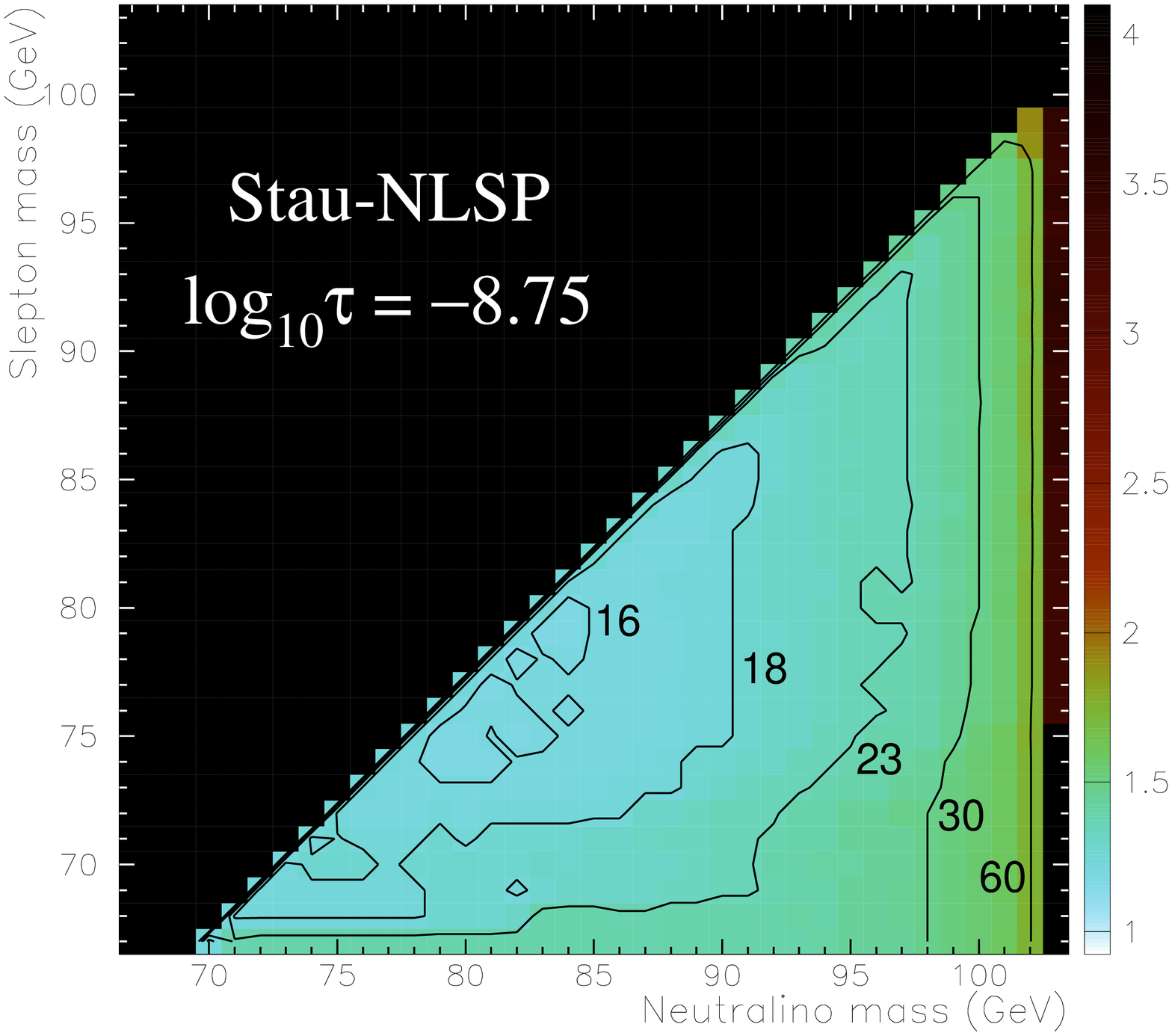}}
\resizebox{\textwidth}{!}{\epsfig{figure=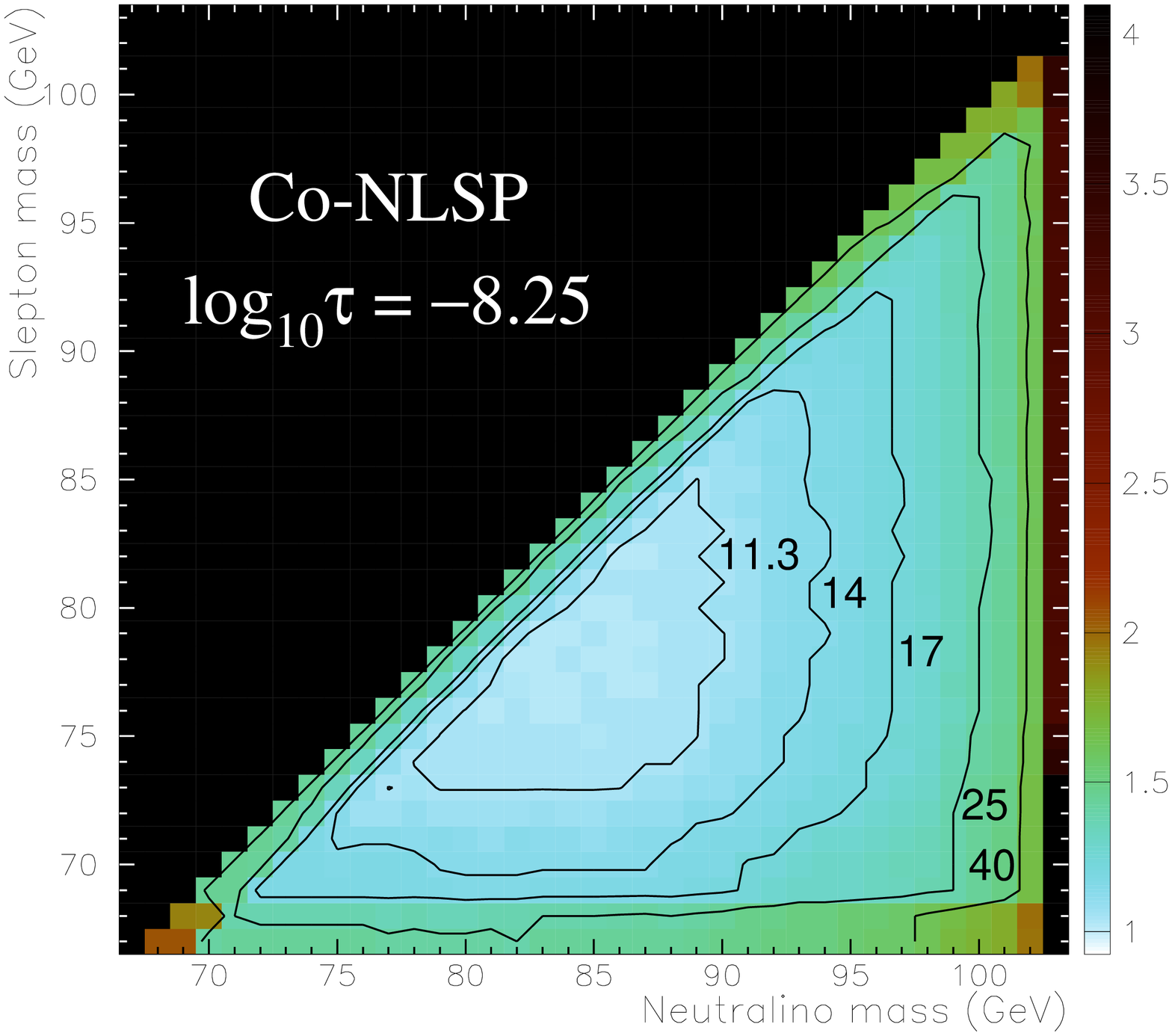}\epsfig{figure=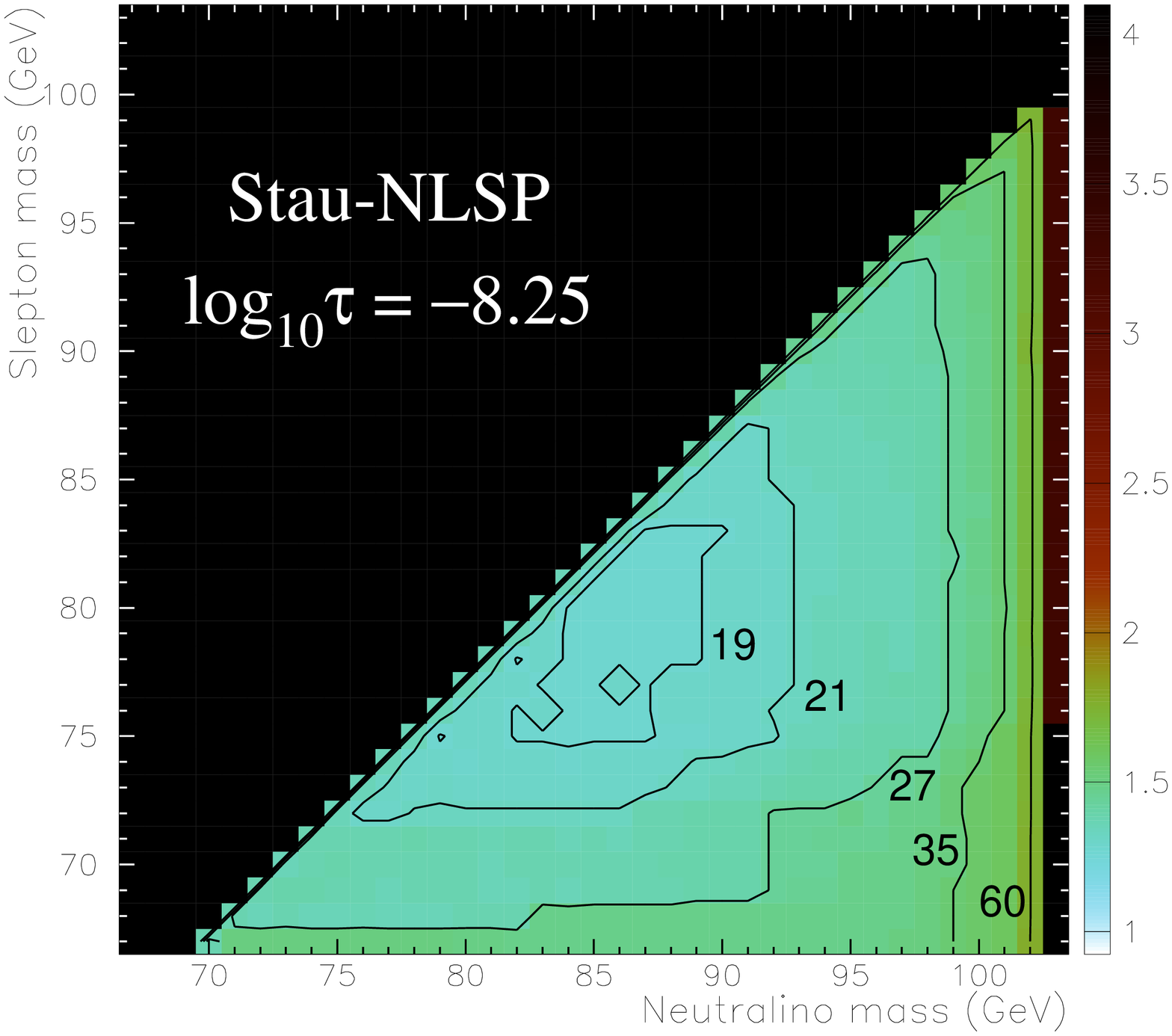}}
\caption{
\protect\iforig
{Model independent limits on the neutralino production cross-section at a LEP energy of 208\GeV\ as a function of neutralino and slepton mass. \ladd{Limits are given in units of nanobarns for the co-NLSP and stau-NLSP scenarios and a range of slepton lifetimes.}}
{Model independent limits on the neutralino production cross-section at a LEP energy of 208\GeV\ as a function of neutralino and slepton mass. \ladd{Limits are given at 95\% confidence level in units of nanobarns for the co-NLSP and stau-NLSP scenarios and a range of slepton lifetimes.}}
}
\label{xslim2}
\end{figure}
\begin{figure}[H]
\resizebox{\textwidth}{!}{\epsfig{figure=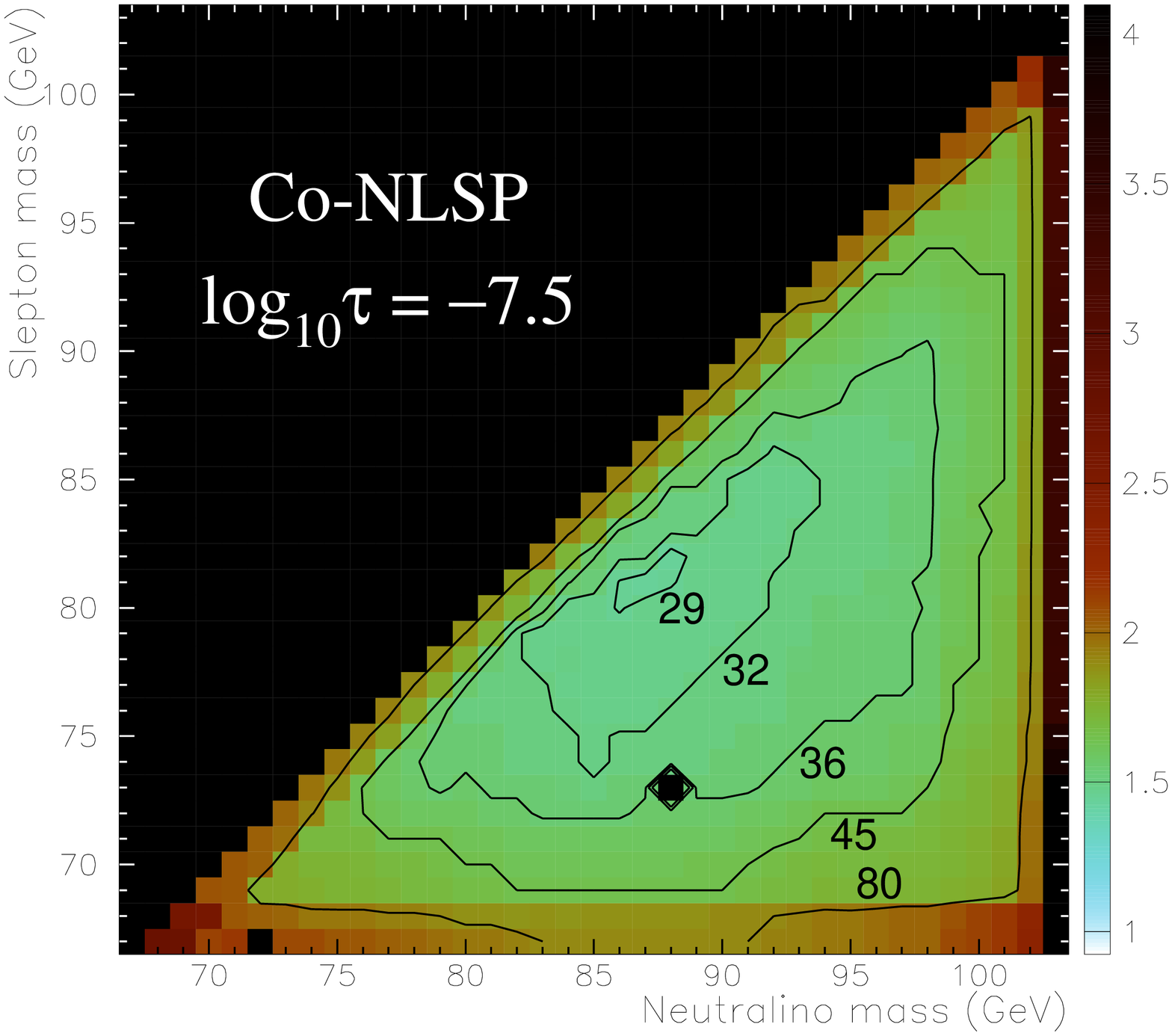}\epsfig{figure=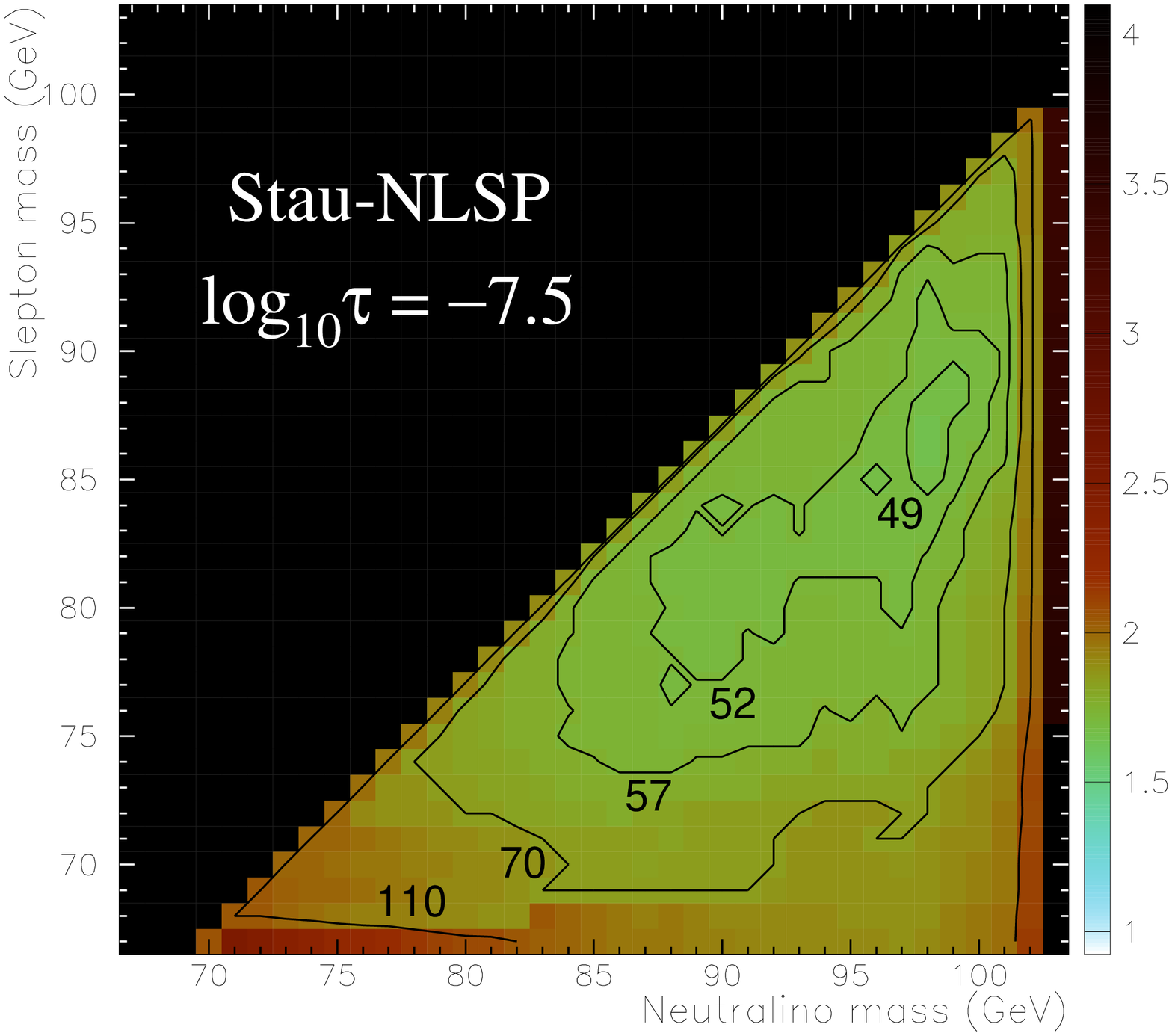}}
\resizebox{\textwidth}{!}{\epsfig{figure=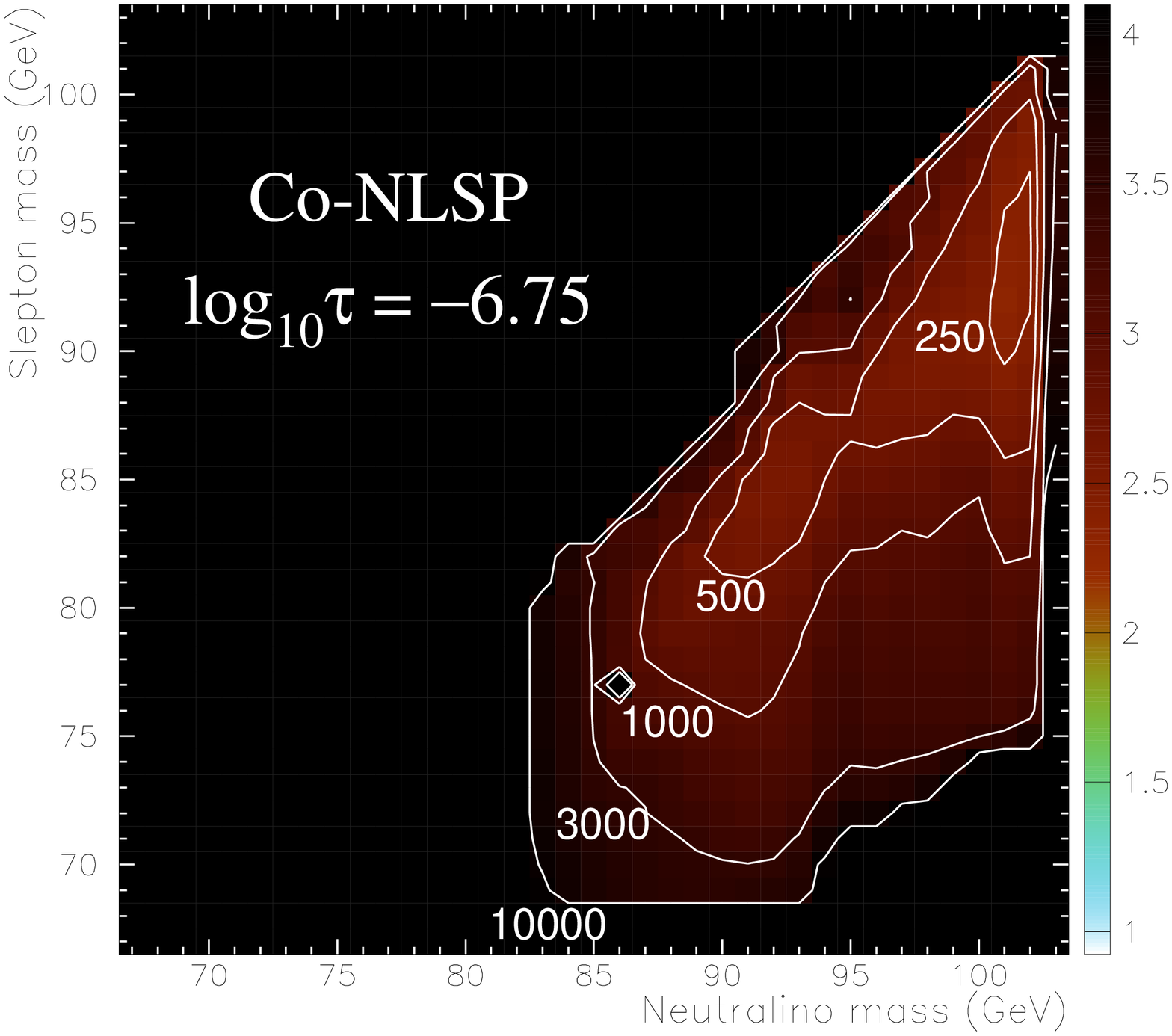}\epsfig{figure=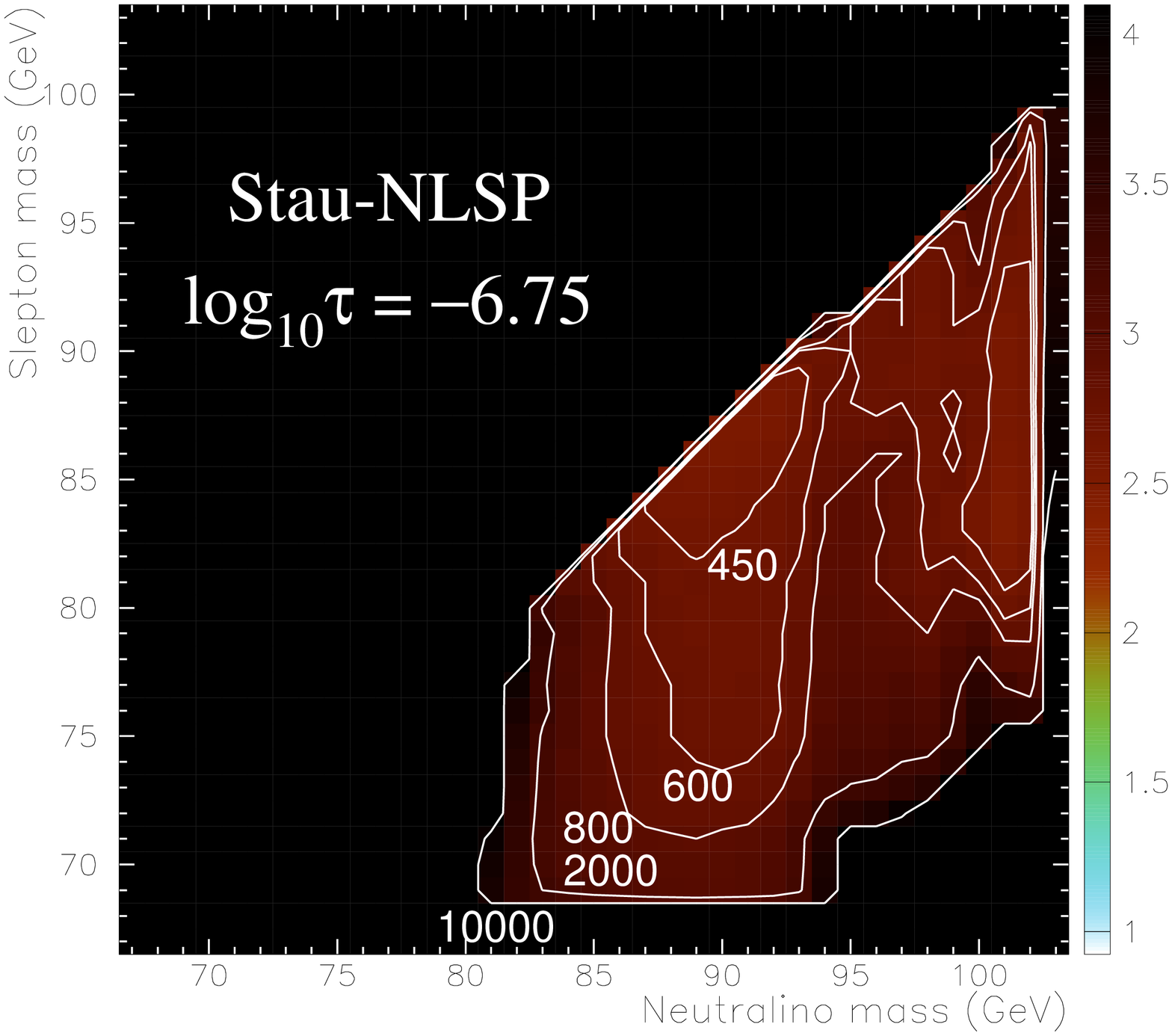}}
\caption{
\protect\iforig
{Model independent limits on the neutralino production cross-section at a LEP energy of 208\GeV\ as a function of neutralino and slepton mass. \ladd{Limits are given in units of nanobarns for the co-NLSP and stau-NLSP scenarios and a range of slepton lifetimes.}}
{Model independent limits on the neutralino production cross-section at a LEP energy of 208\GeV\ as a function of neutralino and slepton mass. \ladd{Limits are given at 95\% confidence level in units of nanobarns for the co-NLSP and stau-NLSP scenarios and a range of slepton lifetimes.}}
}
\label{xslim3}
\end{figure}

\begin{figure}[htpb]
\resizebox{\textwidth}{!}{\epsfig{figure=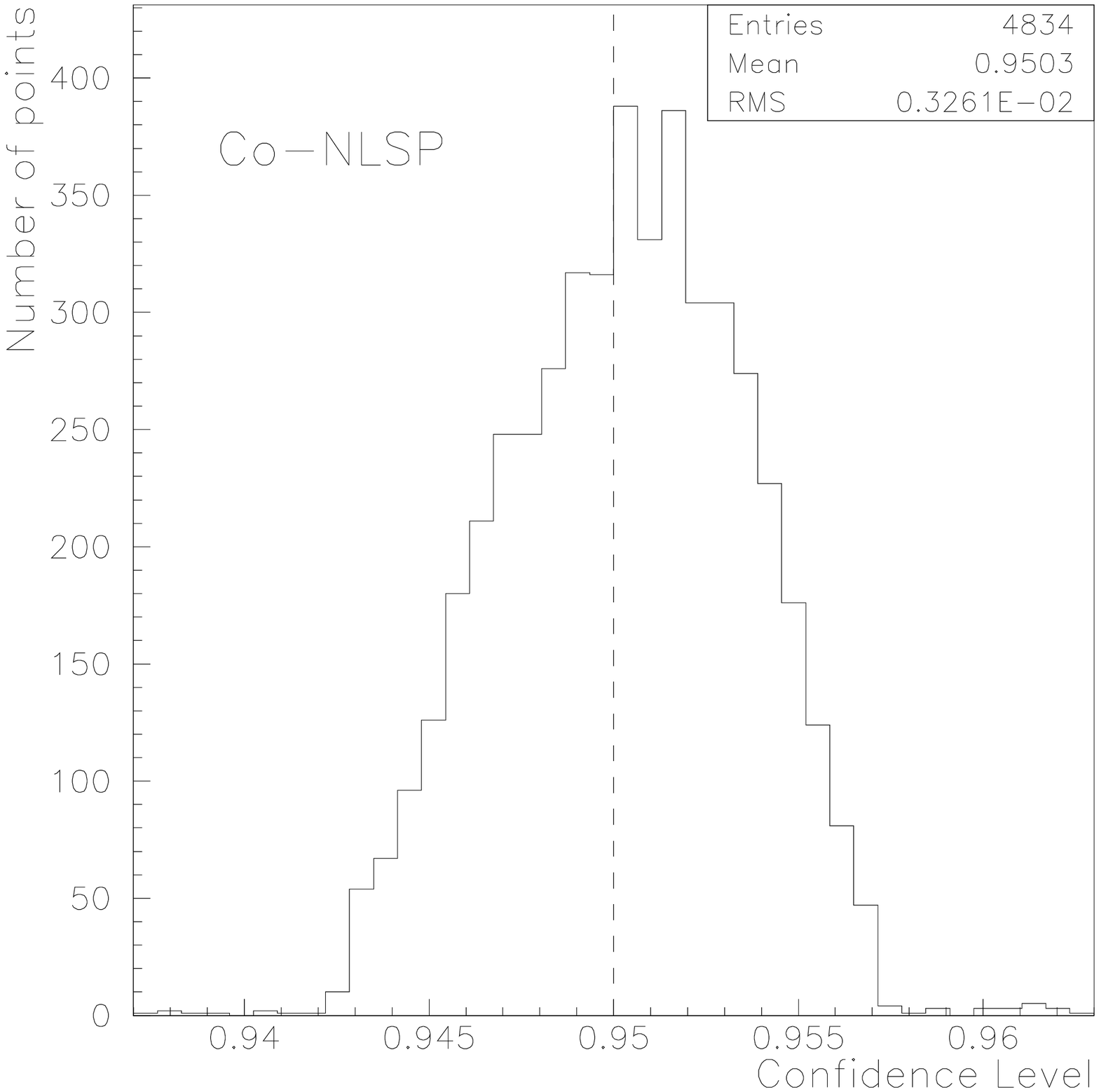}\hspace{0.5cm}\epsfig{figure=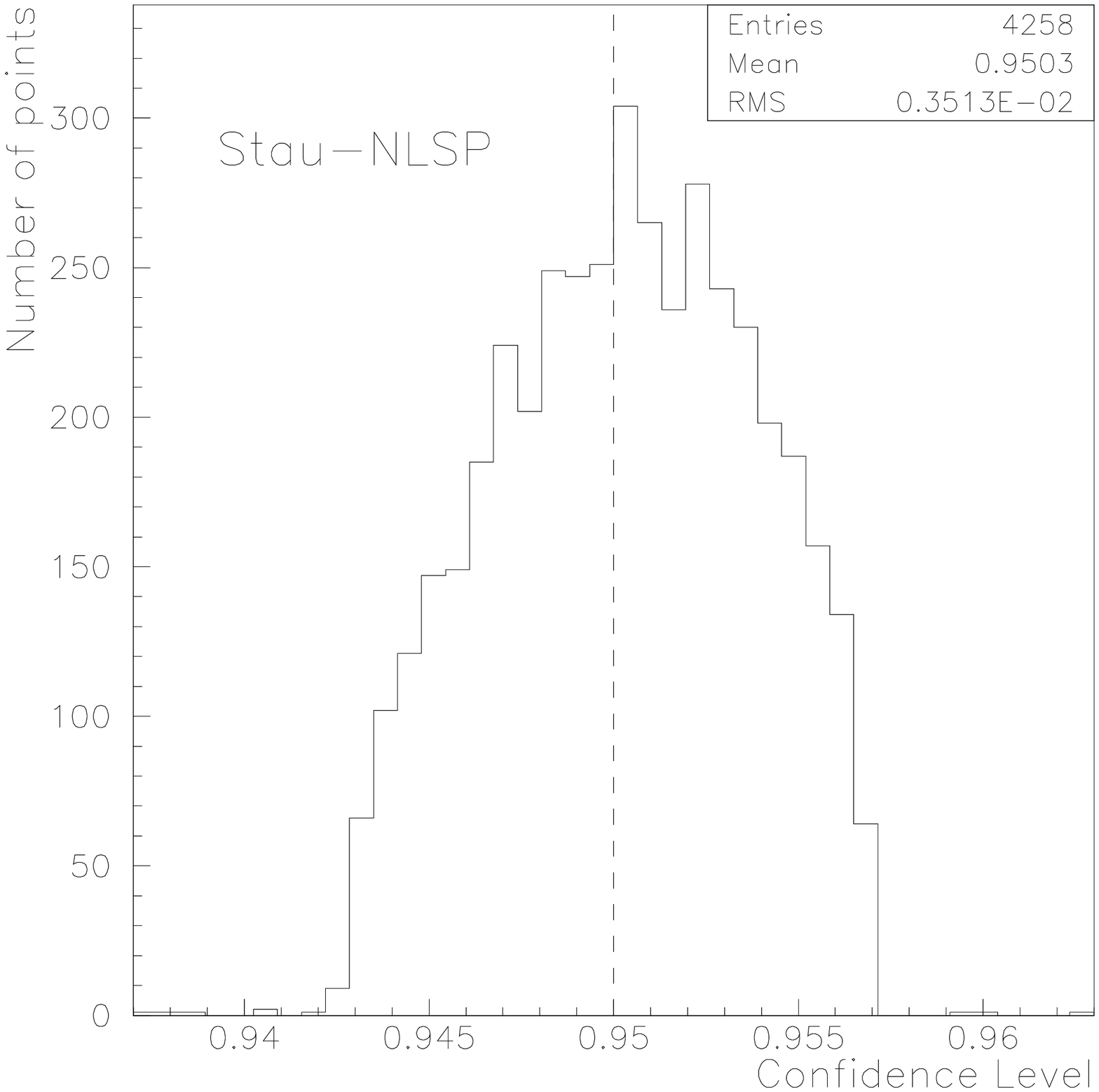}}
\caption{The distributions of the confidence levels by which the cross-section limits are excluded. \ladd{They are centred on 0.95 with a small spread.}}
\label{cldist}
\end{figure}

\begin{figure}[htpb]
\resizebox{\textwidth}{!}{\epsfig{figure=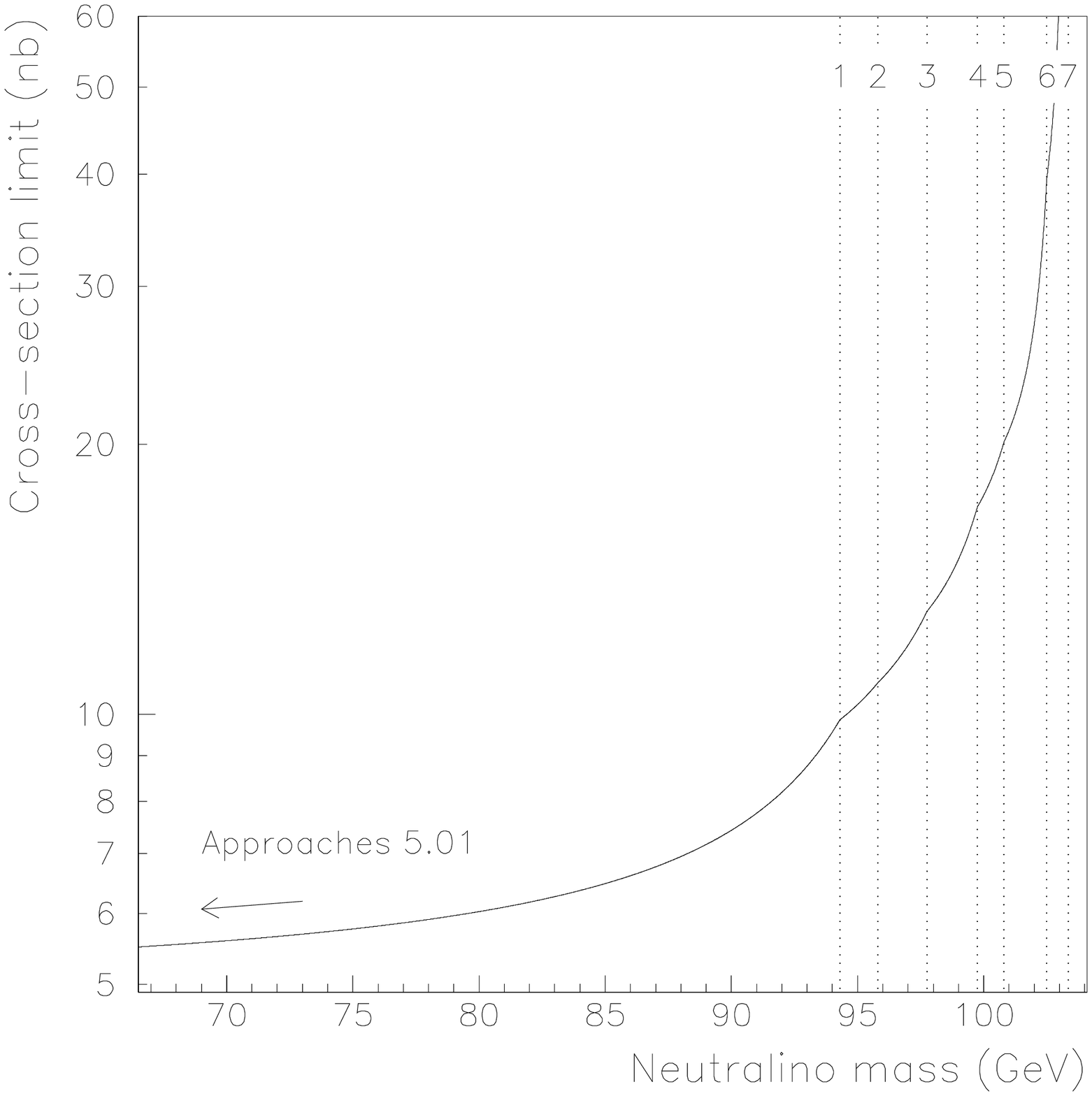}\hspace{1cm}\epsfig{figure=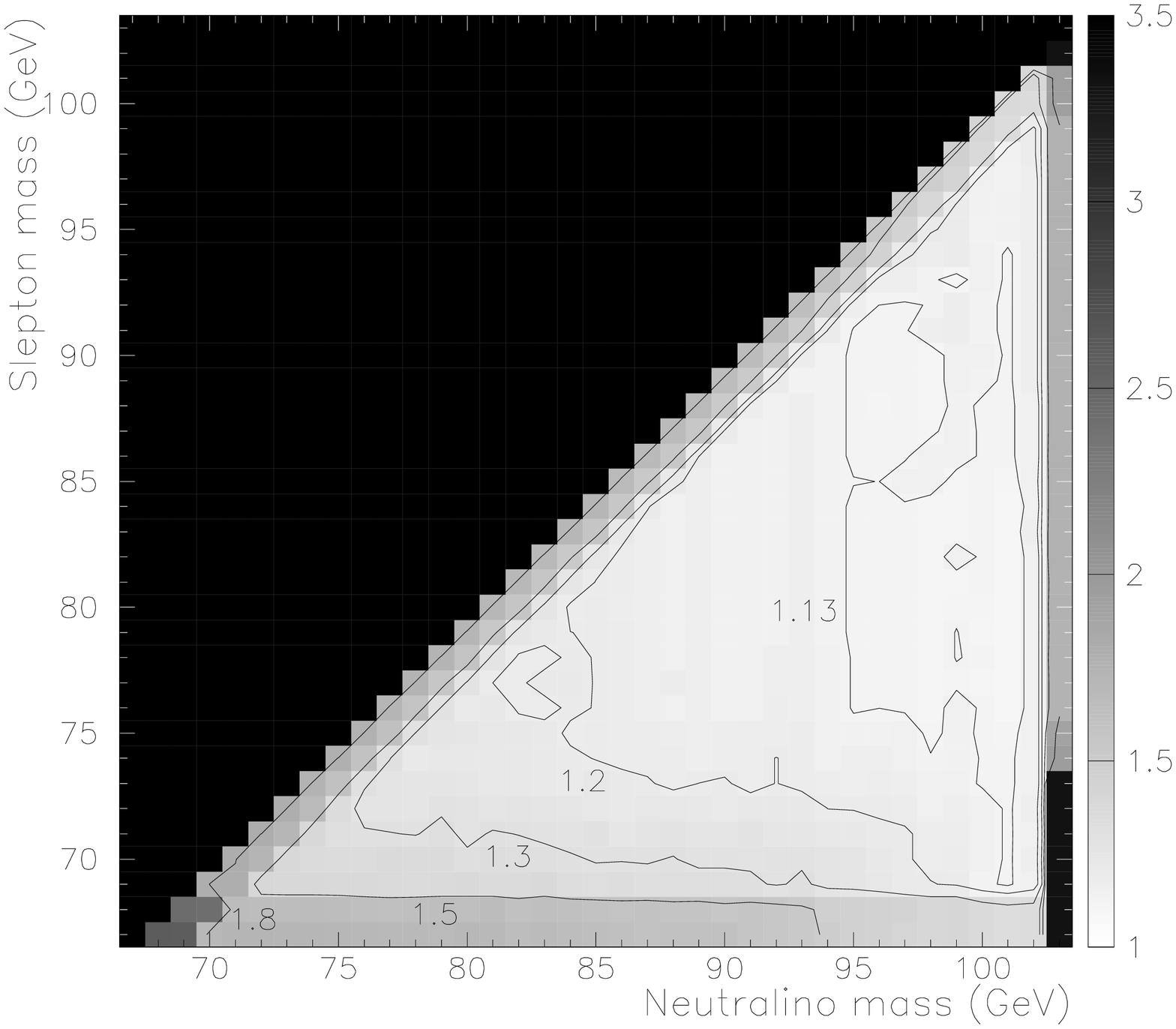}}
\parbox{\textwidth}{\HS[0.5] a).\HS b). \HS[0.5]\hspace{0mm}}
\caption{\lscap{a). The cross-section limit at 208\GeV, as a function of \mc, that would result from a search with 100\% efficiency and which gave zero observed events. The limit increases smoothly with \mc\ because, for a given cross-section at 208\GeV, a higher neutralino mass causes a lower cross-section at the lower energies (according to the linear evolution scheme described earlier). The gradient has discontinuities however, corresponding to the neutralino mass exceeding the beam energies of successive energy bins. b). A plot of the ratio of the limit for the co-NLSP scenario with a $\log\tl=-8.75$ to the ideal limit shown in the plot on the left. It shows that the limit obtained is close to the theoretical minimum.}{The cross-section limit at 208\GeV, as a function of \mc, that would result from a search with 100\% efficiency and which gave zero observed events, and a comparison with quoted limits.}}
\label{xslimperf}
\end{figure}

\section{Model specific parameter space exclusion}
\label{parspace}
A scan over the parameter space of the model described in
\cite{themodel} was performed by Aran Garcia-Bellido in order to
translate the results of all the ALEPH GMSB searches into excluded
regions of the model's parameter space. This was an update of the scan
described in \cite{gmsb189} using the new highest energy LEP data and
including the analysis described in this thesis. The scan ranges for
the parameters (which are described in Section~\ref{gmsb}) and the
number of points used in each are given in Table~\ref{scan}. The five
values for the upper limit on \Lam\ correspond to increasing values of
\Nfive, \ie\ for $\Nfive=1$ the upper limit on \Lam\ is 75\TeV, for
$\Nfive=2$ the upper limit is 60\TeV, etc. At each point the ISAJET
7.51 program \cite{isajet} was used to calculate the sparticle masses,
branching ratios and lifetimes, together with their production
cross-sections at the seven LEP energies given in Table~\ref{result1}.
\renewcommand{\arraystretch}{1.3}
\begin{table}
\centerline{
\begin{tabular}{|c|>{$}c<{$}|>{$}c<{$}|c|}
\hline
Parameter & $Lower limit$ & $Upper limit$          & No. of points \\
\hline
\Mmess    & 10^4\GeV      & 10^{12}\GeV            & 5             \\
\Mgrav    & 10^{-9.5}\GeV & 10^{-4}\GeV            & 12            \\
\Lam      & 2\TeV         & 75, 60, 45, 45, 45\TeV & 151           \\
\tanb     & 1.5           & 40                     & 9             \\
\Nfive    & 1             & 5                      & 5             \\
\signmu   & -             & +                      & 2             \\
\hline
\end{tabular}
}
\caption{The ranges for the scan parameters and the number of points in each. \ladd{815,400 points were considered in total.}}
\label{scan}
\end{table}
\par
These were then passed to the same routine that calculated the
confidence level for the cross-section limit scan, and the same method
was used for judging whether the confidence level was greater than,
less than, or not distinguishable from 95\%. Points with a confidence
level greater than 95\% were considered excluded, points with a
confidence level less than, or indistinguishable from 95\% were
considered unexcluded.
\par
One problem arose in this scan which did not arise in the
cross-section limit scan however. In the cross-section limit scan only
the case of perfect slepton degeneracy, and the case of a stau so much
lighter than the selectron and smuon it was effectively the sole NLSP
were considered. In this scan, points in-between these two extremes
exist, where the stau is significantly lighter but the neutralino
branching ratios to the selectron and smuon are still quite
significant. This does not pose a problem for the \selsel, \smusmu,
\staustau\ and \selsmu\ channels since the two sleptons involved in 
each channel are still degenerate with each other. But events in the
\selstau\ and \smustau\ channel will involve sleptons of different
properties in the same event -- a case for which signal Monte Carlo data
has not been generated. However, since the efficiencies (partial and
total) depend almost solely on the slepton decay length, it can be
expected that the true efficiencies will lie somewhere between those
calculated based on both sleptons having the stau's properties, and
those based on both having the other slepton's properties. Although it
cannot be expected that the true efficiencies can be obtained by any
kind of simple interpolation between the two, it can be expected that
the true confidence level will lie between the confidence levels
calculated assuming both sleptons have the one set of properties and
both having the other. So two confidence level calculations are run in
parallel for both optimisations, and the lowest is chosen as the real
confidence level to be conservative.
\par
Significant regions of excluded parameter space in the (\mc,
$m_{\Stau}$) plane were only found for four of the twelve gravitino
masses used in the scan, and these regions are shown in
Figure~\ref{exclude}. The sudden rise in the lower edges around
$\mc\sim90\GeV$ corresponds to the threshold for neutralino production
beginning to exceed the energies at which data was taken. The
positions of the lower edges of the regions for $\Mgrav=\!10^{-8}$,
$10^{-7.5}$ and $10^{-7}$\GeV\ are predominantly defined by the lowest
slepton mass for which Monte Carlo data was generated rather than any
physical effect\footnote{Note that this does not correspond exactly to
the actual lowest slepton mass for which Monte Carlo data was
generated. The technique for interpolation in $\sqrt{s}$ allows
efficiencies to be calculated for masses down to $\sim189/208 \times$
the lowest slepton mass.}. For $\Mgrav=\!10^{-6.5}$\GeV\ the lower edge
is higher. This is because the slepton lifetime is proportional to
($\Mgrav^2/\ms^5$). So for a high gravitino mass and decreasing
slepton mass the slepton decay length can rise beyond the maximum
value for which there is sensitivity before the slepton mass drops
below the minimum defined by Monte Carlo data. A similar effect
accounts for the fact that the right hand edge of the
$\Mgrav=\!10^{-8}$\GeV\ region is at a lower value of \mc\ than for the
other regions. Here the slepton lifetime is short, and a high
neutralino mass exacerbates this to produce a decay length below the
minimum value for which there is sensitivity.
\begin{figure}[pthb]
\epsfig{figure=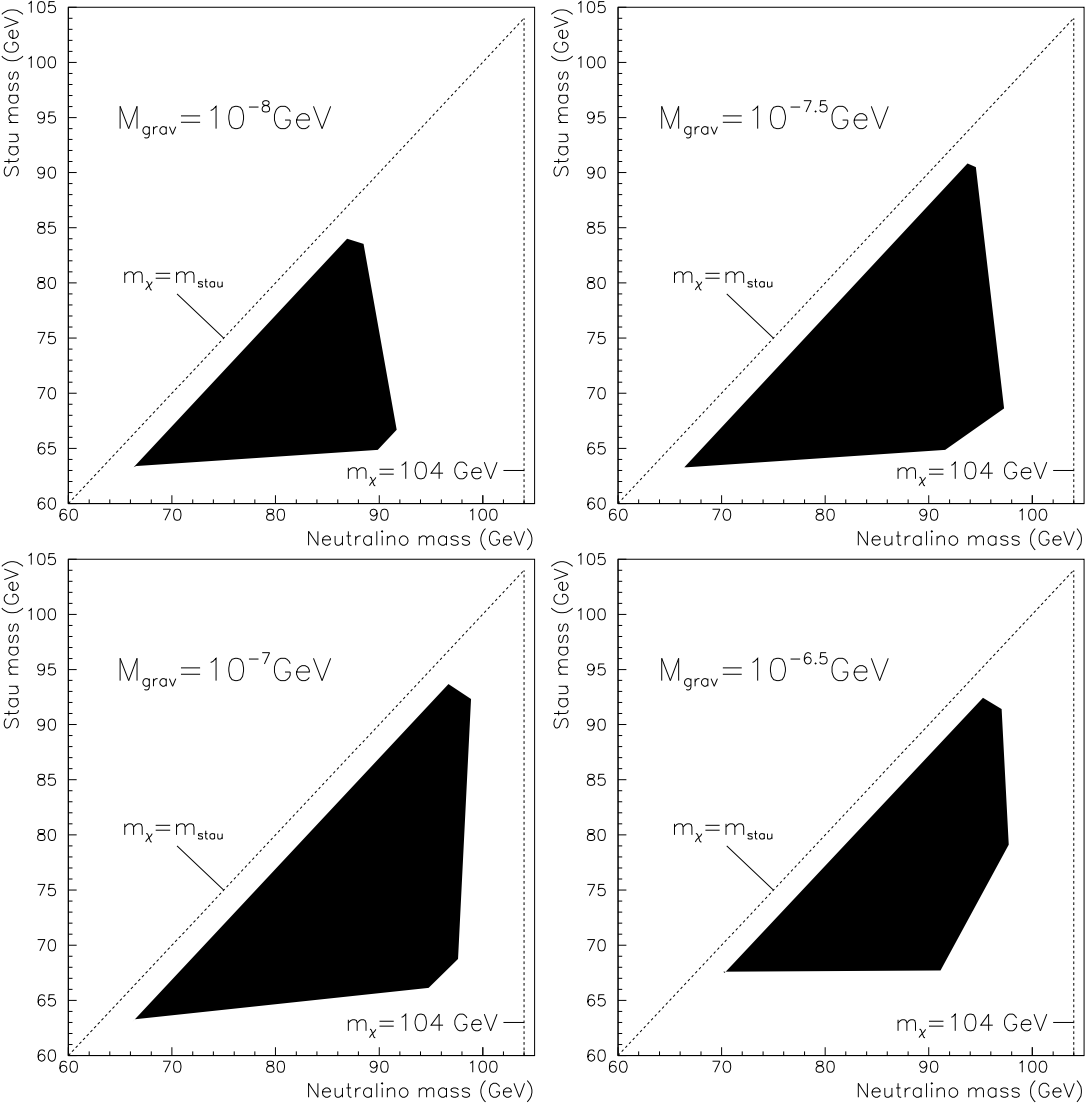,width=\textwidth}
\caption{Excluded regions of (\mc, $\m_{\tilde{\tau}}$) space for four different gravitino masses.}
\label{exclude}
\end{figure}

\section{The results in a wider context}
\label{currentlims}
The analysis described in this thesis was incorporated in an update of
the common interpretation of all ALEPH GMSB work in
\cite{gmsb208}. This includes a description of a scan essentially
identical to that described in Section~\ref{parspace}, and contains
plots similar to those in Figure~\ref{exclude} but showing the regions
excluded by every search. The search described in this thesis did not
exclude any significant amount of parameter space that the search for
direct slepton pair production over the same energy range
(\cite{slep208}) did not, since although the neutralino cross-section
was often higher, the slepton cross-section was never so low that
points could not be excluded by the slepton pair-production search
when they could be by this search. The search for pair-produced
sleptons is also not limited by the neutralino mass and so covers a
wider range of the parameter space. This search then serves to confirm
the exclusion of the relevant region of parameter space under the
model considered, but still could prove independently useful should
models be hypothesised in the future which predict a greater disparity
between the neutralino and slepton cross-sections. It also brings the
overall ALEPH GMSB search effort close to complete.
\par
The current ALEPH limits are given in \cite{gmsb208} as:\\
\par\noindent
\begin{centering}
\setlength{\xa}{1.5cm}\setlength{\xb}{12.3cm}
\begin{tabular}{p{\xa}p{\xb}} \hline
\multicolumn{2}{c}{In the neutralino-NLSP scenario:} \\ \hline
99\GeV & on \mc\ for short lifetime \\
60\GeV & on \mc\ for $c\tau_\chi<100\m$ \\
54\GeV & on \mc\ for long lifetime (indirect limit from MSSM chargino searches)
\end{tabular} \\ \par\noindent\vspace{\intextsep}\centering
\begin{tabular}{p{\xa}p{\xb}} \hline
\multicolumn{2}{c}{In the stau-NLSP scenario:} \\ \hline
74.7\GeV & on $m_{\Stau_1}$ for any lifetime \\
99\GeV & on \mc\ for no stau lifetime \\
\end{tabular} \\ \par\noindent\vspace{\intextsep}\centering
\begin{tabular}{p{\xa}p{\xb}} \hline
\multicolumn{2}{c}{In general:} \\ \hline
54\GeV & on the NLSP mass for any NLSP lifetime \\
10\TeV & on \Lam\  for any NLSP lifetime \\
0.024\eV & on \Mgrav\ \\ \hline
\end{tabular}
\end{centering}
\par
LEP and the associated detectors have now been dismantled in
preparation for the installation of the new LHC, and so \cite{gmsb208}
constitutes the final word on \mbox{GMSB} detection with the ALEPH
detector. A paper combining the GMSB results of all the LEP
experiments is being planned. No evidence for supersymmetry was found
by LEP in any guise. The Tevatron, LHC and FLC present the next
opportunities for supersymmetry to make itself known, and their
findings are awaited with anticipation.


\begin{appendix}
\chapter{Signal Monte Carlo}
\label{mc}

The exact points in (\mc, \ms, \dl) space at which signal Monte Carlo
was generated for each LEP energy are given here.
\par
At 189\GeV\ the slepton decay lengths (in units of cm) were chosen for each channel as follows:
\par
\setlength{\topsep}{0mm}
\setlength{\parskip}{0mm}
\setlength{\parsep}{0mm}
\newlength{\lja}\newlength{\ljb}\newlength{\ljc}\newlength{\ljd}\newlength{\lje}
\settowidth{\ljb}{\smusmu}
\setlength{\ljc}{20mm}
\setlength{\ljd}{2mm}
\settowidth{\lje}{2000}
\addtolength{\ljd}{\lje}
\setlength{\lja}{\textwidth}
\addtolength{\lja}{-\ljb}\addtolength{\lja}{-\ljc}\addtolength{\lja}{-7\ljd}\addtolength{\lja}{-\lje}
\setlength{\lja}{0.5\lja}
\begin{tabbing}
\hspace*{\lja}\=\hspace{\ljb}\=\hspace{\ljc}\=\hspace{\ljd}\=\hspace{\ljd}\=\hspace{\ljd}\=\hspace{\ljd}\=\hspace{\ljd}\=\hspace{\ljd}\=\hspace{\ljd}\=\kill
\> \selsel  \>:	\> 0.2, \> 0.5, \> 2, \> 5, \> 30, \> 80, \> 150, \> 500.  \\
\> \smusmu  \>:	\> 0.2, \> 0.5, \> 2, \> 5, \> 30, \> 80, \> 300, \> 2000. \\
\> \staustau\>:	\> 0.2, \> 0.5, \> 2, \> 5, \> 30, \> 80, \> 300, \> 2000. \\
\> \selsmu  \>:	\> 0.2, \> 0.5, \> 2, \> 5, \> 30, \> 80, \> 300, \> 500.  \\
\> \selstau \>:	\> 0.2, \> 0.5, \> 2, \> 5, \> 30, \> 80, \> 300, \> 500.  \\
\> \smustau \>:	\> 0.2, \> 0.5, \> 2, \> 5, \> 30, \> 80, \> 300, \> 500.
\end{tabbing}
At 208\GeV\ the decay lengths
\begin{tabbing}
\hspace*{\lja}\=\hspace{\ljb}\=\hspace{\ljc}\=\hspace{\ljd}\=\hspace{\ljd}\=\hspace{\ljd}\=\hspace{\ljd}\=\hspace{\ljd}\=\hspace{\ljd}\=\hspace{\ljd}\=\kill
\> \> \> 0.2, \> 0.5, \> 2, \> 5, \> 30, \> 80, \> 300, \> 500,
\end{tabbing}
were used for all channels.
\par
The mass points at 189\GeV, in units of GeV and in the format
\mc-\ms, were chosen for each channel as follows (the points in brackets were only generated at the two highest respective decay lengths):
\setlength{\lja}{1.4cm}
\setlength{\ljc}{0.5cm}
\settowidth{\ljd}{99-99 99-99 99-99}
\addtolength{\ljd}{5mm}
\begin{table}[H]
\begin{tabbing}
\hspace*{\lja}\=\hspace{\ljb}\=\hspace{\ljc}\=\kill
\> \selsel \>: \> 	69-68 74-68 79-68 84-68 89-68 94-68 \\
\> \> \>		74-73 79-73 84-73 89-73 94-73 \\
\> \> \>		79-78 84-78 89-78 94-78 \\
\> \> \>		84-83 89-83 94-83 \\
\> \> \>		89-88 94-88 \\
\> \> \>		94-93 
\end{tabbing}
\end{table}
\begin{table}[H]
\begin{tabbing}
\hspace*{\lja}\=\hspace{\ljb}\=\hspace{\ljc}\=\hspace{\ljd}\=\hspace{\ljb}\=\hspace{\ljc}\=\hspace{\ljd}\=\hspace{\ljb}\=\hspace{\ljc}\=\kill
\> \smusmu \>: \> 70-69 82-69 94-69 \> \staustau \>: \> 70-67 82-69 94-69 \> \selsmu\ \>: \> 70-69 82-69 94-69\\
\> \> \>	  82-81 88-81 94-81 \> \> \>            82-79 88-81 94-81 \> \> \>           82-81 88-81 94-81\\
\> \> \>	  94-93             \> \> \>            94-91             \> \> \>           94-93 \\
\> \> \>	  (94-87 94-75)	    \> \> \>            (94-87 94-75)     \> \> \>           (94-87 94-75)
\end{tabbing}
\end{table}
\begin{table}[H]
\begin{tabbing}
\hspace*{\lja}\=\hspace{\ljb}\=\hspace{\ljc}\=\hspace{\ljd}\=\hspace{\ljb}\=\hspace{\ljc}\=\kill
\> \selstau \>: \> 70-67 82-67 94-67\> \smustau\ \>: \>     70-67 82-67 94-67\\	
\> \> \>           82-79 88-79 94-79\> \> \>                82-79 88-79 94-79\\
\> \> \>           94-91        \> \> \>                94-91\\      
\> \> \>           (94-85 94-73)\> \> \>                (94-85 94-73)
\end{tabbing}
\end{table}
At 208\GeV\ the mass points were the same amongst the three channels that
did not involve a stau, and amongst the three that did. They were as
follows:
\settowidth{\ljb}{No stau}
\addtolength{\lja}{-\ljb}
\settowidth{\ljlen}{Stau}
\setlength{\ljlenb}{\ljb}
\addtolength{\ljlenb}{-\ljlen}
\setlength{\ljc}{1cm}
\begin{table}[H]
\begin{tabbing}
\hspace*{\lja}\=\hspace{\ljb}\=\hspace{\ljc}\=\hspace{\ljd}\=\hspace{\ljb}\=\hspace{\ljc}\=\kill
\> No stau \>: \> 	70-69 87-69 103-69 \> \hspace{\ljlenb}Stau \>: \> 70-67 87-67 103-67 \\
\> \> \>		87-86 95-86 103-86 \> \> \>       87-84 95-84 103-84 \\
\> \> \>		103-102            \> \> \>       103-100\\
\> \> \>		(103-94 103-77)	   \> \> \>       (103-92 103-75)
\end{tabbing}
\end{table}

\end{appendix}

\begin{spacing}{1.5}
\bibliographystyle{rgo}
\bibliography{references}

\end{spacing}

\end{document}